\documentclass[11pt,a4paper]{article}

\pdfoutput=1 

\usepackage{jheppub}                             
\usepackage{bm}        
\usepackage{graphicx}                            
\usepackage{soul}                                
\usepackage{microtype}                           
\usepackage[sharp]{easylist}                     
\usepackage{amsmath}                             
\usepackage{amssymb}                             
\usepackage{tensor}                              
\usepackage{booktabs}                            
\usepackage{wrapfig}                             
\usepackage[T1]{fontenc}                        
\usepackage{titlesec}                            

\usepackage{diffcoeff}
\usepackage{tikz}
\usepackage{amsthm}                              
\usepackage{mathtools}
\usepackage{dsfont}  
\usepackage{derivative}

\usepackage[section]{placeins}

\graphicspath{{figures/}} 

\usepackage{todonotes}
\presetkeys{todonotes}{color=white, linecolor=gray, size=tiny}{}

\makeatletter
\if@todonotes@disabled
  
\else
  
\fi
\makeatother

\usepackage{color}
    \definecolor{darkgreen}{rgb}{0,0.5,0}
    \definecolor{darkred}{rgb}{0.5,0,0}
    \definecolor{darkblue}{rgb}{0,0,0.6}
    \definecolor{purple}{rgb}{0.4,.2,0.7}

\newcommand{\eg}{{\it e.g.\,}\ }
\newcommand{\ie}{{\it i.e.\,}\ }

\def\yp2{{y_+}^2}
\def\ym2{{y_-}^2}

\numberwithin{theorem}{section}

\begin{document}
\title{Double-trace instability of BTZ black holes}
\author[a]{\'Oscar J.~C.~Dias,}
\author[a]{David Sola Gil,}
\author[b]{Jorge E. Santos}

\affiliation[a]{STAG research centre and Mathematical Sciences, Highfield Campus, University of Southampton, Southampton SO17 1BJ, UK}
\affiliation[b]{Department of Applied Mathematics and Theoretical Physics, University of Cambridge, Wilberforce Road, Cambridge CB3 0WA, UK} 

\emailAdd{ojcd1r13@soton.ac.uk}
\emailAdd{d.sola-gil@soton.ac.uk}
\emailAdd{jss55@cam.ac.uk}
\subheader{\today}

\abstract{
We perform a comprehensive study of the linear stability of rotating BTZ black holes under massive scalar field perturbations with double-trace boundary conditions. While BTZ black holes are stable under standard Dirichlet and Neumann boundary conditions, we demonstrate that they can develop instabilities when subjected to double-trace boundary conditions. Our key findings are threefold. First, we show that BTZ black holes exhibit instabilities not only for non-axisymmetric modes—previously the only known unstable sector—but crucially also for axisymmetric modes. Second, we prove that the axisymmetric instability is the dominant and most fundamental: configurations unstable to any non-axisymmetric mode are already unstable to the axisymmetric one. Third, we identify regions in the BTZ parameter space where these black holes are unstable while global AdS$_3$ remains stable, and we map the complete onset curves that determine the corresponding stability boundaries.
Unlike conventional superradiant instabilities, the BTZ double-trace instability occurs for angular velocities always satisfying the Hawking-Reall bound. We trace the physical origin of these instabilities to the influx of energy and angular momentum through the asymptotic boundary permitted by double-trace deformations for a particular sign of the coupling, rather than to near-horizon effects. Our results provide a prototype for understanding double-trace instabilities in higher-dimensional rotating AdS black holes and suggest the existence of rotating hairy black hole solutions with scalar condensates, which we construct in a companion paper.
}

\maketitle
\flushbottom

\section{Introduction and summary of main results}
One of the most interesting features of $d$-dimensional asymptotically 
Anti-de Sitter (AdS$_d$) rotating black holes is their potential 
superradiant instability under bosonic perturbations $-$ whether these are  perturbations of a test field with spin $0$ or $1$, or perturbations of the  gravitational background itself 
\cite{Cardoso:2004hs,Uchikata:2009zz,Dias:2013sdc,Cardoso:2013pza,
Wang:2015fgp,Brito:2015oca,BarraganAmado:2018zpa,Ishii:2022lwc}. 
This instability can occur only when the angular velocity $\Omega_+$ of  the black hole violates the Hawking-Reall bound \cite{Hawking:1999dp},
\begin{equation}
    \Omega_+ L > 1,
\end{equation}
where $L$ is the AdS length scale \cite{Cardoso:2004hs,Dias:2013sdc,Cardoso:2013pza}.

When this bound is not violated ($\Omega_+ L < 1$), there exists a Killing 
vector field $-$ specifically, the horizon generator
\begin{equation}
    K = \partial_t + \Omega_+\,\partial_\phi ,
\end{equation}
which remains timelike throughout the entire outer domain of communication,  from the horizon out to the conformal boundary (the Einstein static universe  $\mathbb{R}_t \times S^{d-2}$). Here $t$ and $\phi$ denote the time and  azimuthal coordinates. However, when $\Omega_+ L > 1$, no Killing vector field (including $K$) is timelike everywhere outside the horizon.

Superradiant scattering can be viewed as the wave analogue of the Penrose process: energy and angular momentum are extracted from the ergoregion, where negative-energy states are allowed. When $\Omega_+ L > 1$, repeated superradiant amplification together with reflections at the AdS conformal boundary can lead to an instability. Throughout this discussion we assume Dirichlet boundary conditions at the boundary, ensuring vanishing energy and angular-momentum flux. These boundary conditions make AdS behave as an effectively confining ``box'', returning any null particle or wave that reaches the boundary back toward the central region of the spacetime.

Rotating black holes in Einstein-AdS$_d$ gravity can satisfy 
$\Omega_+ L > 1$ for spacetime dimension $d \geq 4$. For example, 
Kerr-AdS$_4$ black holes \cite{Carter:1968ks} may either satisfy or violate the Hawking-Reall bound, and black resonators $-$ whether in pure AdS gravity \cite{Dias:2015rxy} or supported by a spinning scalar field \cite{Dias:2011at,Stotyn:2011ns,Ishii:2021xmn} $-$ exist only when 
$\Omega_+ L > 1$.

The situation changes significantly when we turn to asymptotically 
three-dimensional AdS spacetimes. The asymptotically AdS$_3$ rotating 
black hole $-$ the Ba\~nados-Teitelboim-Zanetti (BTZ) solution \cite{Banados:1992wn,Banados:1992gq} $-$ is special in several respects. One particularly important feature is that, unlike its higher-dimensional counterparts, the BTZ black hole is not superradiantly unstable under the conditions considered above. This remains true for spin-0 and spin-1 test fields (since three-dimensional gravity has no propagating gravitational degrees of freedom) and for both Dirichlet and Neumann boundary conditions 
\cite{Birmingham:2001hc,Dappiaggi:2017pbe,Dias:2019ery}.

This absence of superradiant instabilities is striking because BTZ black holes still possess an ergoregion and behave as a confining box when Dirichlet (reflecting) boundary conditions are imposed at the conformal boundary. The underlying reason is ultimately simple: unlike rotating AdS black holes in $d \geq 4$, the BTZ geometry always satisfies the Hawking-Reall bound. And that is the end of the story.

AdS spacetimes have a particularly distinctive property (among many others) when one considers scalar fields propagating in their interior. A massive scalar field in AdS$_d$ can remain normalizable $-$ meaning its excitations have finite energy $-$ even when its dimensionless mass squared, $\mu^2 L^2$, is negative, provided it lies above the Breitenlohner-Freedman (BF) bound~\cite{Breitenlohner:1982jf,Breitenlohner:1982bm,Mezincescu:1984ev},
\begin{equation}
    \mu_{\scriptscriptstyle \mathrm{BF}}^2 L^2 = -\frac{(d-1)^2}{4}.
\end{equation} 
This contrasts sharply with asymptotically flat spacetimes, where 
Klein-Gordon fields are normalizable only if their mass squared is 
non-negative.

To describe this more explicitly, consider Fefferman-Graham coordinates \cite{Fefferman:1985}, in which the AdS$_d$ boundary lies at $z=0$, and the metric satisfies $g_{zz}=L^2/z^2$ and $g_{za}=0$ for $a\ne z$. Near the boundary, a scalar field $\Phi$ behaves as
\begin{equation}
    \Phi|_{z\to 0} \sim 
        \alpha\, z^{\Delta_-} + \cdots 
        + \beta\, z^{\Delta_+} + \cdots ,
\end{equation}
where
\begin{equation}
    \Delta_\pm = \frac{d-1}{2} 
        \pm \sqrt{\frac{(d-1)^2}{4} + \mu^2 L^2}
\end{equation}
are the conformal dimensions, and $\alpha$ and $\beta$ are arbitrary 
functions of the boundary coordinates. The ellipses denote terms fixed by the equations of motion and may include logarithmic terms, depending  on the dimensions and conformal dimension.

If the scalar mass lies at or above the unitarity bound,
\begin{equation}
    \mu_{\scriptstyle \mathrm{unit}}^2 
        = \mu_{\scriptscriptstyle \mathrm{BF}}^2 + \frac{1}{L^2},
\end{equation}
only the faster-decaying mode is normalizable, and therefore one must impose a homogeneous Dirichlet boundary condition, $\alpha = 0$. As discussed above, with this Dirichlet condition rotating black holes in $d\ge4$ can exhibit superradiant instabilities, whereas the rotating BTZ black hole in $d=3$ cannot.

However, when the scalar mass lies strictly between the BF and unitarity bounds,
\begin{equation}
    \mu_{\scriptscriptstyle \mathrm{BF}}^2 
        < \mu^2 
        < \mu_{\scriptscriptstyle \mathrm{BF}}^2 + \frac{1}{L^2},
\end{equation}
both asymptotic modes are normalizable, and one may impose arbitrary mixed boundary conditions. This more general class includes Dirichlet and Neumann conditions as special cases but extends beyond them 
\cite{Wald:1980jn,Klebanov:1999tb,Witten:2001ua,Amsel:2007im,
Sever:2002fk,Berkooz:2002ug,Ishibashi:2003jd,Ishibashi:2004wx,
Henneaux:2004zi,Henneaux:2006hk,Hertog:2004dr,Martinez:2004nb,
Hertog:2004ns,Amsel:2006uf,Faulkner:2010fh,Faulkner:2010gj}.

However, the fact that mixed boundary conditions (BCs) allow normalizable scalar modes does not imply that the corresponding finite-energy solutions are necessarily stable. One must therefore analyse the stability of both AdS$_d$ and asymptotically AdS$_d$ black holes when mixed BCs are imposed (within the scalar-mass window described above). For pure AdS$_d$, this question was studied in detail by Ishibashi and Wald \cite{Ishibashi:2004wx}. They showed that for any $d \ge 3$, AdS$_d$ is stable under scalar perturbations with Dirichlet or Neumann BCs, but can be unstable under \emph{certain} mixed BCs.

This instability can be understood as follows. Mixed BCs $-$ unlike homogeneous Dirichlet BCs $-$  break the conformal invariance of the system and therefore no longer enforce vanishing flux of energy or angular momentum at the asymptotic boundary. In effect, the AdS boundary ceases to behave as a perfectly reflecting confining box: mixed BCs allow energy and angular momentum to leak out of or enter the spacetime. It is thus plausible that some mixed BCs lead to instabilities, as demonstrated in \cite{Ishibashi:2004wx}. Remarkably, such instabilities can arise even for non-spinning perturbations.

To see this, consider expanding scalar perturbations in Fourier modes using 
the AdS Killing vectors $\partial_t$ and $\partial_\phi$,
\begin{equation}
    \Phi \sim e^{-i\,\omega\, t}\, e^{i\, m\, \phi}\psi(z),
\end{equation}
where $\omega$ is the frequency and $m$ the azimuthal number (with $m>0$ and $\text{Re}(\omega L)>0$ corresponding to modes co-rotating with $\Omega_+$). Ishibashi and Wald showed that AdS$_d$ can be unstable to mixed BCs even for modes with $m=0$ \cite{Ishibashi:2004wx}.

One may therefore wish to restrict attention (or not!) to mixed BCs for which AdS$_d$ itself is stable. After doing so, one can proceed to analyse the stability of AdS$_d$ black holes under perturbations satisfying the same mixed BCs.

Within the AdS/CFT correspondence $-$  which has been one of the main motivations for studying AdS black holes $-$ the boundary condition
\begin{equation}
    \beta = \kappa\, \alpha^{\,p-1}, \qquad p \in \mathbb{Z}_{\ge 2},
\end{equation}
is commonly referred to as a \emph{multi-trace boundary condition}, with real parameter $\kappa \in \mathbb{R}$. In this language, the limit $\kappa \to \pm \infty$ reduces to the homogeneous Dirichlet condition $\alpha = 0$, while $\kappa = 0$ yields the homogeneous Neumann condition $\beta = 0$. This terminology reflects the fact that such boundary conditions correspond to adding a multi-trace deformation to the boundary theory, which without the deformation is a conformal field theory (CFT) \cite{Klebanov:1999tb,Witten:2001ua,Amsel:2007im,Sever:2002fk,Berkooz:2002ug,Hertog:2004dr,Martinez:2004nb,Hertog:2004ns,Amsel:2006uf,Faulkner:2010fh,Faulkner:2010gj}.\footnote{Although not considered here, multi-trace boundary conditions can also be imposed on vector and gauge fields in asymptotically AdS$_d$ spacetimes, without restriction, and are relevant in the context of 
AdS/CFT; see, e.g.,\cite{Witten:2003ya,Ishibashi:2004wx,Marolf:2006nd}. 
Robin boundary conditions for the gravitational field in AdS$_d$ 
($d\ge4$) are also of interest; see, e.g., \cite{Compere:2008us}.}

The case $p = 2$ $-$ the double-trace deformation $-$  is particularly important as the simplest nontrivial example. The analysis of \cite{Ishibashi:2004wx} shows that for any $d \ge 3$, scalar fields in pure AdS$_d$ with a \emph{positive} 
double-trace parameter $\kappa$ are stable, whereas fields with $\kappa$ below a certain critical \emph{negative} value are unstable. Intuitively, a double-trace boundary condition with $\kappa>0$ allows energy to leak out at infinity, which reinforces the stability of Dirichlet perturbations. In contrast, $\kappa<0$ allows for incoming energy flux from infinity, which, as discussed above, can trigger an instability. 

In addition to these stability results, \cite{Faulkner:2010fh} proved a \emph{positive energy theorem} for AdS$_d$ ($d\ge3$) equipped with 
double-trace boundary conditions.

These observations naturally motivate the study of AdS$_d$ black holes and their stability under double-trace boundary conditions, a topic with a growing literature. Double-trace deformations have been extensively investigated in the context of the holographic superconductor program (the AdS/condensed matter correspondence) 
\cite{Allais:2010qq,Faulkner:2010gj,Basu:2013soa,Dias:2013bwa,Ren:2022qkr,Kinoshita:2023iad,Auzzi:2025nxx}. A key point is that a \emph{negative} double-trace parameter $\kappa$ can induce a scalar condensation instability $-$ eventually even for a neutral scalar field around a neutral planar Schwarzschild-AdS spacetime $-$ when the temperature falls below a critical value. This mechanism yields a dominant planar AdS hairy black hole phase, interpreted as the \emph{holographic superconductor phase}, below the critical temperature. In this sense, a neutral double-trace perturbation can effectively mimic Gubser's original mechanism of 
charged scalar condensation in charged AdS black holes that launched the superconductor program \cite{Gubser:2008px,Hartnoll:2008vx,Hartnoll:2008kx}.

Beyond the planar case, the linear stability of global AdS$_4$-Reissner-Nordstr\"om black holes under neutral or charged complex scalar perturbations with Robin boundary conditions was studied in \cite{Katagiri:2020mvm}. The presence of unstable modes in that work indicates the existence of corresponding hairy black hole solutions featuring a double-trace  induced scalar condensate floating above the central RN-AdS black hole, which were indeed constructed in \cite{Harada:2023cfl}. Similarly, double-trace perturbations of global AdS$_4$-Schwarzschild black holes were likewise analysed in \cite{Kinoshita:2023iad}.

Finally, we turn to the case that is central to our work: the rotating BTZ black hole \cite{Banados:1992wn,Banados:1992gq,Carlip:1995qv} perturbed by a scalar field subject to double-trace boundary conditions. Recall that for Dirichlet and Neumann boundary conditions $-$  which are the only choices that preserve the full AdS asymptotic symmetries \cite{Hertog:2004rz} and therefore enforce vanishing asymptotic flux $-$  the BTZ black hole is stable against scalar perturbations. This stability ultimately follows from the fact that, 
under such boundary conditions, the condition $\Omega_+ L \leq 1$ prevents the onset of any superradiant-type instability 
\cite{Hawking:1999dp,Birmingham:2001hc,Iizuka:2015vsa,Dappiaggi:2017pbe,Dias:2019ery}.

With double-trace boundary conditions, however, the situation is more subtle. Although BTZ always satisfies $\Omega_+ L \leq 1$, this no longer guarantees stability because double-trace deformations permit a non-vanishing asymptotic flux. In particular, a negative double-trace parameter $\kappa$ allows an incoming flux of energy, which can drive an instability independently of rotational effects. Indeed, it was shown in \cite{Iizuka:2015vsa,Dappiaggi:2017pbe,Ferreira:2017cta} that BTZ becomes unstable for sufficiently negative $\kappa$, although those works restricted their analysis to perturbations with azimuthal number $m \geq 1$ and did not address the $m = 0$ sector.

One of the main goals of this paper is to complete that analysis by including the $m=0$ modes. This is strongly motivated by earlier results showing that both pure AdS$_d$ \cite{Ishibashi:2004wx} and global AdS$_4$ black holes \cite{Katagiri:2020mvm,Harada:2023cfl} can be unstable to $m=0$ perturbations with double-trace boundary conditions. As we demonstrate, the same phenomenon also occurs for the rotating BTZ black hole.

As we will show, the $m=0$ instability is in fact the most significant 
instability of the BTZ black hole. There are several reasons for this. 
First, only those BTZ black holes that are already unstable to $m=0$ 
perturbations can exhibit instabilities in the $m \geq 1$ sectors. 
That is, whenever BTZ is unstable for some $m \geq 1$, it must already be unstable for $m=0$. One of our goals is to map the regions in the BTZ two-parameter space, for various values of the double-trace parameter $\kappa$, where each $m$-mode becomes unstable.

Second, for any $\kappa$ for which BTZ is unstable to the $m=1$ mode (and hence to $m=0$), the instability is \emph{strongest} in the $m=0$ sector, in the sense that
\begin{equation}
   {\rm Im}(\omega L)\big|_{m=0}
   > {\rm Im}(\omega L)\big|_{m=1}
   > {\rm Im}(\omega L)\big|_{m>1}.
\end{equation}
Establishing this hierarchy is another central aim of this paper.

Third, all BTZ black holes that are unstable for $m \geq 1$ necessarily correspond to values of $\kappa$ for which AdS$_3$ is already unstable to the $m=0$ perturbation. However, there is a window of $\kappa$ for which AdS$_3$ is stable (for all $m$), while BTZ can be either stable or unstable to $m=0$ perturbations. Some of these BTZ solutions are then stable for $m \geq 1$, whereas others also develop instabilities in the $m=1$ or higher-$m$ sectors. One of the aims of this paper is to identify the structure of these regions in the BTZ parameter space.

Ultimately, the instabilities of BTZ for both $m=0$ and $m \geq 1$ 
double-trace modes suggest that the phase diagram of asymptotically 
AdS$_3$ stationary solutions should contain a family of hairy black holes with a rotating horizon and a scalar condensate floating or corotating above it. In a companion paper \cite{DSS:2026}, we construct these BTZ hairy black holes and analyse their properties. At fixed mass and angular momentum, they possess higher entropy than BTZ, suggesting that unstable BTZ black holes evolve dynamically toward the hairy solutions of \cite{DSS:2026}. In the companion paper  \cite{DSS:2026} we also construct double-trace hairy boson stars.\footnote{Several AdS$_3$ studies are relevant in this context. AdS$_3$ perfect-fluid stars were constructed in \cite{Cruz:1994ar,Lubo:1998ue}. AdS$_3$ boson stars with scalar self-interaction potentials appeared in \cite{Sakamoto:1998hq,Sakamoto:1999zt,Sakamoto:1998aj}, while boson stars sourced by scalar fields were obtained in \cite{Astefanesei:2003qy,Astefanesei:2003rw}. AdS$_3$ boson stars with homogeneous Dirichlet boundary conditions and AdS$_3$ hairy black holes with Robin boundary conditions were constructed in \cite{Stotyn:2013spa,Stotyn:2012ap} and \cite{Iizuka:2015vsa}, 
respectively, providing three-dimensional analogues of the solutions in \cite{Dias:2011at,Stotyn:2011ns,Dias:2015rxy,Ishii:2021xmn}. 
AdS$_3$ hairy black holes with non-homogeneous Dirichlet boundary conditions were constructed in \cite{Gao:2023rqc}. Finally, AdS$_4$ solitons with Robin boundary conditions were found in \cite{Bizon:2020yqs}.}

The plan of this paper is as follows. In section~\ref{sec:BTZ+KG} we begin with a brief review of the rotating BTZ black hole solution 
(section~\ref{sec:BTZbh}), focusing on the properties relevant for later sections; complete reviews can be found in \cite{Banados:1992gq,Carlip:1995qv}. This is a two-parameter family of solutions, which we take to be the dimensionless mass $\hat{M}$ and angular momentum $\hat{J}$. We then discuss the Klein-Gordon equation in BTZ, emphasizing the scalar field mass range where double-trace boundary conditions can be imposed while keeping the energy of scalar excitations finite (section~\ref{sec:BTZ-KG2xTr}).

In section~\ref{sec:Methods} we describe the two complementary methods used to solve the eigenfrequency problem for double-trace regular modes. One method (section~\ref{sec:Findroot-Method}) is largely analytical, except for a final numerical root-finding step; the eigenfunctions are, however, fully analytic. The second method is a fully numerical approach that efficiently scans the BTZ and double-trace parameter space for the corresponding frequencies (section~\ref{sec:Numerical-Method}). We also review the AdS$_3$ 
and BTZ mode spectra under Dirichlet or Neumann boundary conditions.

Section~\ref{sec:Ad3} revisits the AdS$_3$ instability originally identified by Ishibashi and Wald \cite{Ishibashi:2004wx}. In section~\ref{sec:AdS3w}, we explore additional properties of the instability not discussed in \cite{Ishibashi:2004wx}, which focused on the existence and onset conditions for the instability. Our results (section~\ref{sec:AdS3wald}) are fully consistent with their findings.

Section~\ref{sec:BTZresults} focuses on double-trace perturbations of the BTZ black hole. Subsection~\ref{sec:BTZonset1} studies how the solutions vary with $\kappa$ for $m=0$ and $m \geq 1$ modes. As expected, the double-trace frequency families interpolate between the Dirichlet ($\kappa = \pm \infty$) and Neumann ($\kappa = 0$) frequencies, though not always in a straightforward way (see Figs.~\ref{fig:BTZw-kappa:m0}-\ref{fig:BTZw-kappa:m1} for BTZ, 
and Figs.~\ref{fig:AdS3m0-wk}-\ref{fig:AdS3m1-wk} for AdS$_3$). In 
particular, we find that the fundamental radial overtone ($n=0$) of $m=0$ modes, and the fundamental $n=0$ overtone of $m \geq 1$ modes, are unstable in BTZ. Both AdS$_3$ and BTZ are stable for $\kappa \geq 0$ and become unstable when $0 > \kappa^{\rm AdS}_{m,\mu^2 L^2} > \kappa$ or $0 > \kappa^{\rm BTZ}_{m,\mu^2 L^2}(\hat{M},\hat{J}) > \kappa$, respectively.

Subsection~\ref{sec:BTZonset2} identifies the regions of the 2-dimensional BTZ parameter space and $\kappa$ where instabilities occur in both AdS$_3$ and BTZ, where only one of them occurs, or where both are absent. We compare $\kappa^{\rm AdS}_{m,\mu^2 L^2}$ with $\kappa^{\rm BTZ}_{m,\mu^2 L^2}(\hat{M},\hat{J})$ (see Fig.~\ref{fig:3D_kappa_onset} and Figs.~\ref{fig:m0_kappa_onset_curves}-\ref{fig:m1_kappa_onset_curves}). We find regions where both AdS$_3$ and BTZ are stable, regions where AdS$_3$ is stable while BTZ is unstable to the $m=0$, $n=0$ mode, and regions where BTZ develops instabilities for $m=0$, $n=0$ and increasingly higher $m$, $n=0$ 
modes. The negative onset $\kappa$ decreases as $m$ increases 
(Figs.~\ref{fig:BTZw-kappa:m1_m5}-\ref{fig:m0tom5_btz_VS_ads_instability}), so that whenever BTZ is unstable to any $m \geq 1$ mode, it is also unstable to $m=0$, $n=0$. In this sense, the $m=0$, $n=0$ mode effectively determines 
the stability properties of BTZ. Furthermore, the strength of the instability follows ${\rm Im}(\omega L)|_{m=0} > {\rm Im}(\omega L)|_{m=1} > {\rm Im} (\omega L)|_{m>1}$ (see Fig.~\ref{fig:m0tom5_btz_VS_ads_instability}).

In subsection~\ref{sec:BTZorigin}, we discuss the physical origin of the AdS$_3$ and BTZ instabilities. These cannot be purely superradiant, as the $m=0$ mode is also unstable, and even in AdS$_3$. Double-trace boundary conditions allow energy and angular momentum influx at the asymptotic boundary, which feeds the instability.

Finally, in subsection~\ref{sec:3Dspectrum}, we fix $\kappa$ and study the BTZ frequency spectrum as a function of $\hat{M}$ and $\hat{J}$ for $m=0$, $n=0$ modes and co-rotating $m \geq 1$, $n=0$ modes that can become unstable. We usually choose $\kappa$ such that AdS$_3$ is stable, but also illustrate cases where AdS$_3$ is already unstable. For $m \geq 1$, $n=0$, only co-rotating modes can become unstable, though we discuss main properties of counter-rotating modes as well. For concreteness, plots in the main text use $\mu^2 L^2 = -15/16$, but we have performed an exhaustive analysis for all masses in the double-trace range $-1 < \mu^2 L^2 < 0$, and the plots illustrate universal qualitative properties. Additional plots for $\mu^2 L^2 = \{-8/9, 
-3/4, -1/10\}$ are provided in Appendix~\ref{sec:AppOtherMasses}.
 
\section{Scalar perturbations in BTZ with double-trace boundary conditions}\label{sec:BTZ+KG}

\subsection{The BTZ black hole}\label{sec:BTZbh}
The BTZ black hole is a 3-dimensional solution of Einstein's gravity with a negative cosmological constant $\Lambda = -1/L^2$, with $L$ being the AdS$_3$ radius. In the Schwarzschild gauge, the solution is described by the metric \cite{Banados:1992wn, Banados:1992gq, Carlip:1995qv}
\begin{align}\label{eqn:BTZ_metric}
    {\rm d}s^2 = -f(r) {\rm d}t^2 + \frac{{\rm d}r^2}{f(r)} + r^2 \Big[{\rm d}\phi - \Omega(r)\,{\rm d}t \Big]^2,
\end{align}
where 
\begin{align}\label{eqn:BTZ_metric_functions}
    f(r) = \frac{(r^2- r_+^2)(r^2-r_-^2)}{L^2 r^2}, \qquad \Omega(r) = \frac{r_+ r_-}{L \,r^2}.
\end{align}
Here $\{t, r, \phi\}$ are the usual time, radial and azimuthal coordinates with $\phi \sim \phi + 2 \pi$.  The function $f$ has two positive roots $r_{\pm}$, with $0\leq r_- \leq r_+$, which describe the location of the Cauchy and event horizons, respectively. 

The mass, angular momentum, horizon angular velocities and  surface gravities can be expressed in terms of the horizons via ($G$ is Newton's constant)
\begin{align}\label{eqn:BTZ_physical_quantities}
   8 G \,M = \frac{r_+^2+r_-^2}{L^2}, \quad 8 G\, J = \frac{2 r_+ r_-}{L}, \quad \Omega_\pm \equiv \Omega(r_\pm) = \frac{r_{\mp}}{L r_\pm},\quad \text{and}\quad \kappa_\pm = \frac{r_+^2-r_-^2}{L^2r_\pm}.
\end{align}
The BTZ black hole is extremal when $\kappa_\pm= 0$, \ie $r_+ = r_-$ or $|J| = M L$. Consequently, BTZ black holes must have  $0\leq |J| \leq M L$.  BTZ has an ergosphere at $r=r_{\hbox{\tiny ergo}}$ (circle of infinite redshift), where the norm of $\partial/ \partial t$ vanishes,  given by
 \begin{align}\label{BTZ:ergo}
   r_{\hbox{\tiny ergo}}=(8 G M)^{1/2}L\,
\end{align}
and thus it has an ergoregion in the region $r_+< r \leq r_{\hbox{\tiny ergo}}$. Assuming that $J>0$ (as we do), timelike curves in the ergoregion necessarily have $\mathrm{d}\phi/\mathrm{d}t >0$ and thus all such observers are
dragged along by the rotation of the black hole. 

The BTZ family is well known to exhibit a mass gap. Although these geometries are asymptotically AdS$_3$, recovering global AdS$_3$ requires setting $ 8G\,M = -1$ and $J = 0$, which corresponds to $r_+ = L$ and $r_- = 0$. This places global AdS$_3$ one unit of mass below the ``massless'' or ``vacuum'' BTZ solution with $M = J = 0$, for which the horizon radius vanishes. For intermediate values $-1 < 8G\,M < 0$, the metric \eqref{eqn:BTZ_metric} does not describe a black hole but rather the naked conical defect geometries originally studied in \cite{Deser:1983tn,Deser:1983nh,Staruszkiewicz:1963zza,Gott1984GeneralRI}; we will not consider such solutions here. At the special point $8G\,M = -1$ and $J = 0$, and only there among negative-mass configurations, the conical defect disappears and the metric reduces smoothly to global AdS$_3$ \cite{Banados:1992wn,Banados:1992gq}. 

To make the solution regular across the future event horizon, $r= r_+$, we need to resort to a different coordinate system. For non-extremal BTZ black holes, the first step is to define a radial tortoise coordinate via ${\rm d}r_*={\rm d}r/f(r)$, which upon integration gives \cite{Dias:2019ery}
\begin{align}
    r_* = \frac{1}{2\kappa_+}\log|F_+(r)|, \quad \text{where} \quad F_+(r) = \frac{r-r_+}{r+r_+}\left(\frac{r+r_-}{r-r_-}\right)^{r_-/r_+}. 
\end{align}
Notice that $r_*\to -\infty$ as $r \to r_+$ and $r_* \to \infty$ as $r \to \infty$. We then define ingoing Eddington-Finkelstein coordinates $(v,r, \phi')$ as
\begin{align}\label{ingoingEF}
    {\rm d}t = {\rm d}v - {\rm d}r_* \quad\text{and}\quad  {\rm d}\phi = {\rm d}\phi' - \Omega(r){\rm d}r_*\,,
\end{align}
which allow us to analytically extend the BTZ solution from the exterior region $r>r_+$ into the region $r_-< r \leq r_+$. 

At first sight, the BTZ geometry is specified by three dimensionful parameters, for instance $\{r_-,r_+,L\}$ or, equivalently, $\{M,J,L\}$. These parameters are indeed independent, but the metric admits a simple scaling symmetry that allows one to remove an overall scale. Indeed, under the transformation $\{t,r,\phi\} \to \{\lambda t,\lambda r,\phi\}$ together with $\{r_-,r_+,L\} \to \{\lambda r_-,\lambda r_+,\lambda L\}$, the metric rescales homogeneously as $g \to \lambda^2 g$. Thus one may, if convenient, use this symmetry to fix $L$ (or any one of the other length scales) without loss of generality.

As a result, the BTZ family can be parametrized by two independent \emph{dimensionless} quantities, which may be taken to be the mass and angular momentum in units of $L$, or equivalently the ratios $r_\pm/L$ (this corresponds to fixing $L \equiv 1$ using the scaling symmetry). Accordingly, in what follows we often work with the dimensionless coordinates $(T,R) = (t/L, r/L)$ and scan the parameter space using $(0 \leq y_- \leq y_+)$, with
\begin{align}\label{def:y+y-}
    y_+ \equiv \frac{r_+}{L}, \qquad y_- \equiv \frac{r_-}{L}\,.
\end{align}
Additionally, we will typically present our results in terms of the dimensionless mass, $\hat{M} \equiv 8 G M = y_+^2 + y_-^2$, and the dimensionless angular momentum, $\hat{J} \equiv 8 G J / L = 2 y_+ y_-$. This means that $y_{\pm}$ can be written as a function of  $\{\hat{M},\hat{J}\}$ as
\begin{align}\label{ypmFromMJ}
  y_{\pm}= \frac{1}{2} \left( \sqrt{\hat{M} +\hat{J}} \pm \sqrt{ \hat{M} -\hat{J}} \right)\,.
  \end{align}
Similarly, we will often use a hat to denote some dimensionless physical quantities, \eg
\begin{align}
\hat{T}_{\pm}\equiv T_{\pm}\,L,
   \quad\text{and}\quad \hat{\omega}\equiv \omega \, L.
\end{align}
Note that the black hole mass is dimensionless in $d=3$; the BTZ black hole has a horizon because the AdS radius provides the necessary length scale.

The outer event horizon temperature and angular velocity can be written as 
 \begin{align}\label{TangVMJ}
 \hat{T}_+= \frac{1}{\pi} \frac{\sqrt{\hat{M}^2 - \hat{J}^2}}{\sqrt{\hat{M}+\hat{J}}+\sqrt{\hat{M}-\hat{J}}} \quad\text{and}\quad
 \hat{\Omega}_+= \frac{\hat{M} - \sqrt{\hat{M}^2 - \hat{J}^2}}{\hat{J}}.
  \end{align}
All BTZ black holes have $\hat{\Omega}_+ \leq 1$ (with the upper bound reached at extremality), \ie they have angular velocities that never exceed the Hawking–Reall bound \cite{Hawking:1999dp}.
   
\subsection{Klein-Gordon equation and double-trace boundary conditions}\label{sec:BTZ-KG2xTr}
The study of scalar field perturbations of the BTZ black hole reduces to solving the Klein-Gordon equation on this background, subject to the appropriate physical boundary conditions.
For a neutral complex scalar field $\Phi$ with mass $\mu$ the Klein-Gordon equation reads
\begin{align}\label{eqn:BTZ_KG_equation}
    \Big(\nabla^\mu\nabla_\mu - \mu^2 \Big)\Phi(t,r,\phi) = 0,
\end{align}
where $\nabla$ is the covariant derivative with respect to the BTZ metric \eqref{eqn:BTZ_metric}. As BTZ is stationary and axisymmetric, we can perform a Fourier decomposition of the scalar field, solving for each Fourier mode separately $\Phi(t,r,\phi) = e^{-i\omega t}e^{i\,m\phi}\psi(r)$. This decomposition introduces the frequency $\omega$ and the azimuthal quantum number $m$ (regularity requires it to be an integer). Consequently, \eqref{eqn:BTZ_KG_equation} reduces to a radial ODE given by 
\begin{align}\label{eqn:BTZ_radial_QNM_eqn}
    \psi''(r)+\left[\frac{f'(r)}{f(r)}+\frac{1}{r}\right]\psi'(r)+\frac{1}{f(r)}\left[\frac{\big[\omega -m \,\Omega (r)\big]^2}{f(r)}-\frac{m^2}{r^2}-\mu ^2\right]\psi(r) = 0,
\end{align} 
where $f(r)$ and $\Omega(r)$ are the BTZ functions defined in \eqref{eqn:BTZ_metric_functions} and $^\prime$ denotes derivatives with respect to $r$. The BTZ background has a $t-\phi$ symmetry, $\{t,\phi\}\to\{-t,-\phi\}$, and thus the symmetry $\{m,\omega\}\to \{-m,-\omega^*\}$ (we are keeping the BTZ  angular velocity $\Omega_+>0$ fixed in this discussion). This symmetry provides two equivalent options: we can either consider only modes with $\text{Re}\,\omega > 0$, allowing for $m \ge 0$ (co-rotating or prograde) and $m < 0$ (counter-rotating or retrograde) modes, or we can consider only $m \ge 0$ modes, in which case perturbations with $\text{Re}\,\omega > 0$ correspond to co-rotating modes with the BTZ black hole, while perturbations with $\text{Re}\,\omega < 0$ correspond to counter-rotating modes with respect to the BTZ angular velocity. In this paper, we adopt the latter option.

Note, however, that the terms ``co-/counter-rotating'' involve a slight abuse of language for two reasons: one arises when $m = 0$, and the other occurs when the BTZ spin vanishes ($\hat{J} = 0$). Nevertheless, this nomenclature does not pose any issues in the presentation of our results.

One expects the frequencies that can propagate on the BTZ background to be quantized and to have an imaginary part, since the energy of the scalar field can dissipate, at least partially, through the horizon. As in any eigenvalue problem, the quantization of (eigen)frequencies depends on the boundary conditions imposed. While physical considerations uniquely fix the boundary condition at the horizon, different physical motivations allow more freedom in the choice of asymptotic boundary conditions. In the following, we discuss the boundary conditions we impose and their physical justification.

The event horizon (located at $r = r_+$) and the asymptotic timelike AdS boundary ($r \to \infty$) are both regular singular points of \eqref{eqn:BTZ_radial_QNM_eqn}. Therefore, we can perform a Frobenius analysis to determine the behaviour of the two independent solutions near these boundaries. 

A Frobenius expansion near the event horizon $r = r_+$ yields the two linearly independent solutions:
\begin{equation}
\psi(r) \sim (r - r_+)^{\pm \frac{i}{2\kappa_+}(\omega - m\Omega_+)}.
\end{equation}
Our perturbations must be smooth at the horizon, which requires that the radial field $\psi(r)$ is regular in ingoing Eddington-Finkelstein coordinates \eqref{ingoingEF} at the future event horizon $\mathcal{H}^+$. This condition selects, as the boundary condition, the ingoing behaviour:
\begin{align}\label{BChorizon}
\psi\big|_{r=r_+} \sim (r - r_+)^{-\frac{i}{2\kappa_+}(\omega - m\Omega_+)}.
\end{align}

Consider now the asymptotic boundary, $r \to \infty$. It is convenient to discuss the boundary conditions using the Fefferman-Graham (FG) coordinates $\{t, z, \phi\}$ \cite{Fefferman:1985, Graham:1999jg, Skenderis:2002wp, Anderson:2004yi}, which are defined by imposing
\begin{equation}
g_{zz} = \frac{L^2}{z^2}, \quad g_{za} = 0,
\end{equation}
where $z$ is the radial FG coordinate such that the AdS boundary lies at $z = 0$, and $a = \{t, \phi\}$ are the boundary coordinates. For generic spacetimes, these conditions can be solved order by order in a Taylor expansion in powers of $z$ around $z = 0$. However, for the BTZ black hole, the solution is simple enough that the full spacetime can be expressed in FG coordinates without much effort (note that $z$ is dimensionless):
\begin{align}\label{eqn:BTZ_rtozFGcoordinate}   
    r =\frac{L}{z} \sqrt{1+\frac{1}{2}\left(\frac{r_+^2+r_-^2}{L^2}\right)z^2+\frac{1}{16} \left(\frac{r_+^2-r_-^2}{L^2}\right)^2 z^4}\approx \frac{L}{z}+\frac{1}{4}\left(r_+^2+r_-^2\right)\frac{z}{L} +\mathcal{O}\left(z^3\right). 
\end{align}
Performing a Frobenius expansion of \eqref{eqn:BTZ_KG_equation} in Fefferman-Graham coordinates at $z = 0$ yields the standard result (see, e.g., \cite{Skenderis:2002wp, Dias:2015nua}):
\begin{align}\label{FGasympExp}
    \psi(z)\big|_{z \to 0} = \alpha \, z^{\Delta_-} + \dots + \beta \, z^{\Delta_+} + \mathcal{O}(z^{\Delta_+ + 1}), \quad \text{where} \quad
    \Delta_\pm &= 1 \pm \sqrt{1 + \hat{\mu}^2},
\end{align}  
are the conformal dimensions of the system, \emph{i.e.}, the solutions of $\Delta(\Delta - 2) = \mu^2 L^2 \equiv \hat{\mu}^2$ (note that $\Delta_- = 2 - \Delta_+$). Here, $\alpha$ and $\beta$ are the amplitudes of the two linearly independent decays, and the dots represent terms that are fixed as functions of $\alpha$ and $\beta$ by solving the Klein-Gordon equation order by order in the $z$-expansion.\footnote{Generally, in arbitrary dimensions or beyond linear order $-$ where the scalar and metric back-react $-$ logarithmic terms may appear in the Fefferman-Graham expansion \eqref{FGasympExp} for very special values of the conformal dimensions $\Delta_{\pm}$, typically reflecting \eg conformal anomalies \cite{deHaro:2000vlm,Kanitscheider:2006zf,Grumiller:2008qz,Skenderis:2009nt,Bzowski:2015pba}. However, at linear order in a fixed BTZ background or fixed AdS$_3$, the Klein–Gordon equation yields no logarithmic terms for any mass $\mu^{2}>\mu^{2}_{\text{\tiny BF}}$ \cite{deHaro:2000vlm,Bzowski:2015pba}.}
 
To have a well-posed initial-boundary value problem, one must impose a boundary condition that fixes either $\alpha$, $\beta$, or a relation between them. For physical relevance, such a choice should be motivated by underlying physics. In particular, studies of scalar fields in asymptotically AdS backgrounds are of interest in the context of the AdS/CFT correspondence, and we will motivate our choice(s) accordingly.

Stability of asymptotically AdS$_3$ geometries under scalar perturbations requires that the mass of the scalar field satisfies the Breitenlohner-Freedman (BF) bound, $\mu^2 \geq \mu^2_{\rm BF}$, where $\mu^2_{\rm BF} L^2 = -\frac{(d-1)^2}{4} = -1$ for $d = 3$ \cite{Breitenlohner:1982jf, Mezincescu:1984ev}. As long as this bound is satisfied, one can impose Dirichlet boundary conditions by fixing $\alpha$. Such a mode is normalizable, and solving the bulk equation of motion with the inner boundary condition then yields the expectation value (VEV)
\begin{equation}
\big\langle \mathcal{O}_{\Phi}^{(\Delta_+)} \big\rangle = (\Delta_+ - \Delta_-) \, \beta
\end{equation}
of the dual CFT operator $\mathcal{O}_{\Phi}^{(\Delta_+)}$ with conformal dimension $\Delta_+$.  

If a non-homogeneous Dirichlet boundary condition is imposed ($\alpha \neq 0$), the source $\alpha$ induces a VEV proportional to $\beta$. Alternatively, one can impose a homogeneous Dirichlet BC ($\alpha = 0$), in which case a VEV is generated spontaneously, even in the absence of a source.

However, if the scalar mass lies in the range
\begin{align}
\mu^2_{\text{BF}} L^2 < \mu^2 L^2 < \mu^2_{\text{BF}} L^2 + 1, \quad \text{with} \quad \mu^2_{\text{BF}} L^2 = -1,
\label{2xTrace:rangeMass}
\end{align}
and only in this case, both modes $z^{\Delta_\pm}$ are normalizable (the upper bound is equivalent to the condition that $\Delta_-$ is above the unitarity bound, $\Delta_- \geq (d-3)/2$). It follows that, in this mass range, one can still impose a Dirichlet BC that fixes $\alpha$ (e.g., $\alpha = 0$; this is known as the \emph{standard quantization}). In addition, one is also allowed to impose a Neumann BC (e.g., $\beta = 0$; the \emph{alternative quantization}) or even mixed BCs that express $\beta$ as a function of $\alpha$. In the last two cases, the conformal dimension of the dual operator $\mathcal{O}_\Phi^{(\Delta_-)}$ is $\Delta_-$.

Of particular importance is the case of a mixed BC given by the double-trace condition
\begin{align}
\beta = \kappa \, \alpha,
\label{2xTrace:BC}
\end{align}
where $\kappa \in \mathbb{R}$. The most interesting case for our purposes is $\kappa \leq 0$ as discussed below.

In the context of the AdS/CFT duality, this Robin BC corresponds to a double-trace deformation of the dual CFT,
\begin{align}
S_{\rm CFT} \to S_{\rm CFT} - \int \mathrm{d}^2 x \, \kappa \, \mathcal{O}_{\Phi}^{(\Delta_-)} \mathcal{O}_{\Phi^\dagger}^{(\Delta_-)},
\end{align}
where the dual operator has an expectation value $\langle \mathcal{O}_{\Phi}^{(\Delta_-)} \rangle$ proportional to $\alpha$ \cite{Witten:2001ua,Berkooz:2002ug,Mueck:2002gm,Sever:2002fk,Papadimitriou:2007sj,Faulkner:2010fh,Faulkner:2010gj}.\footnote{The double-trace BC is the special case of a multi-trace BC, $\beta = \kappa \, \alpha^{p-1}$, for integer $p$ obeying $2 \leq p \leq (d-1)/\Delta_-$ (with $d=3$ in our case), which describes multi-trace deformations of the CFT.} 

Henceforth, we will work in the scalar mass range \eqref{2xTrace:rangeMass} and impose the double-trace BC \eqref{2xTrace:BC} because:  
1) this BC includes, as special cases, the homogeneous Dirichlet BC ($\alpha=0$, effectively corresponding to the limit $\kappa \to \pm \infty$) and the Neumann BC ($\beta=0$, corresponding to $\kappa = 0$); and  
2) except for the Dirichlet and Neumann limits, the BTZ background can be unstable to linear scalar field perturbations obeying double-trace boundary conditions. 

We will consider both signs of $\kappa$ for completeness, but the most interesting case occurs when $\kappa \leq 0$ in the double-trace BC \eqref{2xTrace:BC}.  

The reason is that it is now well established $-$ mainly from the stability study of AdS$_3$ \cite{Ishibashi:2004wx} and from investigations in the context of the holographic superconductor program $-$ that for finite $\kappa > 0$, the double-trace deformation only reinforces the stability of the bald black hole. Indeed, we explicitly checked that BTZ does not develop linear mode instabilities when \eqref{2xTrace:BC} is imposed with $\kappa \geq 0$ (more later).  

On the other hand, for $\kappa < 0$, the double-trace deformation can induce a $U(1)$-symmetry breaking that effectively leads to the existence of unstable (i.e., with $\mathrm{Im}(\omega) > 0$) linear mode perturbations in planar AdS$_d$ systems, as shown in \cite{Faulkner:2010gj} (see also \cite{Klebanov:1999tb,Witten:2001ua,Amsel:2007im,Sever:2002fk,Berkooz:2002ug,Hertog:2004dr,Martinez:2004nb,Hertog:2004ns,Amsel:2006uf,Hartnoll:2008kx,Faulkner:2010fh,Faulkner:2010gj}) using results from \cite{DeWolfe:1999cp,Freedman:2003ax}, and, as we will confirm in the following sections, also for the BTZ black hole.  

Consequently, the energy of the system acquires new minima, and when backreaction is included, one expects the scalar field to condense around the horizon of the original bald black hole, forming a hairy black hole.

\section{Methods to find the frequencies of unstable modes in BTZ}\label{sec:Methods} 
We solve the eigenvalue problem defined by the Klein-Gordon equation
\eqref{eqn:BTZ_KG_equation}, subject to the ingoing boundary condition at the
horizon \eqref{BChorizon} and the double-trace asymptotic boundary condition
\eqref{2xTrace:BC}. Our goal is to determine the quantized frequencies that can
propagate in the BTZ black hole when the scalar mass lies in the window
\eqref{2xTrace:rangeMass}. To this end, we employ two complementary methods,
each with its own advantages, which also provide an independent check of our results.

The first method is the most natural. It is well known that the Klein-Gordon
equation \eqref{eqn:BTZ_KG_equation} can be solved analytically in the BTZ
black hole, yielding the general solution as a linear combination of two
independent wavefunctions. For homogeneous Dirichlet ($\alpha=0$) or Neumann
($\beta=0$) boundary conditions, the horizon and asymptotic conditions
\eqref{BChorizon} and \eqref{2xTrace:BC} can be imposed straightforwardly to
obtain the quantized frequencies analytically. One boundary condition fixes the
relative amplitudes of the two independent wavefunctions, while the other reduces
to a simple gamma-function condition, $\Gamma(W)=\infty$, where $W=W(\omega)$
is linear in the frequency $\omega$. This condition has the simple solution
$W(\omega)=-n$ for non-negative integers $n$, corresponding to the radial
overtone, which directly gives the exact quantized frequencies in closed form.
However, this simplification no longer holds for double-trace boundary
conditions. In that case, the final quantization condition cannot be solved
analytically and must instead be addressed numerically, for example using the
\texttt{FindRoot} routine in \textit{Mathematica}.

The second method is fully numerical. We solve the eigenvalue problem directly,
determining eigenvalues and eigenfunctions using either the
\texttt{Eigensystem}/\texttt{Eigenvalue} routines in \textit{Mathematica} or a
Newton-Raphson algorithm. Although this approach may seem unnecessary, in
practice it allows efficient exploration of the two-dimensional parameter space of
BTZ solutions. In the following subsections, we describe the formulation and
setup of both methods, while also revisiting the analytical expressions for
AdS$_3$ and BTZ frequencies in the Dirichlet and Neumann limits.

\subsection{Method 1: Analytical eigenfunctions and double-trace frequencies}\label{sec:Findroot-Method}
To solve the Klein-Gordon equation \eqref{eqn:BTZ_KG_equation}, it is
convenient to introduce the coordinate
\begin{align}
\label{def:y}
y \equiv \frac{r^2 - r_+^2}{r^2 - r_-^2},
\end{align}
so that the event horizon corresponds to $y = 0$ and the conformal boundary to
$y = 1$. We then redefine the mode function as
\begin{align}\label{eqn:Psi_decomposition}
\psi(y) = y^{\frac{i}{2\hat{\kappa}_+}(\hat{\omega} - m \hat{\Omega}_+)}
(1-y)^{\frac{\Delta_+}{2}} F(y),
\end{align}
where the surface gravity $\hat{\kappa}_+$ and angular velocity
$\hat{\Omega}_+$ are defined in \eqref{eqn:BTZ_physical_quantities}.

Not by accident, the prefactors in this redefinition capture the outgoing
behaviour at the horizon - alternatively, one could have chosen the ingoing
behaviour \eqref{BChorizon} - as well as the leading asymptotic decay in
\eqref{FGasympExp}. This redefinition is convenient because it transforms the
Klein-Gordon equation \eqref{eqn:BTZ_KG_equation} into the hypergeometric
equation:
\begin{align}
y(1-y)F''(y) + [c- y(a+b+1)]F'(y) - ab F(y) =0, 
\end{align}
where the parameters $a,b,c$ are given by 
\begin{align}\label{eqn:BTZ_abc_parameter}
a &= \frac{1}{2}\left(1+ \sqrt{1+\hat{\mu}^2} + \frac{i(\hat{\omega}+m)}{y_+ + y_-}\right),  \nonumber\\
 b &= \frac{1}{2}\left(1+ \sqrt{1+\hat{\mu}^2} + \frac{i(\hat{\omega} - m)}{y_+ - y_-}\right), \\
c &= 1 + \frac{\hat{\omega}- m \, \hat{\Omega}_+}{\hat{\kappa}_+}. \nonumber
\end{align}
Near the event horizon, the most general solution is a linear combination,
$\psi = \mathcal{A}\, \psi_{\rm in,+} + \mathcal{B}\, \psi_{\rm out,+}$, of the
two linearly independent solutions~\cite{abramowitz1968handbook}:
\footnote{Recall that we are assuming $-1 < \hat{\mu}^2 < 0$ and $m \geq 0$.
Since $c-a-b = -\sqrt{1+\hat{\mu}^2}$ and $a-b = - i (\omega- \Omega_-)/\kappa_-$,
these quantities are generally not integers. For other specific parameter choices,
one must verify this explicitly~\cite{abramowitz1968handbook}.}
\begin{align}\label{HorizonF}
     \psi_{\rm in,+} (y)&= y^{-\frac{1}{2}(c-1)}(1-y)^{\frac{1}{2}(1+a+b-c)}{}_2F_1(1+a-c,1+b-c,2-c;y), \\
     \psi_{\rm out,+} (y)&= y^{\frac{1}{2}(c-1)}(1-y)^{\frac{1}{2}(1+a+b-c)}{}_2F_1(a,b,c;y), \nonumber
\end{align}
where ${}_2F_1(a,b,c;y)$ is the hypergeometric function.
To impose the ingoing horizon BC, we first expand both linearly independent solutions near $y=0$ (\ie $r = r_+$). Using the property ${}_2F_1(\rho,\sigma,\gamma;0) =1$ (valid for arbitrary parameters with non-integer $\gamma$) near $y = 0$ we have 
\begin{align}
 \psi_{\rm in,+}\big|_{y = 0} &\sim y^{-i\frac{\hat{\omega}- m \, \hat{\Omega}_+}{2\hat{\kappa}_+}}\tilde{\psi}_{\rm in,+}(\hat{\omega}, m, y), \\
     \psi_{\rm out,+}\big|_{y = 0} &\sim y^{i\frac{\hat{\omega}- m \, \hat{\Omega}_+}{2\hat{\kappa}_+}}\tilde{\psi}_{\rm out,+}(\hat{\omega}, m, y),
\end{align}
where $\tilde\psi_{\rm in,+}(\hat{\omega}, m, y)$ and $\tilde\psi_{\rm out,+}(\hat{\omega}, m, y)$ are both analytic at $y = 0$, with  $\tilde\psi_{\rm in,+}(\hat{\omega}, m, 0) = \tilde\psi_{\rm out,+}(\hat{\omega}, m, 0)\equiv 1$. Converting the field $\psi$ to ingoing Eddington-Finkelstein coordinates $(v,r,\phi')$ $-$ as defined in \eqref{ingoingEF} $-$ and substituting in \eqref{eqn:Psi_decomposition} while also recalling that $\Phi(t,r,\phi) = e^{-i\omega t}e^{i\,m\phi}\psi(r)$, we find near $y = 0$ that 
\begin{align}
  \Phi_{\rm in, +} &= e^{-i\omega (v - v_0)}e^{im(\phi'-\phi'_0)}\big[1+ \mathcal{O}(y)\big], \nonumber \\
    \Phi_{\rm out, +} &= e^{-i\omega (v - v_0)}e^{im(\phi'-\phi'_0)}y^{i\frac{\hat{\omega}- m \, \hat{\Omega}_+}{\hat{\kappa}_+}}\big [1+ \mathcal{O}(y) \big], 
\end{align}
where $v_0$ and $\phi_0'$ are real constants depending only on black hole parameters. Then, we see that $\Phi_{\rm in, +}$ is smooth at the future event horizon, while $\Phi_{\rm out, +}$ is not. Hence, we must pick the  $\psi_{\rm in,+}$ behaviour to find regular modes at the horizon, \ie we must impose the boundary condition $\mathcal{B}\equiv 0$. 

It remains to impose the double-trace boundary condition \eqref{2xTrace:BC} at $y=1$ $(r \to \infty)$.  For that, we start by applying the $y\to 1-y$ hypergeometric transformation law, namely  (15.3.6) of \cite{abramowitz1968handbook}, to rewrite $\psi_{\rm in,+}(y)$ in \eqref{HorizonF} as
\begin{multline} \label{Fasymptotic}
     \psi_{\rm in,+}(y)= y^{-\frac{1}{2}(c-1)}(1-y)^{\frac{1}{2}(1-a-b+c)}\frac{\Gamma(2-c)\Gamma(a+b-c)}{\Gamma(1+a-c)\Gamma(1+b-c)}{}_2F_1(1-a,1-b,1-a-b+c,1-y) \\ + y^{-\frac{1}{2}(c-1)}(1-y)^{\frac{1}{2}(1+a+b-c)}\frac{\Gamma(2-c)\Gamma(c-b-a)}{\Gamma(1-a)\Gamma(1-b)}{}_2F_1(1+a-c,1+b-c,1+a+b-c,1-y)  
     ,
\end{multline}
where $\Gamma(x)$ is the gamma function.
Next, recall that in the double-trace boundary condition \eqref{2xTrace:BC}, namely $\beta=\kappa\, \alpha$,  the coefficients $\alpha$ and $\beta$ are the FG amplitudes as introduced in \eqref{FGasympExp}. Therefore, to read the asymptotic FG amplitudes $\alpha$ and $\beta$ of \eqref{Fasymptotic} one needs to first use \eqref{eqn:BTZ_rtozFGcoordinate} and \eqref{def:y} to find the FG transformation $y(z)=1-\left( y_+^2 - y_-^2 \right)z^2 + \frac{1}{2}\left( y_+^2 - y_-^2 \right)^2 z^4 + \mathcal{O}\left( z^6 \right)$. Under this transformation, the asymptotic decay of \eqref{Fasymptotic} about $z=0$ is: 
\begin{align}\label{FGdecay}
     \psi_{\rm in,+}\big|_{z\to 0} \simeq &\:
   \frac{\Gamma(2-c)\Gamma(a+b-c)}{\Gamma(1+a-c)\Gamma(1+b-c)}(y_+^2-y_-^2)^{\frac{1}{2}\,\Delta_-} z^{\Delta_-}\Big[1+\mathcal{O}(z^2)\Big]  \nonumber  \\
      &  +\frac{\Gamma(2-c)\Gamma(c-b-a)}{\Gamma(1-a)\Gamma(1-b)} (y_+^2-y_-^2)^{\frac{1}{2}\,\Delta_+} z^{\Delta_+}\Big[1+\mathcal{O}(z^2)\Big], 
\end{align}
where the conformal dimensions $\Delta_{\pm}= 1 \pm \sqrt{1 + \hat{\mu}^2} $  were introduced in \eqref{FGasympExp}. Comparing the FG decay \eqref{FGdecay} with \eqref{FGasympExp} immediately yields the FG amplitudes for our system:
\begin{align} \label{def:alpha-beta}
    \alpha &= \frac{\Gamma(2-c)\Gamma(a+b-c)}{\Gamma(1+a-c)\Gamma(1+b-c)} (y_+^2-y_-^2)^{\frac{1}{2}\,\Delta_-}, \\
    \beta &= \frac{\Gamma(2-c)\Gamma(c-b-a)}{\Gamma(1-a)\Gamma(1-b)}  (y_+^2-y_-^2)^{\frac{1}{2}\,\Delta_+},
\end{align}
where $a,b,c$ are defined in \eqref{eqn:BTZ_abc_parameter}. These relations, when inserted into the double-trace boundary condition  $\beta=\kappa\, \alpha$, give the condition that quantizes the double-trace frequencies, namely:
\begin{subequations}
\begin{equation}
\label{eqn:BTZ_Gamma_criterion}
     \frac{\Gamma  \left(-\sqrt{1+\hat{\mu}^2}\right)}{\Gamma \left(\Delta^+_- \right) \Gamma \left(\Delta_-^-\right)}\\ 
     -\frac{\kappa}{(y_+^2-y_-^2)^{\sqrt{1+\hat{\mu}^2}}}\frac{\Gamma \left(\sqrt{1+\hat{\mu}^2}\right)}{\Gamma \left(\Delta_+^+\right) \Gamma \left(\Delta_+^-\right)}=0,
\end{equation}
with
\begin{equation}
\Delta_-^{\pm}\equiv \frac{1}{2}\left[\Delta_-\pm\frac{i (m\mp\hat{\omega})}{y_+ - y_-}\right]\quad\text{and}\quad \Delta_+^{\pm}\equiv \frac{1}{2}\left[\Delta_+\pm\frac{i (m\mp\hat{\omega})}{y_+ - y_-}\right]\,,
\end{equation}
\end{subequations}
which is the main result of this subsection and is valid within the scalar mass window \eqref{2xTrace:rangeMass}. As a first outcome of our analysis, note that imposing $\alpha = 0$ in \eqref{def:alpha-beta} $-$ equivalently, $\kappa = 0$ in \eqref{eqn:BTZ_Gamma_criterion} $-$ yields the well-known BTZ frequencies for Dirichlet boundary conditions. Similarly, imposing $\beta = 0$ in \eqref{def:alpha-beta} $-$ equivalently, taking $\kappa \to \infty$ in \eqref{eqn:BTZ_Gamma_criterion} $-$ recovers the well-known BTZ frequencies for Neumann boundary conditions. The frequencies are then given by
\begin{subequations} 
\begin{align}
    \label{BTZ:wDir}\hat{\omega}_{\pm}\big|_{\hbox{\tiny Dir}} = \pm m - i(y_+ \mp y_-)\left(2n + \Delta_+ \right)\,, \\
     \label{BTZ:wNeu}\hat{\omega}_{\pm}\big|_{\hbox{\tiny Neu}}  = \pm m - i(y_+ \mp y_-)\left(2n + \Delta_- \right)\,,
\end{align}
\end{subequations}
where $\hat{\omega}_{\pm}$ applies to co-rotating/counter-rotating modes, respectively (recall that $m\geq 0$) and $n$ is the radial overtone, \ie a non-negative integer that gives the number of radial zeroes of the mode function $\psi$. These Dirichlet/Neumann quasinormal mode frequencies agree with those obtained in \cite{Birmingham:2001hc,Dappiaggi:2017pbe,Dias:2019ery}. Note that \eqref{BTZ:wDir} holds for any $\hat{\mu}^2>-1$, whereas normalizable Neumann modes satisfying \eqref{BTZ:wNeu} exist only for $-1<\hat{\mu}^2<0$. An analysis of \eqref{BTZ:wDir}–\eqref{BTZ:wNeu} shows that BTZ is always stable (\ie ${\rm Im}\,\hat{\omega}<0$) for any $y_+$ and $0<y_-\leq y_+$. For generic double-trace BCs (\ie finite $\kappa\neq 0$), the gamma-function condition \eqref{eqn:BTZ_Gamma_criterion} cannot be solved analytically, but it can be treated numerically, as we do in Sec.~\ref{sec:BTZresults}. There we find that BTZ can become unstable for double-trace BCs when $\kappa$ and/or $y_{\pm}$ take particular values.

Before closing this section, we briefly analyse the mode frequencies in the absence of a BTZ black hole, \ie the frequencies that propagate in global AdS$_3$. Because of the mass gap between BTZ and AdS$_3$, this limit is not obtained by sending $y_{\pm}\to 0$ in \eqref{eqn:BTZ_Gamma_criterion}; that limit is not finite. Instead, one must redo the analysis from scratch, the only change being that the inner boundary condition now requires regularity (\ie finiteness) of the AdS$_3$ modes at the origin $r=0$. Imposing the double-trace boundary condition $\beta - \kappa\,\alpha = 0$ then yields the quantization condition for the global AdS$_3$ frequencies: 
\begin{multline}
\label{eqn:AdS3_Gamma_criterion}\frac{\Gamma  \left(-\sqrt{1+\hat{\mu}^2}\right)}{\Gamma \left(\frac{1}{2} \left(\Delta_-+ m - \hat{\omega}^{\hbox{\tiny AdS}}\right)\right) \Gamma \left(\frac{1}{2} \left(\Delta_-+ m +\hat{\omega}^{\hbox{\tiny AdS}}\right)\right)} \\
   - \frac{\kappa \, \Gamma \left(\sqrt{1+\hat{\mu}^2}\right)}{\Gamma \left(\frac{1}{2} \left(\Delta_++ m - \hat{\omega}^{\hbox{\tiny AdS}}\right)\right) \Gamma \left(\frac{1}{2} \left(\Delta_++m+\hat{\omega}^{\hbox{\tiny AdS}}\right)\right)}=0.
\end{multline}
As before, this condition can be solved analytically only in the Dirichlet limit $\kappa=0$ and in the Neumann limit $\kappa\to\infty$; for generic double-trace coefficients ($-\infty<\kappa<\infty$) it must be solved numerically, which we do in Sec.~\ref{sec:AdS3w}. In these two special cases, the global AdS$_3$ frequencies are, respectively (with $n$ the radial overtone),
\begin{subequations}\label{AdS3:wDirichletNeumann}
\begin{align}
    \label{AdS3-wDir}\hat{\omega}_{\pm}^{\hbox{\tiny AdS}}\big|_{\hbox{\tiny Dir}} = \pm \left(m +2n + \Delta_+ \right), \\
     \label{AdS3-wNeu}\hat{\omega}_{\pm}^{\hbox{\tiny AdS}}\big|_{\hbox{\tiny Neu}}  = \pm \left(m +2n + \Delta_- \right).
\end{align}
\end{subequations}  
which agrees with the values first found in \cite{Burgess:1984ti}.\footnote{Recall that $m\ge 0$ and that the symmetries of the system imply that the normal-mode frequencies of global AdS always come in trivial pairs $\{\omega,-\omega^*\}$ (the first are co-rotating and the second counter-rotating modes, but in AdS there is no distinction between them). Hence, we may focus only on the $+$ solutions of \eqref{AdS3:wDirichletNeumann}.}  Note that \eqref{AdS3-wDir} holds for any $\hat{\mu}^2>-1$, while normalizable modes with Neumann BC satisfying \eqref{AdS3-wNeu} exist only for $-1<\hat{\mu}^2<0$. Analysing \eqref{AdS3-wDir}–\eqref{AdS3-wNeu} shows that global AdS$_3$ is always stable for any allowed value of $\hat{\mu}$, since ${\rm Im}\,\hat{\omega}=0$ (for either Dirichlet or Neumann boundary conditions). In Sec.~\ref{sec:Ad3} we will see that this need not hold for certain double-trace perturbations.

As noted above, for generic double-trace coefficients (\ie finite $\kappa\neq 0$), the BTZ frequency condition \eqref{eqn:BTZ_Gamma_criterion} and the AdS$_3$ condition \eqref{eqn:AdS3_Gamma_criterion} can only be solved numerically. This can be done using, for example, the \verb|FindRoot| routine in {\it Mathematica}. Since the $\kappa=0$ (Dirichlet) solutions \eqref{BTZ:wDir} and \eqref{AdS3-wDir} are known analytically, we start from the $\kappa=0$ mode (for BTZ with fixed $y_{\pm}$) and slowly march to smaller or larger values of $\kappa$ to obtain the eigenfrequencies for the desired double-trace boundary condition. We carry out this analysis and present the results in Sec.~\ref{sec:BTZonset1}. The conditions \eqref{eqn:BTZ_Gamma_criterion} and \eqref{eqn:AdS3_Gamma_criterion} will be especially useful for mapping the regions of the two-dimensional BTZ parameter space and double-trace parameter in which both (AdS$_3$ and BTZ) instabilities occur, where neither occurs, or where only one of them is present. This is done in Sec.~\ref{sec:BTZonset2} (see Fig.~\ref{fig:3D_kappa_onset} below).

\subsection{Method 2: Full numerical solution of the eigenvalue problem}\label{sec:Numerical-Method}
An alternative way to obtain the linear-mode frequencies in BTZ or AdS$_3$ with double-trace boundary conditions is to use a fully numerical approach. One may employ the {\tt Eigensystem}/{\tt Eigenvalue} routines in {\it Mathematica}, or a Newton–Raphson algorithm combined with a pseudo-spectral discretization of the grid. These methods exhibit excellent convergence and accuracy in similar problems in the literature (see, \eg \cite{Dias:2010eu,Dias:2010ma,Dias:2010maa,Dias:2010gk,Dias:2011jg,Dias:2013sdc,Cardoso:2013pza,Dias:2022mde,Davey:2023fin,Davey:2024xvd}). A review of such numerical techniques can be found in \cite{Dias:2015nua}.

To solve numerically the eigenvalue problem defined by the Klein-Gordon ODE \eqref{eqn:BTZ_KG_equation} with horizon \eqref{BChorizon} and double-trace boundary conditions \eqref{2xTrace:BC}, we first perform a field redefinition of the wavefunction $\psi$ to factor out the leading decays at the horizon and at the asymptotic boundary, as given in \eqref{BChorizon}–\eqref{FGasympExp}. Namely, one redefines the scalar mode function as 
\begin{align}\label{eqn:scalar_peeledoff_factors}
    \Phi(r) = \left(1 - \frac{r_+}{r} \right)^{-\frac{i}{2\kappa_+}(\omega-m\,\Omega_+)}r^{-\Delta_-}Q(r),
\end{align}
so that the ingoing horizon condition is automatically satisfied by seeking smooth wavefunctions $Q(r)$ that approach a constant at infinity.

One still needs to translate the double-trace boundary condition \eqref{FGasympExp} (or, in the FG context, \eqref{2xTrace:BC}) into a condition for the redefined mode function $Q(r)$, which depends on the scalar mass. For numerical purposes, it is also convenient to introduce a compact radial coordinate so that the boundary condition for $Q$ reduces to a simple Robin condition; this choice of coordinate likewise depends on $\mu L$. Rather than discussing the general case for any $\mu L$ in the range \eqref{2xTrace:rangeMass}, we will henceforth take $\mu^2 L^2 = -\frac{15}{16}$.
\footnote{For $-1<\mu^2L^2<-3/4$, one can define $\rho = 1 - (r_+/r)^{\Delta_+-\Delta_-}$ to obtain integer powers in the expansion of $Q(\rho)$ at $\rho = 1$, so that $\Phi|_{\rho\to 1} = y_+^{\Delta_-} (1-\rho)^{\frac{\Delta_-}{\Delta_+-\Delta_-}} Q|_{\rho\to 1}$ with $Q|_{\rho\to 1} = \alpha + \frac{\beta}{y_+^{\Delta_-}} (1-\rho) + \dots$. For example, if $\mu^2L^2=-8/9$, then $\rho = 1 - (r_+/r)^{2/3}$ and $\Phi|_{\rho\to 1} = y_+^{-2/3} (1-\rho) Q|_{\rho\to 1}$ with $Q|_{\rho\to 1} = \alpha + \frac{\beta}{y_+^{2/3}} (1-\rho) + \dots$. For $\mu^2L^2=-3/4$, after $\rho = 1 - r_+/r$, the first term in \eqref{eqn:scalar_peeledoff_factors} modifies the $\Phi|_{\rho\to 1}$ expansion at order $(1-\rho)$, so the boundary condition becomes $Q^\prime(1) + \frac{\kappa}{\sqrt{y_+}} Q(1) - \frac{i(\omega - m \Omega_+)}{2\kappa_+} Q(1) = 0$. For $-3/4 < \mu^2L^2 < 0$, the same coordinate change makes the first term in \eqref{eqn:scalar_peeledoff_factors} non-smooth near $\rho=1$, requiring either a Gamma-function computation or a different setup. Hence, this approach is not fully general.} 

For $\mu^2L^2=-\frac{15}{16}$, and thus $\{ \Delta_-,\Delta_+\}=\{\frac{3}{4},\frac{5}{4}\}$, we introduce the field redefinition \eqref{eqn:scalar_peeledoff_factors} and the compact radial coordinate 
\begin{align}\label{def:rho}
    \rho =  1- \sqrt{\frac{r_+}{r}}
\end{align}
so that  $0 \leq \rho \leq 1$ with the horizon being at $\rho=0$ and the AdS$_3$ boundary at $\rho=1$. From \eqref{eqn:BTZ_rtozFGcoordinate} and \eqref{def:rho}, we find that the relation between $\rho$ and the FG coordinate $z$ is 
\begin{align}\label{eqn:BTZ_rtozFGcoordinateRHO}
    \rho = 1- \sqrt{y_+}\,z^{1/2}+\frac{1}{8}\sqrt{y_+}\left( y_+^2 +y_-^2 \right) z^{5/2}+\mathcal{O}\left(z^{9/2}\right). 
\end{align}
It follows that the asymptotic decay \eqref{FGasympExp} can be written as 
\begin{align}
    \Phi\big|_{\rho \to 1} = y_+^{-3/4}(1-\rho)^{3/2} Q\big|_{\rho \to 1} \quad \hbox{with} \quad Q\big|_{\rho \to 1}= \alpha + \frac{\beta}{\sqrt{y_+}} (1-\rho) + \dots.
\end{align}
where we have used \eqref{eqn:scalar_peeledoff_factors} and \eqref{def:rho}.
Hence, in the context of \eqref{FGasympExp}, the double-trace boundary condition \eqref{2xTrace:BC} for $\Phi$ becomes a Robin boundary condition for $Q$ at the asymptotic boundary $\rho=1$:
\begin{align}\label{eqn:scalar_mixedbcs_final}
   Q^\prime(1) + \frac{\kappa}{\sqrt{y_+}} Q(1)=0.
\end{align}
At the horizon ($\rho=0$), as discussed above, the field redefinition \eqref{eqn:scalar_peeledoff_factors} already ensures that smooth wavefunctions $Q$ will only capture ingoing modes at the horizon. That is to say, at $\rho=0$, we impose a derived boundary condition that is simply the equation of motion for $Q$ evaluated at the horizon.  

The quasinormal modes can be computed using one of two methods. The first employs Mathematica’s built-in {\tt Eigensystem} routine with a pseudo-spectral discretization; details can be found in \cite{Dias:2010eu,Dias:2010maa,Dias:2011jg,Dias:2015nua}. This approach yields many modes simultaneously, allowing for a comprehensive determination of the spectrum. The second method applies a Newton–Raphson root-finding algorithm, also with a pseudo-spectral grid, as described in \cite{Dias:2013sdc,Cardoso:2013pza,Dias:2015nua}. While it computes one family of modes at a time and requires a seed close to the exact solution, it is faster and more efficient for high-resolution grids or extreme regions of parameter space.

\section{Instability of global \texorpdfstring{AdS$_3$}{AdS3} to double-trace perturbations}\label{sec:Ad3}
Our main goal is to study the stability of BTZ black holes under scalar perturbations with double-trace boundary conditions. Before addressing this, however, it is important to ask whether global AdS$_3$ itself is stable under scalar perturbations, particularly with double-trace boundary conditions. This question has been studied previously, so this section will necessarily reproduce known results. At the same time, we will highlight properties of the AdS$_3$ instability that may not be clearly identified in earlier studies and that are relevant for the BTZ analysis.

Global AdS$_3$ is known to be stable against scalar modes satisfying Dirichlet or Neumann boundary conditions at the asymptotic boundary. This was first established in \cite{Burgess:1984ti}, and the expression \eqref{AdS3:wDirichletNeumann} for the normal mode frequencies confirms that they are purely real for these boundary conditions.
For scalar fields with masses in the range \eqref{2xTrace:rangeMass}, the case of double-trace boundary conditions has also been addressed. In \cite{Ishibashi:2004wx}, the authors showed, without explicitly computing the (quasi)normal mode frequencies, that global AdS$_3$ can become unstable when the double-trace parameter $\kappa$ (defined in \eqref{FGasympExp}-\eqref{2xTrace:BC}) falls below a critical negative value. More generally, they proved that global AdS$_d$ can be unstable to double-trace perturbations for any dimension $d\geq 3$.

In the remainder of this section, we present the relevant physical properties of the unstable modes (subsection~\ref{sec:AdS3w}) and explicitly verify that our analysis quantitatively agrees with \cite{Ishibashi:2004wx} (subsection~\ref{sec:AdS3wald}).

\subsection{Global \texorpdfstring{AdS$_3$}{AdS3} frequency spectrum}\label{sec:AdS3w}
We find that the \emph{qualitative} behaviour of the modes is independent of the scalar field mass $\hat{\mu}\equiv\mu L$. For concreteness, we therefore focus on a representative mass $\mu^2 L^2 = -15/16$ to illustrate the properties of scalar fields in the range \eqref{2xTrace:rangeMass}, $-1 < \mu^2 L^2 < 0$, for which the double-trace boundary condition \eqref{2xTrace:BC} is allowed. In Appendix~\ref{sec:AppOtherMasses}, we also present results for $\mu^2L^2 = \{-8/9, -3/4, -1/10\}$.

Recall that the double-trace boundary condition, $\beta = \kappa\, \alpha$ with $\kappa \in \mathbb{R}$, includes Neumann and Dirichlet boundary conditions as the special limits $\kappa = 0$ and $\kappa \to \pm \infty$, respectively.

In AdS$_3$, modes with azimuthal number $m=0$ and $m \ge 1$ exhibit similar qualitative behaviour. This similarity does not hold for the BTZ black hole. Therefore, for comparison with BTZ, we will typically present results for $m=0$ and $m=1$ (the latter being representative of $m>0$), while occasionally showing higher-$m$ results when they are relevant to illustrate specific physical properties.

\begin{figure}[ht]
    \centering
    \includegraphics[width=0.49\linewidth]{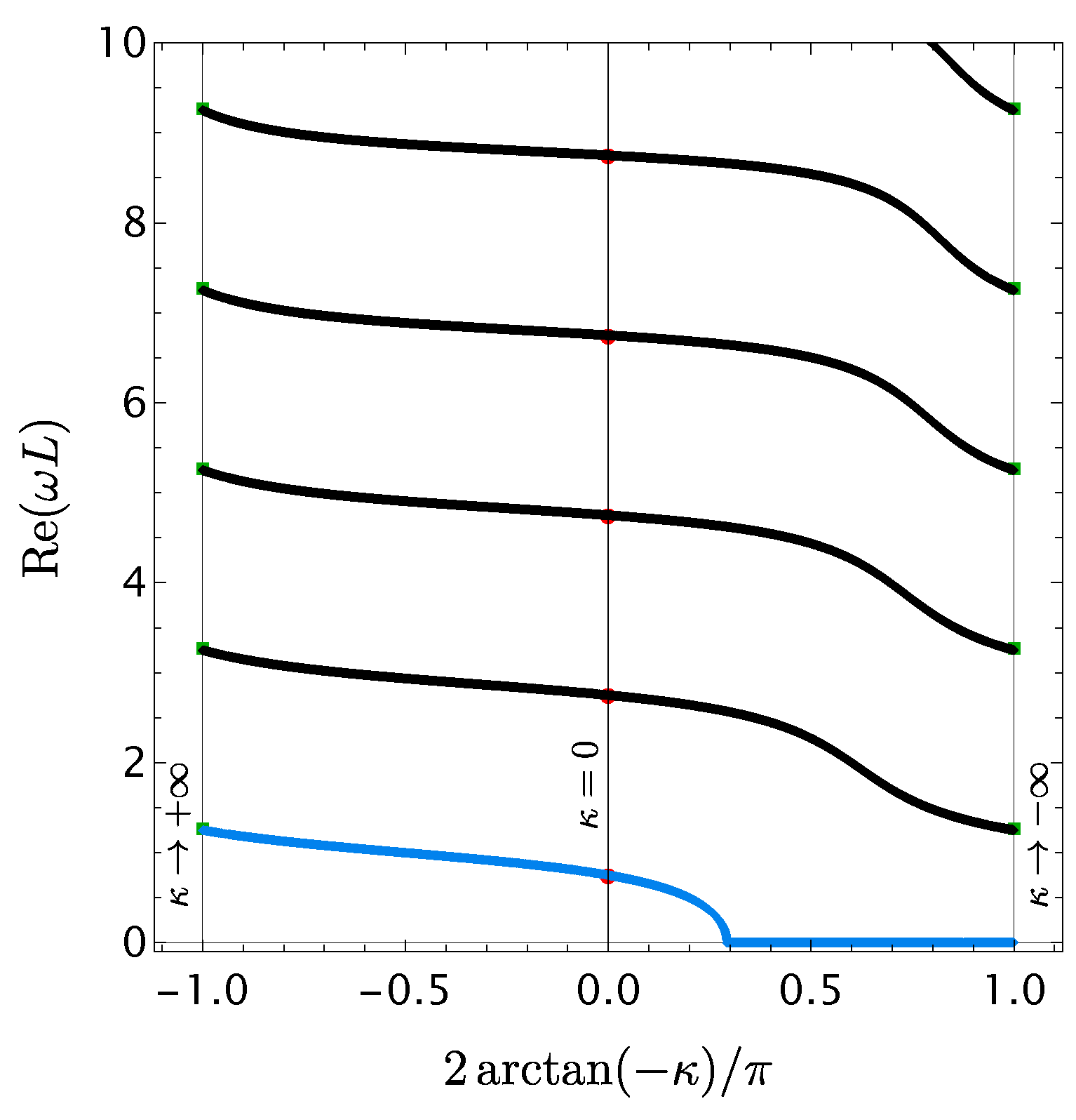}\\
  \includegraphics[width=0.48\linewidth]{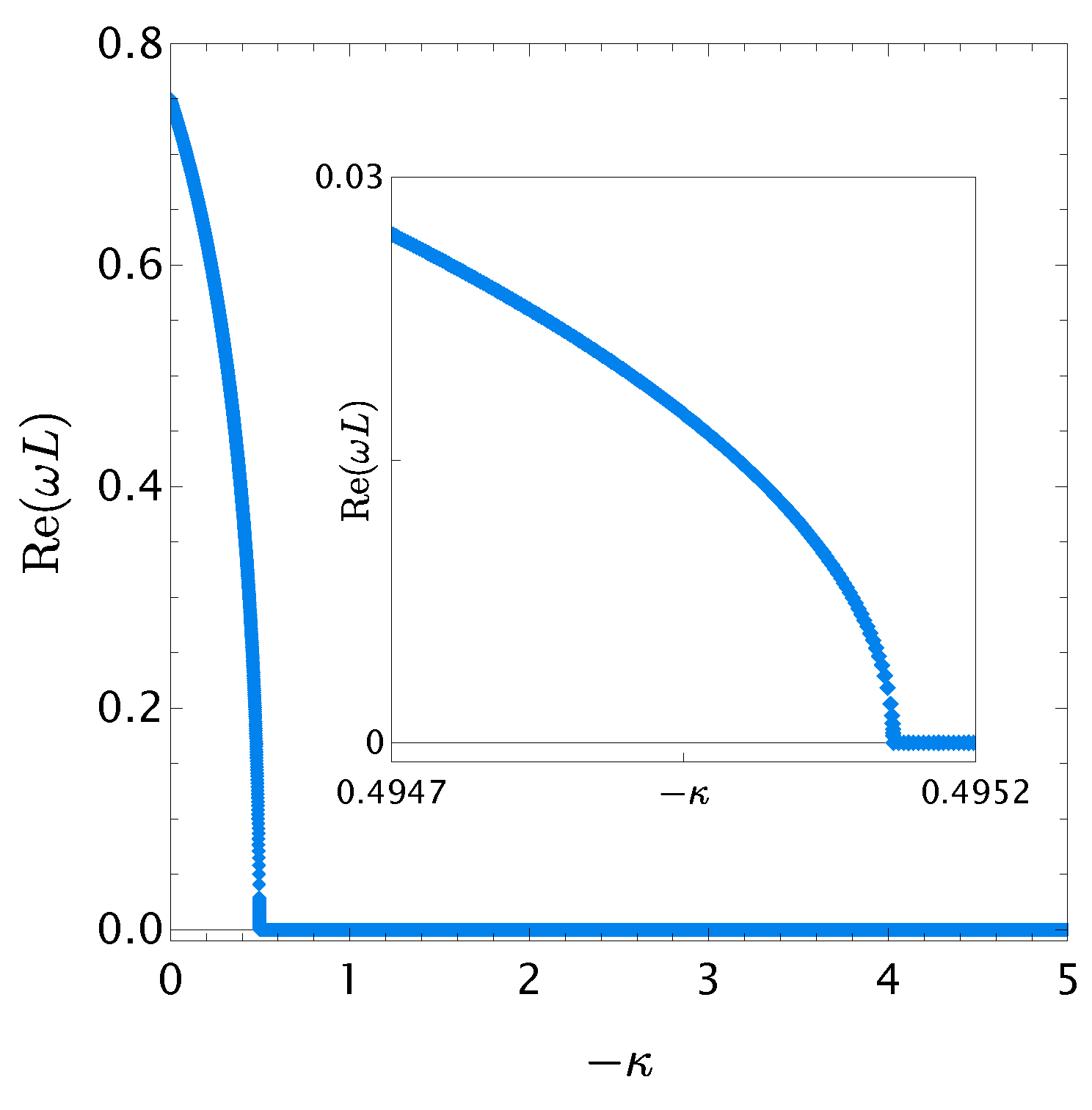}
        \hspace{0.2 cm}
     \includegraphics[width=0.48\linewidth]{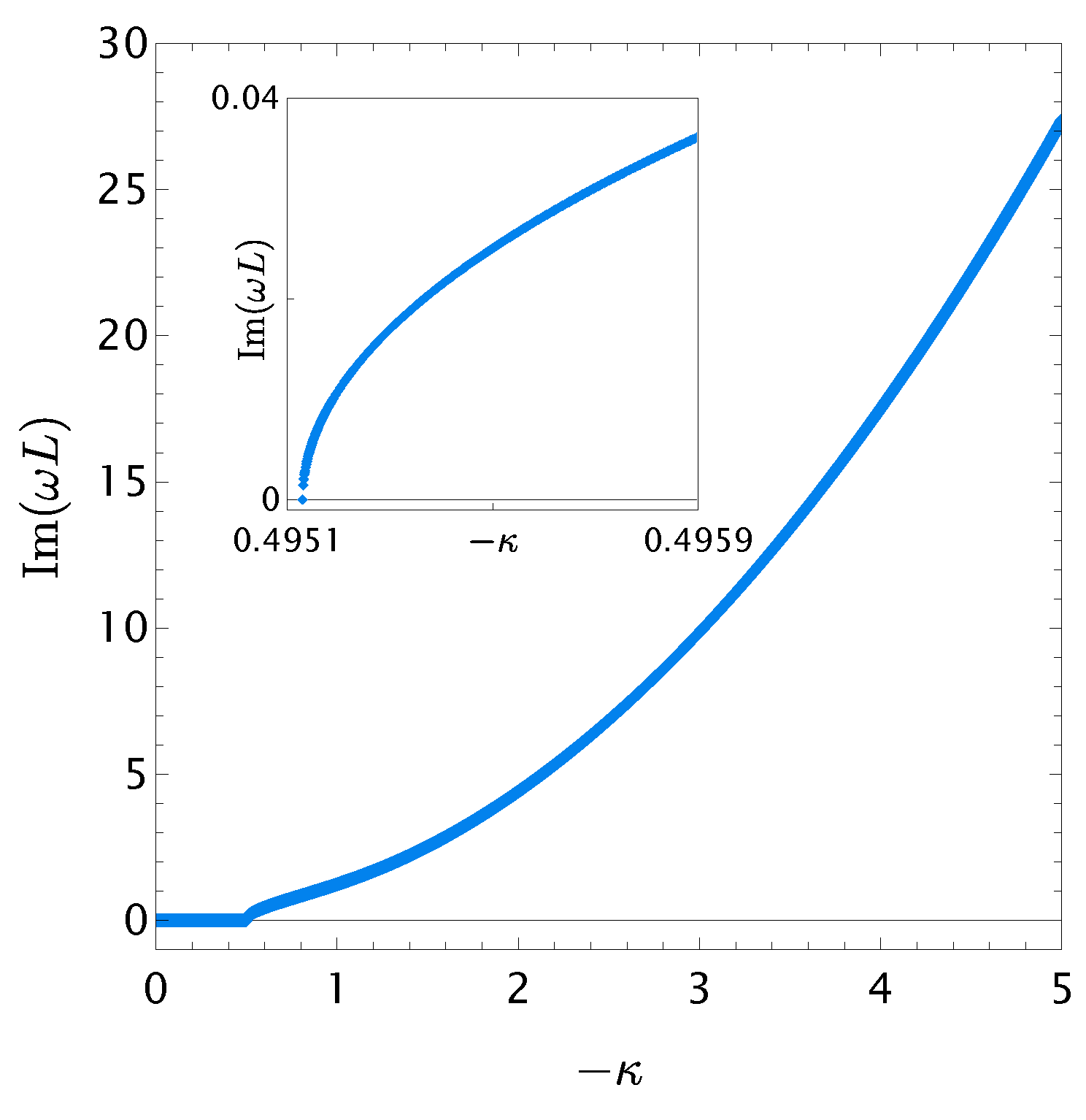}
    \caption{\textbf{Top panel:} Real part of the frequency, $\text{Re}(\omega L)$, as a function of $\theta = \frac{2}{\pi}\arctan(-\kappa)$ for scalar modes of \textbf{global AdS$_3$} with $\mu^2 L^2 = -15/16$ and $m=0$, for the first five radial overtones $n=0,1,2,3,4$ (bottom to top curves). All overtones start at the Dirichlet green square on the left ($\kappa=+\infty$, i.e., $\theta=-1$), connect continuously to the Neumann red dot ($\kappa=0$), and for $n\ge 1$ (black curves) end at the Dirichlet green square at $\kappa\to -\infty$ ($\theta=+1$). These overtones have ${\rm Im}\,\hat{\omega}^{\rm AdS}_{n\ge1} = 0$ for all $\kappa$. (Frequencies $-\hat{\omega}$ are also solutions, not shown.)  \textbf{Bottom panel:} Detailed view of $n=0$ (blue curve, also in the top panel). Left: $\text{Re}(\omega L)$; right: $\text{Im}(\omega L)$, both as functions of $-\kappa$ (restricted to $\kappa \in \mathbb{R}^-$). For $0 \ge \kappa \gtrsim -0.49$, $\text{Re}(\omega L) >0$ and $\text{Im}(\omega L) =0$. For $\kappa \lesssim -0.49$, $\text{Re}(\omega L) =0$ and $\text{Im}(\omega L) >0$, indicating that this AdS$_3$ mode becomes unstable, with $\underset{{\kappa \to -\infty}}{\lim} {\rm Im}(\omega L) \to +\infty$.
}
    \label{fig:AdS3m0-wk}
\end{figure}

We analyse how the stability of global AdS$_3$ evolves as the double-trace parameter $\kappa$ is varied. Starting from $\kappa \to +\infty$ (Dirichlet BC), we continuously decrease $\kappa$ through $0$ (Neumann BC) and into negative values, down to $\kappa \to -\infty$, which again corresponds to Dirichlet BC. AdS$_3$ is stable for Dirichlet and Neumann boundary conditions \cite{Burgess:1984ti}, as confirmed by \eqref{AdS3:wDirichletNeumann}, but becomes unstable below a critical negative $\kappa$ \cite{Ishibashi:2004wx}. This raises the question: how can the solution be stable in the Dirichlet and Neumann limits, yet unstable for sufficiently negative $\kappa$?

To address this question, we consider the $m=0$ case. To represent the infinite range $-\infty < \kappa < +\infty$ compactly, we introduce the auxiliary parameter
\begin{equation}\label{def:theta}
\theta= \frac{2}{\pi}\arctan (-\kappa)
\qquad \Rightarrow
\begin{cases}
 \theta= -1  \quad & \longleftrightarrow \quad  \kappa\to +\infty \,,\\
\theta  = 0 \quad   & \longleftrightarrow \quad  \kappa =0  \,,\\
\theta = +1  \quad  & \longleftrightarrow \quad  \kappa\to -\infty \,.
\end{cases}
 \end{equation} 
In the top panel of Fig.~\ref{fig:AdS3m0-wk}, we plot the real part of the frequency, ${\rm Re}(\omega L)$, as a function of the double-trace parameter $\theta(\kappa)$ for the first five radial overtones, $n=0,1,2,3,4$; overtones with $n\ge5$ are qualitatively similar but with higher ${\rm Re}(\omega L)$. The fundamental mode ($n=0$) is shown with blue points, while $n=1,2,3,4$ are black. The horizontal axis runs from $\kappa = +\infty$ ($\theta=-1$) on the left, decreasing through $\kappa=0$ ($\theta=0$) to $\kappa \to -\infty$ ($\theta=+1$) on the right (in practice, $+10^3 > \kappa > -10^3$). Green squares at $\kappa = \pm \infty$ ($\theta = \mp 1$) indicate Dirichlet frequencies, $\hat{\omega}^{\rm AdS}_n |_{\rm Dir}$, from \eqref{AdS3-wDir}, while red circles at $\kappa=0$ ($\theta=0$) show Neumann frequencies, $\hat{\omega}^{\rm AdS}_n |_{\rm Neu}$, from \eqref{AdS3-wNeu}. Using $\kappa=0$ to label overtones, the curves naturally split into $n=0$ and $n\ge1$.  

For $n\ge1$, each double-trace curve $\hat{\omega}^{\rm AdS}_n$ connects the $n^{\rm th}$ Neumann frequency at $\kappa=0$ with the $n^{\rm th}$ Dirichlet frequency at $\kappa \to +\infty$ and the $(n-1)^{\rm th}$ Dirichlet frequency at $\kappa \to -\infty$, i.e., $\hat{\omega}^{\rm AdS}_n|_{\kappa\to +\infty} = \hat{\omega}^{\rm AdS}_n|_{\rm Dir}$, $\hat{\omega}^{\rm AdS}_n|_{\kappa=0} = \hat{\omega}^{\rm AdS}_n|_{\rm Neu}$, and $\hat{\omega}^{\rm AdS}_n|_{\kappa \to -\infty} = \hat{\omega}^{\rm AdS}_{n-1}|_{\rm Dir}$. These modes always have vanishing imaginary part, ${\rm Im}\, \hat{\omega}^{\rm AdS}_{n\ge1} = 0$.  

The $n=0$ mode (blue) behaves differently. Starting at $\hat{\omega}^{\rm AdS}_{0}|_{\rm Dir}$ for $\kappa \to +\infty$, it reaches $\hat{\omega}^{\rm AdS}_{0}|_{\rm Neu}$ at $\kappa=0$, but for $\kappa \to -\infty$, ${\rm Re}(\omega L)$ decreases to zero and does not connect to the Dirichlet value. In the bottom panel of Fig.~\ref{fig:AdS3m0-wk}, plotted as a function of $-\kappa$ for $\kappa < 0$, we see that ${\rm Re}\,\hat{\omega}^{\rm AdS}_{0}$ decreases monotonically from $\hat{\omega}^{\rm AdS}_{0}|_{\rm Neu}$ at $\kappa=0$ to zero at a critical $\kappa = \kappa^{\rm AdS}_{m,\hat{\mu}^2}<0$, which for $\mu^2L^2=-15/16$ and $m=0$ is $\kappa^{\rm AdS}_{m,\hat{\mu}^2} \sim -0.495$.  

For $\kappa > \kappa^{\rm AdS}_{m,\hat{\mu}^2}$, the $n=0$ mode is purely real, ${\rm Im}\,\hat{\omega}^{\rm AdS}_{0} = 0$, but for $\kappa < \kappa^{\rm AdS}_{m,\hat{\mu}^2}$ it becomes purely imaginary, ${\rm Im}\,\hat{\omega}^{\rm AdS}_{0} > 0$, indicating instability. As $\kappa \to -\infty$, the instability grows without bound, $\underset{{\kappa \to -\infty}}{\lim} {\rm Im}\,\hat{\omega}^{\rm AdS}_{0} \to +\infty$. Notably, $\underset{{\kappa \to -\infty}}{\lim} \hat{\omega}^{\rm AdS}_{0} \neq \hat{\omega}^{\rm AdS}_{0}|_{\rm Dir}$; instead, the first overtone matches the Dirichlet $n=0$ frequency, $\underset{{\kappa \to -\infty}}{\lim} \hat{\omega}^{\rm AdS}_1 = \hat{\omega}^{\rm AdS}_{0}|_{\rm Dir}$. This “discontinuity” in the unstable $n=0$ mode is similar to behaviours observed in AdS$_d$ with $d>3$ \cite{Ishibashi:2004wx,Katagiri:2020mvm,Kinoshita:2023iad}.

\begin{figure}[ht]
    \centering
    \includegraphics[width=0.49\linewidth]{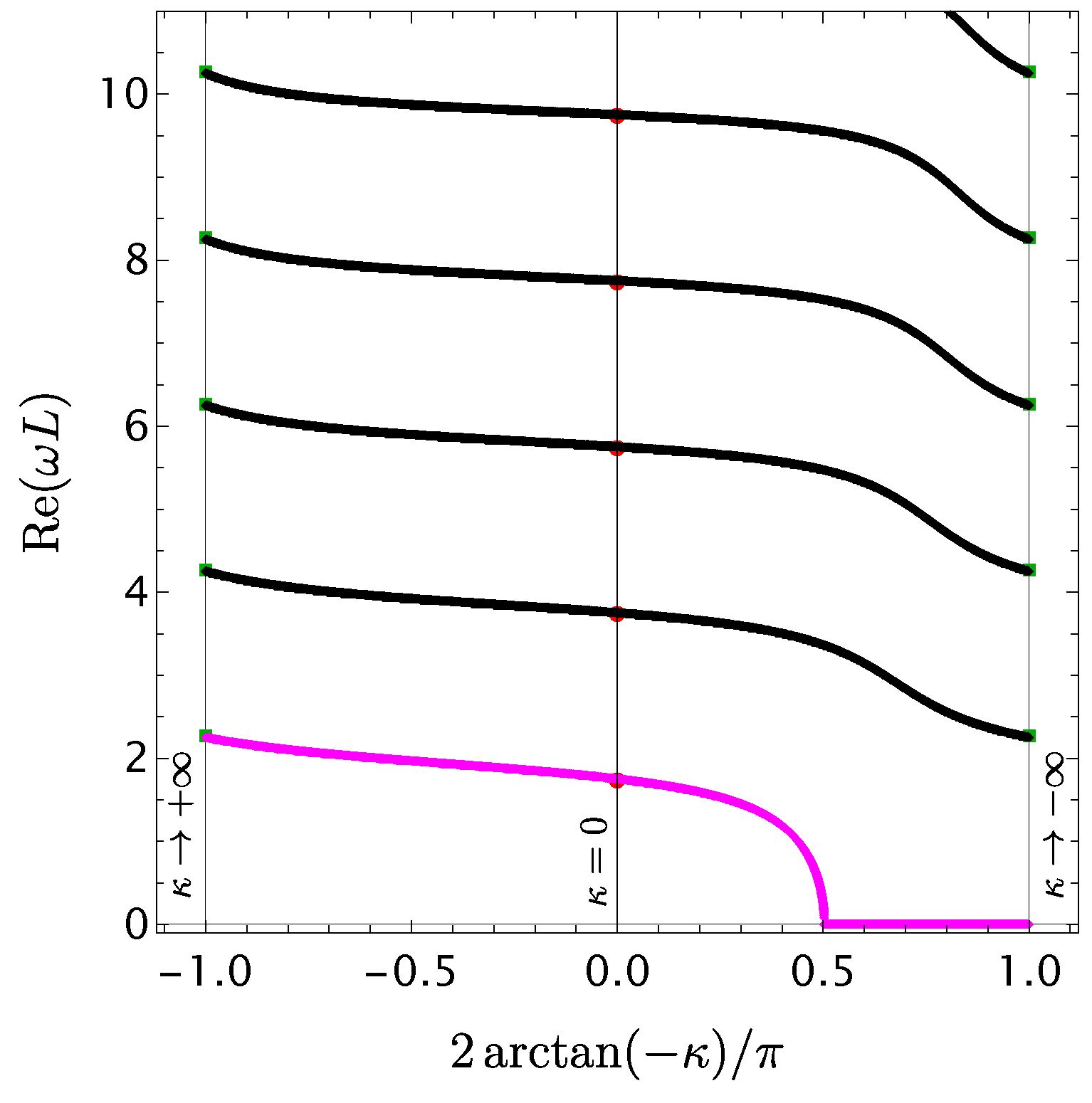}\\
     \includegraphics[width=0.48\linewidth]{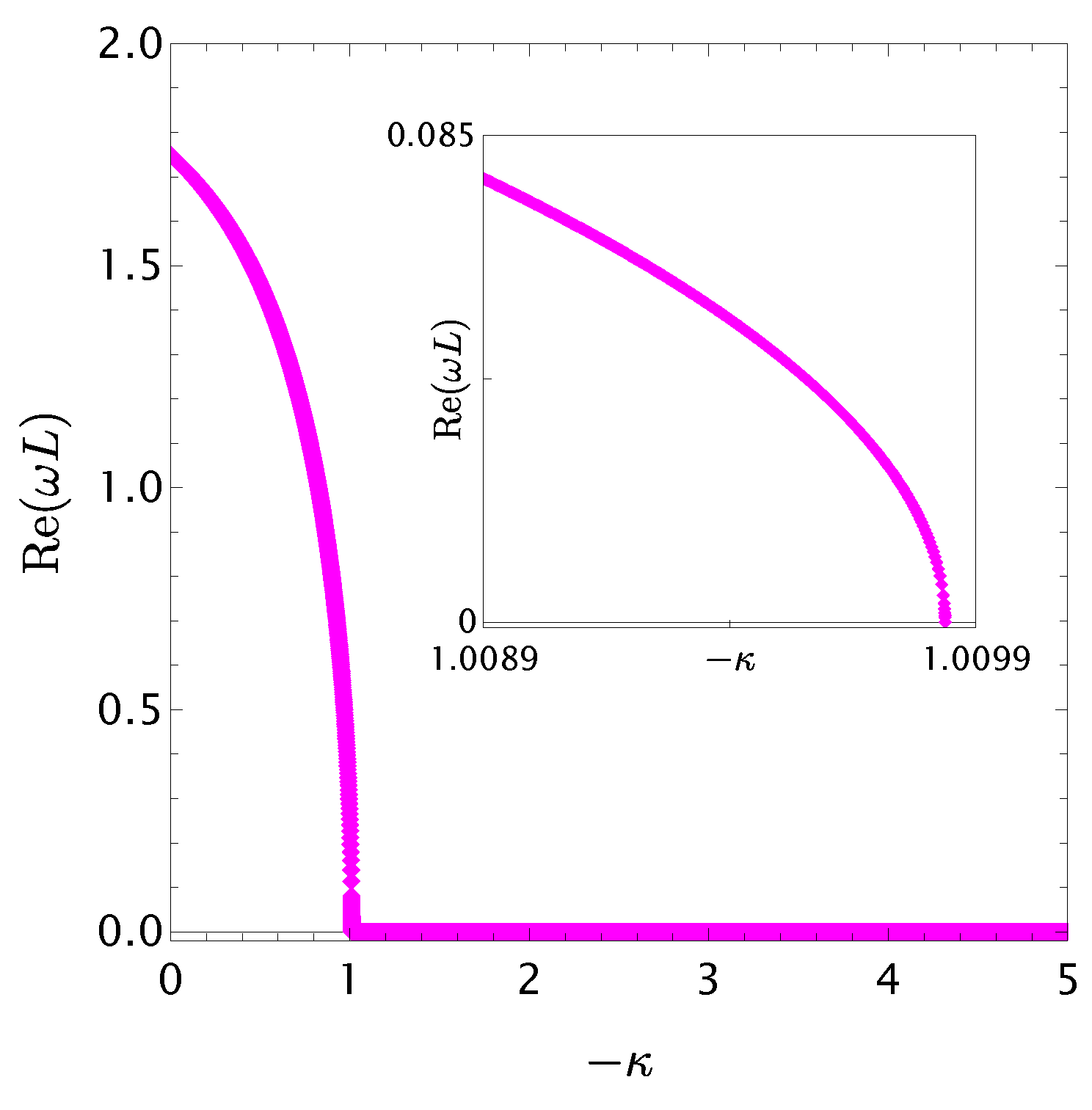}
      \hspace{0.2 cm}
     \includegraphics[width=0.48\linewidth]{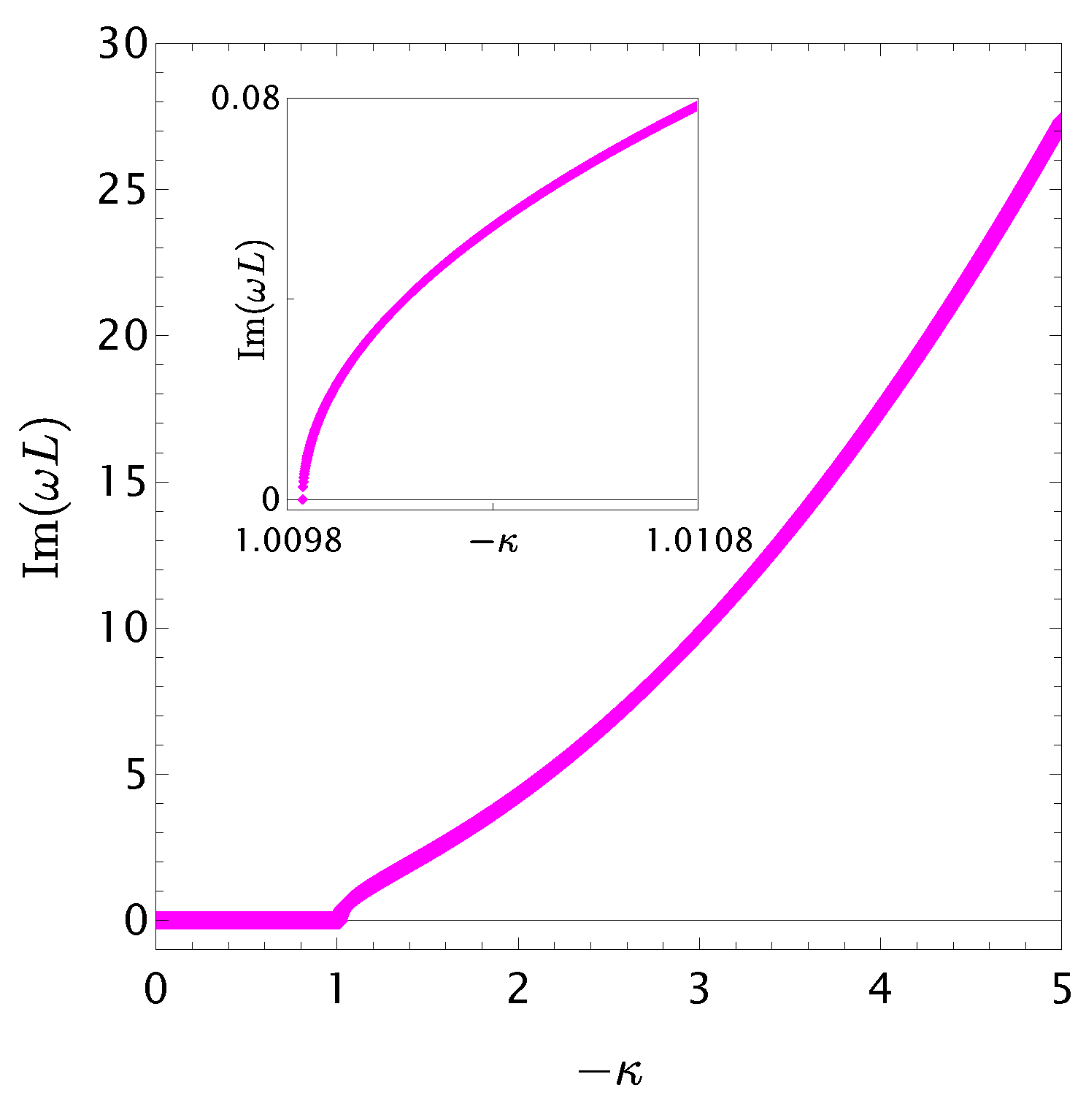}
    \caption{\textbf{Top panel:} Real part of the frequency, $\text{Re}(\omega L)$, as a function of $\theta = \frac{2}{\pi}\arctan(-\kappa)$ for scalar modes of \textbf{global AdS$_3$} with $\mu^2 L^2 = -15/16$ and $m=1$, for the first five radial overtones $n=0,1,2,3,4$ (bottom to top curves). All overtones start at the Dirichlet green square on the left ($\kappa=+\infty$, $\theta=-1$), connect continuously to the Neumann red dot ($\kappa=0$), and for $n\ge1$ (black curves, excluding the purple $n=0$ curve) end at the Dirichlet green square at $\kappa \to -\infty$ ($\theta=+1$). Overtones with $n\ge1$ have vanishing imaginary part, ${\rm Im}\, \hat{\omega}^{\rm AdS}_{n\ge1} = 0$ (frequencies $-\hat{\omega}$ are also solutions, not shown). \textbf{Bottom panel:} Detailed view of the $n=0$ mode (purple curve, also in the top panel). Left: $\text{Re}(\omega L)$; right: $\text{Im}(\omega L)$, as a function of $-\kappa$ (restricted to $\kappa \in \mathbb{R}^-$). For $0 \ge \kappa \gtrsim -1.01$, the mode is stable with $\text{Re}(\omega L) >0$ and $\text{Im}(\omega L)=0$, while for $\kappa \lesssim -1.01$ it becomes purely imaginary, $\text{Re}(\omega L) =0$, $\text{Im}(\omega L) >0$, indicating an $m=1$ instability. The instability grows as $\kappa \to -\infty$, $\underset{{\kappa \to -\infty}}{\lim} {\rm Im}(\omega L) \to +\infty$.}\label{fig:AdS3m1-wk}
\end{figure}

Consider now modes with $m\ge1$ in global AdS$_3$. Their qualitative behaviour is similar to the $m=0$ case (unlike in BTZ). This is illustrated in the top panel of Fig.~\ref{fig:AdS3m1-wk}, which shows ${\rm Re}(\omega L)$ versus $\theta(\kappa)$ \eqref{def:theta} for $m=1$ and $n=0,1,2,3$, and in the bottom panel, which plots ${\rm Re}(\omega L)$ and ${\rm Im}(\omega L)$ versus non-positive $\kappa$ for $m=1$, $n=0$.  

As for $m=0$, overtones with $n\ge1$ always have purely real frequencies. Each $n\ge1$ double-trace curve (black) connects Dirichlet$_{\rm left}$, Neumann, and Dirichlet$_{\rm right}$ points as $\theta$ runs from $-1$ ($\kappa \to +\infty$) through $0$ ($\kappa=0$) to $+1$ ($\kappa \to -\infty$). Neumann frequencies, $\hat{\omega}^{\rm AdS}_n|_{\rm Neu}$ (red circles), and Dirichlet frequencies, $\hat{\omega}^{\rm AdS}_n|_{\rm Dir}$ (green squares), are given by \eqref{AdS3-wNeu} and \eqref{AdS3-wDir} with $m=1$.  

The fundamental mode $n=0$ (purple) becomes unstable for $\kappa < \kappa^{\rm AdS}_{m,\hat{\mu}^2}$, with $\mu^2L^2=-15/16$ and $m=1$ giving $\kappa^{\rm AdS}_{m,\hat{\mu}^2} \sim -1.01$. For $\kappa > \kappa^{\rm AdS}_{m,\hat{\mu}^2}$, ${\rm Im}\,\hat{\omega}^{\rm AdS}_{0}=0$, while for $\kappa < \kappa^{\rm AdS}_{m,\hat{\mu}^2}$, ${\rm Im}\,\hat{\omega}^{\rm AdS}_{0} >0$, signalling instability (bottom-right panel of Fig.~\ref{fig:AdS3m1-wk}). The instability grows as $\kappa \to -\infty$ with $\underset{{\kappa \to -\infty}}{\lim} {\rm Im}\,\hat{\omega}^{\rm AdS}_{0} \to +\infty$.  

As in the $m=0$ case, $\underset{{\kappa \to -\infty}}{\lim} \hat{\omega}^{\rm AdS}_{0} \neq \hat{\omega}^{\rm AdS}_{0}|_{\rm Dir}$; instead, the first overtone satisfies $\underset{{\kappa \to -\infty}}{\lim} \hat{\omega}^{\rm AdS}_{1} = \hat{\omega}^{\rm AdS}_{0}|_{\rm Dir}$, making the $n=0$ mode “discontinuous” in the $\kappa\to -\infty$ limit. This behaviour occurs for $m=0$, $m=1$, and all $m>1$ studied.  

Since the AdS$_3$ instability (and the BTZ instability studied later) appears only for negative $\kappa$, in the following we often show plots restricted to $0 \ge \kappa$. 
 
It is notable that only the $n=0$ overtone is unstable, unlike many other instabilities (e.g., superradiant, ultraspinning, scalar condensation, Gregory-Laflamme) where multiple radial overtones are unstable. While the precise reason is unclear, higher overtones must remain stable to connect the stable Neumann frequencies $\hat{\omega}^{\rm AdS}_{n}|_{\rm Neu}$ with the stable Dirichlet frequencies $\hat{\omega}^{\rm AdS}_{n-1}|_{\rm Dir}$ for $n\ge1$ as $\kappa \to +\infty$.  

The qualitative behaviour of $m>1$ modes is similar to $m=0$ and $m=1$. The critical value of the double-trace parameter, $\kappa = \kappa^{\rm AdS}_{m,\hat{\mu}^2}$, below which only the $n=0$ mode becomes unstable, occurs when $\omega^{\rm AdS}=0$ (both real and imaginary parts vanish, as seen in the bottom panels of Figs.~\ref{fig:AdS3m0-wk}-\ref{fig:AdS3m1-wk}). Inserting $\omega^{\rm AdS}=0$ into the AdS frequency quantization condition \eqref{eqn:AdS3_Gamma_criterion} gives  
\begin{align}\label{kAdS-onset} 
\kappa^{\rm AdS}_{m, \hat{\mu}^2} = \frac{\Gamma\big(-\sqrt{1+\mu^2L^2}\big)}{\Gamma\big(\sqrt{1+\mu^2L^2}\big)} 
\frac{\Gamma\big(\frac{1}{2}(\Delta_+ + m)\big)^2}{\Gamma\big(\frac{1}{2}(\Delta_- + m)\big)^2}.
\end{align}
This threshold is always negative for any $m$ and $-1 < \hat{\mu} < 0$. Global AdS$_3$ is unstable to $n=0$ modes for $\kappa < \kappa^{\rm AdS}_{m, \hat{\mu}^2}$ (the instability is absent in the strict $\kappa \to -\infty$ Dirichlet limit, which is “discontinuous” as discussed above).  
 
For $\mu^2L^2=-15/16$, the critical values of $\kappa$ for the instability onset are  

\begin{align}\label{kAdS-onset:mValues}
\kappa_{m, -15/16}^{\rm AdS} \simeq 
\begin{cases} 
-0.4951294\,, & m = 0, \\
-1.0098371\,, & m = 1, \\
-1.3753594\,, & m = 2,\\
-1.6693225\,, & m = 3, \\
-2.1441520\,, & m = 5, \\
-3.0254135\,, & m = 10. \\
\end{cases}
\end{align}

Recall that global AdS$_3$ is unstable to $n=0$ perturbations for $\kappa < \kappa_{m, \hat{\mu}^2}^{\rm AdS}$. These examples illustrate a universal feature of the AdS$_3$ instability: for fixed $\mu L$ and $n$, higher $m$ modes become unstable at more negative $\kappa$. Consequently, if $\kappa$ is below the instability onset of a given $m$, it is automatically unstable to all lower $m$ modes, in particular to $m=0$. This is non-standard compared to typical black hole instabilities, where instability in a higher $m$ mode does not imply instability in lower $m$ modes. Hence, the $m=0$, $n=0$ mode governs the overall stability of AdS$_3$.  

For completeness, one can also fix $m$ and solve \eqref{kAdS-onset} to find $\kappa_{m, \hat{\mu}^2}^{\rm AdS}$ for different scalar masses in the double-trace range \eqref{2xTrace:rangeMass}, $-1 < \mu^2 L^2 < 0$. For example, for $m=0$ and $m=1$ (to six decimal places)  

\begin{align}\label{AdS:CritKappa2}
\kappa_{0, \hat{\mu}^2}^{\rm AdS} &\simeq 
\begin{cases} 
-0.916039\,, & \mu^2 L^2=-999/1000, \\
-0.495129\,, & \mu^2 L^2=-15/16, \\
-0.387438\,, & \mu^2 L^2=-8/9, \\
-0.228473\,, & \mu^2 L^2=-3/4, \\
-0.101910\,, & \mu^2 L^2=-1/2, \\
-0.0825216\,, & \mu^2 L^2=-7/16, \\
-0.0135220\,, & \mu^2 L^2=-1/10, \\
\end{cases} \\
\kappa_{1, \hat{\mu}^2}^{\rm AdS} &\simeq 
\begin{cases} 
-1.00002\,, & \mu^2 L^2=-999/1000, \\
-1.00984\,, & \mu^2 L^2=-15/16, \\
-1.02429\,, & \mu^2 L^2=-8/9, \\
-1.09422\,, & \mu^2 L^2=-3/4, \\
-1.38147\,, & \mu^2 L^2=-1/2, \\
-1.51476\,, & \mu^2 L^2=-7/16, \\
-5.32880\,, & \mu^2 L^2=-1/10. \\
\end{cases}
\end{align}

This illustrates that, for $m=0$ ($m\ge1$), $\kappa_{m, \hat{\mu}^2}^{\rm AdS}$ increases (decreases) as $\hat{\mu}$ increases.  

\subsection{Comparison with Ishibashi-Wald's stability study of global \texorpdfstring{AdS$_d$}{AdSd}}\label{sec:AdS3wald}
As discussed in the previous subsection, one can quickly find the critical double-trace parameter $\kappa=\kappa_{m, \hat{\mu}^2}^{\hbox{\tiny AdS}}$ for the onset of the global AdS$_3$ instability (in the $n=0$ overtone sector) by setting $\omega^{\hbox{\tiny AdS}}=0$ in the AdS$_3$ frequency quantization condition \eqref{eqn:AdS3_Gamma_criterion} and solving it to find the onset $\kappa$. This yields $\kappa=\kappa_{m, \hat{\mu}^2}^{\hbox{\tiny AdS}}$ given by \eqref{kAdS-onset}.
Ishibashi and Wald \cite{Ishibashi:2004wx}  first found that global AdS$_d$ (for $d\geq 3$) can be unstable to perturbations with double-trace boundary conditions. Therefore, we should compare our results against those of \cite{Ishibashi:2004wx} to check our computations and to confirm that the nature of the instability is the same. Since \cite{Ishibashi:2004wx} did not compute directly the frequency spectrum, this comparison is not straightforward and in this subsection we map the two studies.\footnote{Notice that \cite{Ishibashi:2004wx} provided a formal proof of AdS stability conditions while we are just doing a linear mode analysis that finds the AdS unstable modes and AdS normal modes.} 

In \cite{Ishibashi:2004wx}, a detailed mathematical study of the dynamics of scalar, electromagnetic and gravitational perturbations of global AdS$_d$ ($d\geq 3$) spacetime was carried out. In particular, they studied all possible boundary conditions that can be imposed at the AdS boundary in every case, and established criteria to determine whether each choice yields stable or unstable dynamics. For the scalar field (the only of interest here), the cases are split by studying the variable (see equation (44) in \cite{Ishibashi:2004wx} where $m_0\equiv \mu$ and $n_d \equiv d-2$)
\begin{align}
    \nu = \sqrt{\frac{d(d-2)}{4} + \mu^2L^2 + \frac{1}{4} } \qquad  \overset{d=3 \:\:}{\hbox{\Large $\longrightarrow$}}  \qquad \nu = \sqrt{1+ \mu^2L^2}. 
\end{align}
This means that for the range $-1<\mu^2 L^2<0$  where double-trace BCs are permitted $-$ see \eqref{2xTrace:rangeMass} $-$ we have $0< \nu^2<1$  which corresponds to case (ii) for the scalar field as introduced in section 3.1 of \cite{Ishibashi:2004wx}. For such case, AdS$_3$ is stable if and only if equation (169) in \cite{Ishibashi:2004wx} holds, namely:
\begin{align}\label{eqn:Wald_Ishibashi_criterion}
    \frac{b_\nu}{a_\nu} \geq  - \left|\frac{\Gamma(-\nu)}{\Gamma(\nu)}\right| \frac{\Gamma(\zeta_{\nu, \sigma}^0)^2}{\Gamma(\zeta_{-\nu, \sigma}^0)^2},
\end{align}
where $-$ see discussion of  equation (168) of \cite{Ishibashi:2004wx} $-$  the ratio $b_\nu/a_\nu$ is effectively equivalent to our $\kappa$ introduced in \eqref{2xTrace:BC} in the context of the FG expansion \eqref{FGasympExp}. 
From equation (127) in \cite{Ishibashi:2004wx}, their $\sigma$ is effectively our azimuthal number $m$:
\begin{align}
    \sigma = \ell + \frac{d-3}{2} \equiv m,
\end{align}
as we are in $d=3$. Finally, from equation (145) in \cite{Ishibashi:2004wx} one has
\begin{align}
    \zeta_{\nu, \sigma}^\omega = \frac{\nu + \sigma + \omega^{\hbox{\tiny AdS}} +1}{2}.
\end{align}
For \eqref{eqn:Wald_Ishibashi_criterion}, we  need to evaluate this for $\omega^{\hbox{\tiny AdS}}=0$ which yields (after translation into our notation):  
\begin{align}
\begin{cases} 
   & \!\! \zeta_{\nu, \sigma}^0 = \frac{\nu + \sigma +1}{2} \equiv \frac{1+m + \sqrt{1+ \mu^2L^2}}{2}\,, \\
    & \!\! \zeta_{-\nu, \sigma}^0 = \frac{-\nu + \sigma +1}{2} \equiv  \frac{1+m - \sqrt{1+ \mu^2L^2}}{2}\,.
  \end{cases} 
\end{align}
Therefore, rewriting \eqref{eqn:Wald_Ishibashi_criterion} for our case and in our notation, for global AdS$_3$ to be stable under double-trace boundary conditions one must have \cite{Ishibashi:2004wx}: 
\begin{align} 
   \kappa \geq  -\left|\frac{\Gamma\left(-\sqrt{1+ \mu^2L^2}\right)}{\Gamma\left(\sqrt{1+ \mu^2L^2}\right)}\right| \,\frac{\Gamma\left(\frac{1}{2}\left(1+m+\sqrt{1+\mu^2L^2}\right)\right)^2}{\Gamma\left(\frac{1}{2}\left(1+m-\sqrt{1+\mu^2L^2}\right)\right)^2}.
\end{align}
For $-1 < \mu^2L^2<0$, the quotient inside the absolute value is always negative, and hence we can rewrite this stability condition as 
\begin{align} \label{IshibashiWaldStabilityCriterion}
      \kappa \geq \frac{\Gamma\left(-\sqrt{1+ \mu^2L^2}\right)}{\Gamma\left(\sqrt{1+ \mu^2L^2}\right)}\,\frac{\Gamma\left(\frac{1}{2}\big(1+m+\sqrt{1+\mu^2L^2}\big)\right)^2}{\Gamma\left(\frac{1}{2}\big(1+m-\sqrt{1+\mu^2L^2}\big)\right)^2}  \qquad \Leftrightarrow   \qquad \kappa \geq \kappa^{\hbox{\tiny AdS}}_{m, \hat{\mu}^2}\,,
\end{align}
where $\kappa^{\hbox{\tiny AdS}}_{m, \hat{\mu}^2}$ is exactly given by the very same critical value \eqref{kAdS-onset}
that emerges from our independent analysis. 

Thus, our conclusion that AdS$_3$ is linearly unstable to double-trace boundary conditions when $ \kappa < \kappa_{m, \hat{\mu}^2}^{\hbox{\tiny AdS}}$  matches the stability criterion \eqref{IshibashiWaldStabilityCriterion} of \cite{Ishibashi:2004wx}. The physical origin of this AdS$_3$ instability will be discussed in subsection~\ref{sec:BTZorigin}.

\section{Double-trace instabilities of BTZ black hole}\label{sec:BTZresults}
In this section, we study instabilities of the BTZ black hole under perturbations by a massive scalar field with mass in the range $-1 < \hat{\mu}^2 < 0$, for which double-trace boundary conditions, $\beta = \kappa \, \alpha$, can be imposed (cf.~\eqref{FGasympExp}-\eqref{2xTrace:BC}).  

In the context of the AdS/CFT correspondence, fixing the boundary theory requires specifying the boundary conditions and thus the double-trace parameter $\kappa$, which we do in subsection~\ref{sec:3Dspectrum}. To identify physically relevant values of $\kappa$, we first analyse how the solutions vary with $\kappa$, finding that BTZ becomes unstable for negative $\kappa < \kappa^{\hbox{\tiny BTZ}}_{m, \hat{\mu}^2}\text{\small $(\hat{M},\hat{J})$}$. Initially, we compare this critical BTZ onset with the AdS$_3$ onset $\kappa^{\hbox{\tiny AdS}}_{m, \hat{\mu}^2}$ from \eqref{kAdS-onset} in subsections~\ref{sec:BTZonset1} and \ref{sec:BTZonset2}, and discuss the physical origin of these instabilities in subsection~\ref{sec:BTZorigin}. In subsection~\ref{sec:3Dspectrum}, we typically fix $\kappa$ so that BTZ can be unstable for some $\{\hat{M},\hat{J}\}$ while global AdS$_3$ remains stable, also briefly illustrating BTZ mode behaviour when AdS$_3$ is unstable.

The qualitative results are largely independent of the scalar mass $\hat{\mu} = \mu L$. For concreteness, we use $\mu^2 L^2 = -15/16$ to illustrate properties in the range \eqref{2xTrace:rangeMass}, $-1 < \mu^2 L^2 < 0$, where the double-trace BC \eqref{2xTrace:BC} is allowed.\footnote{Appendix~\ref{sec:AppOtherMasses} presents additional results for $\mu^2L^2 = -8/9,-3/4,-1/10$ to highlight universal properties and briefly discuss non-universal features for specific masses.} Recall that $\beta = \kappa \, \alpha$ (with $\kappa \in \mathbb{R}$) also encompasses Neumann ($\kappa = 0$) and Dirichlet ($\kappa \to \pm \infty$) BCs. Since positive $\kappa$ never leads to instabilities (see Fig.~\ref{fig:BTZw-kappa:m0} top panel and \cite{Hartnoll:2008kx,Faulkner:2010fh,Faulkner:2010gj,Iizuka:2015vsa,Dappiaggi:2017pbe}), from subsection \ref{sec:BTZorigin} we restrict our discussion to $0\leq \kappa$.

Modes with azimuthal number $m=0$ and $m\geq 1$ behave qualitatively differently in BTZ, though $m\geq 1$ modes share similar features. Therefore, we typically present results for $m=0$ and $m=1$ (representing $m>0$ behaviour) and occasionally include higher $m$ when relevant to illustrate specific physical properties.

\subsection{Varying \texorpdfstring{$\kappa$}{kappa}. Double-trace instabilities \& march from Neumann to Dirichlet}\label{sec:BTZonset1}
\begin{figure}[ht]
    \centering
    \includegraphics[width=0.47\linewidth]{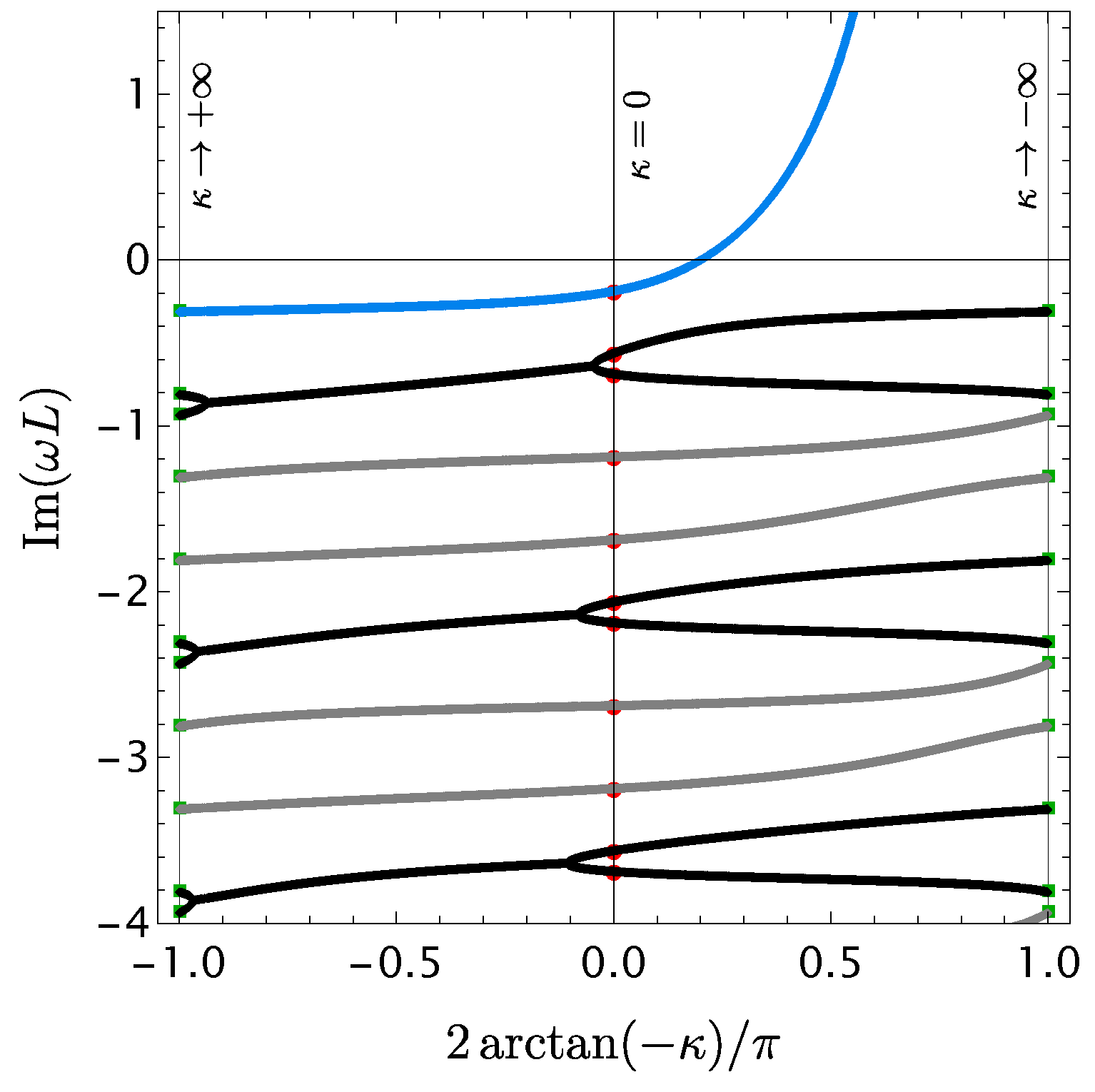}
    \includegraphics[width=0.51\linewidth]{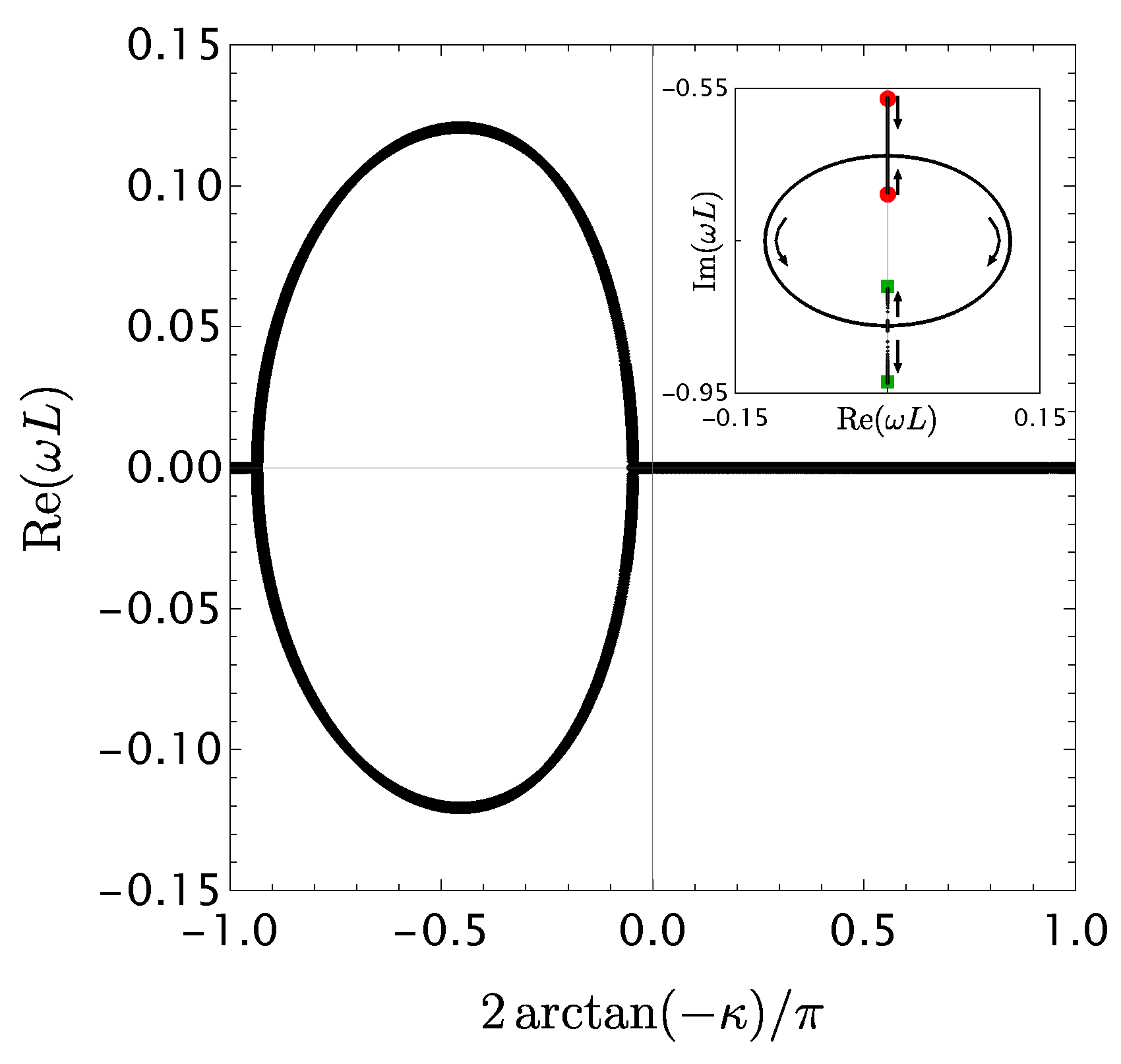}
    \caption{\small 
{\bf Left panel:} $\text{Im}(\omega L)$ vs.~$\theta = \frac{2}{\pi}\arctan(-\kappa)$ for radial overtones $n=0,\dots,10$ of $m=0$ modes ($\mu^2 L^2 = -15/16$) for a BTZ black hole with $\{\hat{M},\hat{J}\} = \{5/16,1/4\}$ ($\{y_-,y_+\}=\{1/4,1/2\}$). Red discs at $\kappa=0$ ($\theta=0$) denote $\hat{\omega}^{\rm BTZ}_n|_{\rm Neu}$ \eqref{BTZ:wNeu}, and green squares at $\kappa = \pm\infty$ ($\theta = \mp 1$) denote $\hat{\omega}^{\rm BTZ}_n|_{\rm Dir}$ \eqref{BTZ:wDir}. The $n=0$ mode (blue curve) becomes unstable at $\kappa \sim -0.33$. Modes $n=0$ (blue) and $n=3,4,7,8,11,12,15,16,\dots$ (grey) have $\text{Re}(\omega L)=0$, unlike $n=1,2,5,6,9,10,13,14,\dots$ (black, see right panel).  {\bf Right panel:} $\text{Re}(\omega L)$ vs.~$\theta(\kappa)$ for the black $n=1,2$ overtones. These modes have $\text{Re}(\omega L)=0$ when $\text{Im}(\omega L)$ is distinct; when $\text{Im}(\omega L)$ coincides, $\text{Re}(\omega L)$ is symmetric.  The $n=0$ blue curve always has $\text{Re}(\omega L)=0$; for $0 \ge \kappa \gtrsim -0.33$, $\text{Im}(\omega L)<0$, while for $\kappa \lesssim -0.33$, $\text{Im}(\omega L)>0$, indicating instability with $\underset{{\kappa \to -\infty}}{\lim} \text{Im}(\omega L) \to +\infty$ (details in top panel of Fig.~\ref{fig:BTZw-kappa:m1_m5}). See Fig.~\ref{fig:BTZw-kappa:m0-OtherMasses} (Appendix~\ref{sec:AppOtherMasses}) for other masses $\mu^2 L^2=-8/9,-3/4,-1/10$.
}\label{fig:BTZw-kappa:m0}
\end{figure}

In this subsection, we scan the parameter space of linear modes by varying the double-trace parameter $\kappa$. To illustrate the system's properties, we consider a BTZ black hole with $\{\hat{M},\hat{J}\} = \{5/16,1/4\}$ (\ie $\{y_-,y_+\} = \{1/4,1/2\}$) and a scalar field with $\mu^2 L^2=-15/16$. We compute the BTZ frequencies as $\kappa$ varies from $\kappa\to +\infty$ (Dirichlet BC) to $\kappa=0$ (Neumann BC) and further to $\kappa\to -\infty$ (Dirichlet BC again), for modes with $m=0$ and $m=1$ (the latter representing $m\geq 1$ modes which behave qualitatively similarly). Recall that the Dirichlet ($\kappa\to\pm\infty$) and Neumann ($\kappa=0$) quasinormal frequencies are given analytically by \eqref{BTZ:wDir}-\eqref{BTZ:wNeu}.  

For $m=0$, the left panel of Fig.~\ref{fig:BTZw-kappa:m0} shows ${\rm Im}(\omega L)$ vs.~$\theta(\kappa)$ \eqref{def:theta} for the first 11 radial overtones ($n=0,\dots,10$), with red discs at $\kappa=0$ marking $\hat{\omega}^{\rm BTZ}_n|_{\rm Neu}$ and green squares at $\kappa=\pm\infty$ marking $\hat{\omega}^{\rm BTZ}_n|_{\rm Dir}$. All $n\geq 1$ modes (black or grey diamonds) interpolate continuously from Dirichlet$_{\rm left}$ to Neumann to Dirichlet$_{\rm right}$ as $\theta$ goes from $-1$ ($\kappa\to+\infty$) to $0$ ($\kappa=0$) to $1$ ($\kappa\to-\infty$). For $\mu^2 L^2=-15/16$, $\hat{\omega}^{\rm BTZ}_n|_{\kappa=0}=\hat{\omega}^{\rm BTZ}_n|_{\rm Neu}$ and $\underset{{\kappa \to -\infty}}{\lim} \hat{\omega}^{\rm BTZ}_n = \hat{\omega}^{\rm BTZ}_{n-1}|_{\rm Dir}$, reproducing the $n|_{\rm Neu} \to (n-1)|_{\rm Dir}$ mapping similar to AdS$_3$ (cf.~Fig.~\ref{fig:AdS3m0-wk}), though BTZ frequencies for $0\le\theta\le 1$ are purely imaginary.  

The black overtone pairs ($n=1,2$, $n=5,6$, $n=9,10$, etc.) interact nontrivially: they have $\text{Re}(\omega L)=0$ when $\text{Im}(\omega L)$ differs, and symmetric $\text{Re}(\omega L)$ when $\text{Im}(\omega L)$ coincides (right panel, Fig.~\ref{fig:BTZw-kappa:m0}). Other scalar masses show similar behaviour (Appendix~\ref{sec:AppOtherMasses}, Fig.~\ref{fig:BTZw-kappa:m0-OtherMasses}) and similar features were noted in other black holes \cite{Tanabe:2015isb,Dias:2020ncd,Davey:2024xvd,Davey:2022vyx}.  

Modes $n=0$ (blue curve) and $n=3,4,7,8,11,12,15,16,\dots$ (grey) always have $\text{Re}(\omega L)=0$. The fundamental mode $n=0$ (blue) has ${\rm Im}\,\hat{\omega}_{n=0}<0$ from $\theta=-1$ ($\kappa\to +\infty$) to $\theta=0$ ($\kappa=0$), with $\underset{{\kappa \to +\infty}}{\lim}\hat{\omega}^{\rm BTZ}_{n=0}=\hat{\omega}^{\rm BTZ}_{n=0}|_{\rm Dir}$ and $\hat{\omega}^{\rm BTZ}_{n=0}|_{\kappa=0}=\hat{\omega}^{\rm BTZ}_{n=0}|_{\rm Neu}$. For $\kappa \lesssim \kappa^{\rm BTZ}_{m=0,\hat{\mu}^2=-15/16}\sim -0.33$, ${\rm Im}\,\hat{\omega}_{n=0}>0$ and grows as $\kappa\to -\infty$, indicating that only this mode becomes unstable. Details of this blue curve are better seen in the top panel of Fig.~\ref{fig:BTZw-kappa:m1_m5} (restricted to $\kappa\le0$).  

For other scalar masses in $-1<\hat{\mu}^2<0$, the $m=0$, $n=0$ mode remains the only unstable mode, while higher overtones can have qualitative features that differ from Fig.~\ref{fig:BTZw-kappa:m0}; see Appendix~\ref{sec:AppOtherMasses}, Fig.~\ref{fig:BTZw-kappa:m0-OtherMasses} for examples with $\mu^2 L^2=-8/9,-3/4,-1/10$.

\begin{figure}[ht]
    \centering
    \includegraphics[width=0.49\linewidth]{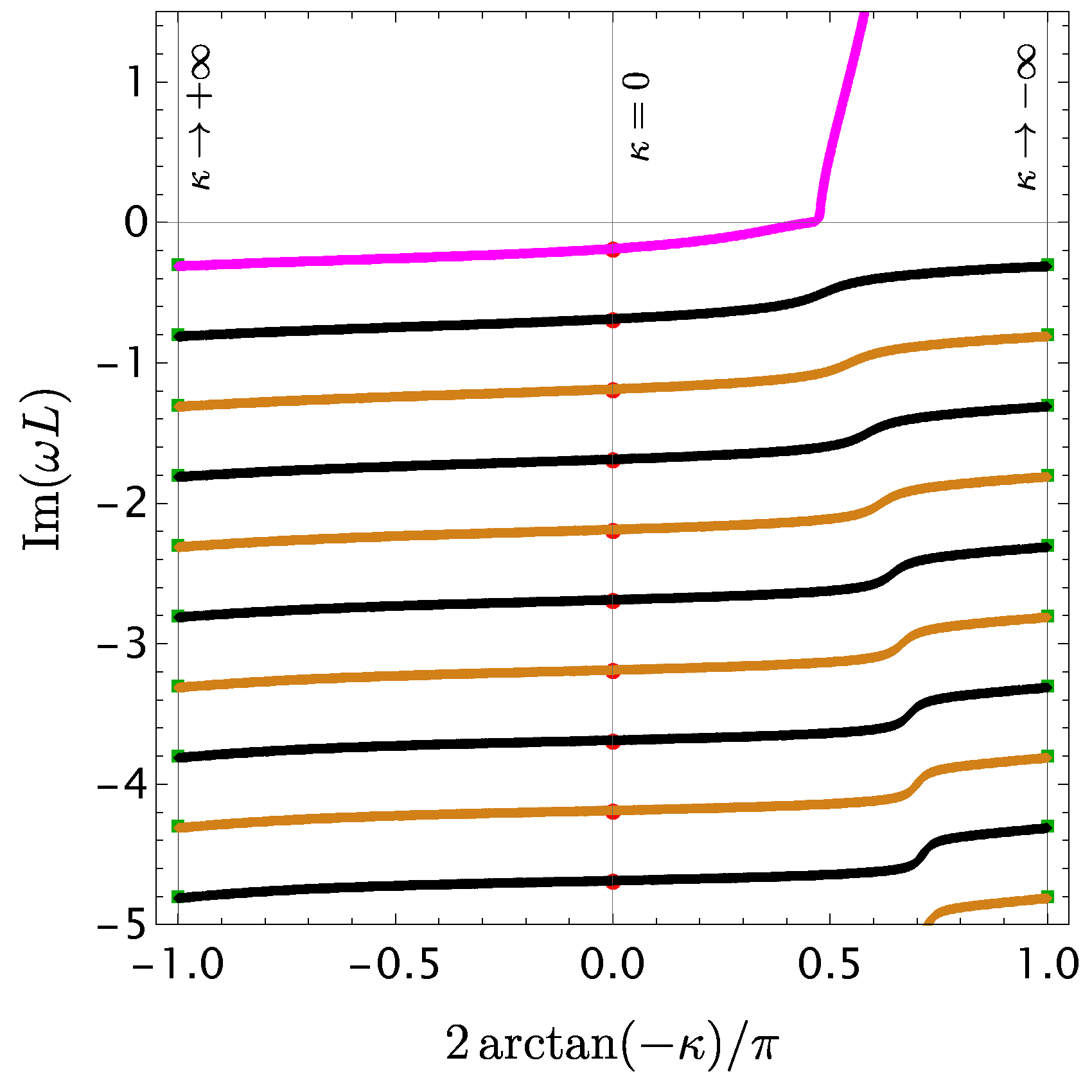}
     \includegraphics[width=0.49\linewidth]{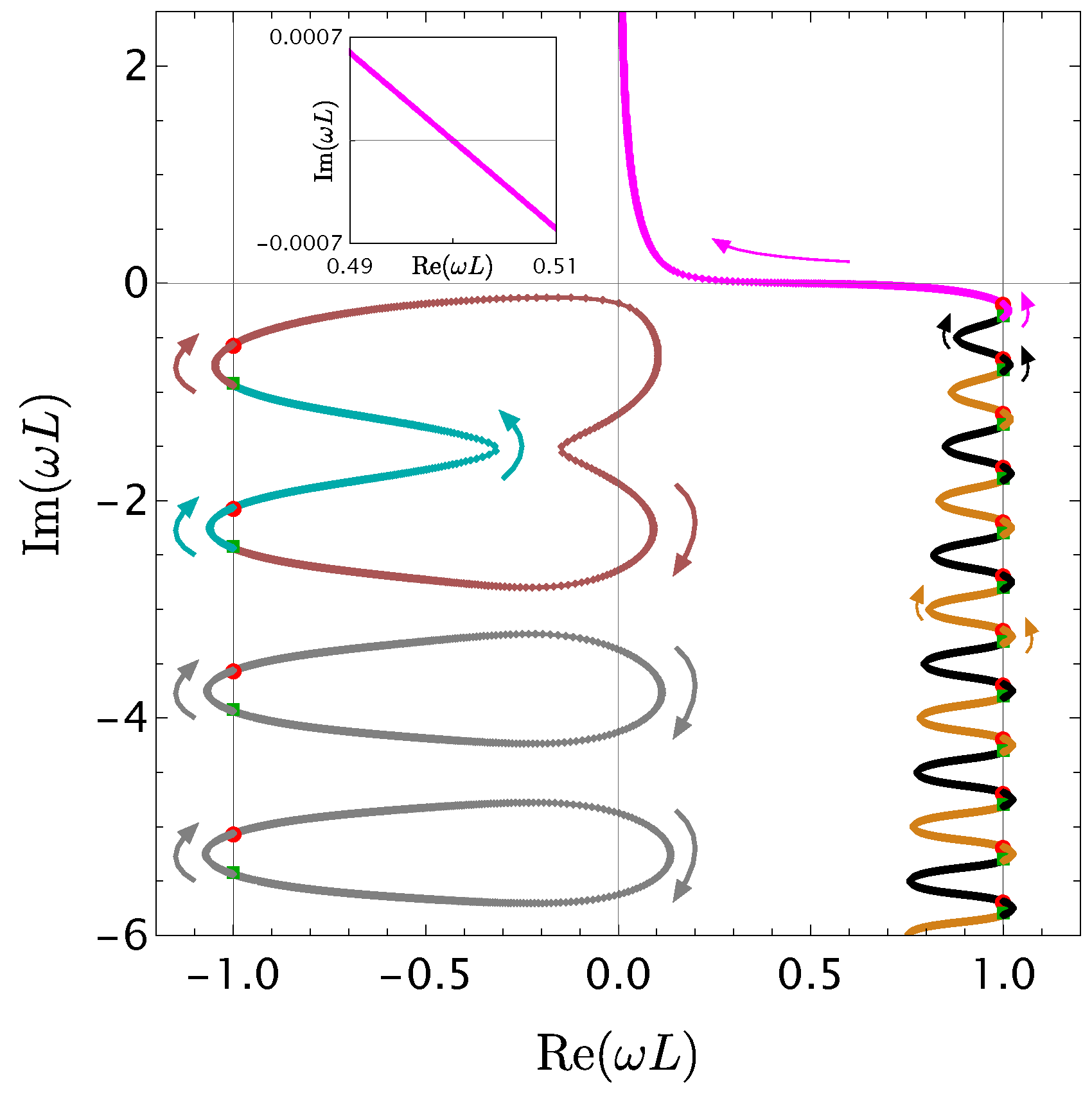}
    \\
    \includegraphics[width=0.45\linewidth]{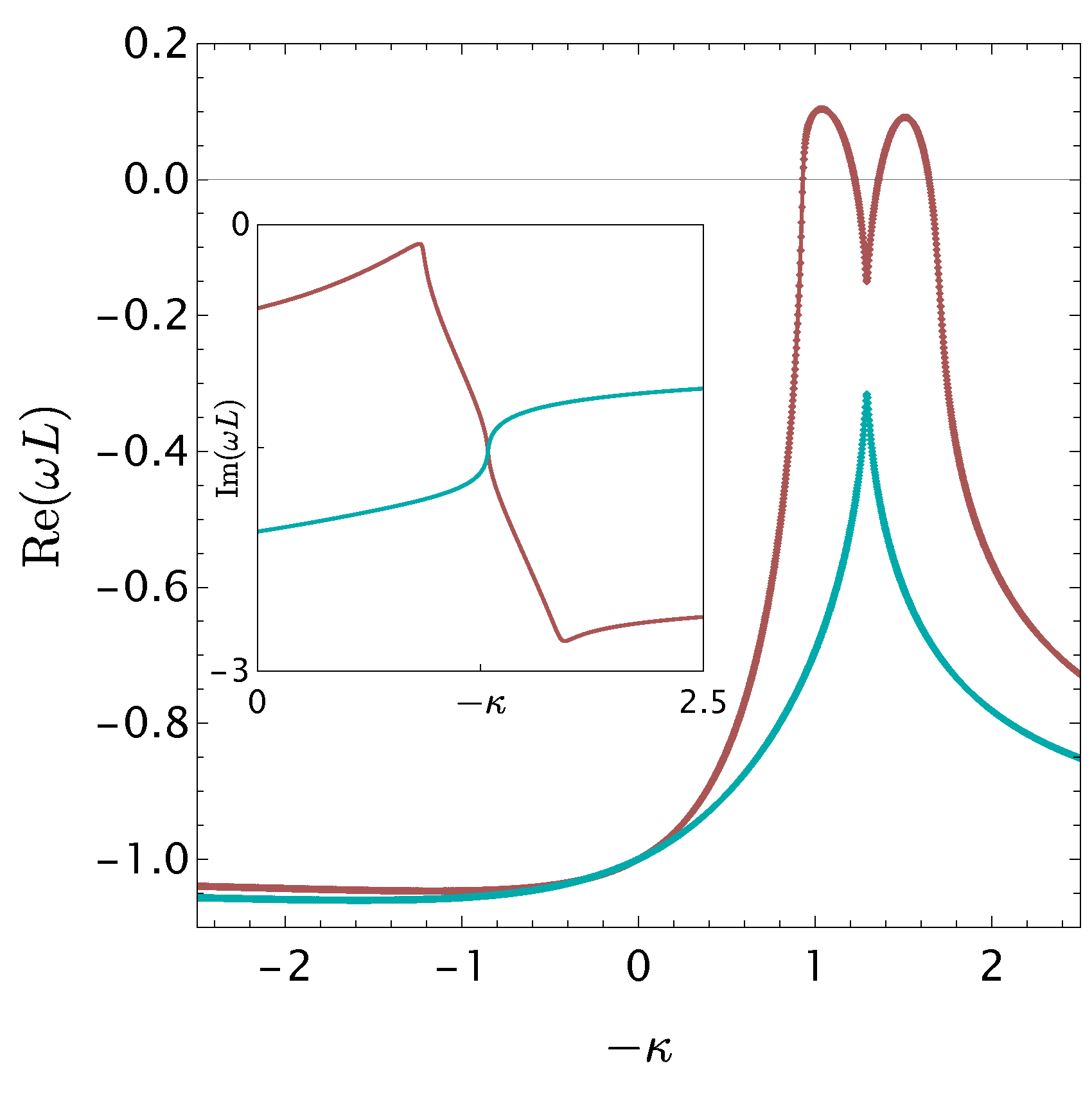}
    \hspace{0.5cm}
    \includegraphics[width=0.46\linewidth]{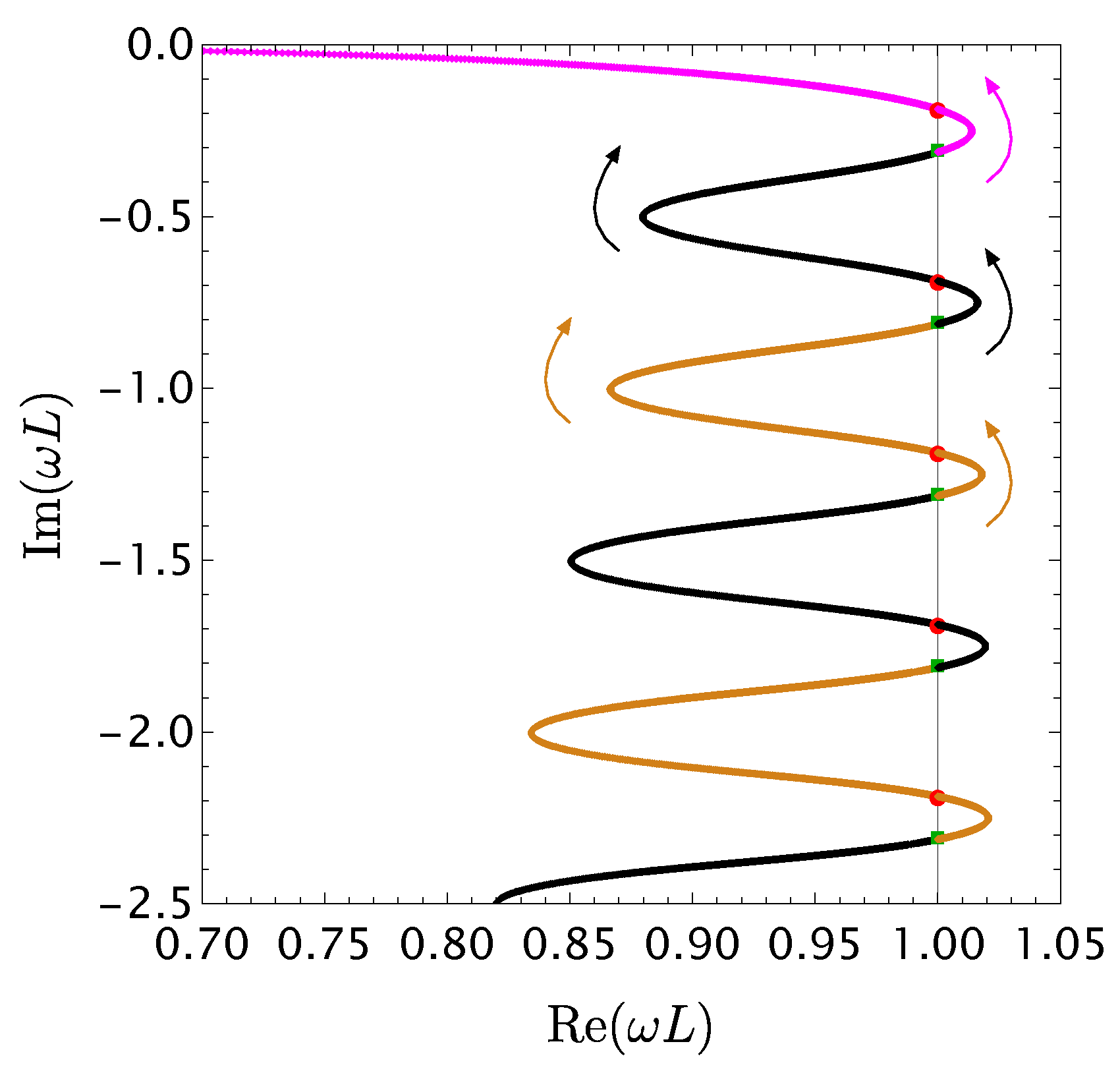}    
    \caption{\small {\bf Top-left:} $\text{Im}(\omega L)$ vs.~$\theta(\kappa)$ for radial overtones $n=0,\dots,9$ of {\it co-rotating} $m=1$ modes ($\mu^2 L^2=-15/16$) for a BTZ black hole with $\{\hat{M},\hat{J}\} = \{5/16,1/4\}$.  {\bf Top-right:} Double-trace BTZ modes in the complex plane, $\text{Im}(\omega L)$ vs.~$\text{Re}(\omega L)$, including co-rotating $n=0,\dots,12$ (curves pass through red Neumann solutions with ${\rm Re}\,\hat{\omega}=+1$ on the right) and counter-rotating $n=0,1,2,3$ (curves pass through red Neumann solutions with ${\rm Re}\,\hat{\omega}=-1$ on the left). All curves are parametrized by $\kappa$; arrows indicate decreasing $\kappa$. Green squares at $\kappa\to \pm \infty$ mark $\hat{\omega}^{\rm BTZ}_n|_{\rm Dir}$ \eqref{BTZ:wDir}, and red discs at $\kappa=0$ mark $\hat{\omega}^{\rm BTZ}_n|_{\rm Neu}$ \eqref{BTZ:wNeu}.  {\bf Bottom-left:} Detail of the interaction/repulsion between the counter-rotating $n=0,1$ modes (dark-red and cyan) near $\kappa \sim -1.3$, where ${\rm Im}(\omega L)$ coincides but ${\rm Re}(\omega L)$ repels.  {\bf Bottom-right:} Zoom-in on the top-right panel showing the co-rotating $n=0,1,2$ curves.  See Fig.~\ref{fig:BTZw-kappa:m1-OtherMasses} (Appendix~\ref{sec:AppOtherMasses}) for results with $\mu^2 L^2=-8/9,-3/4,-1/10$.}\label{fig:BTZw-kappa:m1}
\end{figure}

Next, we ask whether such an instability occurs for $m=1$ (modes with $m>1$ behave similarly). For $m=1$, both co-rotating and counter-rotating modes exist, and any instability arises in the co-rotating sector. In the top-left panel of Fig.~\ref{fig:BTZw-kappa:m1} we plot $\text{Im}(\omega L)$ vs.~$\theta=\frac{2}{\pi}\arctan(-\kappa)$ for co-rotating $m=1$ modes of a BTZ black hole with $\{\hat{M},\hat{J}\} = \{5/16,1/4\}$ and $\hat{\mu}^2=-15/16$, analogous to the $m=0$ case (Fig.~\ref{fig:BTZw-kappa:m0}). Red discs at $\kappa=0$ mark $\hat{\omega}^{\rm BTZ}_n|_{\rm Neu}$ \eqref{BTZ:wNeu}, and green squares denote $\hat{\omega}^{\rm BTZ}_n|_{\rm Dir}$ \eqref{BTZ:wDir}, with $n=0,\dots,9$ from top to bottom. The $n=0$ (magenta) co-rotating mode becomes unstable for sufficiently negative $\kappa$, while all $n\ge 1$ modes (black/brown diamonds) interpolate continuously between Dirichlet$_{\rm left}$, Neumann, and Dirichlet$_{\rm right}$ points as $\theta$ goes from $-1$ ($\kappa\to+\infty$) to $0$ ($\kappa=0$) to $1$ ($\kappa\to-\infty$). For $n\ge 1$, ${\rm Im}\,\hat{\omega}_n<0$, $\underset{{\kappa \to +\infty}}{\lim}\hat{\omega}_n=\hat{\omega}^{\rm BTZ}_n|_{\rm Dir}$, $\hat{\omega}_n|_{\kappa=0}=\hat{\omega}^{\rm BTZ}_n|_{\rm Neu}$, and $\underset{{\kappa \to -\infty}}{\lim}\hat{\omega}_n=\hat{\omega}^{\rm BTZ}_{n-1}|_{\rm Dir}$.

Unlike $m=0$, all $m=1$ modes initially have complex frequencies. Thus, in the top-right panel of Fig.~\ref{fig:BTZw-kappa:m1}, we plot ${\rm Im}\,\hat{\omega}$ vs.~${\rm Re}\,\hat{\omega}$ for co-rotating ($n=0,\dots,11$) and counter-rotating ($n=0,\dots,5$) modes. Green squares and red discs denote Dirichlet and Neumann limits, respectively, and arrows indicate decreasing $\kappa$ ($+\infty \to 0 \to -\infty$).  

Co-rotating modes revolve around ${\rm Re}\,\hat{\omega}=+1$. For $n\ge1$, each $n^{\rm th}$ curve starts at the $(n-1)^{\rm th}$ Dirichlet point at $\kappa\to-\infty$, passes through the $n^{\rm th}$ Neumann point at $\kappa=0$, and terminates at the $n^{\rm th}$ Dirichlet point at $\kappa\to+\infty$, forming a continuous zig-zag sequence (black/light-brown curves). However, for some scalar masses, these curves can form disconnected loops (Appendix~\ref{sec:AppOtherMasses}, Fig.~\ref{fig:BTZw-kappa:m1-OtherMasses}).  

The $n=0$ co-rotating mode (magenta) is distinct: starting from the upper Dirichlet point, it passes through the first Neumann point with ${\rm Im}\,\hat{\omega}_{n=0}<0$ and becomes unstable at $\kappa=\kappa^{\rm BTZ}_{m=1,\hat{\mu}^2=-15/16}\sim -0.87$, where ${\rm Im}\,\hat{\omega}_{n=0}=0$. For $\kappa<\kappa^{\rm BTZ}_{m,\hat{\mu}^2}$, ${\rm Im}\,\hat{\omega}_{n=0}>0$ grows with decreasing $\kappa$, while ${\rm Re}\,\hat{\omega}_{n=0}\to0$ as $\kappa\to-\infty$, analogous to the $m=0$, $n=0$ instability (Fig.~\ref{fig:BTZw-kappa:m0}). Notably, the $m=1$ instability occurs at a lower $\kappa$ than $m=0$ for this BTZ, a feature consistent across other $\{\hat{M},\hat{J}\}$.

Counter-rotating modes revolve around ${\rm Re}\,\hat{\omega}=-1$ and never become unstable. Initially, counter-rotating modes can develop ${\rm Re}\,\hat{\omega}>0$ in intermediate $\kappa$ windows, indicating co-rotation with the BTZ. Higher counter-rotating overtones ($n\ge2$) form closed loops, passing through Dirichlet and Neumann points, while $n=0,1$ interact nontrivially (bottom-left panel, Fig.~\ref{fig:BTZw-kappa:m1}), repelling in ${\rm Re}\,\hat{\omega}$ near $\kappa\sim-1.3$ while sharing the same ${\rm Im}\,\hat{\omega}$.

For other scalar masses $-1<\hat{\mu}^2<0$, only the fundamental co-rotating mode ($m\ge1$, $n=0$) is unstable; details of damped overtones vary (Appendix~\ref{sec:AppOtherMasses}, Fig.~\ref{fig:BTZw-kappa:m1-OtherMasses}). Our numerical results reproduce prior analyses \cite{Iizuka:2015vsa,Ferreira:2017cta,Dappiaggi:2017pbe} for $m=1$, but these works did not study the $m=0$ sector, which is crucial for the double-trace instability.

Comparing Figs.~\ref{fig:BTZw-kappa:m0} ($m=0$) and \ref{fig:BTZw-kappa:m1} ($m=1$, $m>1$ similar), one finds that for any $m$, only the fundamental (and co-rotating if $m>0$) mode ($n=0$) can become unstable for sufficiently negative $\kappa$, $0>\kappa^{\rm BTZ}_{m,\hat{\mu}^2}>\kappa$, a characteristic of double-trace perturbations in AdS black holes across dimensions, horizon topology, rotation, and charge \cite{Ishibashi:2004wx,Hertog:2004dr,Martinez:2004nb,Hertog:2004ns,Faulkner:2010fh,Faulkner:2010gj,Dias:2013bwa,Iizuka:2015vsa,Dappiaggi:2017pbe,Ferreira:2017cta,Katagiri:2020mvm,Kinoshita:2023iad,Harada:2023cfl}. Consequently, we focus on negative $\kappa$.

Finally, the Dirichlet$-$Neumann$-$Dirichlet cycle of double-trace modes is inevitable due to the boundary condition limits, explaining the similar $\kappa$-evolution observed in global AdS$_4$ Schwarzschild and Reissner-Nordstr\"om black holes \cite{Kinoshita:2023iad,Katagiri:2020mvm}. Figs.~\ref{fig:BTZw-kappa:m0}-\ref{fig:BTZw-kappa:m1} show that this cycle can be traversed directly or in a more intricate way depending on $m$, $n$, and co/counter-rotation.

  \begin{figure}[ht]
    \centering
    \includegraphics[width=0.45\linewidth]{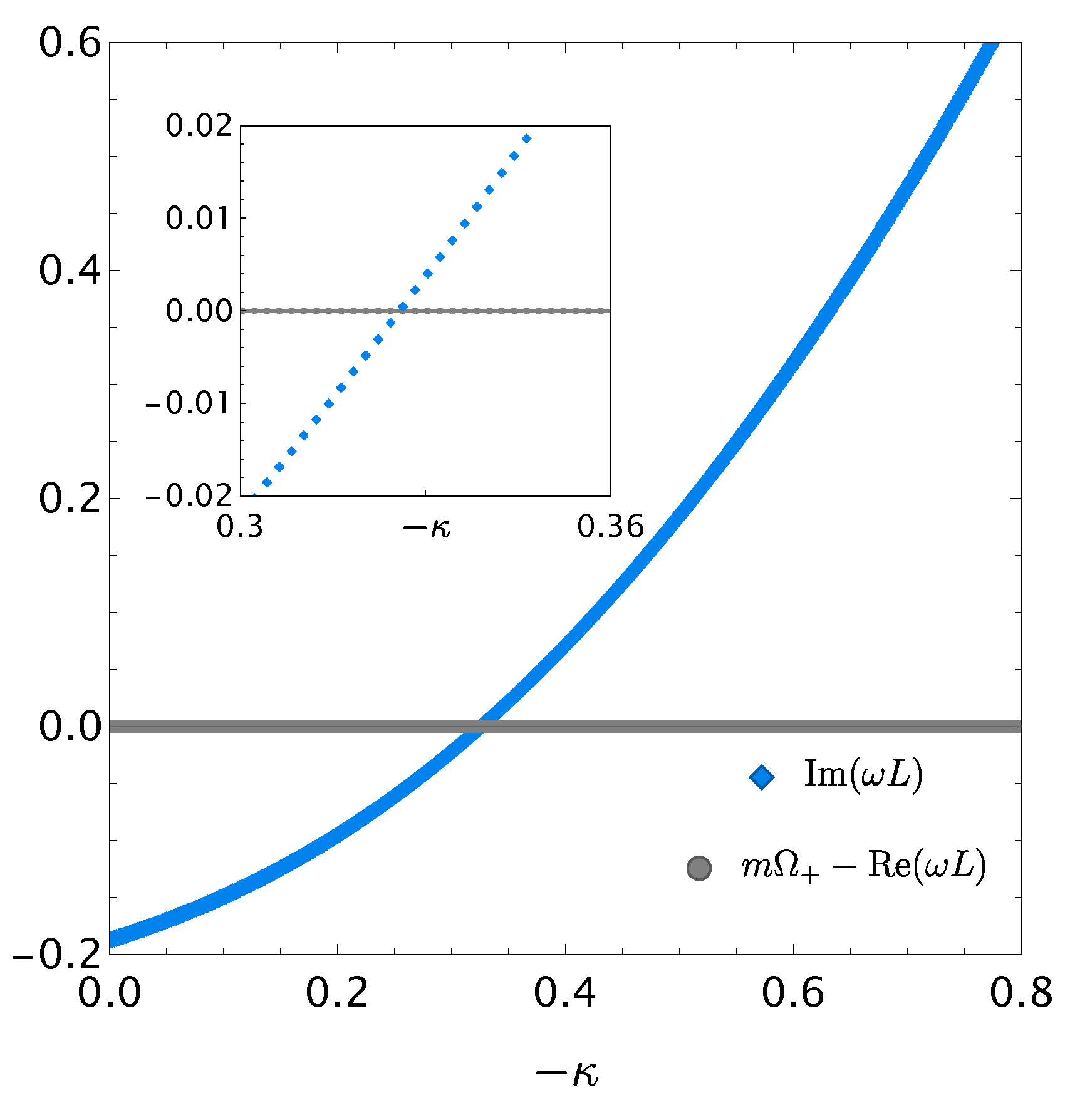} \\
     \includegraphics[width=0.9\linewidth]{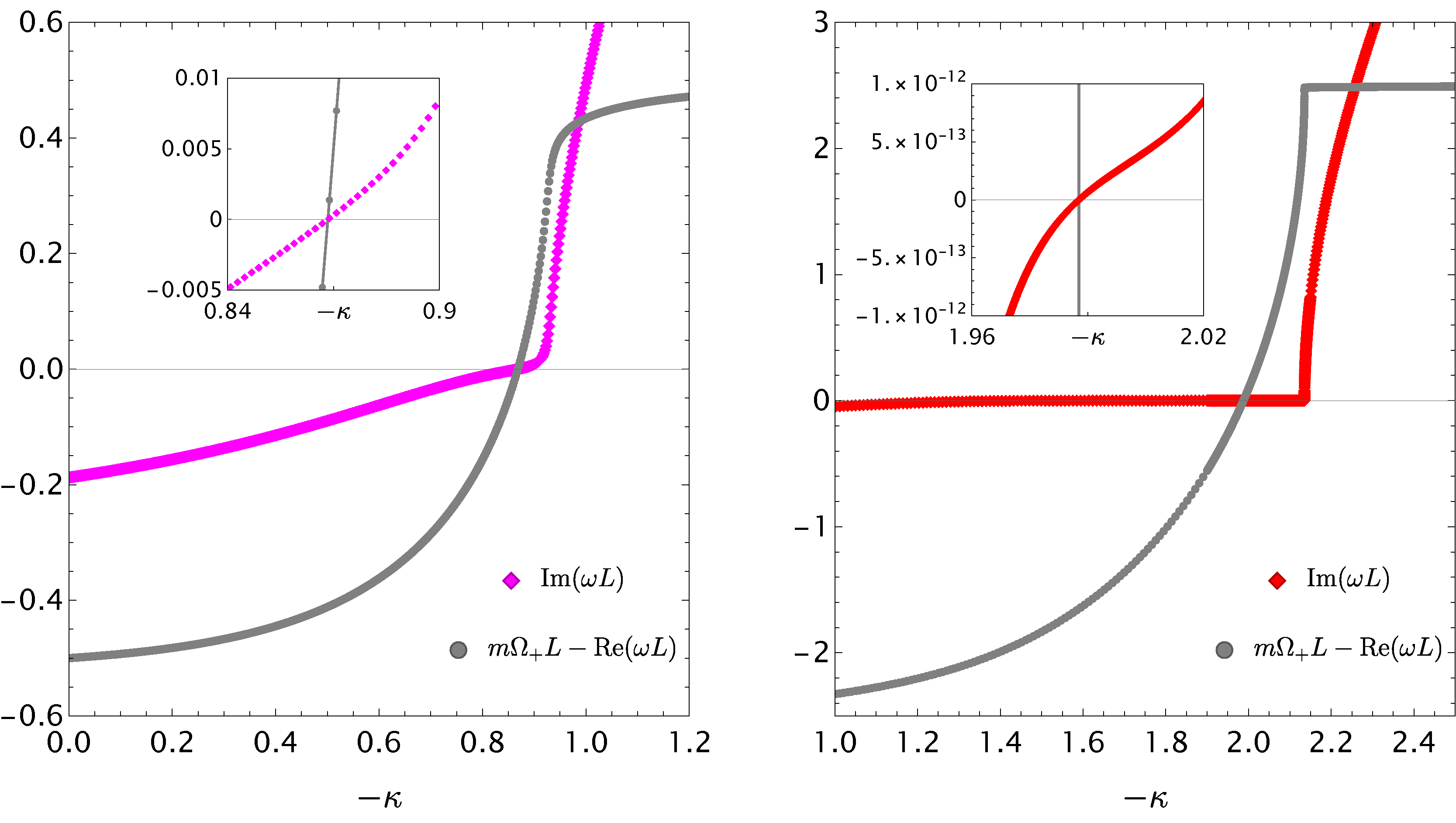}
       \caption{$\text{Im}(\omega L)$ (diamonds) and $m\Omega_+ L - \text{Re}(\omega L)$ (gray circles) as a function of $\kappa$ for the unstable $n=0$ mode of a BTZ black hole with $\{\hat{M},\hat{J}\} = \{5/16,1/4\}$ and $\mu^2L^2=-15/16$. {\bf Top panel:} $m=0$. {\bf Bottom-left panel:} $m=1$. {\bf Bottom-right panel:} $m=5$. The instability onset occurs at $\kappa=\kappa^{\rm BTZ}_{m,\hat{\mu}^2}$, where ${\rm Im}\,\hat{\omega}=0$ and $m\Omega_+ - {\rm Re}\,\hat{\omega}=0$. The magnitude of $|\kappa^{\rm BTZ}_{m,\hat{\mu}^2}|$ increases with $m$, approximately $-0.33$, $-0.87$, and $-1.99$ for $m=0,1,5$, respectively.} 
    \label{fig:BTZw-kappa:m1_m5}
\end{figure}

A third universal aspect is that the onset of the $n=0$ instability occurs at increasingly negative $\kappa^{\rm BTZ}_{m,\hat{\mu}^2}\text{\footnotesize $(\hat{M},\hat{J})$}$ as $m$ increases. That is, if BTZ is unstable to a finite-$m$ mode, it is necessarily unstable to $m=0$; thus, BTZ's stability under double-trace perturbations is controlled by the $m=0$, $n=0$ mode. This is clearly illustrated for the BTZ with $\{\hat{M},\hat{J}\} = \{5/16,1/4\}$ ($\{y_-,y_+\}=\{1/4,1/2\}$) and $\mu^2L^2=-15/16$ in Figs.~\ref{fig:BTZw-kappa:m0}-\ref{fig:BTZw-kappa:m1}, and we have checked it holds universally for a wide range of $\{\hat{M},\hat{J}\}$, $\mu^2L^2\in\{-15/16,-8/9,-3/4,-0.1\}$ and $0\le m\le5$ (see Figs.~\ref{fig:BTZw-kappa:m1_m5},~\ref{fig:3D_kappa_onset} and Appendix~\ref{sec:AppOtherMasses}). For any scalar mass in the double-trace range \eqref{2xTrace:rangeMass}, the negative onset $\kappa=\kappa^{\rm BTZ}_{m,\hat{\mu}^2}\text{\footnotesize $(\hat{M},\hat{J})$}$ decreases with $m$, as shown in Fig.~\ref{fig:BTZw-kappa:m1_m5} for $m=0,1,5$, with approximate onset values $-0.33,-0.87,-1.99$, respectively.  

This property distinguishes double-trace instabilities in BTZ (and more generally AdS) from other black hole instabilities, where typically the unstable region grows with $m$ (e.g., AdS superradiant \cite{Cardoso:2004hs,Dias:2013sdc,Brito:2015oca}, ultraspinning \cite{Dias:2009iu}, Gregory-Laflamme \cite{Dias:2022mde}, and scalar condensation instabilities \cite{Dias:2010ma}). An exception is the superradiant instability of rotating black strings \cite{Dias:2022mde}, where high-$m$ instability implies low-$m$ instability—a behaviour mirrored by BTZ. This has been termed a ``finite-$m$ instability'', with BTZ providing another example. Double-trace instabilities are special because BTZ can also be unstable for $m=0$, i.e., non-spinning modes. This is analogous to planar static AdS black holes, which can be unstable to double-trace scalar condensation even when uncharged \cite{Faulkner:2010gj,Dias:2013bwa}.  

Furthermore, the $m=0$ instability is dominant: Fig.~\ref{fig:BTZw-kappa:m1_m5} shows that when $m\ge1$ modes are unstable, ${\rm Im}\,\hat{\omega}|_{m=0}>{\rm Im}\,\hat{\omega}|_{m\ge1}$.  

A fourth universal property, also visible in Fig.~\ref{fig:BTZw-kappa:m1_m5}, is that at the onset of the $n=0$ instability, $\kappa=\kappa^{\rm BTZ}_{m,\hat{\mu}^2}\text{\footnotesize $(\hat{M},\hat{J})$}$, both ${\rm Im}\,\hat{\omega}=0$ and ${\rm Re}\,\hat{\omega}-m \hat{\Omega}_+=0$ (including for $m=0$). This is expected: the onset of any AdS black hole instability with an ergoregion must satisfy these two conditions \cite{Green:2015kur}. While sometimes associated with superradiance, this is a general feature of AdS instabilities, reflecting the sign change of energy and angular momentum fluxes across the horizon, which are proportional to $\hat{\omega}-m\hat{\Omega}_+ (more in the next subsection)$.

\subsection{Origin of the double-trace instability in BTZ and \texorpdfstring{AdS$_3$}{AdS3}}\label{sec:BTZorigin}
In this subsection, we discuss the mechanisms responsible for the origin of double-trace instabilities in BTZ and AdS$_3$. Ultimately, these instabilities arise from the fact that double-trace boundary conditions with $\kappa < 0$ permit an incoming energy flux across the asymptotic boundary.

Given a Killing vector field $\xi$ and a conserved stress-energy tensor $T$, we define the associated conserved current
\begin{subequations}
\begin{align}
&J^{\xi}_\mu = - T_{\mu\nu} \, \xi^\nu \,,\\
&J^{\xi} \equiv J^{\xi}_\mu \, {\rm d}x^\mu \,,\\
&{\rm d} \star J^{\xi} = 0 \,.
\end{align}
\end{subequations}
The conservation of $J^\xi$ follows from the symmetry and covariant conservation of $T_{\mu\nu}$ together with the Killing property of $\xi$. This allows us to define a conserved energy associated with a hypersurface $\Sigma$ as
\begin{equation}
E^{\xi}[\Sigma] = \int_{\Sigma} \star J^\xi \,.
\end{equation}
Typically, $\Sigma$ is chosen as a hypersurface with a causal normal, well-behaved near the horizon and the conformal boundary, and $\xi$ as a Killing vector field that is causal near the conformal boundary. A natural example is $\xi = \partial/\partial v$ in Eddington–Finkelstein coordinates.

For Dirichlet boundary conditions, where only the standard decay at infinity is allowed, one may take $T_{ab}$ as the usual stress-energy tensor coupling to the Einstein equations:
\begin{equation}
T_{\mu \nu} = \nabla_\mu \Phi \nabla_\nu \Phi^{\dagger} + \nabla_\mu \Phi^{\dagger} \nabla_\nu \Phi - g_{\mu \nu} \nabla_\gamma \Phi \nabla^\gamma \Phi^{\dagger} - g_{\mu\nu} \,\mu^2 |\Phi|^2 \,.
\end{equation}
However, for double-trace deformations, the energy defined with this $T_{\mu \nu}$ fails to converge near the conformal boundary for any causal $\xi$.

Following Breitenlohner and Freedman \cite{Breitenlohner:1982jf}, one can modify the stress-energy tensor to ensure the energy remains finite for the chosen boundary conditions:
\begin{equation}
\widetilde{T}_{\mu \nu} = T_{\mu \nu} + \varpi \Big( g_{\mu \nu} \Box - \nabla_\mu \nabla_\nu + R_{\mu \nu} \Big) |\Phi|^2 \,,
\end{equation}
where $\varpi$ is chosen as
\begin{equation}
\varpi = \frac{\Delta_-}{1 + 2\Delta_-} \,,
\end{equation}
to render the energy finite. For a \emph{test field} propagating on a constant–scalar–curvature spacetime, one can check that $\nabla^\mu \widetilde{T}_{\mu \nu}=0$ for this choice of $\varpi$.

Having a current also allows us to define the energy flux at a given moment of time through the future event horizon and the conformal boundary:
\begin{equation}
\Phi_{\mathcal{H}^+_v} \equiv - \int_{\mathcal{H}_v} \star J^\xi
\,,\qquad
\Phi_{\mathcal{I}_v} \equiv \int_{\mathcal{I}_v} \star J^\xi \,,
\end{equation}
where $\mathcal{H}^+_v$ is the intersection of a partial Cauchy surface of constant $v$ with the future event horizon, and $\mathcal{I}_v$ is defined similarly at the conformal boundary.

Since the BTZ spacetime is regular in Eddington–Finkelstein coordinates (see~\eqref{ingoingEF}), a natural choice is $\xi = \partial/\partial v$, which is timelike near the conformal boundary. The energy flux at a given moment of time through the conformal boundary is then
\begin{equation}
\Phi_{\mathcal{I}_v} = \frac{4 (1 - \Delta_-)}{1 + 2 \Delta_-} \frac{|\alpha|^2}{L^2} \, \kappa \, {\rm Im}\,\omega \,.
\end{equation}
For an instability to exist, we require ${\rm Im}\,\omega > 0$. Instabilities occur only if $\kappa < 0$, which corresponds to a \emph{negative} energy flux through the conformal boundary, thereby explaining the origin of the instability. A similar conclusion can be obtained using the renormalised holographic stress--energy tensor, as in \cite{Kinoshita:2023iad}, but we will not present it here.

Finally, integrating the conservation law ${\rm d} \star J^\xi = 0$ over a spacetime slab
\begin{equation}
\mathcal{M}_{v_-}^{v_+}=\{(v,r,\varphi):\quad v \in [v_-, v_+] \quad \land\quad  r\in[r_+,+\infty)\}
\end{equation}
gives the \emph{energy balance equation}:
\begin{equation}
E^\xi[\Sigma_{v_+}] - E^\xi[\Sigma_{v_-}]
= \int_{v_-}^{v_+} \Phi_{\mathcal{H}^+_v} \, {\rm d}v
+ \int_{v_-}^{v_+} \Phi_{\mathcal{I}_v} \, {\rm d}v \,,
\end{equation}
where the right-hand side is the net flux through the horizon and conformal boundary during the interval $[v_-,v_+]$. This makes explicit that the change in energy in the slab is accounted for entirely by the fluxes through the boundaries, providing a clear interpretation of the role of negative flux at the conformal boundary in driving the instability.

At the onset of the instability, we have ${\rm Im}\,\omega = 0$, so the energy is conserved between slices:
\begin{equation}
E^\xi[\Sigma_{v_+}] = E^\xi[\Sigma_{v_-}] \,,
\end{equation}
and $\Phi_{\mathcal{I}_v}=0$. Hence, the only contribution to the energy balance comes from the flux through the horizon, $\Phi_{\mathcal{H}^+_v}$, which evaluates to
\begin{equation}
\Phi_{\mathcal{H}^+_v} = 4 \pi r_+ \, \left| \Phi(r_+) \right|^2 (\omega - m \Omega_+)\, \omega  \,.
\end{equation}
This immediately shows that, at the onset of the instability, either $\omega = 0$ or $\omega = m \Omega_+$.

For completeness, we should also consider another mechanism often responsible for black hole instabilities. In many AdS black holes with an extremal configuration, certain instabilities can be diagnosed by analysing the perturbation equations (of any spin, not only scalars) in the near-horizon limit~\cite{Gubser:2008px,Hartnoll:2008kx,Dias:2010ma,Durkee:2010ea,Dias:2011tj,Hollands:2014lra,Dias:2016pma,Dias:2019fmz,Dias:2020ncd}. In this limit, the perturbation equation typically reduces to a Klein-Gordon equation on AdS$_2$ with an effective mass. If this effective mass violates the AdS$_2$ Breitenlohner-Freedman bound, one expects the geometry to be unstable. This criterion is sufficient $-$ though not necessary $-$ for an instability in asymptotically AdS spacetimes\footnote{For asymptotically flat spacetimes, one must further restrict to modes that preserve angular momentum and/or electric charge \cite{Hollands:2014lra}.} \cite{Hollands:2014lra}, and when satisfied it often clarifies the physical origin of the instability. It has been particularly useful for understanding instabilities of scalar fields in asymptotically AdS black holes, independent of the charges or angular momenta involved.

Motivated by these successful applications, in Appendix~\ref{sec:AppNHanalysis} we perform a near-horizon analysis of the extremal BTZ geometry. For the present system, however, this approach is inconclusive and provides no additional insight into the nature of the BTZ instability. This further supports the conclusion that the double-trace instability in BTZ $-$ and more generally in AdS$_3$ $-$ originates from the non-trivial flux of energy and angular momentum through the asymptotic AdS boundary.

\subsection{Varying \texorpdfstring{$\kappa$}{kappa}: Double-trace instability onsets in global \texorpdfstring{AdS$_3$}{AdS3} and BTZ}\label{sec:BTZonset2}
In Section~\ref{sec:Ad3} we showed that the fundamental radial overtone $n=0$ (for any $m$) in global AdS$_3$ becomes unstable under double-trace perturbations whenever the parameter $\kappa$ satisfies $\kappa < \kappa^{\hbox{\tiny AdS}}_{m,\hat{\mu}^2}$, where $\kappa^{\hbox{\tiny AdS}}_{m,\hat{\mu}^2}$ is a negative constant depending only on $\hat{\mu}$ and $m$ and given in \eqref{kAdS-onset:mValues}. In Section~\ref{sec:BTZonset1}, we found $-$ after a preliminary analysis $-$ that the $n=0$ overtone of BTZ (again for any $m$) also becomes unstable when $\kappa < \kappa^{\hbox{\tiny BTZ}}_{m,\hat{\mu}^2}(\hat{M},\hat{J})$, where this threshold depends not only on $\hat{\mu}$ and $m$ but also on the BTZ parameters $(\hat{M},\hat{J})$.

This naturally prompts several questions: Is BTZ unstable only in regimes where AdS$_3$ is already unstable? Can BTZ be unstable while AdS$_3$ remains stable? More generally, can BTZ black holes exhibit double-trace instabilities even when global AdS$_3$ does not?

To address these questions, one should compute $\kappa^{\hbox{\tiny BTZ}}_{m,\hat{\mu}^2}\hbox{\footnotesize $(\hat{M},\hat{J})$}$. This can be done analytically. Recall that the double-trace frequency quantization in BTZ is given by $\beta=\kappa\,\alpha$, i.e. by the $\Gamma$-condition \eqref{eqn:BTZ_Gamma_criterion}, which depends on $\kappa,\hat{\omega},m,\hat{\mu},y_{\pm}$. At the onset of instability one has ${\rm Re}\,\hat{\omega}=m\hat{\Omega}_+$ and ${\rm Im}\,\hat{\omega}=0$. Thus, to find $\kappa^{\hbox{\tiny BTZ}}_{m,\hat{\mu}^2}$ we simply insert $\hat{\omega}=m\hat{\Omega}_+=m\,y_-/y_+$ into \eqref{eqn:BTZ_Gamma_criterion} and use the $y_\pm(\hat{M},\hat{J})$ map in \eqref{ypmFromMJ}. We then find that, for any $-1<\hat{\mu}^2<0$, the onset value of the double-trace parameter for the BTZ instability of the co-rotating $m\geq 0$, $n=0$ mode is
\begin{align} \label{kBTZ-onset}
    \kappa^{\hbox{\tiny BTZ}}_{m, \hat{\mu}^2} \hbox{\footnotesize $(\hat{M},\hat{J})$}= & \left( \hat{M}^{2} - \hat{J}^2 \right)^{\!\frac{\sqrt{1 + \hat{\mu}^2}}{2}}
    \!\!\frac{\,\Gamma\!\left[-\sqrt{1 + \hat{\mu}^2}\right]}{\Gamma\!\left[\sqrt{1 + \hat{\mu}^2}\right]} 
    \frac{\,\Gamma\!\left[\frac{1}{2}(\Delta_+- i\,m \,\chi)\right]}{\,\Gamma\!\left[\frac{1}{2}(\Delta_- - i\,m \,\chi)\right]} \, 
    \frac{\,\Gamma\!\left[\frac{1}{2}(\Delta_+ + i\,m \,\chi )\right]}{\,\Gamma\!\left[\frac{1}{2}(\Delta_-+ i\,m \,\chi )\right]}, \nonumber \\
& \hbox{with} \quad \chi\equiv  \hat{J}^{-1}\left( \sqrt{\,\hat{M} - \hat{J}\,} - \sqrt{\,\hat{M} + \hat{J}\,} \right) \,.
\end{align}
This $\kappa^{\hbox{\tiny BTZ}}_{m,\hat{\mu}^2}\hbox{\footnotesize $(\hat{M},\hat{J})$}$ is always negative and finite for any $\{\hat{M},\hat{J}\}$ for which BTZ black holes exist, namely for $J \leq M L$. BTZ is unstable for any co-rotating $n=0$ double-trace boundary condition satisfying $\kappa < \kappa^{\hbox{\tiny BTZ}}_{m,\hat{\mu}^2}\hbox{\footnotesize $(\hat{M},\hat{J})$}$ (and linearly stable otherwise). However, as discussed in detail below (and already mentioned), it should be stressed that BTZ is not unstable in the strict $\kappa \to -\infty$ (Dirichlet) limit, which is “discontinuous,” just as in the AdS$_3$ case.

\begin{figure}[ht]
    \centering
    \includegraphics[width=0.49\linewidth]{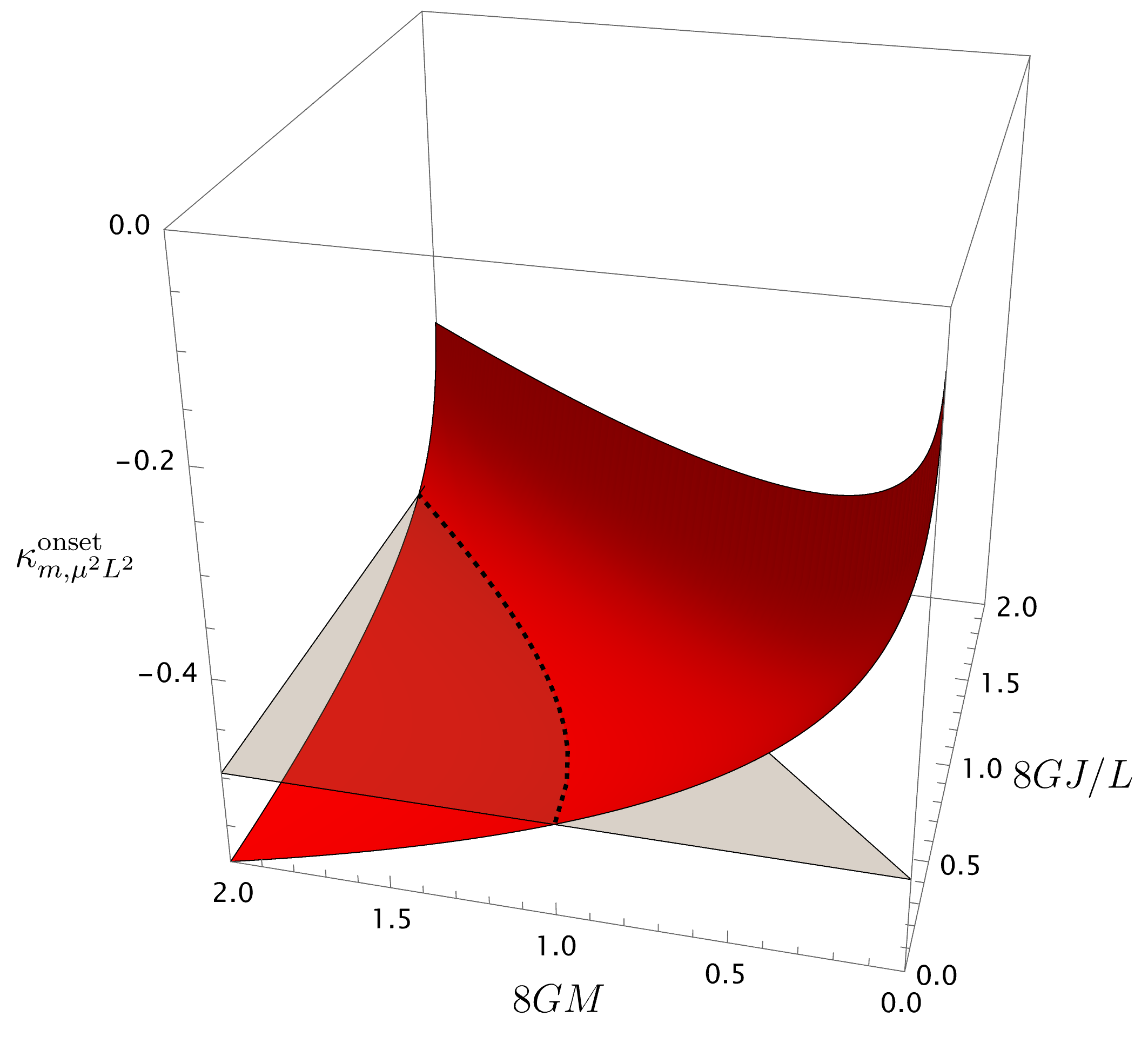}
    \includegraphics[width=0.49\linewidth]{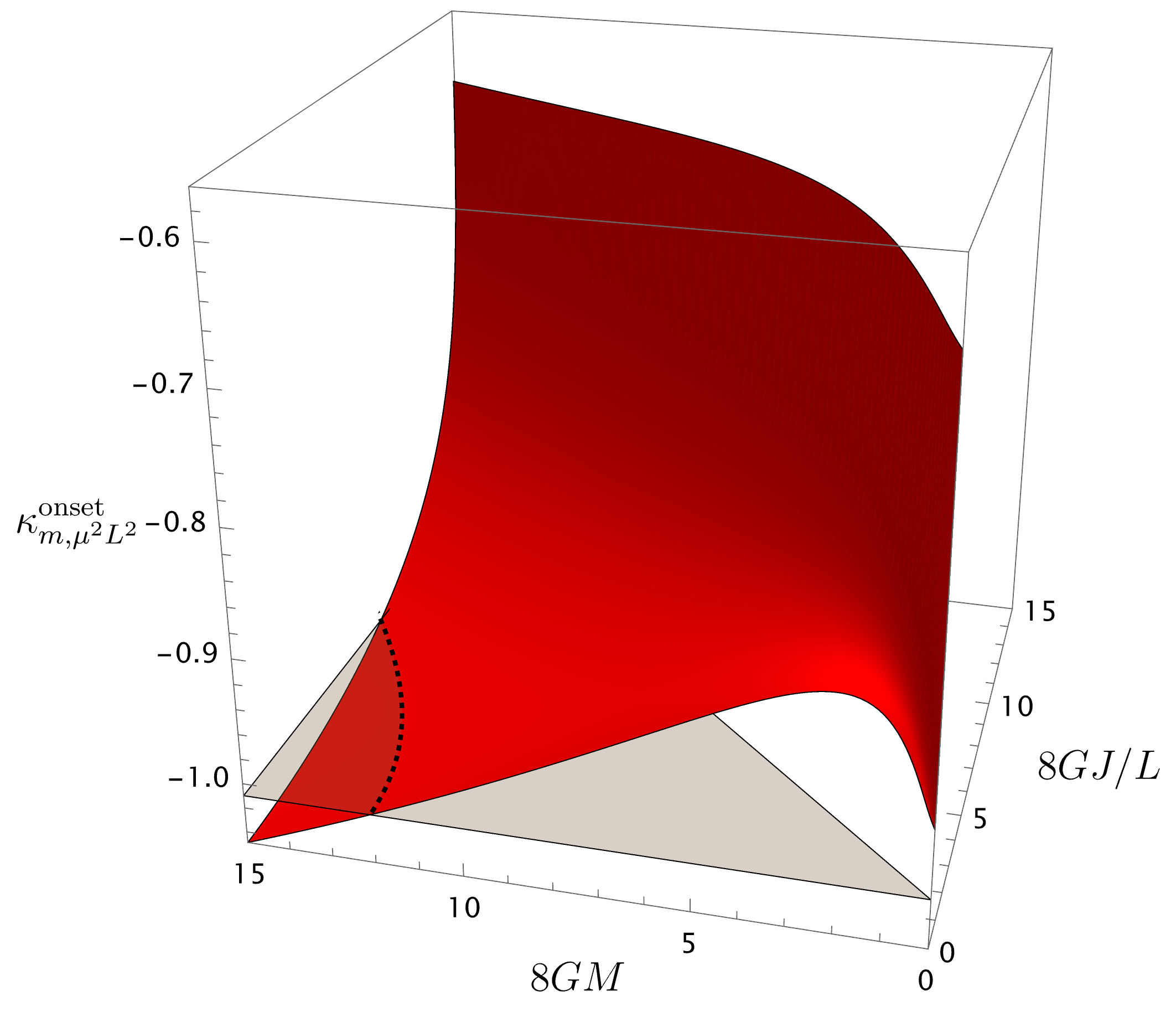}
    \caption{Instability onset values of the double-trace parameters $\kappa^{\hbox{\tiny BTZ}}_{m,\hat{\mu}^2}\hbox{\footnotesize $(\hat{M},\hat{J})$}$ (red surface) and $\kappa^{\hbox{\tiny AdS}}_{m,\hat{\mu}^2}$ (flat grey surface) for $n=0$ double-trace perturbations with $\bm{m=0}$ (left panel) and $\bm{m=1}$ (right panel), both for $\mu^2L^2=-15/16$. Regular BTZ black holes exist for $J\leq M L$. BTZ is unstable for any co-rotating $n=0$ double-trace boundary condition with $\kappa < \kappa^{\hbox{\tiny BTZ}}_{m,\hat{\mu}^2}\hbox{\footnotesize $(\hat{M},\hat{J})$}$ (and linearly stable otherwise). AdS$_3$ is unstable for any $n=0$ double-trace boundary condition with $\kappa < \kappa^{\hbox{\tiny AdS}}_{m,\hat{\mu}^2}$ (and linearly stable otherwise). The $m=0$ BTZ onset surface (left panel) always lies above the $m=1$ (right panel, and in fact any $m\geq 1$) BTZ onset surface. Likewise, the $m=0$ AdS$_3$ onset plane (left panel) always lies above the $m=1$ (right panel, and any $m\geq 1$) AdS$_3$ onset plane; this holds for any scalar mass $-1<\hat{\mu}^2<0$. See Fig.~\ref{fig:3D_kappa_onset-OtherMasses} in Appendix~\ref{sec:AppOtherMasses} for corresponding plots with $\mu^2L^2=\{-8/9,-3/4,-1/10\}$.}
    \label{fig:3D_kappa_onset}
\end{figure}

In Fig.~\ref{fig:3D_kappa_onset} we plot $\kappa^{\hbox{\tiny BTZ}}_{m,\hat{\mu}^2}$ (red surface)  $-$ given by \eqref{kBTZ-onset}  $-$ as a function of the BTZ mass $\hat{M}$ and angular momentum $\hat{J}$ in the region $\hat{J} \leq \hat{M} L$ where BTZ exists as a regular black hole. We again use $\mu^2L^2=-15/16$ to illustrate our results (the qualitative behaviour is similar for $-1<\hat{\mu}^2<0$, and Fig.~\ref{fig:3D_kappa_onset-OtherMasses} in Appendix~\ref{sec:AppOtherMasses} shows analogous plots for $\mu^2L^2=\{-8/9,-3/4,-1/10\}$). We present results for $m=0$ (left panel) and $m=1$ (right panel); cases with $m \geq 2$ are qualitatively similar.

In Fig.~\ref{fig:3D_kappa_onset} we also display the global AdS$_3$ instability onset $\kappa^{\hbox{\tiny AdS}}_{m,\hat{\mu}^2}$ $-$ given by \eqref{kAdS-onset} for $\mu^2L^2=-15/16$ $-$ which is the negative constant \eqref{kAdS-onset:mValues}, represented by the flat grey surface (plotted only in the region $J \leq M L$ where BTZ exists). Recall that AdS$_3$ is unstable for any $n=0$ double-trace boundary condition satisfying $\kappa < \kappa^{\hbox{\tiny AdS}}_{m,\hat{\mu}^2}$ (and linearly stable otherwise).

The BTZ and AdS$_3$ onset surfaces intersect along a curve, dividing the parameter space into two main regions. Away from BTZ extremality $J=M L$ $-$ i.e., at large $\hat{M}$ and small $\hat{J}$ $-$ the BTZ onset surface lies below the AdS$_3$ onset surface, $\kappa^{\hbox{\tiny BTZ}}_{m,\hat{\mu}^2}\hbox{\footnotesize $(\hat{M},\hat{J})$} < \kappa^{\hbox{\tiny AdS}}_{m,\hat{\mu}^2}$. In the region connected to extremal BTZ, the BTZ onset surface lies above the AdS$_3$ onset surface, $\kappa^{\hbox{\tiny BTZ}}_{m,\hat{\mu}^2}\hbox{\footnotesize $(\hat{M},\hat{J})$} > \kappa^{\hbox{\tiny AdS}}_{m,\hat{\mu}^2}$.

Let us analyse Fig.~\ref{fig:3D_kappa_onset} in detail. Consider first the $\{\hat{M},\hat{J}\}$ region far from extremality (large $\hat{M}$, small $\hat{J}$). In both plots, if $\kappa$ lies above both onset surfaces, i.e. if $\kappa>\kappa^{\hbox{\tiny AdS}}_{m,\hat{\mu}^2}>\kappa^{\hbox{\tiny BTZ}}_{m,\hat{\mu}^2}\hbox{\footnotesize $(\hat{M},\hat{J})$}$ or if $\kappa>\kappa^{\hbox{\tiny BTZ}}_{m,\hat{\mu}^2}\hbox{\footnotesize $(\hat{M},\hat{J})$}>\kappa^{\hbox{\tiny AdS}}_{m,\hat{\mu}^2}$, then both BTZ and AdS$_3$ are linearly stable. If $\kappa$ lies between the grey and red surfaces, $\kappa^{\hbox{\tiny AdS}}_{m,\hat{\mu}^2} > \kappa > \kappa^{\hbox{\tiny BTZ}}_{m,\hat{\mu}^2}\hbox{\footnotesize $(\hat{M},\hat{J})$}$, we encounter the somewhat peculiar situation where BTZ is linearly stable to $n=0$ perturbations while AdS$_3$ is unstable (noting, however, that “vacuum BTZ’’ with $M=0=J$ has a mass gap relative to AdS$_3$). Conversely, if $\kappa$ lies below both surfaces, i.e. if $\kappa < \kappa^{\hbox{\tiny BTZ}}_{m,\hat{\mu}^2}\hbox{\footnotesize $(\hat{M},\hat{J})$} < \kappa^{\hbox{\tiny AdS}}_{m,\hat{\mu}^2}$ or if $\kappa < \kappa^{\hbox{\tiny AdS}}_{m,\hat{\mu}^2} < \kappa^{\hbox{\tiny BTZ}}_{m,\hat{\mu}^2}\hbox{\footnotesize $(\hat{M},\hat{J})$}$, then both BTZ and AdS$_3$ are unstable to $n=0$ perturbations.

Finally, consider the remaining scenario in the BTZ parameter region connected to extremality, when $\kappa$ lies between the red and grey surfaces, $\kappa^{\hbox{\tiny BTZ}}_{m,\hat{\mu}^2}\hbox{\footnotesize $(\hat{M},\hat{J})$} > \kappa > \kappa^{\hbox{\tiny AdS}}_{m,\hat{\mu}^2}$. In this case, BTZ is unstable while AdS$_3$ is linearly ($n=0$) stable. This is likely the physically most interesting regime, and it will be the focus of Fig.~\ref{fig:m0_kappam4o10_onset_check} and subsequent figures, unless stated otherwise (e.g. in the right panel of Fig.~\ref{fig:m0_3d_plot}). Double-trace perturbations with $-1<\hat{\mu}^2<0$ produce AdS$_3$ and BTZ onset surfaces with structure very similar to that in Fig.~\ref{fig:3D_kappa_onset}; see Fig.~\ref{fig:3D_kappa_onset-OtherMasses} in Appendix~\ref{sec:AppOtherMasses} for the corresponding plots with $\mu^2L^2=\{-8/9,-3/4,-1/10\}$.

\begin{figure}[ht]
    \centering
    \includegraphics[width=0.49\linewidth]{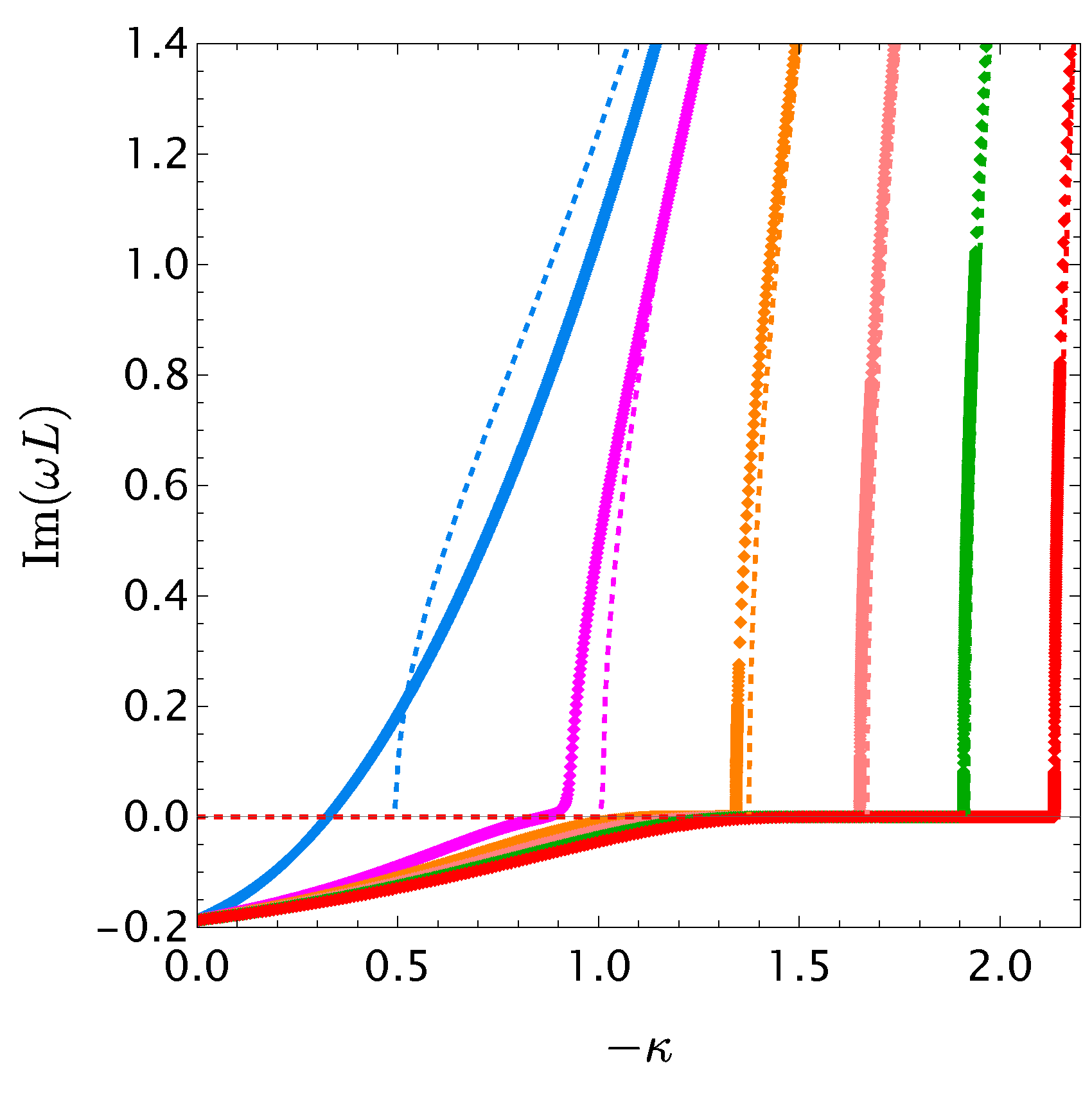}
    \includegraphics[width=0.49\linewidth]{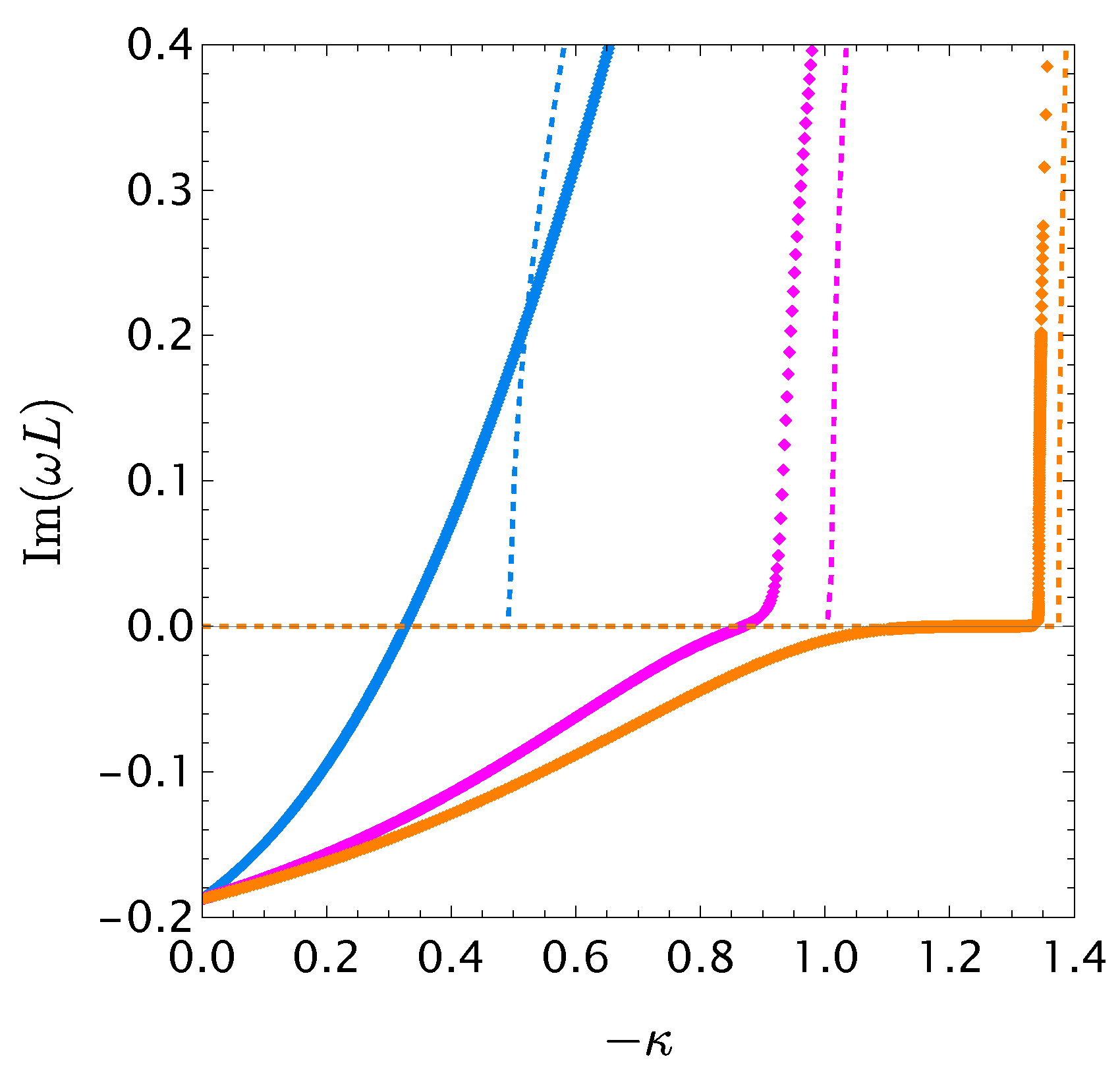}
    \caption{$\text{Im}(\omega L)$ as a function of $\kappa$ for $n=0$ co-rotating modes in global AdS$_3$ (dashed lines) and in a BTZ black hole (diamonds) with $\{\hat{M},\hat{J}\}=\{5/16,1/4\}$ ($\mu^2L^2=-15/16$). The curves correspond to the modes $m=0$ (blue), $m=1$ (purple), $m=2$ (orange), $m=3$ (pink), $m=4$ (green), and $m=5$ (red). The right panel is a zoom of the left panel, showing only $m=0,1,2$. When $\text{Im}(\omega L)>0$, the background is unstable to the corresponding $m$-mode. See Fig.~\ref{fig:m0tom5_btz_VS_ads_instability-OtherMasses} in Appendix~\ref{sec:AppOtherMasses} for the associated plots with $\mu^2L^2=\{-8/9,-3/4,-1/10\}$.}

    \label{fig:m0tom5_btz_VS_ads_instability}
\end{figure}

To analyse the near-extremal BTZ region, in Fig.~\ref{fig:m0tom5_btz_VS_ads_instability} we consider a BTZ black hole with  
$\{\hat{M},\hat{J}\} = \{5/16,1/4\}$, in the $\{\hat{M},\hat{J}\}$ region connected to extremal BTZ in Fig.~\ref{fig:3D_kappa_onset} (the same values used in previous illustrations). For this BTZ, $\kappa$ can lie in the window $\kappa^{\hbox{\tiny BTZ}}_{m,\hat{\mu}^2}\hbox{\footnotesize $(\hat{M},\hat{J})$} > \kappa > \kappa^{\hbox{\tiny AdS}}_{m,\hat{\mu}^2}$, where BTZ is unstable but AdS$_3$ is linearly ($n=0$) stable. We follow the evolution, as $\kappa$ decreases from $0$ to $-\infty$, of the imaginary part of the $n=0$ co-rotating frequency, corresponding to a vertical path at $\{\hat{M},\hat{J}\} = \{5/16,1/4\}$ in Fig.~\ref{fig:3D_kappa_onset}.  

We plot ${\rm Im}\,\hat{\omega}$ for co-rotating $n=0$ modes in BTZ (diamonds) and global AdS$_3$ (dashed lines) with $m = 0$ (blue), $m = 1$ (purple), $m = 2$ (orange), $m = 3$ (pink), $m = 4$ (green), and $m = 5$ (red). The right plot is a zoom of the left, showing only $m=0,1,2$. Focus first on the $m=0$ blue curves: both AdS (dashed) and BTZ (diamonds) start with negative ${\rm Im}\,\hat{\omega}$ at $\kappa=0$ but become unstable as $-\kappa$ increases. Specifically, BTZ becomes unstable at $\kappa^{\hbox{\tiny BTZ}}_{m=0,-15/16}\hbox{\footnotesize $(5/16,1/4)$}\sim -0.33$, while AdS$_3$ becomes unstable only at $\kappa<\kappa^{\hbox{\tiny AdS}}_{m=0,-15/16}\sim -0.495$. Hence, for $\kappa$ in between, BTZ is unstable while AdS$_3$ is stable.

A similar trend occurs for higher $m$ modes: both AdS$_3$ and BTZ become unstable, but the BTZ onset for a given $m$ occurs at smaller $-\kappa$ than the corresponding AdS$_3$ onset. Moreover, the instability onset shifts to higher $-\kappa$ as $m$ increases. For instance, for $|\kappa|>|\kappa^{\hbox{\tiny BTZ}}_{m=0,-15/16}|\sim 0.33$, BTZ is unstable to $m=0$; then there exists a window $|\kappa^{\hbox{\tiny BTZ}}_{m=0,-15/16}| < |\kappa| < |\kappa^{\hbox{\tiny BTZ}}_{m=1,-15/16}|$ where it is stable to $m=1$, becoming unstable to $m=1$ only for $|\kappa| > |\kappa^{\hbox{\tiny BTZ}}_{m=1,-15/16}|\sim 0.87$, and so on. This pattern holds for any scalar mass $-1<\hat{\mu}^2<0$ and follows from Fig.~\ref{fig:3D_kappa_onset} ($\mu^2L^2=-15/16$) and Fig.~\ref{fig:3D_kappa_onset-OtherMasses} ($\mu^2L^2=\{-8/9,-3/4,-1/10\}$), where the $m=0$ BTZ onset surface (left panel) lies above $m=1$ (right panel; and any $m\geq 1$), and similarly for AdS$_3$.

Thus, for a given $\kappa$, the stability of the $n=0$, $m=0$ mode determines the overall stability: $|\kappa^{\hbox{\tiny BTZ}}_{m=0,\hat{\mu}^2}|$ is smaller than the onset for any other $m$, and for $\kappa>|\kappa^{\hbox{\tiny BTZ}}_{m=0,\hat{\mu}^2}|$, ${\rm Im}\,\hat{\omega}$ for $m=0$ is the largest when coexisting with higher-$m$ instabilities. This illustrates one of the main results of our paper. Analogous behaviour for $\mu^2L^2 = \{-8/9,-3/4,-1/10\}$ is shown in Fig.~\ref{fig:m0tom5_btz_VS_ads_instability-OtherMasses}.

Finally, Fig.~\ref{fig:3D_kappa_onset} shows a qualitative difference between the $m=0$ (left panel) and $m=1$ (right panel) onset surfaces near the corner $\{\hat{M},\hat{J}\}=\{0,0\}$. While $\kappa^{\hbox{\tiny BTZ}}_{m=0,\hat{\mu}^2}$ increases monotonically towards the corner, the $m=1$ red surface develops a nontrivial structure: along the $J=0$ curve, $\kappa^{\hbox{\tiny BTZ}}_{m=1,\hat{\mu}^2}$ first rises, reaches a maximum, then drops near $M=0$. This feature will be discussed in detail in Fig.~\ref{fig:m1_kappa_onset_curves}.

Ultimately, fixing the theory requires specifying both the scalar mass and the double-trace parameter $\kappa$. Henceforth, we focus on the dynamics at fixed $\kappa$.
 
To further illustrate the properties of the $n=0$ BTZ instability and its dependence on $\kappa$, we now focus on the $\{\hat{J},\hat{M}\}$ plane and fix $\kappa$ to a constant value, say $\kappa = c$ (for a given $\hat{\mu}$ and $m$). We then ask for the curve $\hat{M} = \hat{M}(\hat{J})$ that describes the instability onset, \ie the curve where $\kappa^{\hbox{\tiny BTZ}}_{m,\hat{\mu}^2}\hbox{\footnotesize $(\hat{M},\hat{J})$} = c$ (a level surface). This corresponds to taking a horizontal cut at $\kappa = c$ in Fig.~\ref{fig:3D_kappa_onset}, finding its intersection with the BTZ onset red surface, and projecting this onset curve down onto the $\hat{J}-\hat{M}$ plane.

Recall that at the onset, $\kappa$ is given by \eqref{kBTZ-onset}, which we now solve for $\hat{M}$. For $m=0$, the $\Gamma$-functions in \eqref{kBTZ-onset} do not depend on $\hat{M}$ or $\hat{J}$, so solving for the onset $M$ is straightforward, yielding
\begin{align}\label{kBTZ-onsetMJ-m0}
    \hat{M}\big|_{\hbox{\tiny $m\! =\! 0$}}^{\hbox{\tiny onset}}  = \sqrt{\hat{J}^2-\left(\frac{\Gamma \left(\sqrt{1+\hat{\mu}^2}\right) \Gamma \left(\frac{1}{2} \Delta_- \right)^2}
    {\Gamma \left(-\sqrt{1+\hat{\mu}^2}\right) \Gamma \left(\frac{1}{2}\Delta_+ \right)^2}
   \, \kappa \right)^{\frac{2}{\sqrt{1+\hat{\mu}^2}}}  } \, \Bigg|_{\kappa=c}\,.
\end{align}

\begin{figure}[ht]
    \centering
    \includegraphics[width=0.55\textwidth]{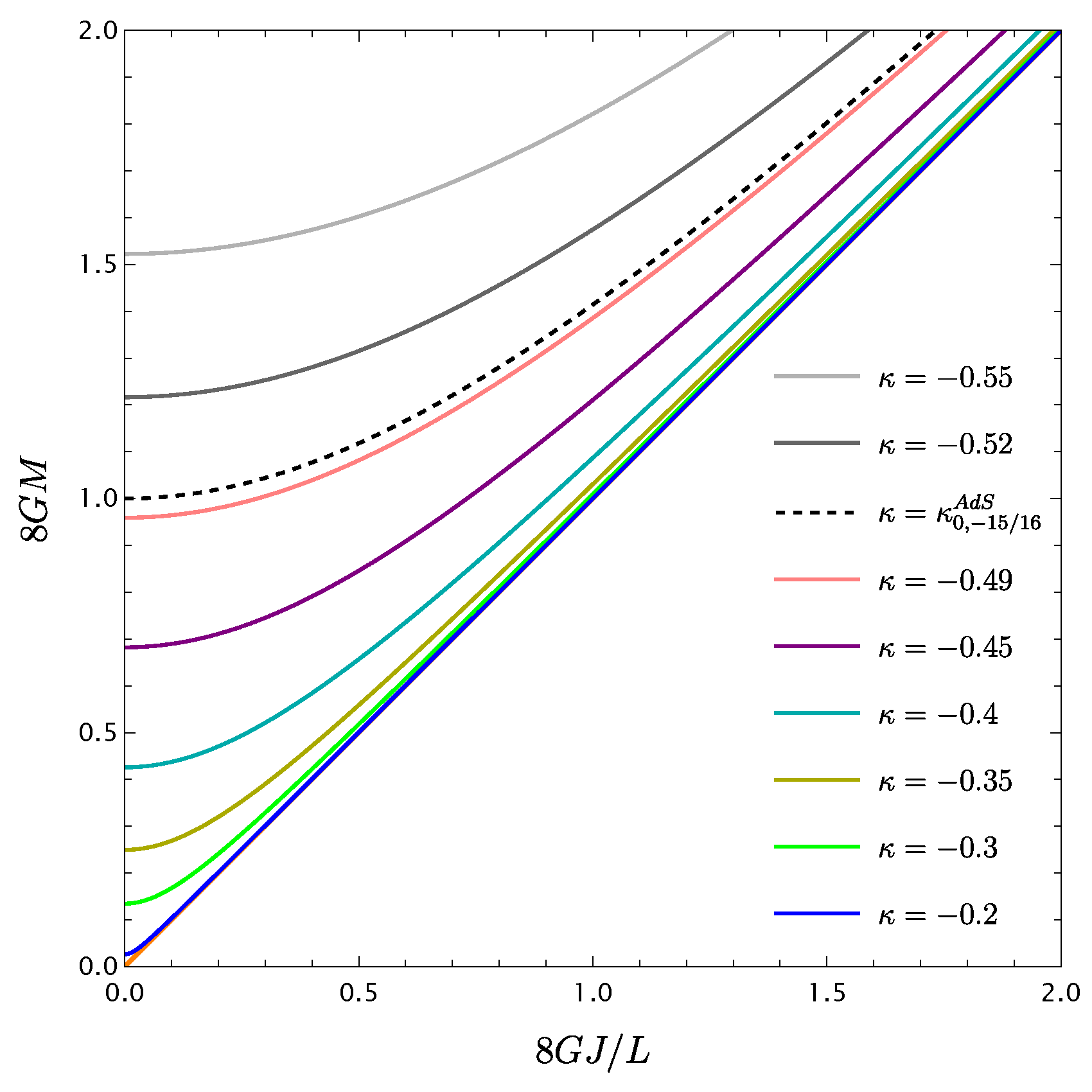}
    \caption{$\bm{m=0}$, $n=0$ instability of {\bf BTZ} for $\mu^2L^2=-15/16$: onset curves in the $\{\hat{J},\hat{M}\}$ plane for different values of the double-trace parameter $\kappa$ (shown in the legend). For a given $\kappa$, BTZ black holes between the associated onset curve and extremal BTZ (orange line, $M=J/L$) are unstable to $m=0$ modes. The dashed black curve indicates the value of $\kappa$ for the onset of the $m=0$ AdS$_3$ instability. For $\kappa$ above this value, AdS$_3$ is linearly stable to $m=0$ (while BTZ can be stable or unstable depending on $\{\hat{J},\hat{M}\}$). For $\kappa$ below $\kappa^{\hbox{\tiny AdS}}_{0,\hat{\mu}^2}$ (e.g., $\kappa=-0.52$ or $\kappa=-0.55$), AdS$_3$ is unstable, and BTZ may be stable or unstable depending on $\{\hat{J},\hat{M}\}$. See the left panels ($m=0$) of Fig.~\ref{fig:m0tom5_btz_VS_ads_instability-OtherMasses} in Appendix~\ref{sec:AppOtherMasses} for corresponding plots with $\mu^2L^2=\{-8/9,-3/4,-1/10\}$.}
    \label{fig:m0_kappa_onset_curves}
\end{figure}

\begin{figure}[ht]
    \centering
    \includegraphics[width=0.48\textwidth]{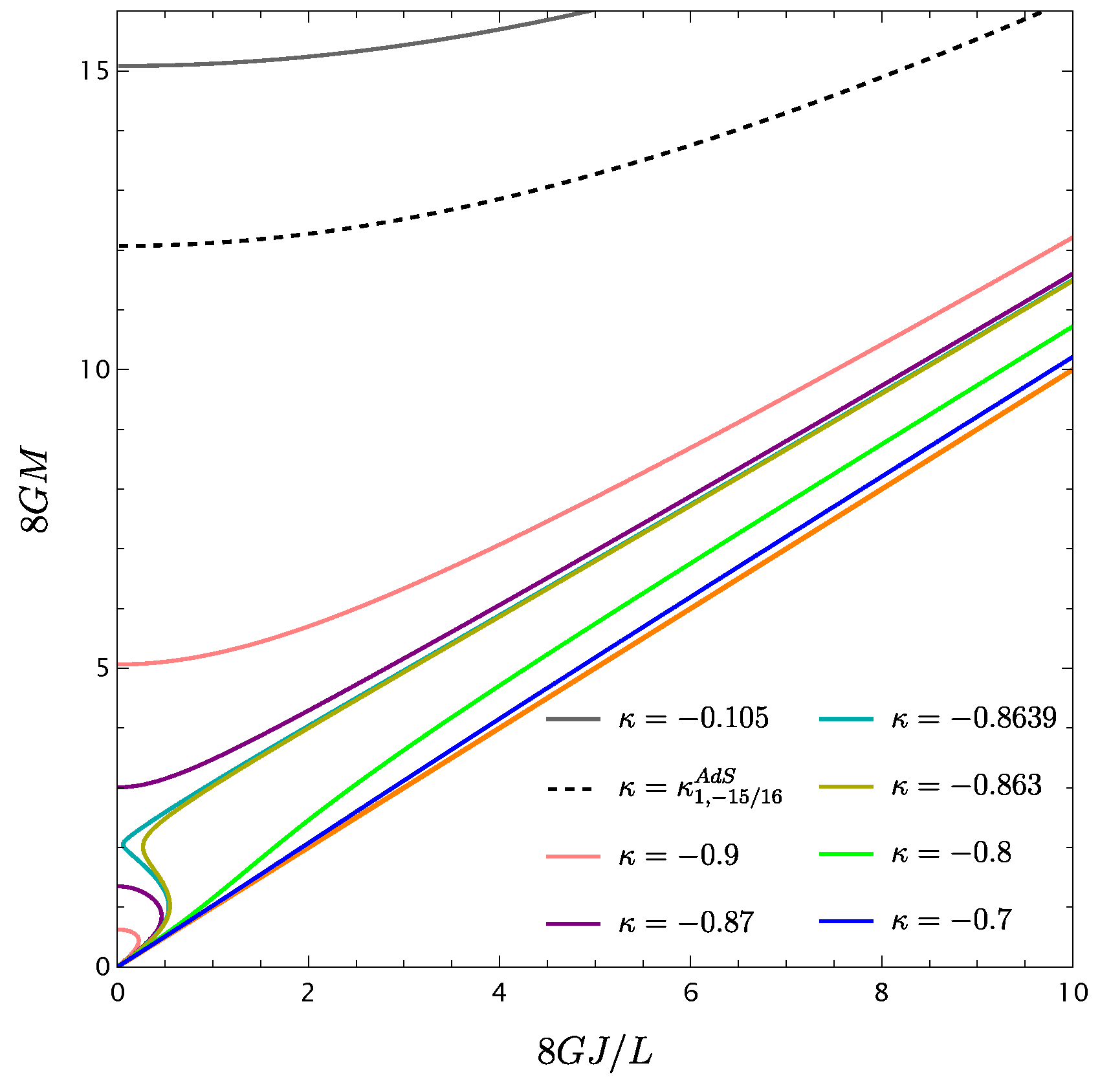}
    \hspace{0.2cm}
    \includegraphics[width=0.48\textwidth]{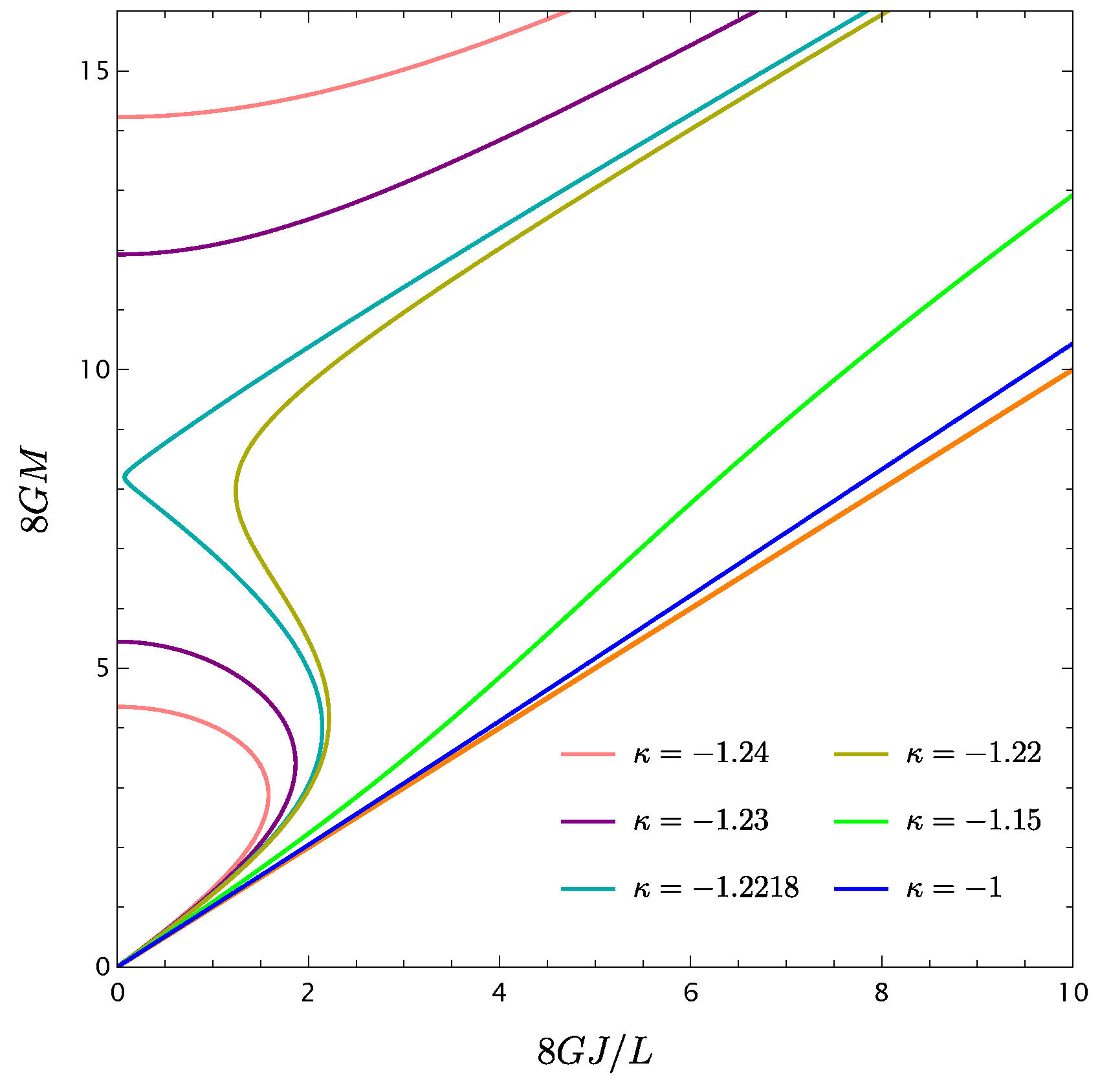}
    \caption{{\bf Left panel:} $\bm{m=1}$, $n=0$ instability of {\bf BTZ} for $\mu^2L^2=-15/16$: onset curves in the $\{\hat{J},\hat{M}\}$ plane for different values of the double-trace parameter $\kappa$ (shown in the legend). For a given $\kappa$, BTZ black holes between the associated onset curve and extremal BTZ (orange line, $M=J/L$) are unstable to $m=1$ modes. The dashed black curve indicates the value of $\kappa$ for the onset of the $m=1$ AdS$_3$ instability. For $\kappa$ above this value, AdS$_3$ is linearly stable to $m=1$ (while BTZ can be stable or unstable depending on $\{\hat{J},\hat{M}\}$). For $\kappa$ below $\kappa^{\hbox{\tiny AdS}}_{1,\hat{\mu}^2}$ (e.g., $\kappa=-1.05$), AdS$_3$ is unstable, and BTZ may be stable or unstable depending on $\{\hat{J},\hat{M}\}$. See the right panels ($m=1$) of Fig.~\ref{fig:m0tom5_btz_VS_ads_instability-OtherMasses} in Appendix~\ref{sec:AppOtherMasses} for corresponding plots with $\mu^2L^2=\{-8/9,-3/4,-1/10\}$. {\bf Right panel:} Similar to the left panel but for $\bm{m=2}$ (the dashed AdS$_3$ onset curve is not shown because it lies well above the displayed region).}
    \label{fig:m1_kappa_onset_curves}  
\end{figure}

\begin{figure}[b]
    \centering
    \includegraphics[width=1.0\textwidth]{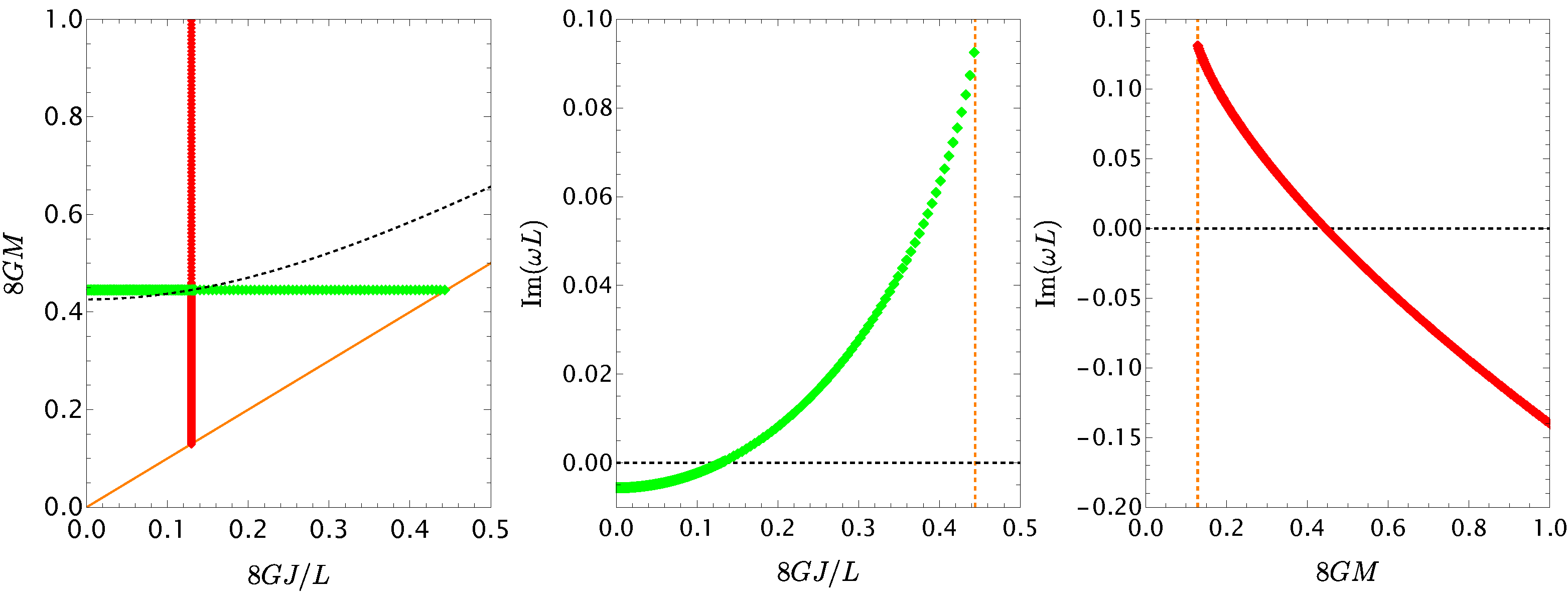}
    \caption{Imaginary part of the frequency for $\bm{m=0}$, $n=0$ modes with $\mu^2L^2=-15/16$ and $\kappa=-4/10$ along a 1-parameter family of BTZ black holes: constant $M=0.28$ (green curve) and constant $J/L=0.18$ (red curve). The orange line indicates extremal BTZ, $M=J/L$. The dashed black line shows the instability onset curve from Fig.~\ref{fig:m0_kappa_onset_curves}.}
    \label{fig:m0_kappam4o10_onset_check}
\end{figure}

Examples of such onset curves for selected values of $\kappa = c$ and $m = 0$ are shown in Fig.~\ref{fig:m0_kappa_onset_curves} for $\mu^2L^2 = -15/16$ (masses in the double-trace range $-1<\hat{\mu}^2<0$ yield qualitatively similar plots; see the left panels of Fig.~\ref{fig:m0tom5_btz_VS_ads_instability-OtherMasses} in Appendix~\ref{sec:AppOtherMasses} for $\hat{\mu}^2=\{-8/9,-3/4,-1/10\}$).

For $m \geq 1$, \eqref{kBTZ-onset} cannot be solved analytically for $\hat{M}$, since it appears inside the $\Gamma$-functions. However, one can instead impose the instability onset condition $\hat{\omega} = m \hat{\Omega}_+ = m\, y_-/y_+$ in the BTZ frequency quantization condition \eqref{eqn:BTZ_Gamma_criterion}. This can then be solved for $y_-$, yielding 
\begin{align}\label{kBTZ-onsetMJ-m1}
    y_-\big|_{\hbox{\tiny $m\! \geq\! 1$}}^{\hbox{\tiny onset}}  =\! \sqrt{
    y_+^2 - \left(\frac{\Gamma\left(\sqrt{1+\hat{\mu}^2}\right)\Gamma\left(\frac{1}{2}\left(\Delta_--\frac{i\,m}{y_+}\right)\right)\Gamma\left(\frac{1}{2}\left(\Delta_-+\frac{i\,m}{y_+}\right)\right)}
    {\Gamma\left(-\sqrt{1+\hat{\mu}^2}\right)\Gamma\left(\frac{1}{2}\left(\Delta_+-\frac{i\,m}{y_+}\right)\right)\Gamma\left(\frac{1}{2}\left(\Delta_++\frac{i\,m}{y_+}\right)\right)}\,
    \kappa \right)^{\!\!\frac{1}{\sqrt{1+\hat{\mu}^2}}}
    }\!\Bigg|_{\kappa=c}\!\!.
\end{align}
Finally, using \eqref{ypmFromMJ}, we map these onset curves $y_-(y_+)$ into $\hat{M}\big|_{\hbox{\tiny $m$}}^{\hbox{\tiny onset}}(\hat{J})$. Examples for selected values of $\kappa = c$ and $m = 1$ are shown in the left panel of Fig.~\ref{fig:m1_kappa_onset_curves} for $\mu^2L^2 = -15/16$ (masses in the double-trace range $-1<\hat{\mu}^2<0$ yield qualitatively similar plots; see the right panels of Fig.~\ref{fig:m0tom5_btz_VS_ads_instability-OtherMasses} in Appendix~\ref{sec:AppOtherMasses} for $\hat{\mu}^2=\{-8/9,-3/4,-1/10\}$). For $m \geq 2$, the onset curves are qualitatively similar to the $m=1$ case: see right panel of Fig.~\ref{fig:m1_kappa_onset_curves} for $m=2$.

 In Fig.~\ref{fig:m0_kappa_onset_curves} ($m=0$), the various curves show the onset, as given by \eqref{kBTZ-onsetMJ-m0}, of the $n=0$ instability for the fixed values of $\kappa$ given in the legend. BTZ black holes exist above the extremal BTZ line $M=J/L$ (red). For a given double-trace parameter $\kappa$, BTZ black holes between the associated onset curve and extremal BTZ are unstable. 

The figure also includes the dashed curve corresponding to the AdS$_3$ onset $\kappa = \kappa^{\hbox{\tiny AdS}}_{m=0,\hat{\mu}^2} \sim -0.495$, which is the dashed black curve in the left panel of Fig.~\ref{fig:3D_kappa_onset} where the grey AdS$_3$ and red BTZ surfaces intersect. In the context of Fig.~\ref{fig:3D_kappa_onset}, any BTZ onset curve below the AdS$_3$ dashed curve represents a double-trace BC setup where BTZ black holes can be stable or unstable to $n=0$ perturbations while AdS$_3$ is linearly stable (these BTZ black holes lie in the region connected to extremal BTZ).

For concreteness, consider $\kappa = -0.45 > \kappa^{\hbox{\tiny AdS}}_{m=0,\hat{\mu}^2}$, whose onset curve is the purple curve in Fig.~\ref{fig:m0_kappa_onset_curves}. In this case, AdS$_3$ is stable to $n=0$ perturbations, while BTZ black holes are unstable or stable depending on whether they lie below or above the purple onset curve. Conversely, for $\kappa = -0.52 < \kappa^{\hbox{\tiny AdS}}_{m=0,\hat{\mu}^2}$, the associated onset curve (dark-grey) lies above the dashed black AdS$_3$ onset curve. Here, BTZ black holes below/above the dark-grey curve are unstable/stable, but AdS$_3$ is now unstable to the same $n=0$ perturbation. Although the ``vacuum BTZ'' ($M=0=J$) has a mass gap to AdS$_3$, BTZ asymptotically approaches AdS$_3$, so cases with the BTZ onset curve above the AdS$_3$ onset curve are (or not!) of less physical interest. Accordingly, in the remainder of this paper (unless otherwise stated, e.g., in the right panel of Fig.~\ref{fig:m0_3d_plot}), we focus on values of $\kappa$ for which the BTZ onset curve is below the AdS$_3$ dashed onset curve, i.e., $0 > \kappa > \kappa^{\hbox{\tiny AdS}}_{m=0,\hat{\mu}^2}$, where $\kappa^{\hbox{\tiny AdS}}_{m=0,\hat{\mu}^2}$ is given by \eqref{kAdS-onset} (e.g., $-0.4951294$ for $\hat{\mu}^2=-15/16$).

Consider now Fig.~\ref{fig:m1_kappa_onset_curves}, the analogue of Fig.~\ref{fig:m0_kappa_onset_curves} for $m=1$ (left panel) and $m=2$ (right panel), as representatives of $m\geq 1$. In the left panel ($m=1$), BTZ black holes exist above the extremal BTZ line, and the dashed black curve represents $\kappa = \kappa^{\hbox{\tiny AdS}}_{m=1,\hat{\mu}^2} \sim -1.01$, the AdS$_3$ onset (corresponding to the right panel of Fig.~\ref{fig:3D_kappa_onset} where the grey and red surfaces intersect). The coloured curves show the onset for $n=0$ co-rotating $m=1$ modes for the $\kappa$ values in the legend. For a given $\kappa$, BTZ black holes between the onset curve and extremal BTZ are unstable, while those above the curve or inside the ``semicircle'' branch of the curve are stable.

The main difference with the $m=0$ case is that for $m\geq 1$, some onset curves have two disconnected pieces. For instance, for $\kappa = -0.87 > \kappa^{\hbox{\tiny AdS}}_{m=1,\hat{\mu}^2} \sim -1.01$, the purple onset curve includes a semicircle piece in the bottom-left corner (small $J/L$) and a standard piece starting at $\{\hat{J},\hat{M}\} = \{0,3\}$. BTZ black holes in the region bounded by the semicircle, by $J=0$, by the standard piece, and by extremal BTZ are unstable to co-rotating $n=0$, $m=1$ perturbations, while those outside these regions are linearly stable.  

 \begin{figure}[ht]
    \centering
    \includegraphics[width=0.99\textwidth]{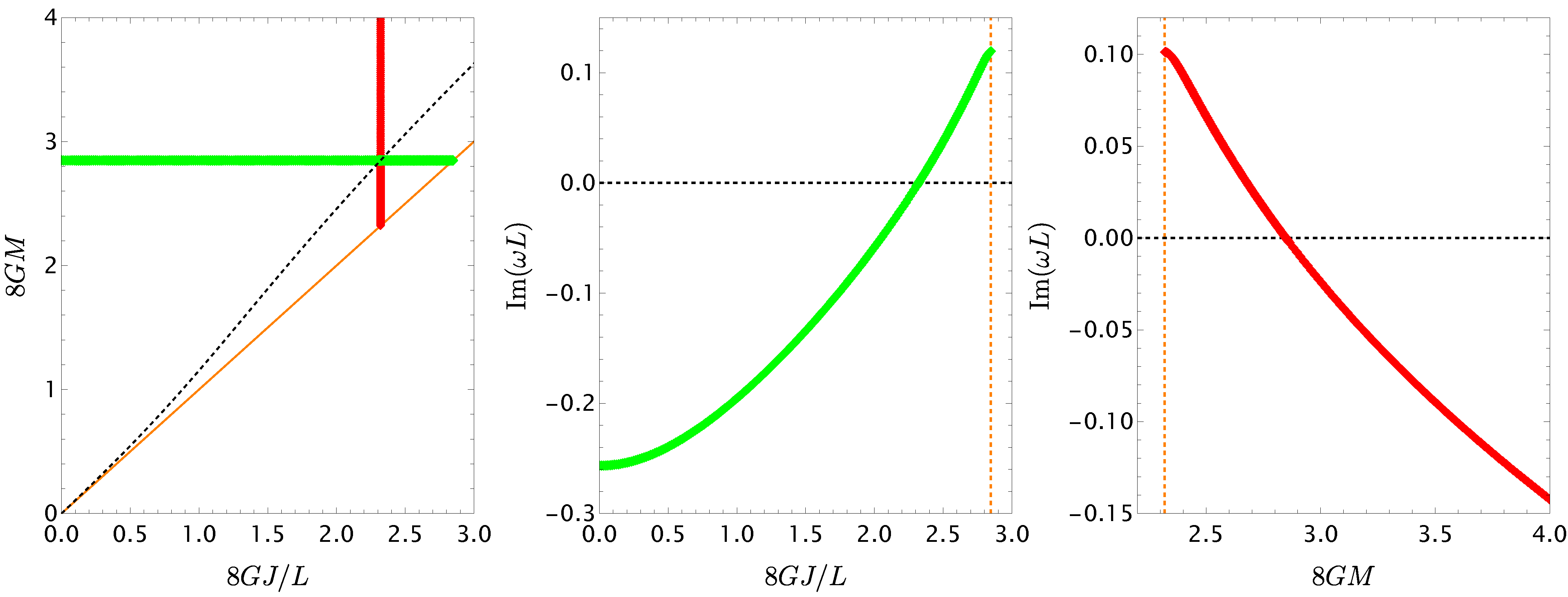}
    \includegraphics[width=0.99\textwidth]{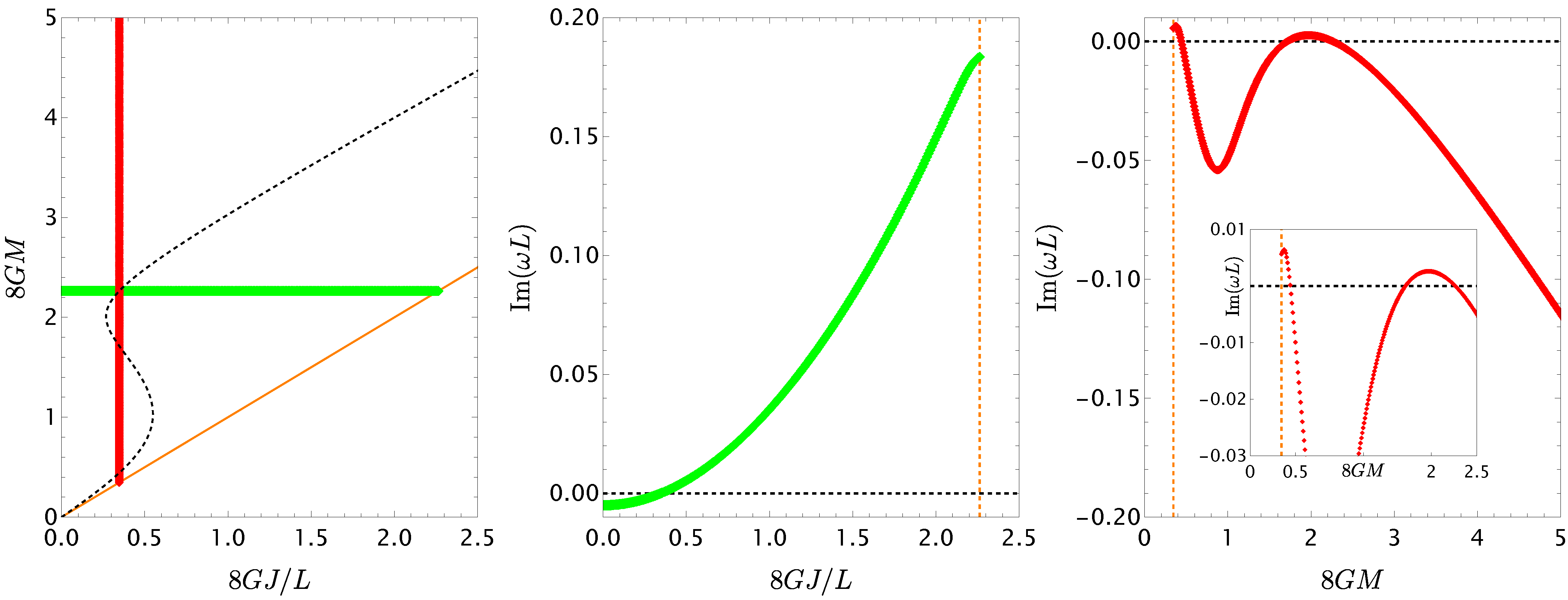}
    \includegraphics[width=0.99\textwidth]{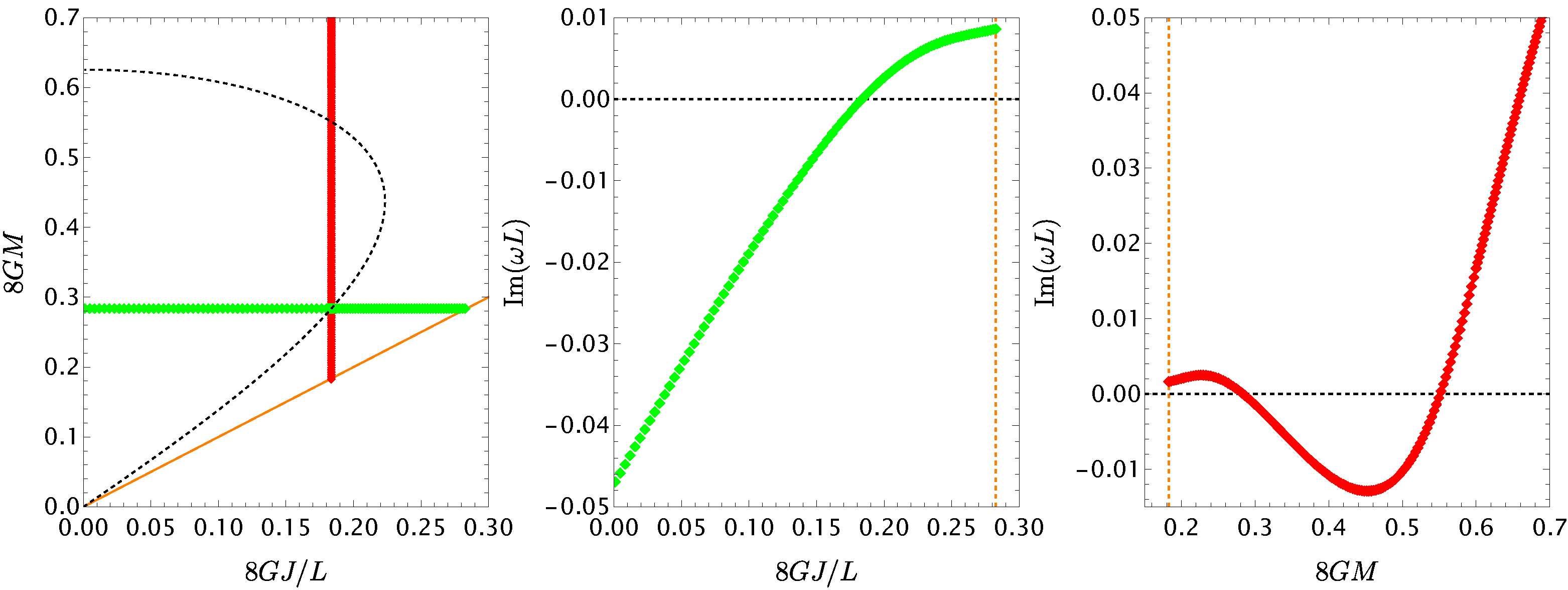}
    \caption{\small Imaginary part of the frequency for $\bm{m=1}$, $n=0$ modes with $\mu^2L^2=-15/16$ and three values of $\bm{\kappa}$ along a 1-parameter family of BTZ black holes: constant $M$ (green curves) and constant $J/L$ (red curves). The orange line indicates extremal BTZ, $M=J/L$. The dashed black line shows the instability onset curve from Fig.~\ref{fig:m1_kappa_onset_curves}.  {\bf Top panel:} $\bm{\kappa=-0.8}$ for BTZ with constant $M=2.84$ (green) and constant $J/L=2.32$ (red).  {\bf Middle panel:} $\bm{\kappa=-0.863}$ for BTZ with constant $M=2.26$ (green) and constant $J/L=0.34$ (red).  {\bf Bottom panel:} $\bm{\kappa=-0.9}$ for BTZ with constant $M=0.45$ (green) and constant $J/L=0.13$ (red). Only the region around the ``semicircle'' branch of the onset curve is shown; see Fig.~\ref{fig:m1_kappa_onset_curves}.}

    \label{fig:m1_kappa_onset_check}
\end{figure}

For larger $\kappa$, the onset curves take the standard single-piece shape with monotonically increasing $\hat{M}$ (e.g., the green curve for $\kappa=-0.8$). Between these cases, curves can be $S$-shaped (e.g., dark-green for $\kappa=-0.863$), with a critical $\kappa$ where the top turning point of the $S$ reaches $J=0$ (cyan curve, $\kappa \sim -0.8639$). For smaller $\kappa$, the onset curves develop the two-piece structure described above. This captures the small neighbourhood of the corner $\{\hat{M},\hat{J}\} = \{0,0\}$ of the 3D plot (right panel, $m=1$) in Fig.~\ref{fig:3D_kappa_onset}. The semicircle shrinks with decreasing $\kappa$ until it disappears at $\{0,0\}$.

As with $m=0$, in the rest of the paper we focus on $\kappa$ for which the BTZ onset curve is below the AdS$_3$ dashed curve, i.e., $0>\kappa > \kappa^{\hbox{\tiny AdS}}_{m=1,\hat{\mu}^2}$ (given by \eqref{kAdS-onset}, e.g., $-1.0098371$ for $\hat{\mu}^2=-15/16$). Moreover, whenever BTZ is unstable to co-rotating $m\geq 1$, $n=0$ modes, it is necessarily unstable to $m=0$ as discussed in Fig.~\ref{fig:m0tom5_btz_VS_ads_instability}. This discussion also applies to $m\geq 2$ (right panel, Fig.~\ref{fig:m1_kappa_onset_curves}); the AdS$_3$ onset curve is not shown because it lies far above the displayed region.

In the discussions of Figs.~\ref{fig:3D_kappa_onset}-\ref{fig:m1_kappa_onset_curves}, we focused on the onset curves of co-rotating $n=0$ modes that can become unstable in BTZ. These curves are determined by the onset conditions ${\rm Re}\,\hat{\omega}=m\hat{\Omega}_+$ and ${\rm Im}\,\hat{\omega}=0$ applied to the BTZ frequency quantization condition \eqref{eqn:BTZ_Gamma_criterion}.

To further check and explore the unstable BTZ system, we fix $\kappa$ (and $\hat{\mu}^2=-15/16$) and compute the $n=0$ mode frequency spectrum numerically, as described in section~\ref{sec:Numerical-Method}. The goal is to confirm that ${\rm Im}\,\hat{\omega}$ changes sign at the onset curve in Fig.~\ref{fig:m0_kappa_onset_curves} ($m=0$) or Fig.~\ref{fig:m1_kappa_onset_curves} ($m=1$). This is illustrated in Fig.~\ref{fig:m0_kappam4o10_onset_check} ($m=0$, $\kappa=-4/10$) and Fig.~\ref{fig:m1_kappa_onset_check} ($m=1$) for $\kappa=-0.8$ (top panel), $\kappa=-0.863$ (middle panel), and $\kappa=-0.9$ (bottom panel). Three panels are needed for $m=1$ because the onset curves can take different shapes (see Fig.~\ref{fig:m1_kappa_onset_curves}).

For this purpose, we pick a 1-parameter family of BTZ black holes at constant $M$ (green curves in Figs.~\ref{fig:m0_kappam4o10_onset_check}-\ref{fig:m1_kappa_onset_check}, values in captions) or constant $J/L$ (red curves in Figs.~\ref{fig:m1_kappa_onset_check}, values in captions). Section~\ref{sec:3Dspectrum} will later allow a full 2-dimensional scan of BTZ parameters.

In the top panels of Figs.~\ref{fig:m0_kappam4o10_onset_check}-\ref{fig:m1_kappa_onset_check}, $8G J/L$ is plotted versus $8G M$: the orange line shows extremal BTZ, $M=J/L$, and the dashed black curve shows the instability onset curve for the chosen $\kappa$, taken from Fig.~\ref{fig:m0_kappa_onset_curves} ($m=0$) or Fig.~\ref{fig:m1_kappa_onset_curves} ($m=1$). As discussed previously, BTZ is unstable when it lies between the onset curve and extremal line in the $\{\hat{M},\hat{J}\}$ plane, and stable otherwise. These figures confirm this: ${\rm Im}\,\hat{\omega}>0$ (unstable) between onset and extremal curves, and ${\rm Im}\,\hat{\omega}<0$ (stable) above the onset curve. 

The numerical results also show ${\rm Im}\,\hat{\omega}=0$ precisely at the values predicted by \eqref{kBTZ-onsetMJ-m0} ($m=0$) or \eqref{kBTZ-onsetMJ-m1} ($m\geq 1$), which were used to generate the dashed onset curves. The bottom panel of Fig.~\ref{fig:m1_kappa_onset_check} further confirms that BTZ black holes inside/outside the ``semicircle'' branch of the $m=1$ onset curve are stable/unstable as discussed in Fig.~\ref{fig:m1_kappa_onset_curves}.

\subsection{Fix the theory (and thus \texorpdfstring{$\kappa$}{kappa}). Frequency spectrum for unstable BTZ}\label{sec:3Dspectrum}
In the previous sections, we have completed our analysis of double-trace perturbations $-$ \ie ingoing horizon perturbations with asymptotic boundary conditions \eqref{2xTrace:BC}, in the context of \eqref{eqn:BTZ_rtozFGcoordinate}-\eqref{FGasympExp}, for scalar field masses in the range \eqref{2xTrace:rangeMass} $-$ for the rotating BTZ black hole \eqref{eqn:BTZ_metric}-\eqref{eqn:BTZ_metric_functions}. In this section, we fix the theory and scan the full 2-dimensional BTZ parameter space to obtain the frequency spectrum of massive scalar field perturbations with double-trace boundary conditions. Fixing the theory entails specifying both the scalar field mass $\hat{\mu} = \mu L$ and the double-trace parameter $\kappa$ in the mixed boundary condition \eqref{2xTrace:BC}, namely $\beta = \kappa \, \alpha$. In the context of AdS$_3$/CFT$_2$, fixing $\kappa$ selects the boundary theory, \ie determines the non-conformal deformation of the original CFT.

For $\hat{\mu}$ in the mass range \eqref{2xTrace:rangeMass}, \ie $-1 < \hat{\mu}^2 < 0$, our numerical experiments indicate that the qualitative features of the frequency spectrum are similar across different $\hat{\mu}$. Therefore, for concreteness, we fix $\mu^2 L^2 = -15/16$ in this section, consistent with the previous discussions. Furthermore, we restrict our attention to modes with the fundamental radial overtone, \ie $n=0$, since these are the only modes that can be unstable both for $m=0$ and $m \geq 1$. More precisely, the $n=0$ mode is continuously connected to the $n=0$ Neumann mode $\hat{\omega}^{\hbox{\tiny BTZ}}_{n=0}|_{\hbox{\tiny Neu}}$ in the limit $\kappa \to 0$, as given by \eqref{BTZ:wNeu}. 

For $m \geq 1$, only co-rotating modes can be unstable, and the results for $m \geq 2$ are qualitatively similar to $m=1$. Hence, we typically discuss co-rotating $n=0$ modes for $m=1$ (except in Fig.~\ref{fig:m1_3d_CoCounter}, where both co-rotating and counter-rotating $n=0$, $m=1$ modes are shown). We also typically select $\kappa$ such that global AdS$_3$ is stable for the mode under consideration, while BTZ can be stable or unstable depending on its location in the $\{\hat{J},\hat{M}\}$ parameter plane (see Fig.~\ref{fig:3D_kappa_onset} and Figs.~\ref{fig:m0_kappa_onset_curves}-\ref{fig:m1_kappa_onset_curves}).  

Finally, as discussed in Fig.~\ref{fig:m0tom5_btz_VS_ads_instability}, it is possible to choose $\kappa$ to satisfy these conditions for $m=0$. However, once a $\kappa$ is selected for which BTZ can be unstable to $m \geq 1$, it must be kept in mind that such BTZ (and, in fact, also AdS$_3$) will necessarily be unstable to lower $m$ modes, in particular to the $m=0$ sector.

With this preamble, we are now ready to present several plots that are largely self-explanatory when viewed alongside their captions. Therefore, we avoid lengthy descriptions. The plots in this section were generally obtained using the numerical approach described in section~\ref{sec:Numerical-Method}.

\begin{figure}[b]
    \centering
    \includegraphics[width=0.49\textwidth]{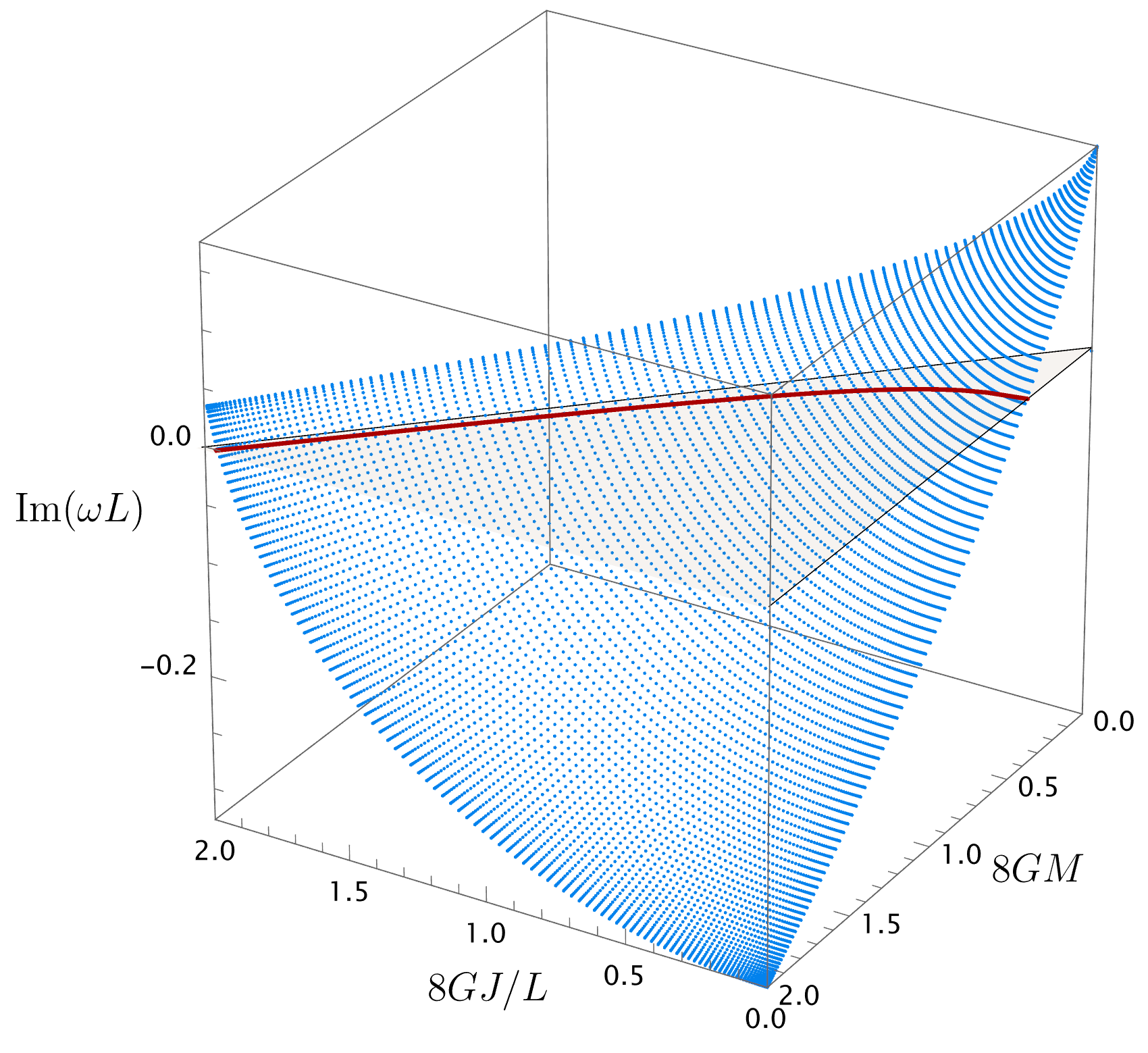}
     \includegraphics[width=0.49\textwidth]{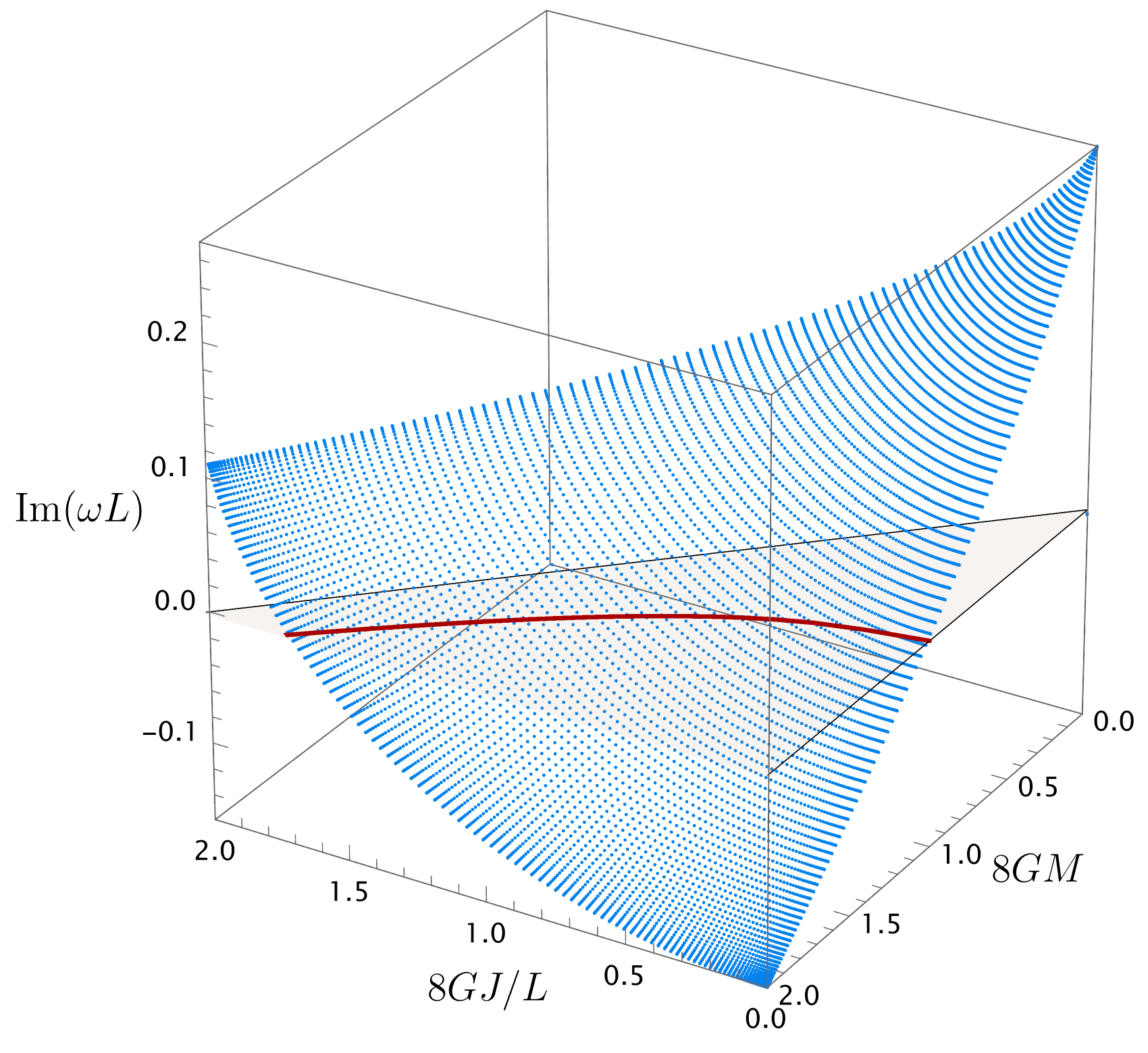}
    \caption{$\text{Im}(\omega L)$ (cyan surface) as a function of the {\bf BTZ} mass $\hat{M}$ and angular momentum $\hat{J}$ for $\bm{m=0}$, $n=0$ modes ($\mu^2L^2=-15/16$). The transparent grey plane corresponds to $\text{Im}(\omega L)=0$ and intersects the cyan surface along the onset of the BTZ instability (red dashed curve). This onset curve agrees with \eqref{kBTZ-onsetMJ-m0} and with the associated curve in Fig.~\ref{fig:m0_kappa_onset_curves}. These modes have ${\rm Re}\,\hat{\omega}=0$. {\bf Left panel:} $\kappa=-0.4$, representing a case where $0 > \kappa > \kappa^{\hbox{\tiny AdS}}_{m=0,\hat{\mu}^2}$ with $\kappa^{\hbox{\tiny AdS}}_{m=0,\hat{\mu}^2} = -0.4951294$. Here, global AdS$_3$ is stable, while BTZ can be either unstable (${\rm Im}\,\hat{\omega}>0$) or stable (${\rm Im}\,\hat{\omega}<0$). {\bf Right panel:} $\kappa=-0.5$, representing a case where $\kappa < \kappa^{\hbox{\tiny AdS}}_{m=0,\hat{\mu}^2}$. In this scenario, BTZ can be unstable or stable, and global AdS$_3$ is unstable.}

    \label{fig:m0_3d_plot}
\end{figure}

Let us start with the $m=0$ case. The real part of the frequency for these modes always vanishes, ${\rm Re}\,\hat{\omega}=0$, regardless of the BTZ parameters $\{\hat{M},\hat{J}\}$. In Fig.~\ref{fig:m0_3d_plot}, we show the imaginary part of the frequency, ${\rm Im}\,\hat{\omega}$, as a function of the BTZ mass $M$ and angular momentum $J/L$. The left panel corresponds to $\kappa=-0.4$, a representative case with $0>\kappa> \kappa^{\hbox{\tiny AdS}}_{m=0,\hat{\mu}^2}$, where $\kappa^{\hbox{\tiny AdS}}_{m=0,\hat{\mu}^2}$ is given by \eqref{kAdS-onset} ($-0.4951294$ for $\hat{\mu}^2=-15/16$). Here, global AdS$_3$ is linearly stable, while BTZ can be either unstable (${\rm Im}\,\hat{\omega}>0$) or stable (${\rm Im}\,\hat{\omega}<0$), as seen in the plot.

The right panel shows $\kappa=-0.5$, representing the case $\kappa^{\hbox{\tiny AdS}}_{m=0,\hat{\mu}^2}>\kappa$. In this scenario, BTZ can still be stable or unstable depending on $\{\hat{M},\hat{J}\}$, but global AdS$_3$ is unstable to the $n=0$, $m=0$ perturbation. These two panels illustrate a general feature: the qualitative behaviour of the BTZ spectrum is similar whether $\kappa$ is above or below $\kappa^{\hbox{\tiny AdS}}_{m=0,\hat{\mu}^2}$.

In Fig.~\ref{fig:m0_3d_plot}, the red dashed curve indicates the instability onset (${\rm Im}\,\hat{\omega}=0$) and agrees with \eqref{kBTZ-onsetMJ-m0} and the associated $\kappa$-onset curve shown in Fig.~\ref{fig:m0_kappa_onset_curves}.

\begin{figure}[ht]
{\vskip -2.3cm}
    \centering
    \includegraphics[width=0.45\textwidth]{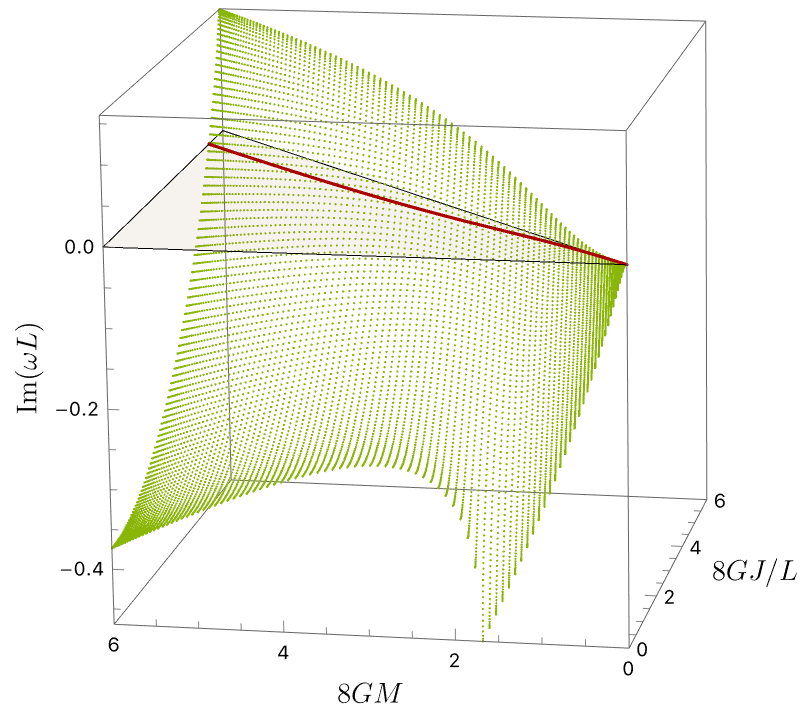}
      \includegraphics[width=0.47\textwidth]{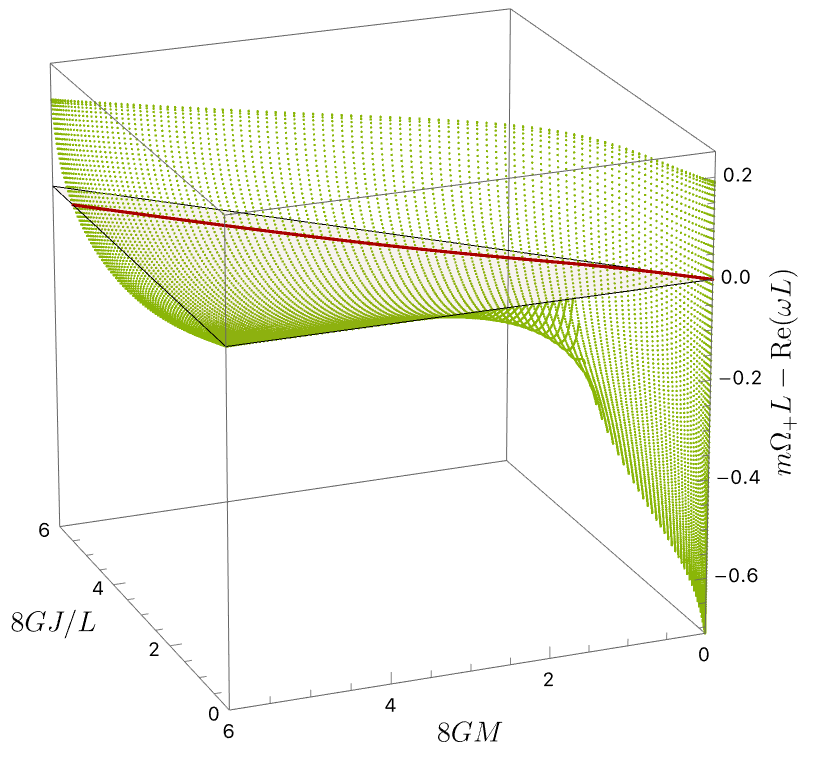}\\
      \includegraphics[width=0.45\textwidth]{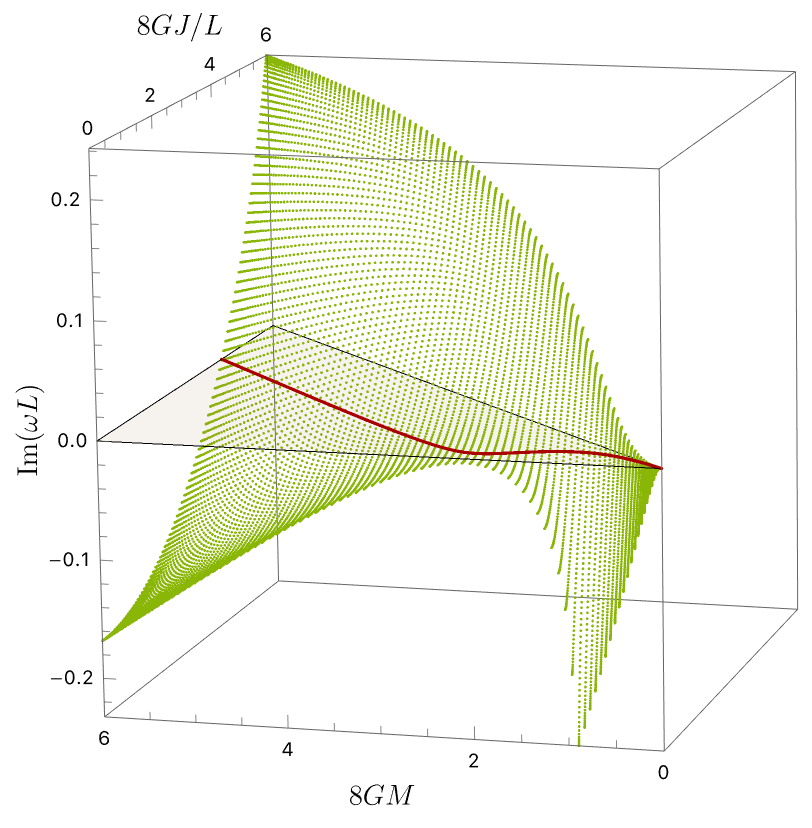}
    \includegraphics[width=0.47\textwidth]{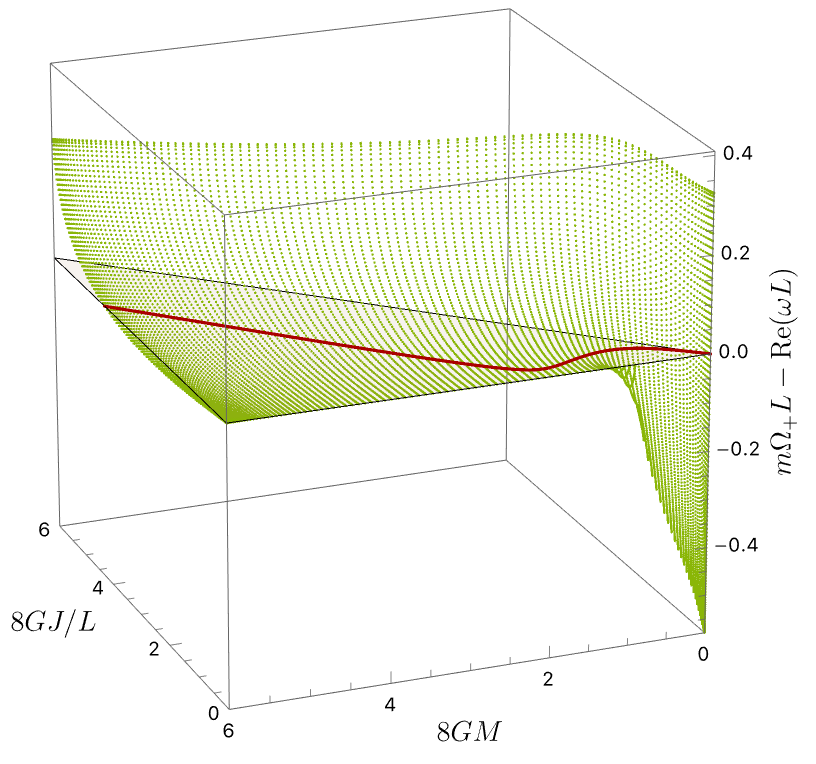}\\
     \includegraphics[width=0.45\textwidth]{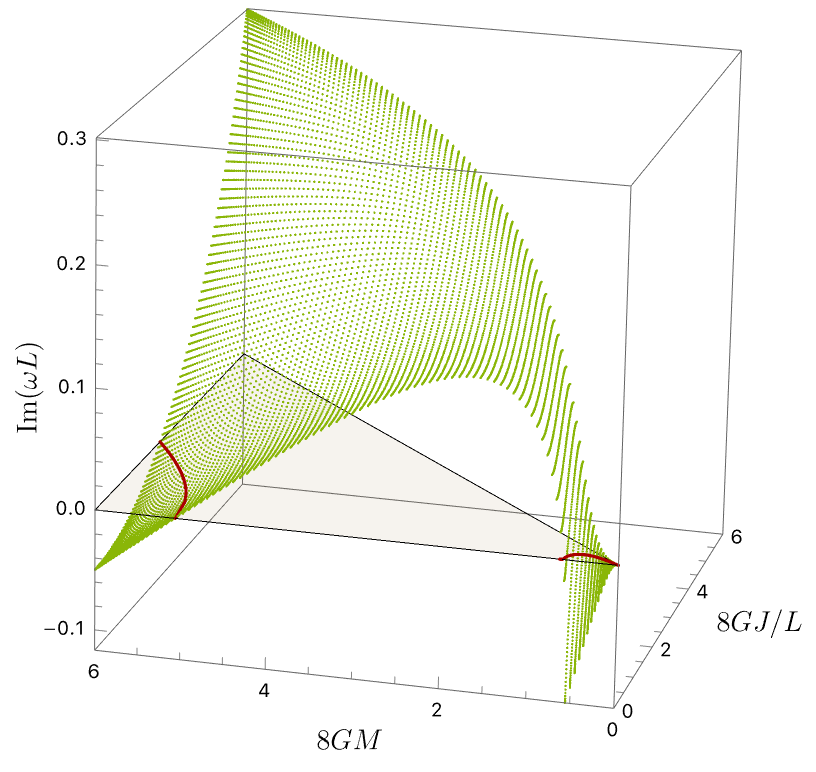}
     \includegraphics[width=0.47\textwidth]{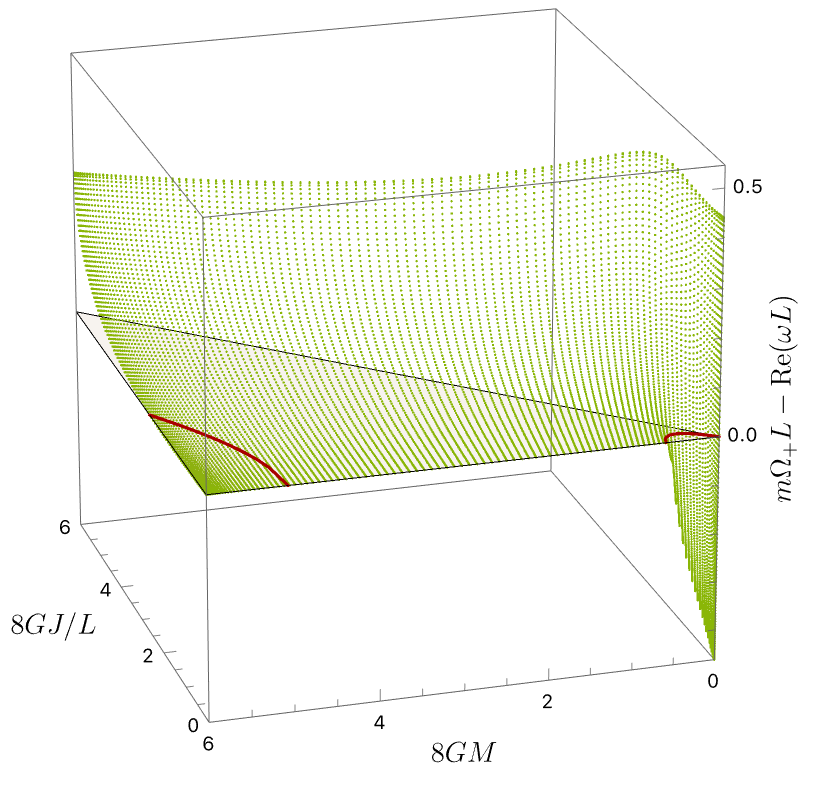}
     {\vskip -0.3cm}
    \caption{\footnotesize $\text{Im}(\omega L)$ (green surface in all left panels) and $m\Omega_+L - \text{Re}(\omega L)$ (green surface in all right panels) as functions of the {\bf BTZ} mass $\hat{M}$ and angular momentum $\hat{J}$ for co-rotating $\bm{m=1}$, $n=0$ modes ($\mu^2L^2=-15/16$). The transparent grey flat triangle corresponds to $\text{Im}(\omega L)=0$ or $\hat{\Omega}_+ - \text{Re}\,\hat{\omega}=0$ (with $J\leq M L$) and intersects the cyan surface at the onset of the BTZ instability (dark-red curve). The {\bf top} ($\bm{\kappa = -0.8}$), {\bf middle} ($\bm{\kappa = -0.863}$), and {\bf bottom} ($\bm{\kappa = -0.9}$) panels correspond to three different values of $\kappa$ within the window $0>\kappa>\kappa^{\hbox{\tiny AdS}}_{m=1,\hat{\mu}^2}$, where $\kappa^{\hbox{\tiny AdS}}_{m=1,\hat{\mu}^2}=-1.0098371$. In this range, global AdS$_3$ is stable to $m=1$ modes, while BTZ can be either stable or unstable, depending on the region in parameter space (see Fig.~\ref{fig:m1_kappa_onset_curves}).}
\label{fig:m1_3d_plot} 
\end{figure}

\begin{figure}[ht]
    \centering
    \includegraphics[width=0.49\textwidth]{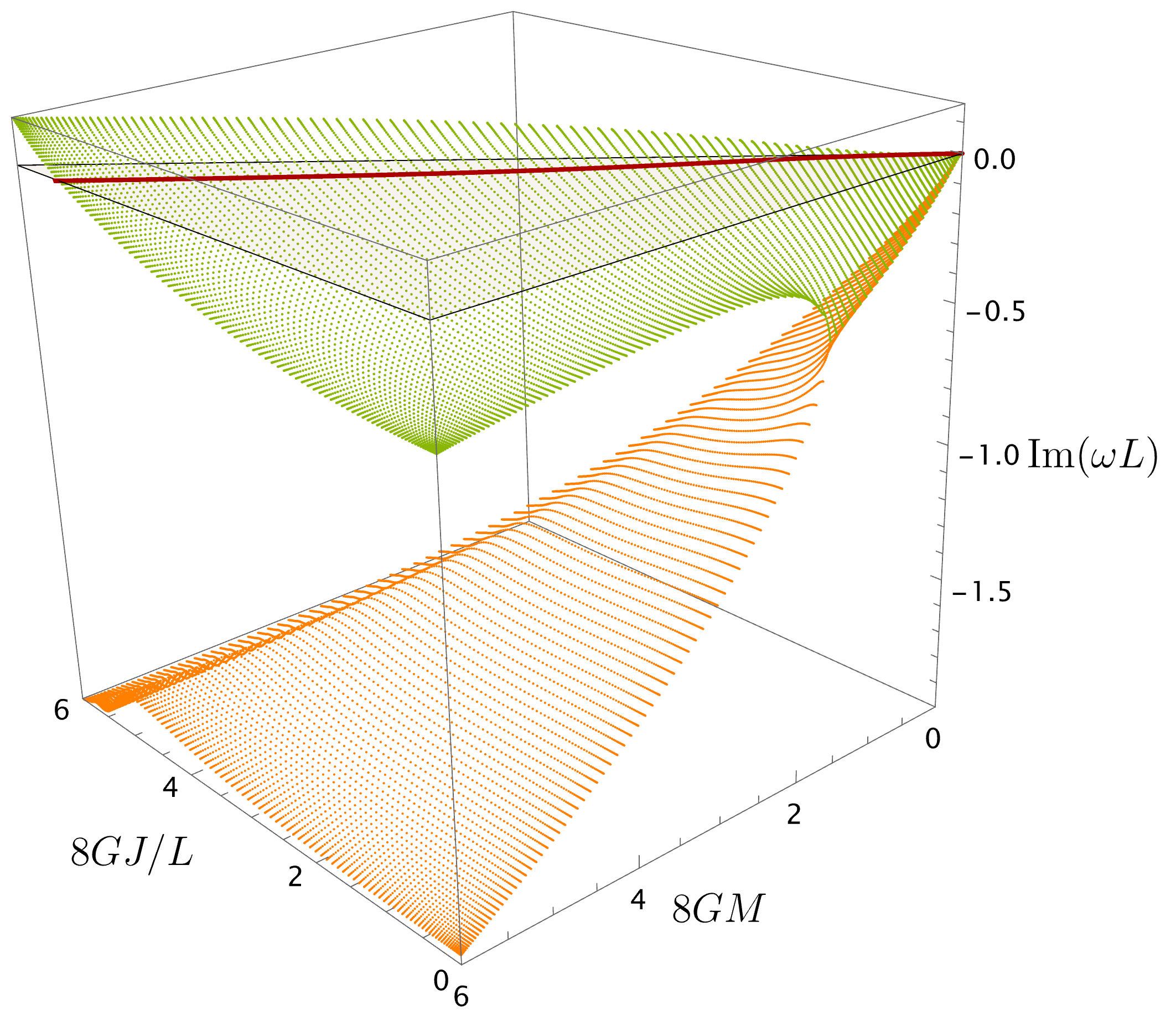}
    \includegraphics[width=0.49\textwidth]{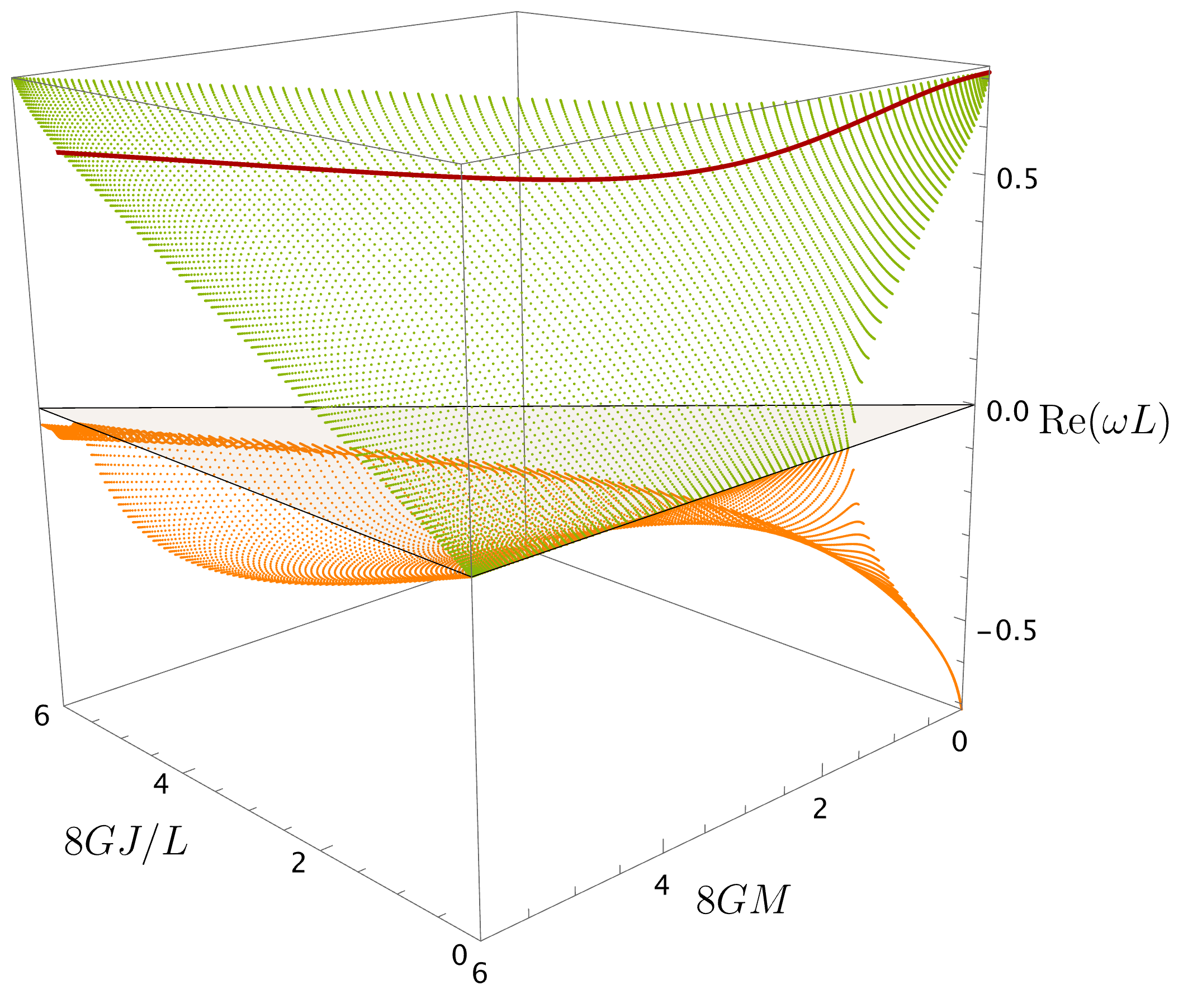}\\
    \includegraphics[width=0.49\textwidth]{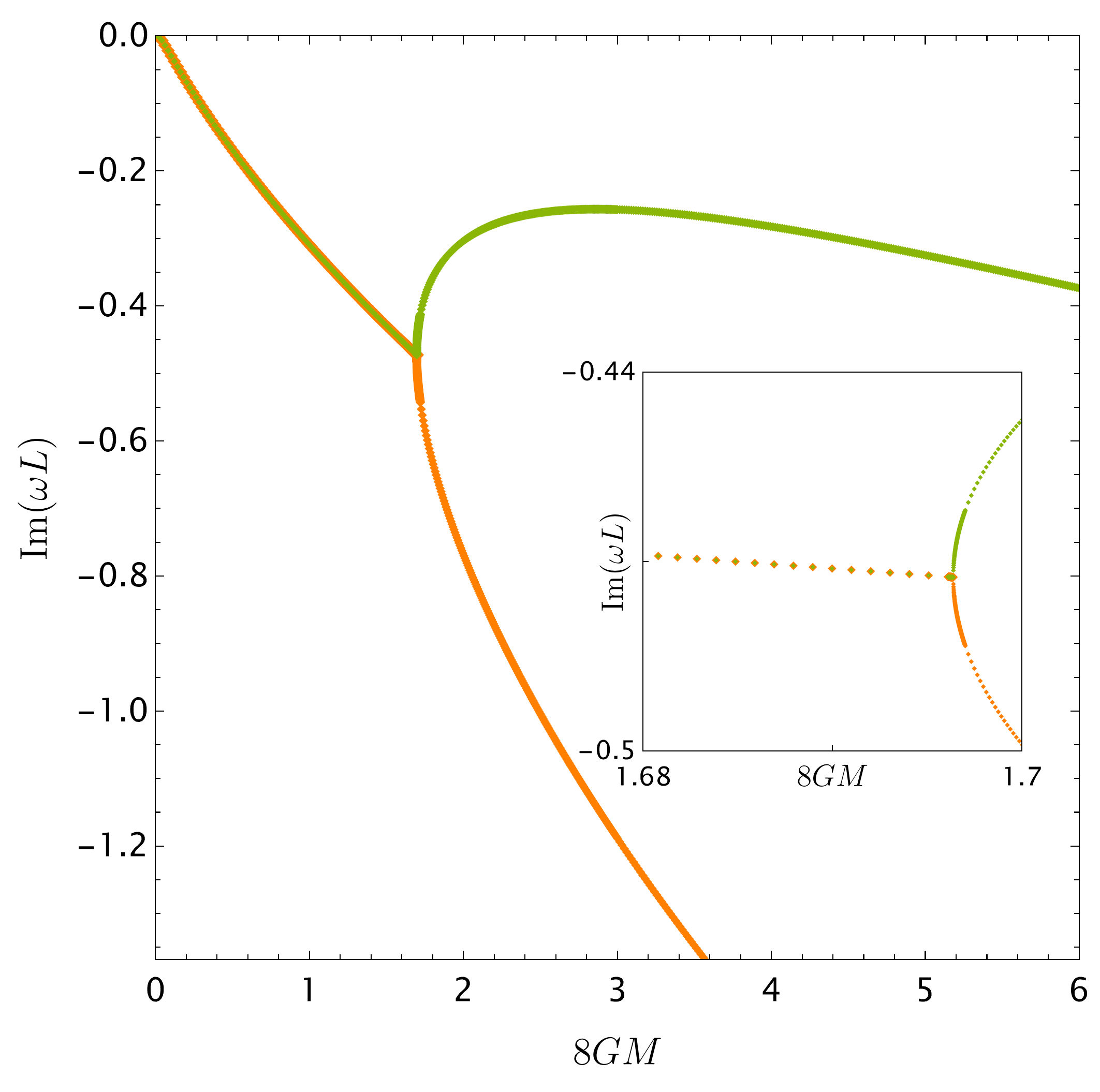}
    \includegraphics[width=0.49\textwidth]{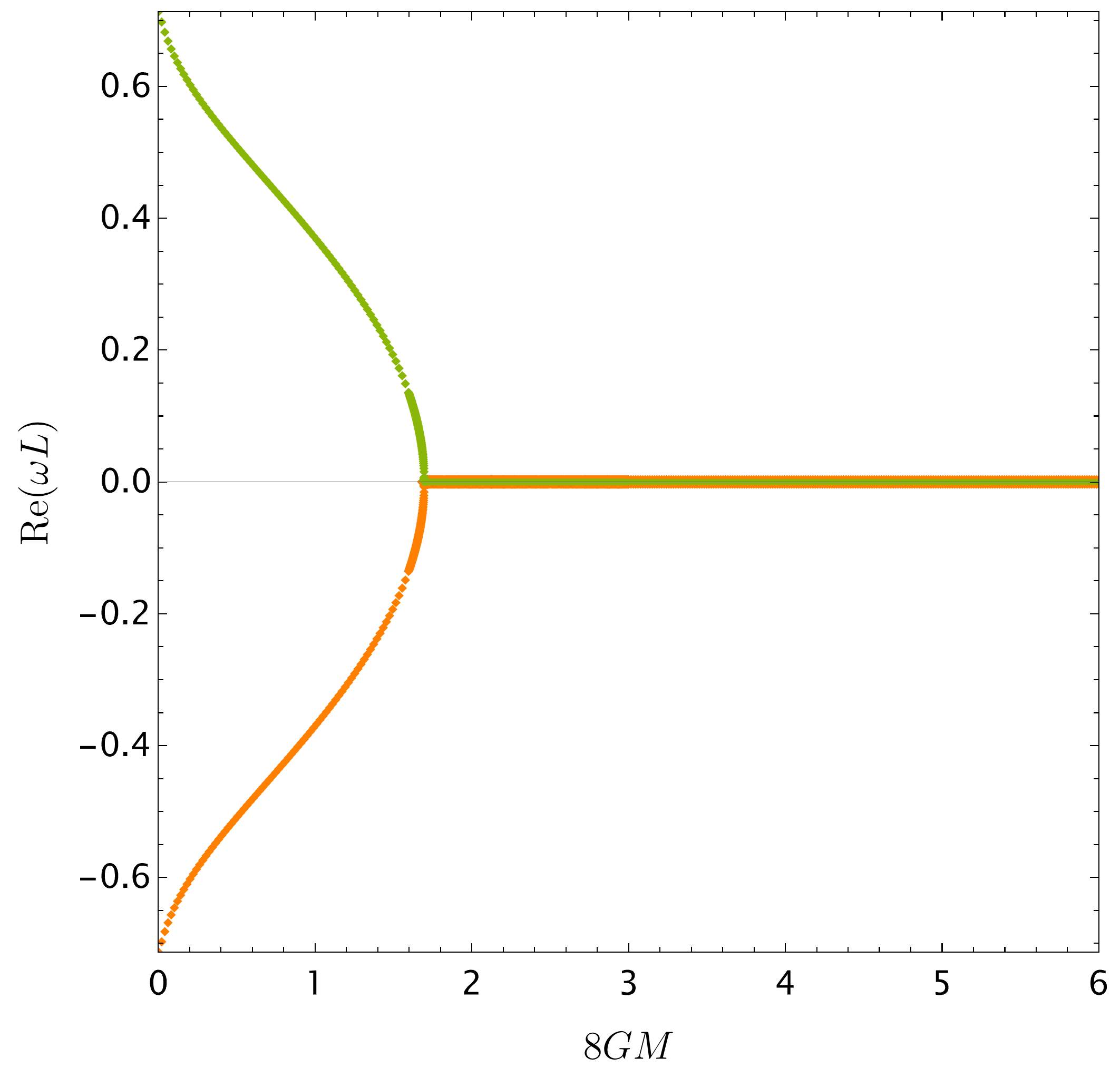}
    \caption{
Understanding the cusp in the top panel of Fig.~\ref{fig:m1_3d_plot} for $\kappa = -8/10$, $\mu^2L^2=-15/16$, and $m=1$, $n=0$ modes. Here we show both the {\it co-rotating} modes (green surface, already in Fig.~\ref{fig:m1_3d_plot}) and the {\it counter-rotating} modes (orange surface).  {\bf Top panel:} $\text{Im}(\omega L)$ (left) and $\text{Re}(\omega L)$ (right) as functions of the BTZ mass $\hat{M}$ and angular momentum $\hat{J}$. The transparent grey flat triangle corresponds to $\text{Im}(\omega L)=0$ or $\text{Re}\,\hat{\omega}=0$ (with $J \leq M L$). The dark-red curve indicates the onset of the BTZ instability. {\bf Bottom panel:} Detail of the $J=0$ plane from the top panel. Shown are $\text{Im}(\omega L)$ (left) and $\text{Re}(\omega L)$ (right) for the co-rotating (green) and counter-rotating (orange) $m=1$, $n=0$ modes of {\it static} BTZ. Starting at $M=0$, as both modes approach $M \sim 1.69$, the absolute value $|\text{Re}(\omega L)|$ decreases until vanishing, during which both modes share the same $\text{Im}(\omega L)$. For $M \gtrsim 1.69$, the imaginary parts split, giving rise to a repulsion observed in the left plot.
}
    \label{fig:m1_3d_CoCounter}
\end{figure}

Consider now the $m=1$ case (illustrative of all $m \geq 1$ cases). Unlike for $m=0$, these modes typically have ${\rm Re}\,\hat{\omega} \neq 0$, and at the instability onset we have ${\rm Re}\,\hat{\omega} - m\hat{\Omega}_+ = 0$ in addition to ${\rm Im}\,\hat{\omega} = 0$. Thus, in Fig.~\ref{fig:m1_3d_plot} we plot both ${\rm Im}\,\hat{\omega}$ (left panels) and $m\hat{\Omega}_+ - {\rm Re}\,\hat{\omega}$ (right panels) as functions of the BTZ mass $\hat{M}$ and angular momentum $\hat{J}$ (with $\Omega_+ = \sqrt{M - J/L} / \sqrt{M + J/L}$).

Unlike the $m=0$ case, the instability onset curves $M(J/L)|_{\rm onset}$ are not always monotonically increasing. As shown in Fig.~\ref{fig:m1_kappa_onset_curves}, the onset curves exhibit three distinct shapes depending on the value of $\kappa$. Accordingly, Fig.~\ref{fig:m1_3d_plot} presents 3D plots for three representative $\kappa$ values that cover these possibilities:
\begin{itemize}
    \item $\kappa=-0.8$ (top panels), where $M(J/L)|_{\rm onset}$ is monotonically increasing (see top panel of Fig.~\ref{fig:m1_kappa_onset_check}),  
    \item $\kappa=-0.863$ (middle panels), where the onset curve has an $S$-shape (middle panel of Fig.~\ref{fig:m1_kappa_onset_check}),  
    \item $\kappa=-0.9$ (bottom panels), where the onset curve has two disconnected branches — one being a roughly semicircular branch at small $M$ and $J/L$ (see bottom panel of Fig.~\ref{fig:m1_kappa_onset_check} and Fig.~\ref{fig:m1_kappa_onset_curves}).
\end{itemize}

All three $\kappa$ values lie within the window $0 > \kappa > \kappa^{\rm AdS}_{m=1,\hat{\mu}^2}$ (with $\kappa^{\rm AdS}_{m=1,\hat{\mu}^2} = -1.0098371$), where global AdS$_3$ is stable to $m=1$ modes, but BTZ can be either stable or unstable to these modes (see Fig.~\ref{fig:m1_kappa_onset_curves}).

In Fig.~\ref{fig:m1_3d_plot}, the dark-red curve indicates the instability onset (${\rm Im}\,\hat{\omega}=0$ and ${\rm Re}\,\hat{\omega}-m\hat{\Omega}_+=0$) and agrees with \eqref{kBTZ-onsetMJ-m1} and the corresponding $\kappa$-curves in Fig.~\ref{fig:m1_kappa_onset_curves}. Regions (in the green surface) where ${\rm Im}(\omega L) > 0$ and $m\hat{\Omega}_+ - {\rm Re}\,\hat{\omega} > 0$ correspond to BTZ black holes unstable to co-rotating $m=1$, $n=0$ double-trace perturbations. It is important to emphasize that BTZ black holes unstable to $m=1$ perturbations are already unstable to $m=0$ perturbations, and some BTZ that are stable to $m=1$ can still be unstable to $m=0$, as discussed in Fig.~\ref{fig:m0tom5_btz_VS_ads_instability}.

The plots of Fig.~\ref{fig:m1_3d_plot} also reveal a cusp in the $J=0$ plane, corresponding to the static BTZ black hole. To understand this feature, Fig.~\ref{fig:m1_3d_CoCounter} shows both co-rotating $m=1$, $n=0$ frequencies (green) and counter-rotating frequencies (orange) for $\kappa=-0.8$ (similar behaviour occurs for other $\kappa$ values). By convention, co-rotating modes have ${\rm Re}\,\hat{\omega} \ge 0$, while counter-rotating modes have ${\rm Re}\,\hat{\omega} \le 0$.

In the static $J=0$ limit, both modes exhibit a critical mass $M = M_\star(m,\kappa)$. For $0 < M < M_\star$, the co- and counter-rotating modes share the same ${\rm Im}\,\hat{\omega} < 0$ and equal $|{\rm Re}\,\hat{\omega}|$ with opposite signs. As $M$ approaches $M_\star$ from below, $|{\rm Re}\,\hat{\omega}|$ decreases monotonically, vanishing exactly at $M=M_\star$ (producing the cusp at $\{\hat{M}, \hat{J}\} = \{M_\star,0\}$ in Fig.~\ref{fig:m1_3d_plot}). For $M > M_\star$, ${\rm Re}\,\hat{\omega} = 0$ for both modes, while the imaginary parts split, giving distinct ${\rm Im}\,\hat{\omega} < 0$ values. The bottom panel of Fig.~\ref{fig:m1_3d_CoCounter} illustrates this evolution along $M$ at fixed $J=0$. For $J>0$, the frequencies of the two modes are always distinct, yielding a smooth frequency surface in Fig.~\ref{fig:m1_3d_plot}.

\section{Final remarks and prospects}\label{sec:Conc}

We have carried out a detailed analysis of rotating BTZ black holes perturbed by a scalar field 
$\Phi(t,r,\phi)$ with mass $-1<\hat{\mu}^2<0$ subject to double-trace boundary conditions. 
Within the AdS/CFT framework, these perturbations correspond to double-trace (non-conformal) 
deformations of the dual boundary theory. 

Starting with the Fourier decomposition 
$\Phi(t,r,\phi)=e^{-i\omega t}e^{im\phi}\psi(r)$, it was previously known that the fundamental 
radial overtone of modes with $m\geq 1$ is unstable 
\cite{Iizuka:2015vsa,Ferreira:2017cta,Dappiaggi:2017pbe}. 
Here we have shown that the BTZ black hole is in fact already unstable in the $m=0$ sector 
(in certain regions where global AdS$_3$ is known to be linearly stable under the same 
double-trace perturbations \cite{Ishibashi:2004wx}). 
This $m=0$ instability is the most relevant one for three main reasons.

First, whenever BTZ is unstable to modes with $m\geq 1$, it is necessarily already unstable 
to $m=0$ modes; however, there are regions of parameter space where BTZ is unstable only to 
$m=0$ perturbations. In this sense, the double-trace instability provides an example of a 
(non-conventional) ``finite-$m$ instability,’’ the first example of which was discovered in the 
context of superradiant instabilities of rotating black strings \cite{Dias:2022mde}. 
This behaviour is unconventional, since in typical black hole instabilities one finds the 
reverse pattern: if a black hole is unstable to a low-$m$ mode, it is already unstable to all 
higher-$m$ perturbations. For example, Kerr–AdS$_4$ black holes with angular velocity 
$\Omega_+L$ just above the Hawking–Reall bound ($\Omega_+L=1$) are unstable to arbitrarily 
large $m$ but not to $m=1$ or $m=2$; only for considerably larger $\Omega_+L$ does one 
encounter the lower-$m$ instabilities 
\cite{Cardoso:2013pza,Dias:2022mde}. 

Second, in regimes where BTZ is unstable to $m=1$ (or higher) and therefore also to $m=0$,  the $m=0$ instability is always the strongest in the sense that
${\rm Im}\,\hat{\omega}|_{m=0} > {\rm Im}\,\hat{\omega}|_{m=1} > {\rm Im}\,\hat{\omega}|_{m>1}$.

Third, every BTZ black hole that is unstable to modes with $m\ge 1$ occurs for a value of the 
double-trace parameter for which AdS$_3$ is already unstable to $m=0$ perturbations 
\cite{Ishibashi:2004wx}. 
Nevertheless, we have identified a window of double-trace parameters for which AdS$_3$ is 
stable (for all $m$) while BTZ may be either linearly stable or unstable to $m=0$ perturbations. 
Some of these BTZ solutions are additionally unstable to $m\geq 1$ modes, while others are not. 
We have mapped the corresponding regions of the BTZ $(M,J)$ parameter space.

In the AdS$_3$/CFT$_2$ duality, a BTZ black hole with temperature $T_+L$ and angular velocity 
$\Omega_+L$ is dual to a thermal state of the boundary CFT with the same temperature and 
chemical potential. Its quasinormal mode frequencies correspond to the characteristic 
oscillation frequencies governing the return to equilibrium of perturbed thermal states. 
The instabilities we found for the $m=0$ and $m\ge 1$ modes must therefore have counterparts in 
the CFT$_2$. These instabilities are especially interesting because, unlike the conventional 
superradiant instabilities of rotating AdS black holes in $d\geq 4$, the BTZ instabilities 
always occur for $\Omega_+L\leq 1$. Thus the Hawking–Reall bound remains satisfied and the 
dual CFT partition function is well-defined.

Although our analysis focused on double-trace perturbations, multi-trace deformations are also of considerable interest in AdS$_3$/CFT$_2$ since they correspond to multi-trace deformations of the boundary CFT \cite{Klebanov:1999tb,Witten:2001ua,Amsel:2007im,Sever:2002fk,Berkooz:2002ug,
Hertog:2004dr,Martinez:2004nb,Hertog:2004ns,Amsel:2006uf,Faulkner:2010fh,Faulkner:2010gj,Witten:2003ya,Ishibashi:2004wx,Marolf:2006nd,Compere:2008us}. 
Our results can serve as a guide for future analyses of multi-trace perturbations, 
which we expect to share qualitative features with the double-trace case.

We believe the rotating BTZ black hole studied here provides a prototype for 
double-trace instabilities of higher-dimensional ($d\geq 4$) rotating AdS black holes. 
Indeed, static charged AdS$_4$ black holes exhibit double-trace instabilities for both 
$m=0$ and $m\ge 1$ modes 
\cite{Faulkner:2010gj,Dias:2013bwa,Katagiri:2020mvm,Harada:2023cfl,Kinoshita:2023iad}, 
much like BTZ. Nevertheless, this conjecture warrants closer inspection, and work in this direction is underway.

Finally, the instability of BTZ to $m=0$ and $m\ge 1$ double-trace modes implies the existence of stationary asymptotically AdS$_3$ solutions with scalar hair: rotating AdS$_3$ hairy black holes in which a scalar condensate co-rotates or “floats’’ above the horizon. Moreover, one expects the existence of AdS$_3$ boson stars or solitons associated with the double-trace instability. In a companion paper we show that such solutions indeed exist and we analyse their properties \cite{DSS:2026}. For fixed mass and angular momentum, the hairy black holes have greater entropy than BTZ, 
making them the natural endpoint of the double-trace instability.

\begin{acknowledgments}

The authors would like to acknowledge Javier Carballo, Kostas Skenderis and Ben Withers for insightful discussions.
O.J.C.D. acknowledges financial support from the  STFC ``Particle Physics Grants Panel (PPGP) 2020" Grant No.~ST/X000583/1. J.~E.~S. has been partially supported by STFC consolidated grant ST/X000664/1 and by Hughes Hall College. The authors also acknowledge the use of the IRIDIS High Performance Computing Facility, and associated support services at the University of Southampton, in the completion of this work.
\end{acknowledgments}


\appendix

\section{Results for scalar field masses \texorpdfstring{$\mu^2L^2=\{-8/9,-3/4,-1/10\}$}{mu2L2={-8/9,-3/4,-1/10}}}\label{sec:AppOtherMasses}
\begin{figure}[ht]
    \centering
    \includegraphics[width=0.48\linewidth]{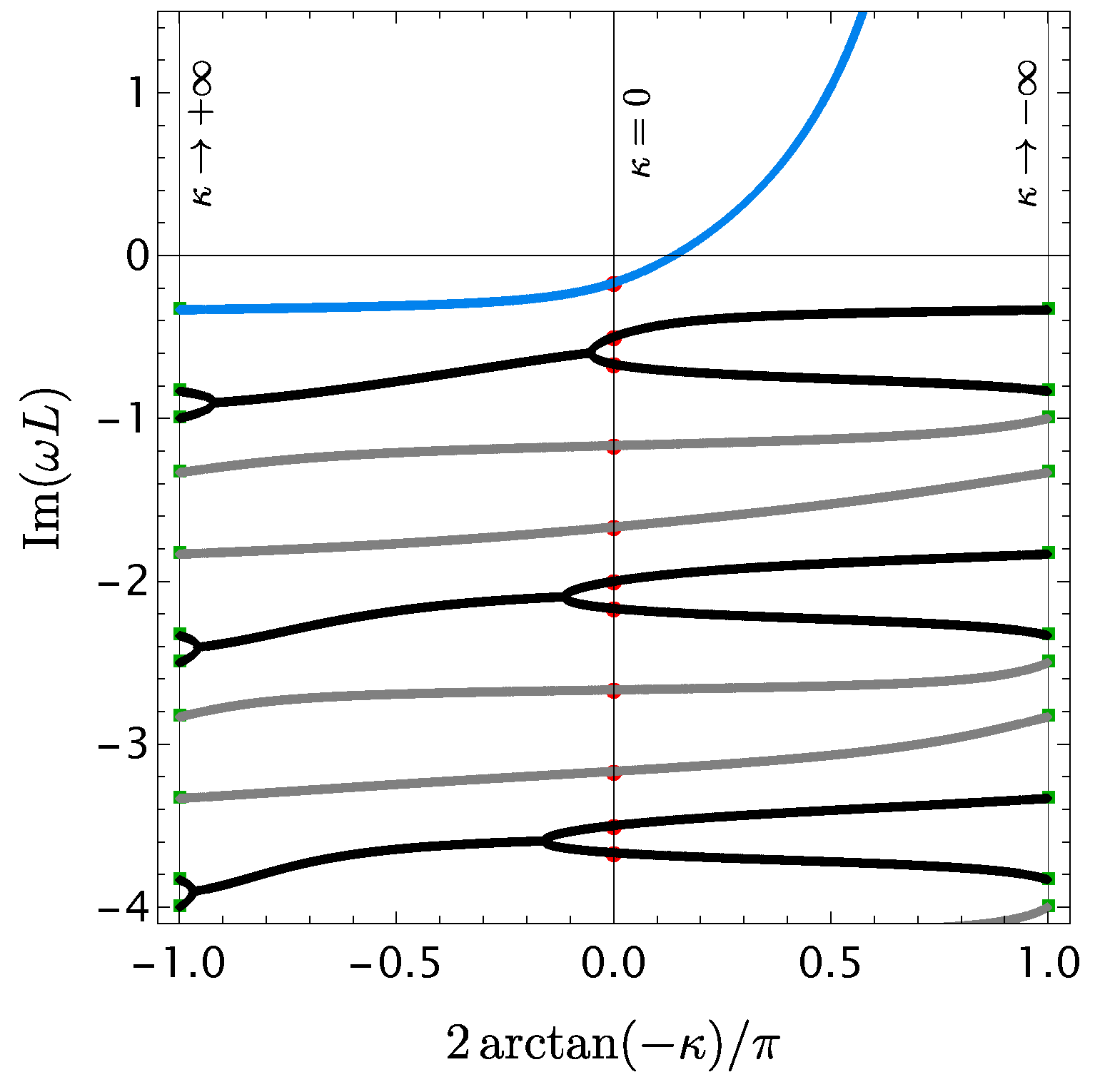}
    \includegraphics[width=0.48\linewidth]{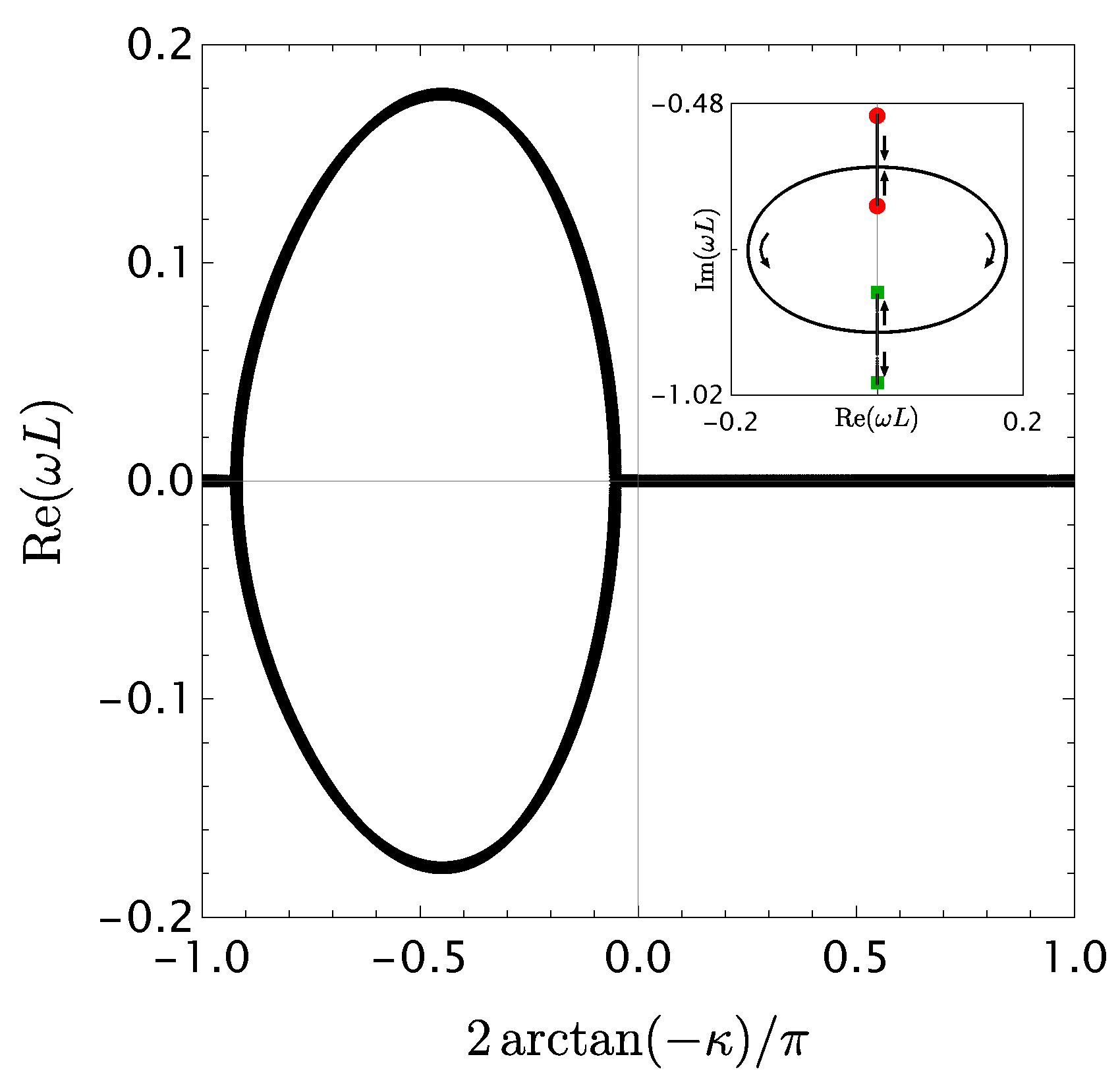}\\
     \includegraphics[width=0.48\linewidth]{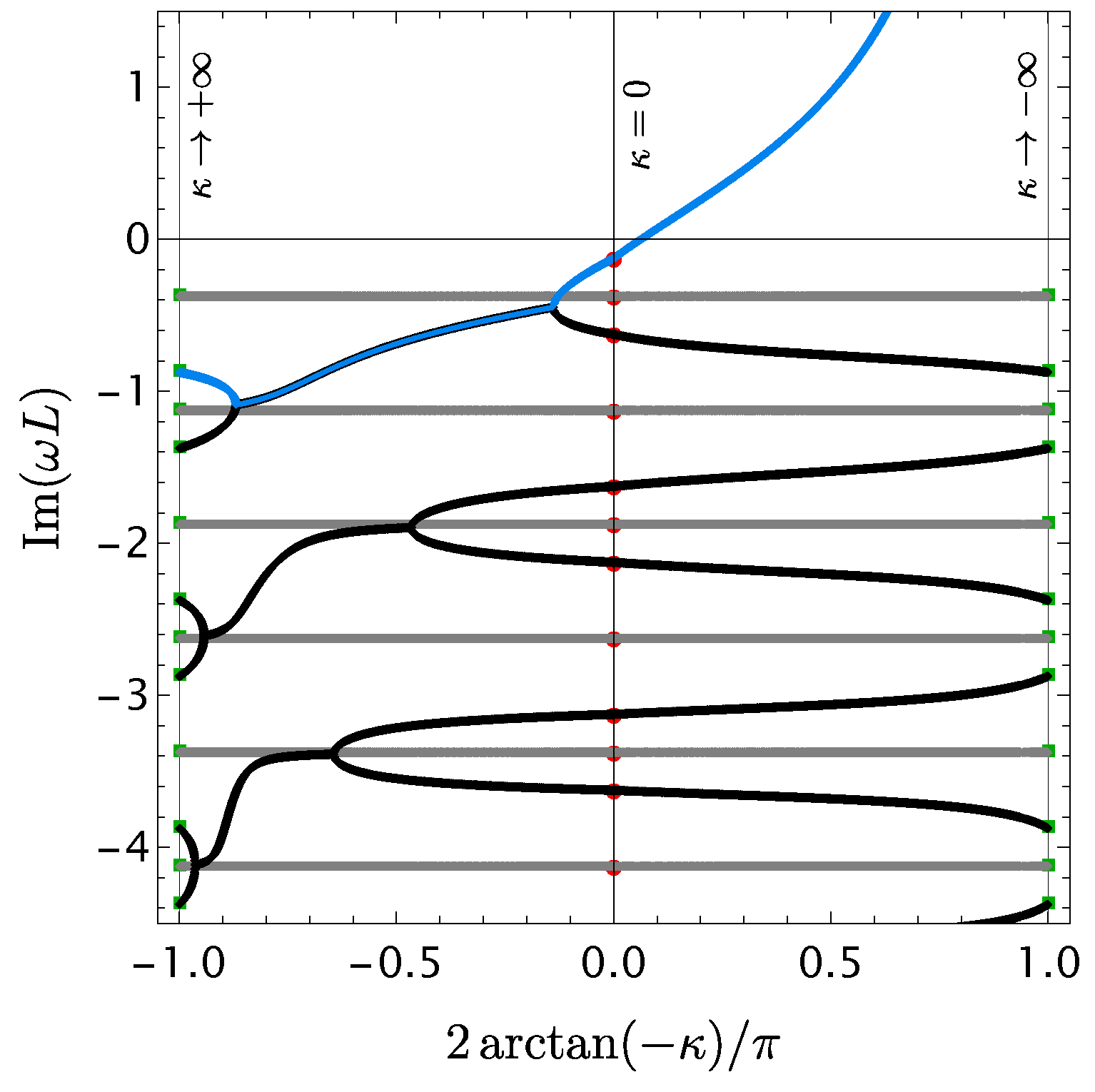}
      \includegraphics[width=0.48\linewidth]{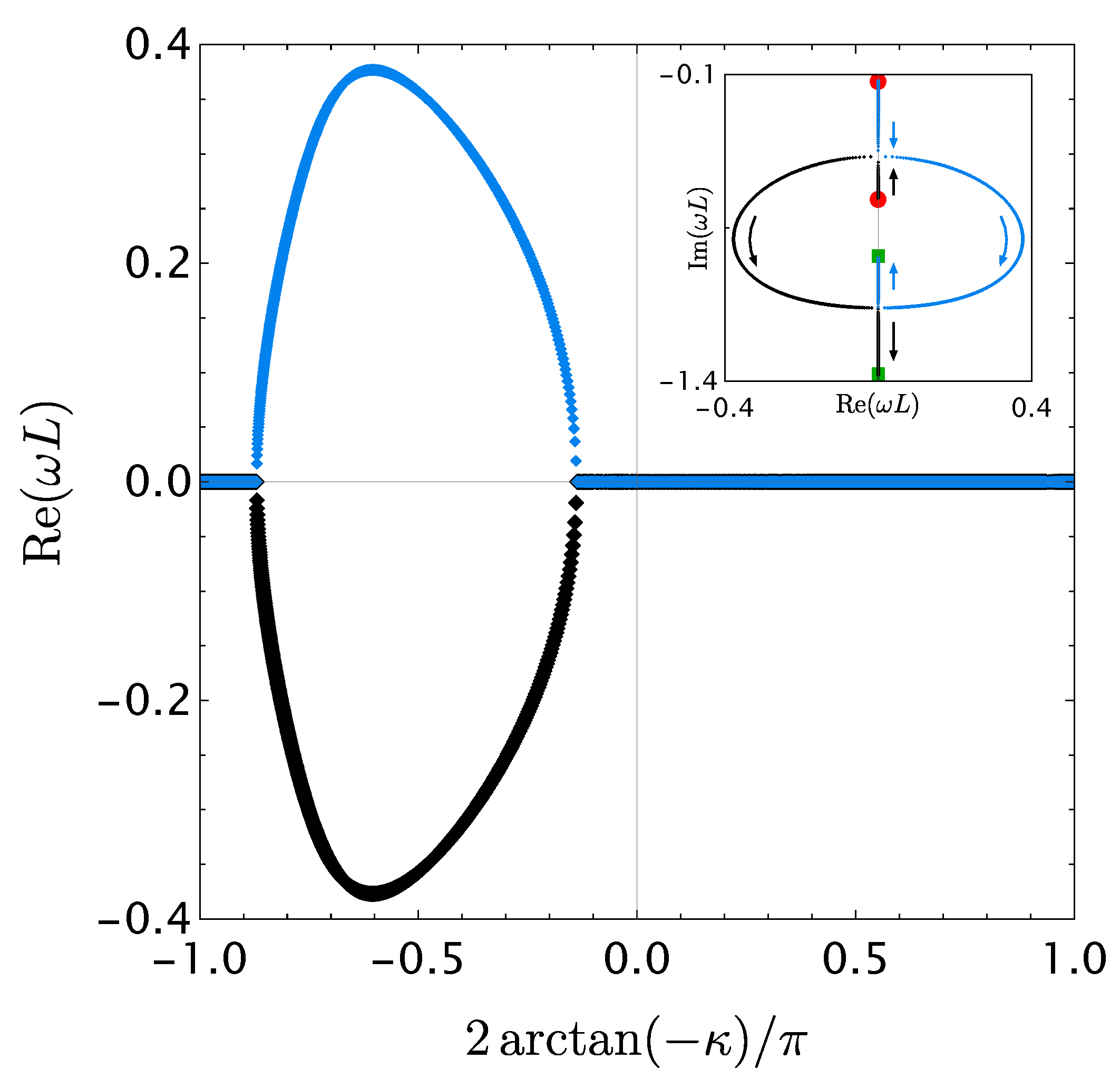} \\
    \includegraphics[width=0.48\linewidth]{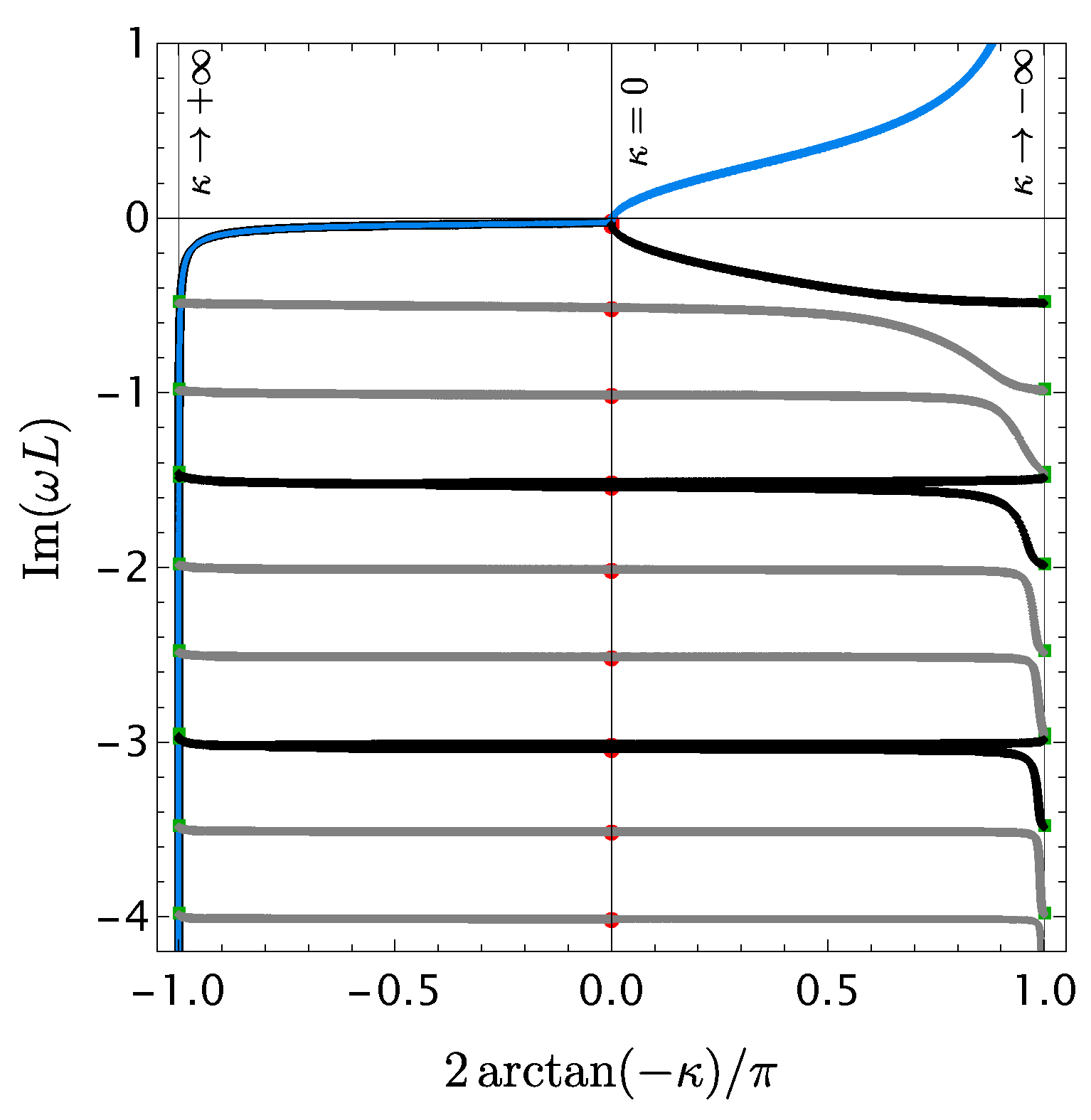}
     \includegraphics[width=0.48\linewidth]{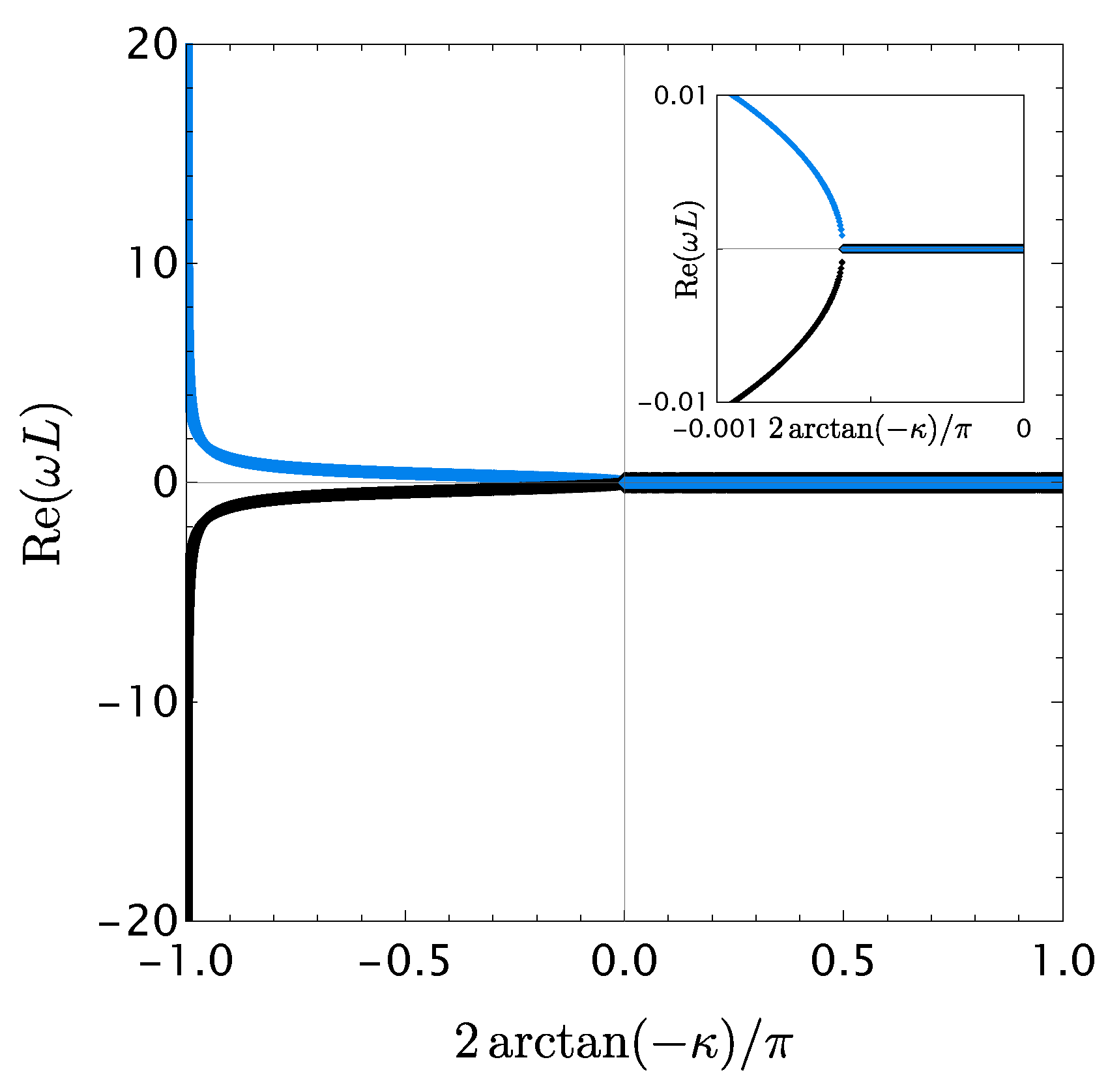}
    \caption{
Analogue of Fig.~\ref{fig:BTZw-kappa:m0}, \ie the $\bm{m=0}$ modes in {\bf BTZ} with 
$\{\hat{M},\hat{J}\}=\{5/16,1/4\}$ (equivalently, $\{y_-,y_+\}=\{1/4,1/2\}$), now shown for scalar masses $\bm{\mu^2L^2=-8/9}$ (top panel), $\bm{\mu^2L^2=-3/4}$ (middle panel), and $\bm{\mu^2L^2=-1/10}$ (bottom panel).
}
    \label{fig:BTZw-kappa:m0-OtherMasses}
 \end{figure}

\begin{figure}[ht]
\vskip -1.3cm
    \centering
    \includegraphics[width=0.48\linewidth]{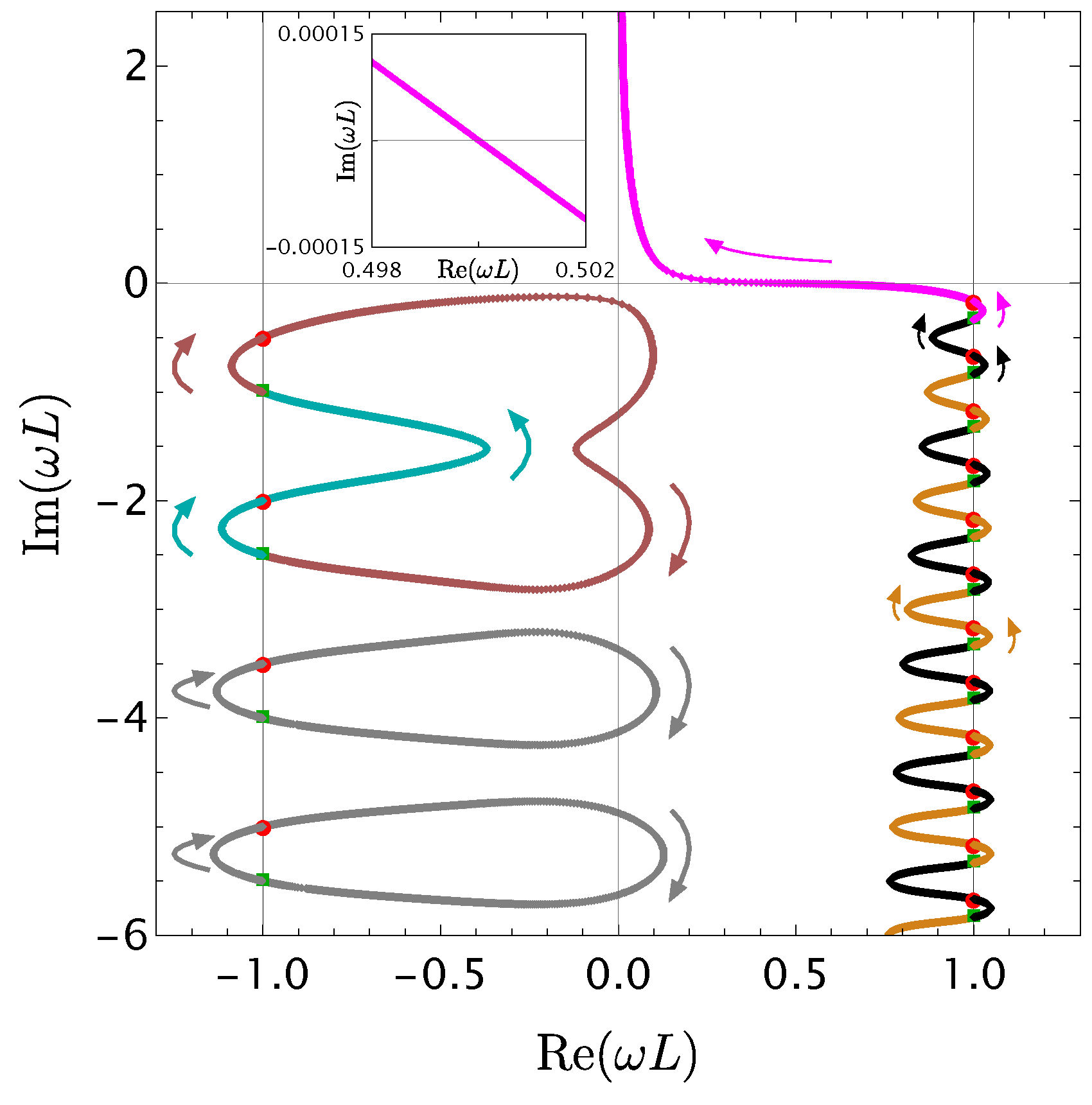}
     \includegraphics[width=0.49\linewidth]{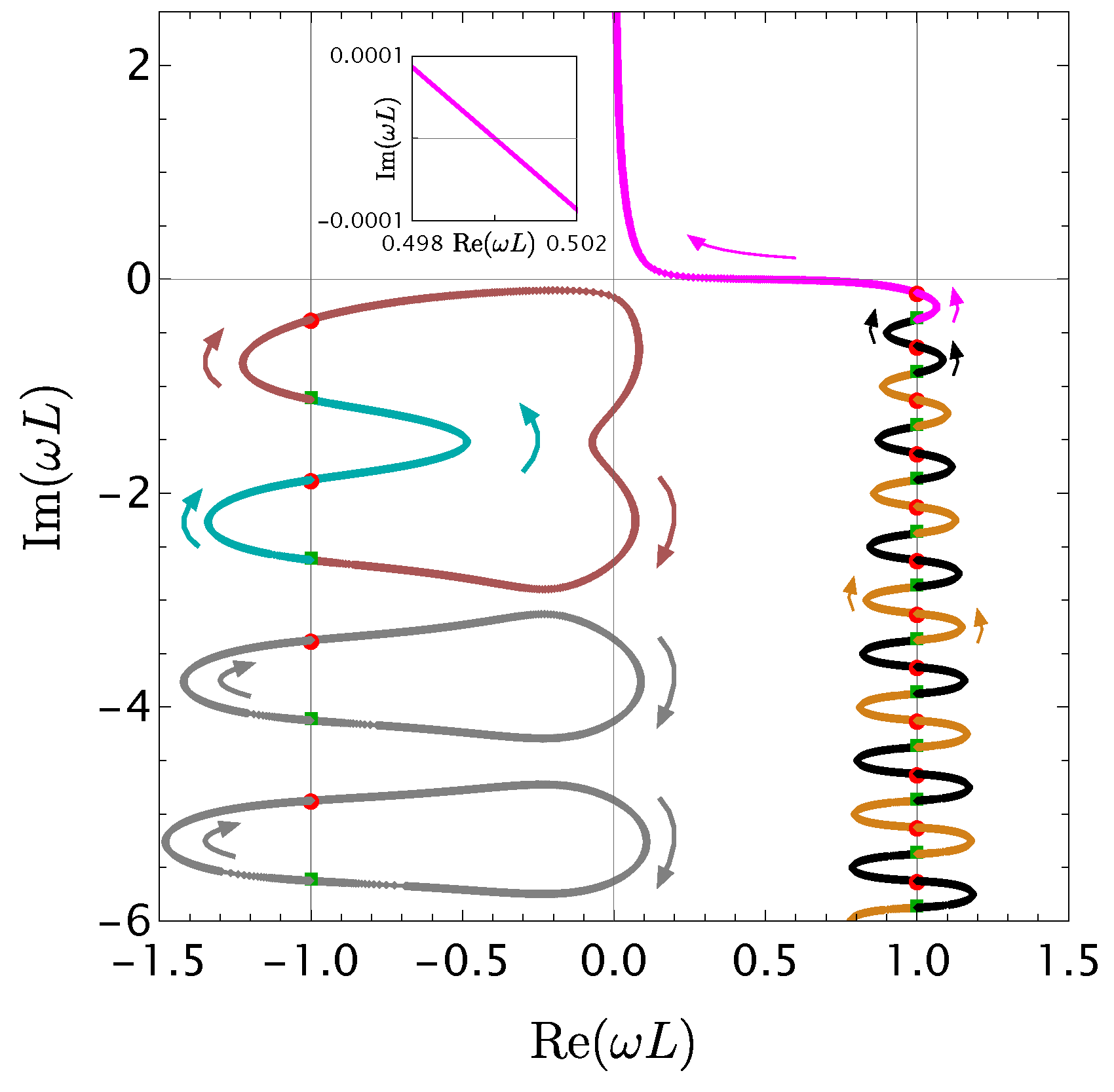}\\
     \includegraphics[width=0.51\linewidth]{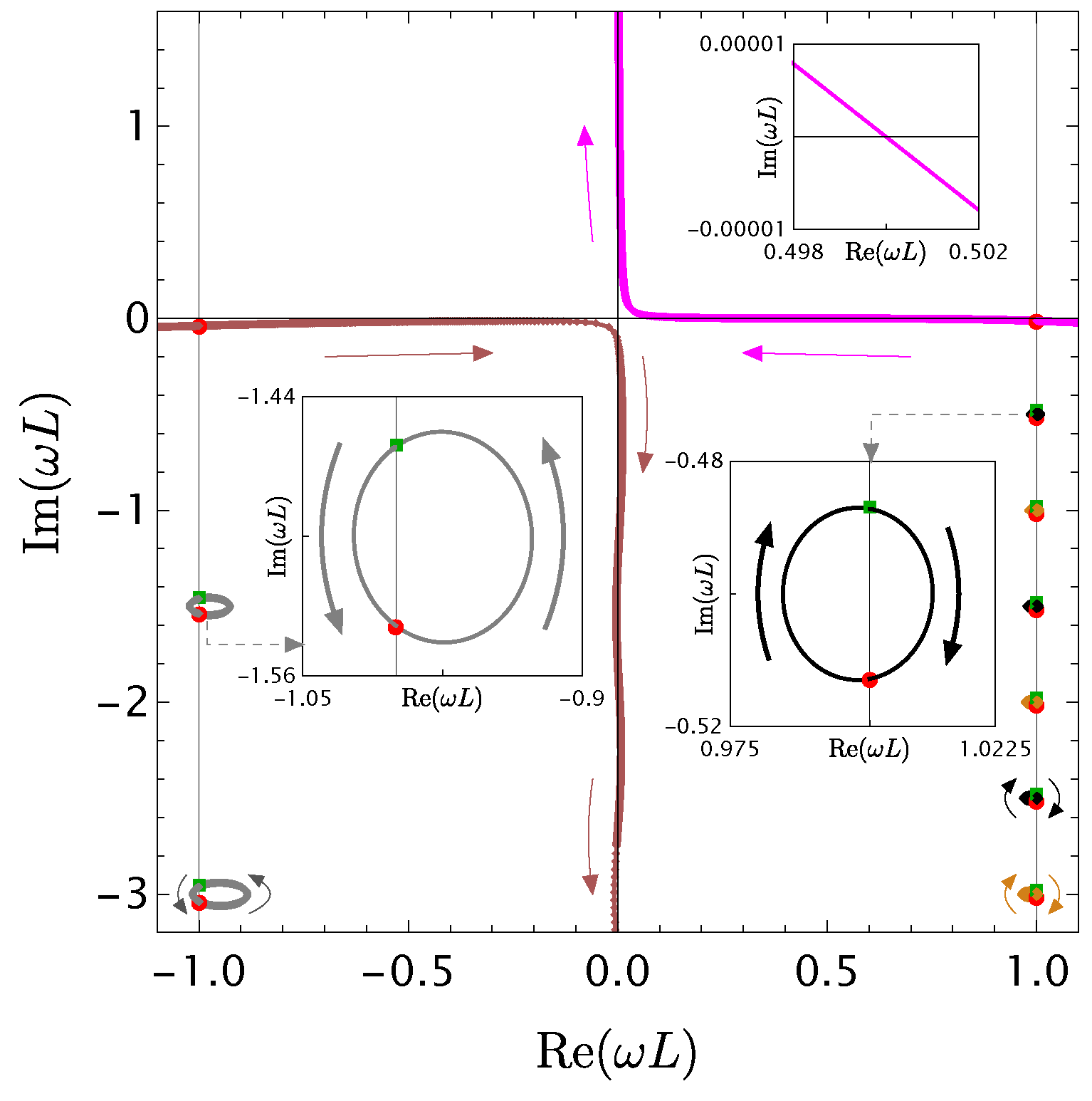}\\
      \includegraphics[width=0.48\linewidth]{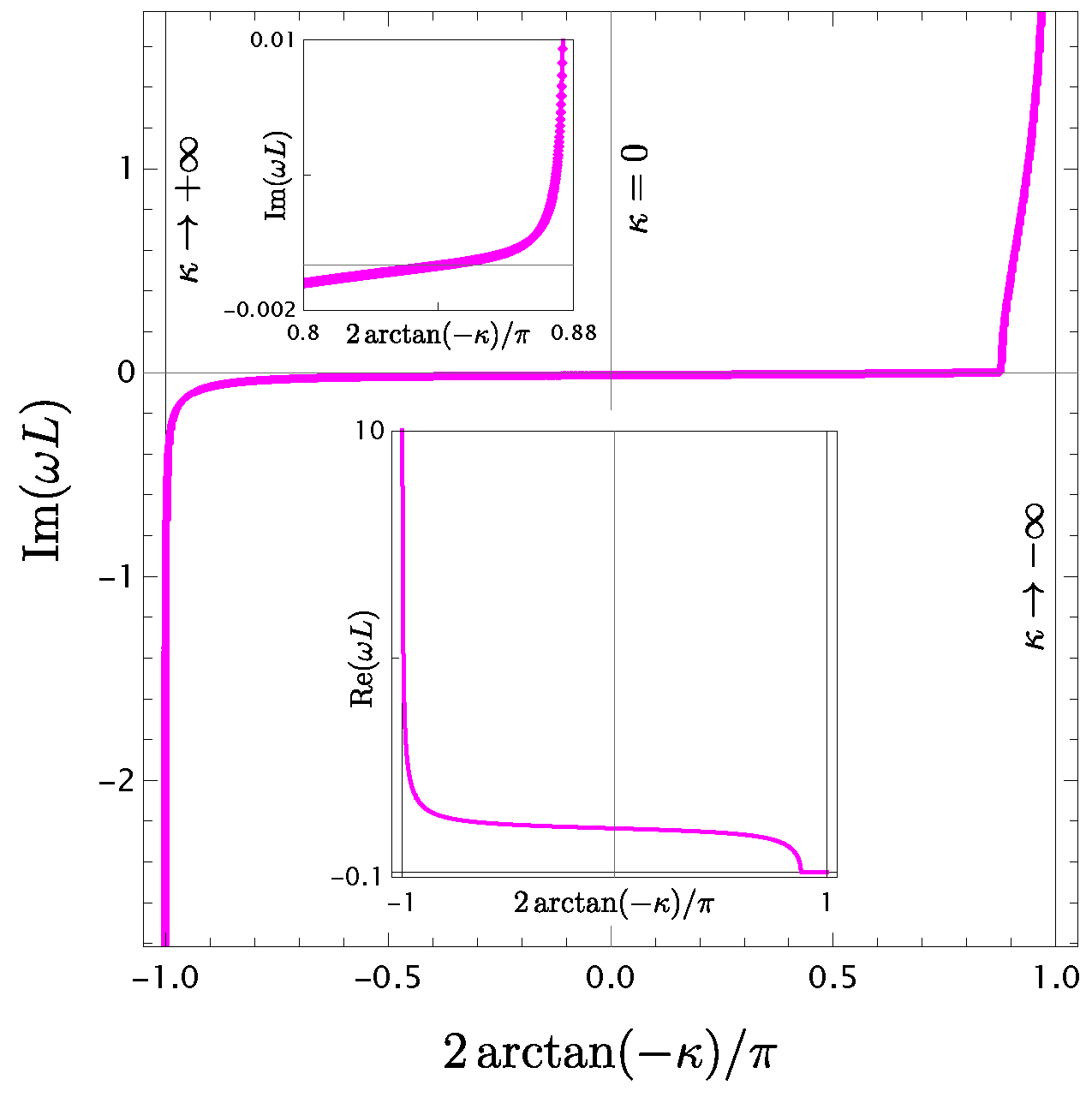}
     \includegraphics[width=0.48\linewidth]{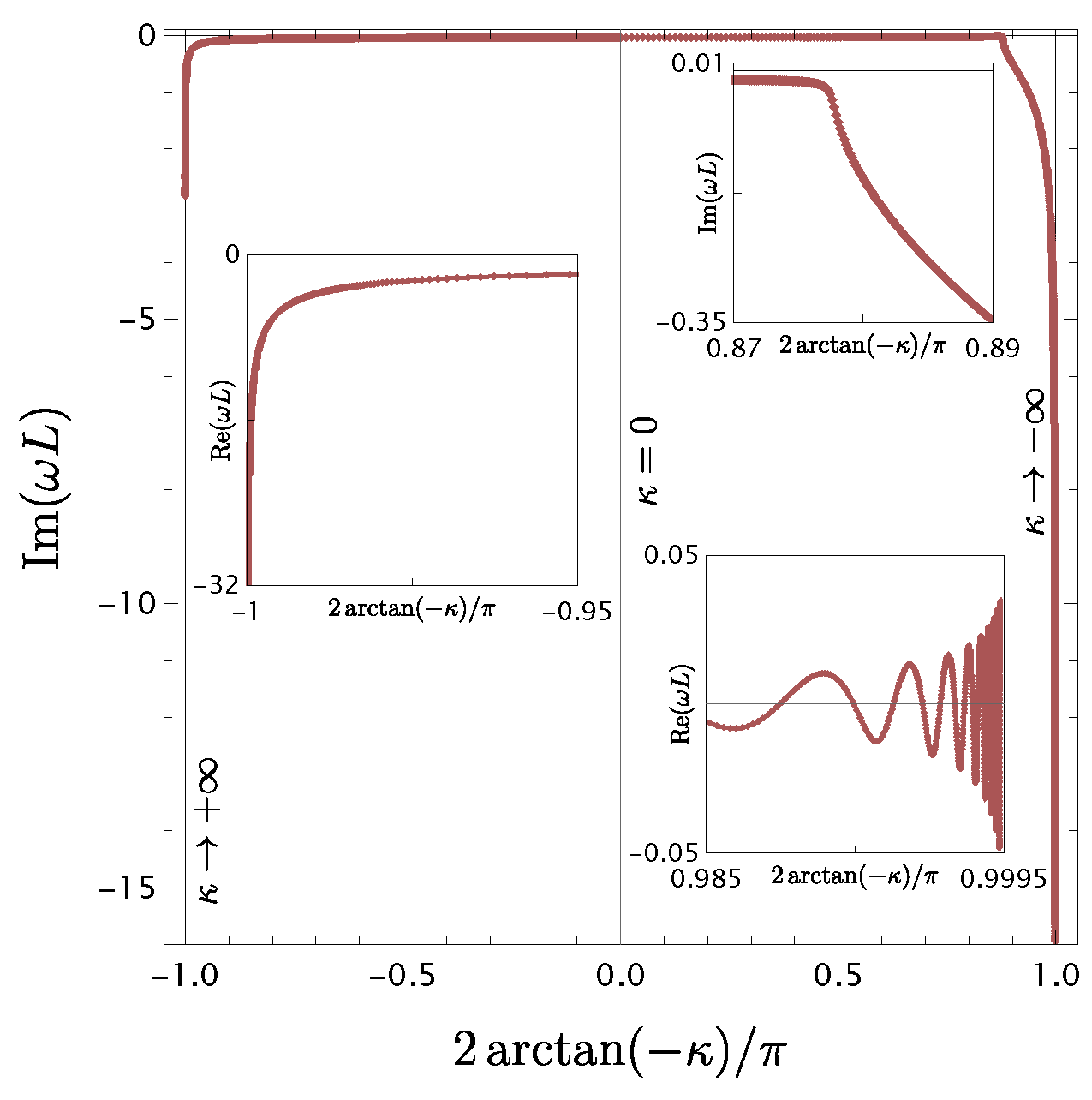}
    \caption{\small Analogue of the top-right panel of Fig.~\ref{fig:BTZw-kappa:m1}, i.e.\ $m=1$ modes in {\bf BTZ} with $\{\hat{M},\hat{J}\} = \{5/16,1/4\}$, but for different scalar masses: $\mu^2L^2= -8/9$ (top-left), $\mu^2L^2= -3/4$ (top-right), and $\mu^2L^2= -1/10$ (middle panel). The bottom panel shows details of the counter-rotating $n=0$ dark-red mode (bottom-left) and the co-rotating $n=0$ magenta mode (bottom-right) for the $\mu^2 L^2=-1/10$ spectrum displayed in the middle panel.}
    \label{fig:BTZw-kappa:m1-OtherMasses}
\end{figure}

\begin{figure}[ht]
    \centering
    \includegraphics[width=0.49\linewidth]{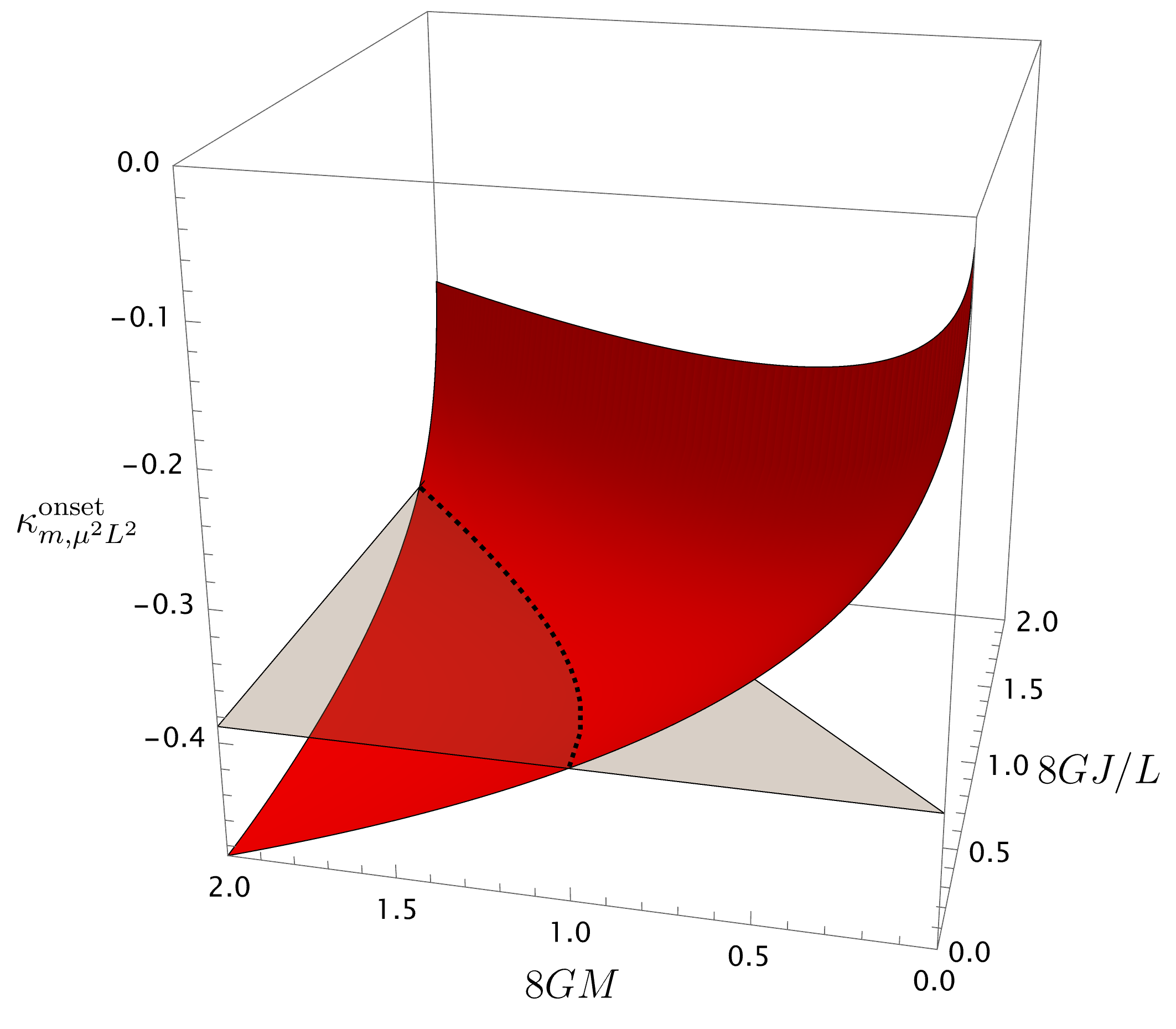}
    \includegraphics[width=0.49\linewidth]{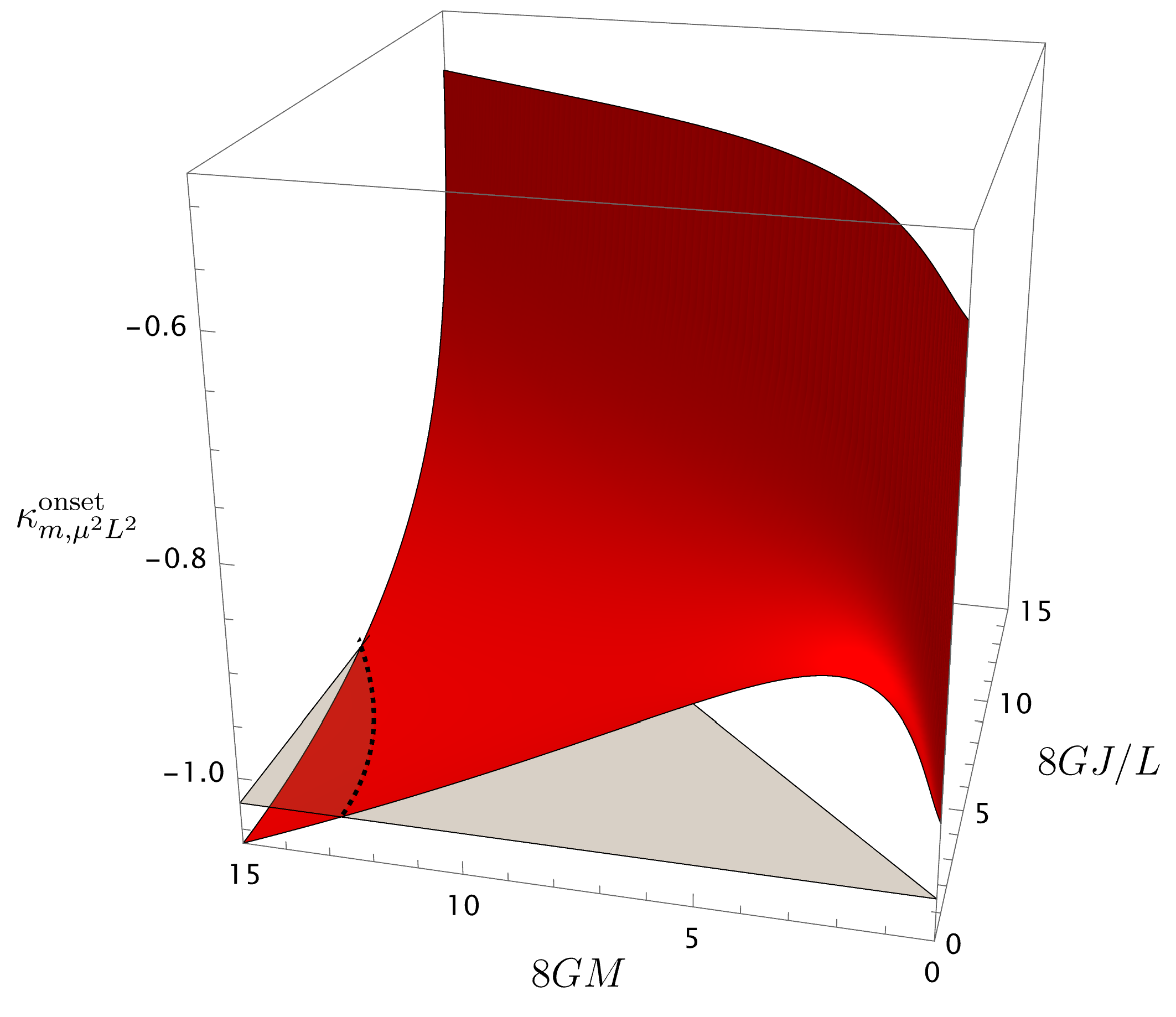}\\
    \includegraphics[width=0.49\linewidth]{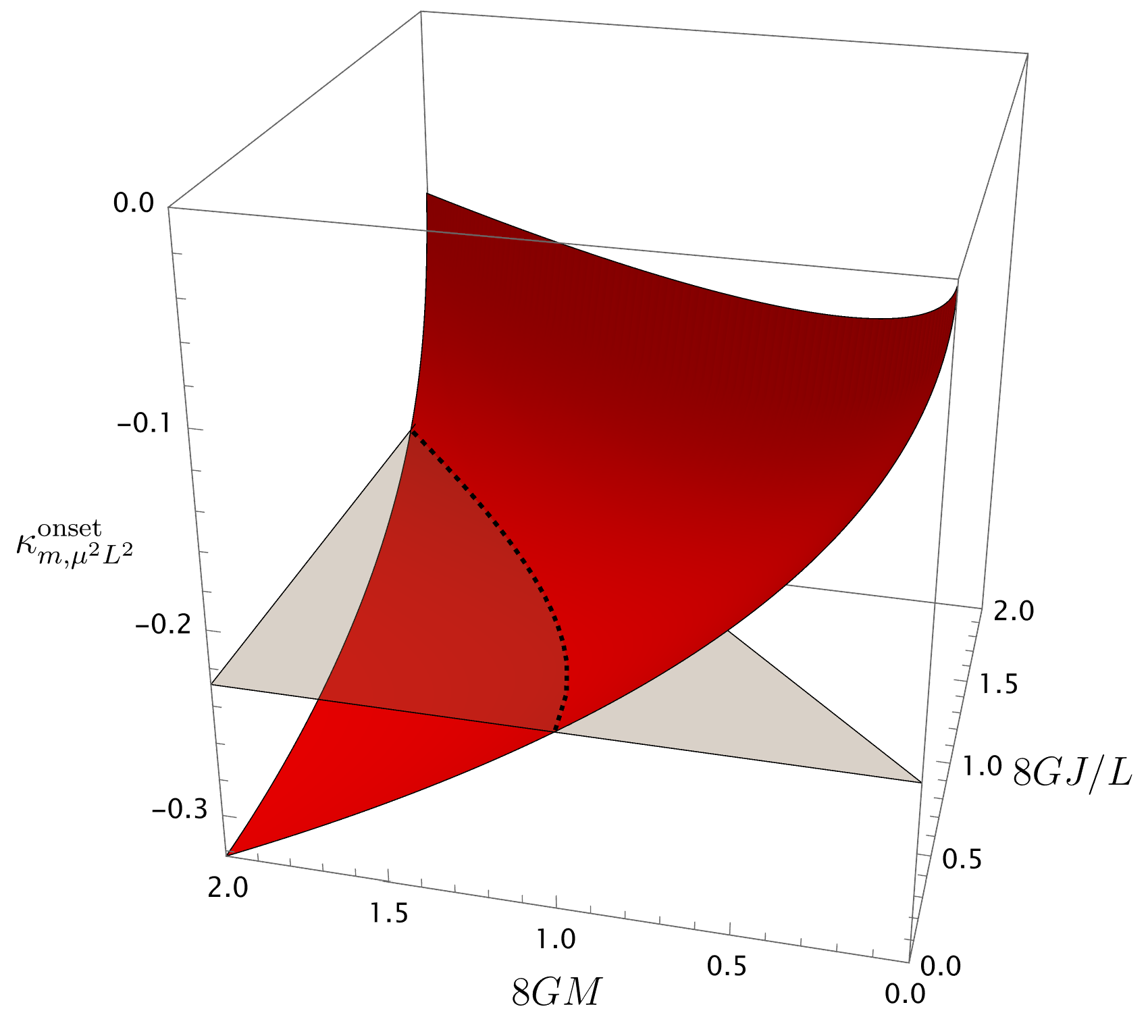}
    \includegraphics[width=0.49\linewidth]{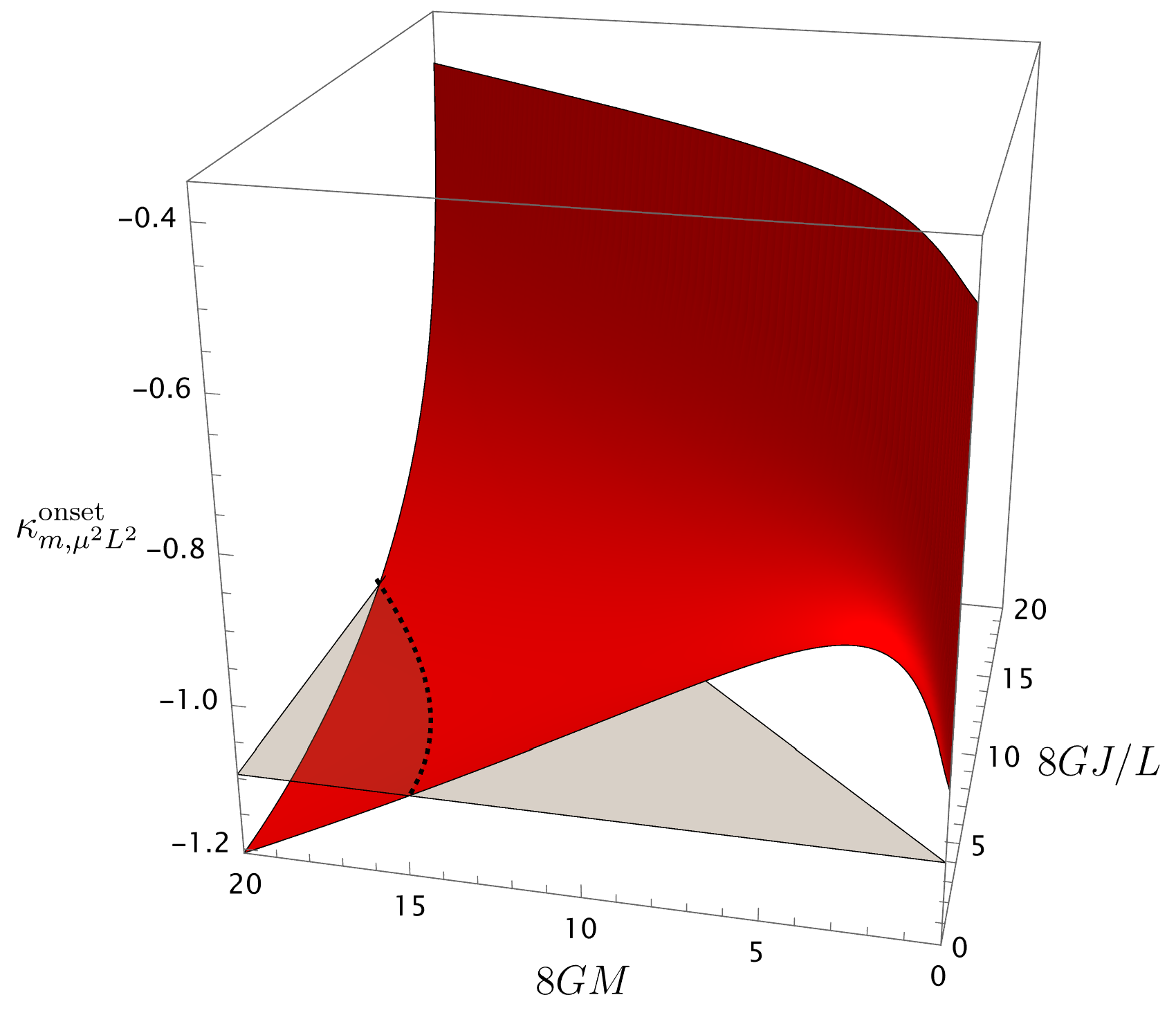}\\
    \includegraphics[width=0.49\linewidth]{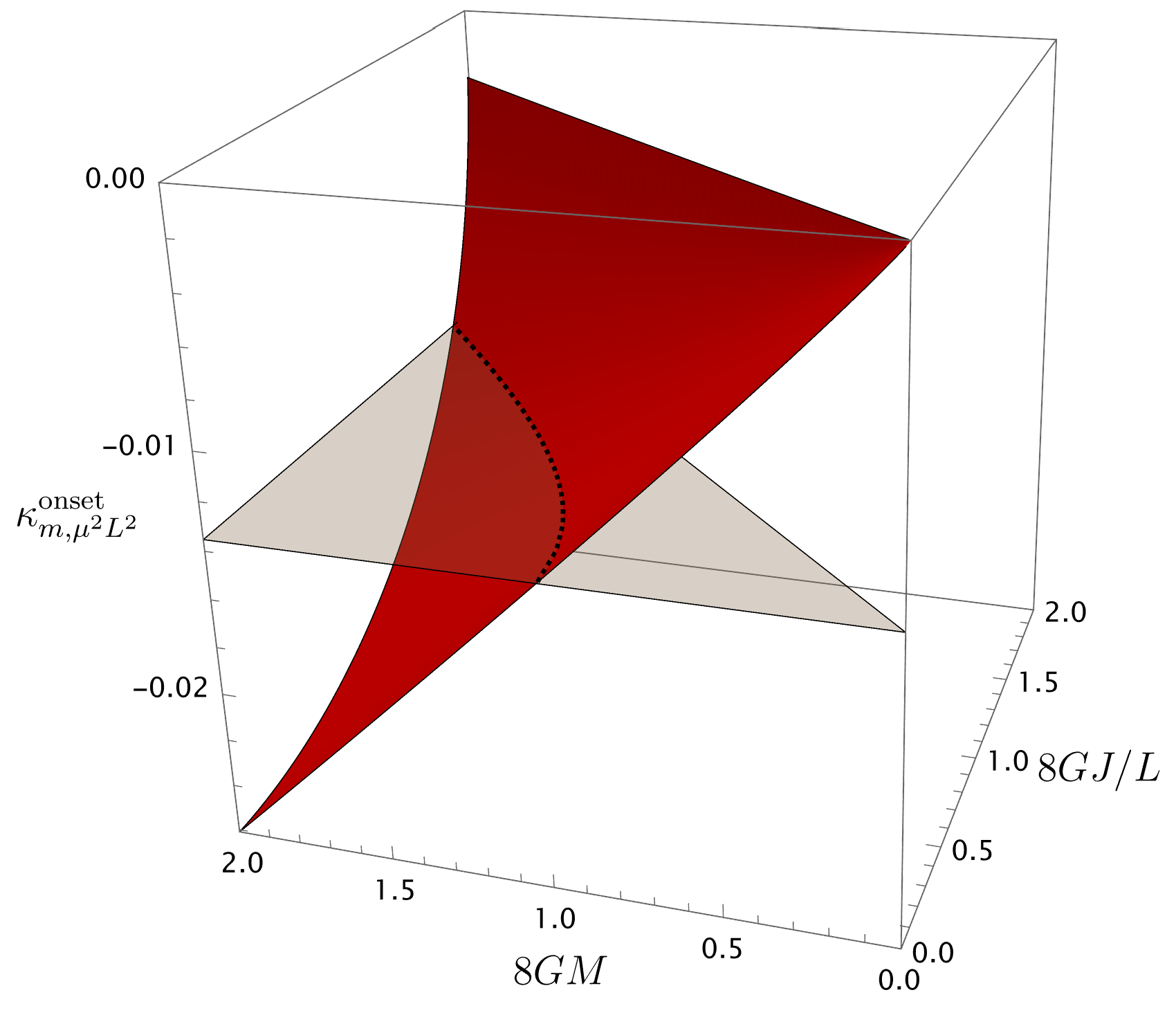}
    \includegraphics[width=0.49\linewidth]{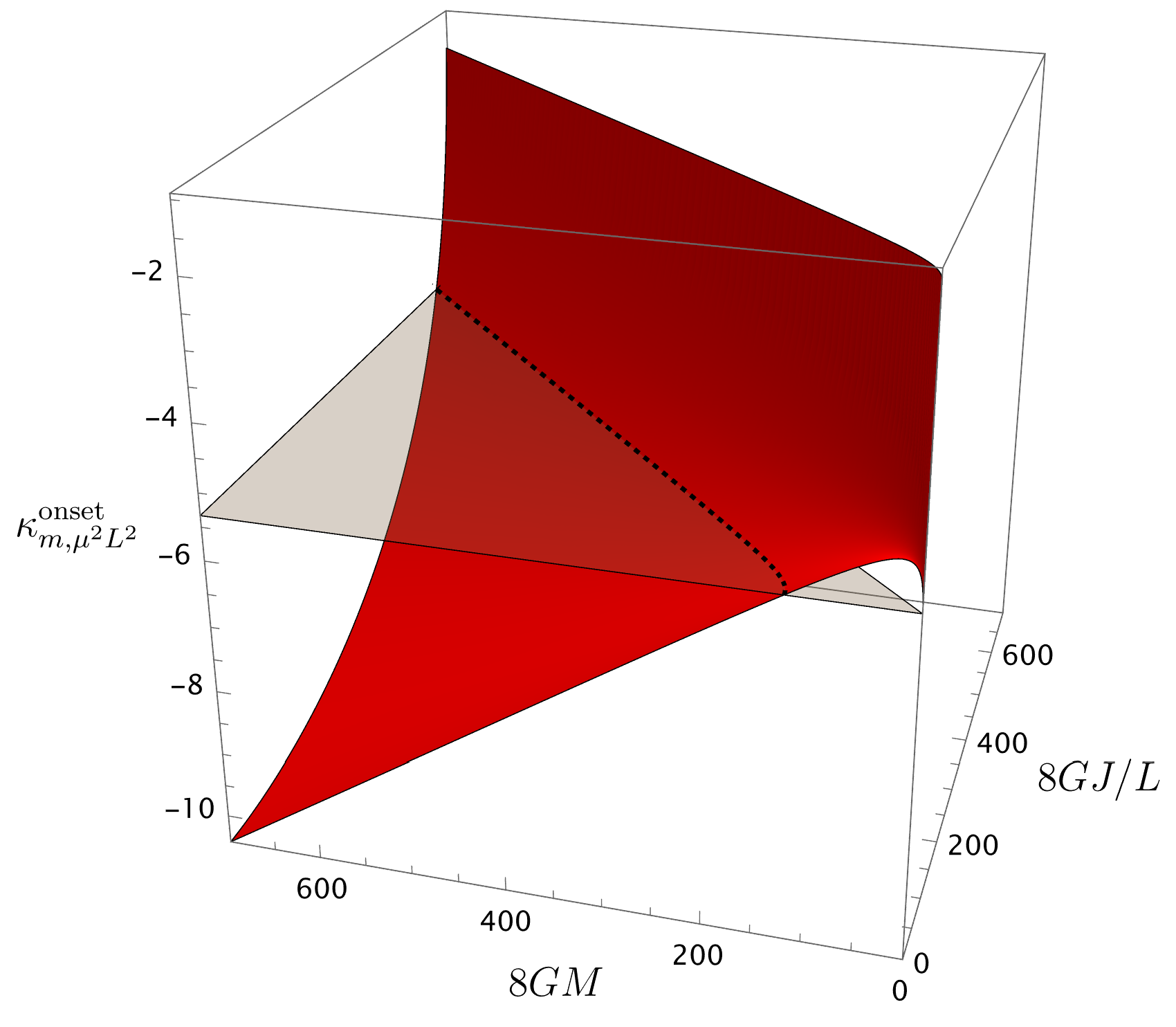}
   \caption{Analogue of Fig.~\ref{fig:3D_kappa_onset}, \ie instability onset values $\kappa^{\hbox{\tiny BTZ}}_{m,\hat{\mu}^2}(\hat{M},\hat{J})$ (red surface) and 
$\kappa^{\hbox{\tiny AdS}}_{m,\hat{\mu}^2}$ (flat grey surface) for $n=0$ double-trace 
perturbations with $m=0$ (left panels) and $m=1$ (right panels), now shown for scalar masses $\bm{\mu^2L^2=-8/9}$ (top), $\bm{\mu^2L^2=-3/4}$ (middle), and $\bm{\mu^2L^2=-1/10}$ (bottom).}
    \label{fig:3D_kappa_onset-OtherMasses}
\end{figure}

\begin{figure}[ht]
    \centering
    \includegraphics[width=0.45\textwidth]{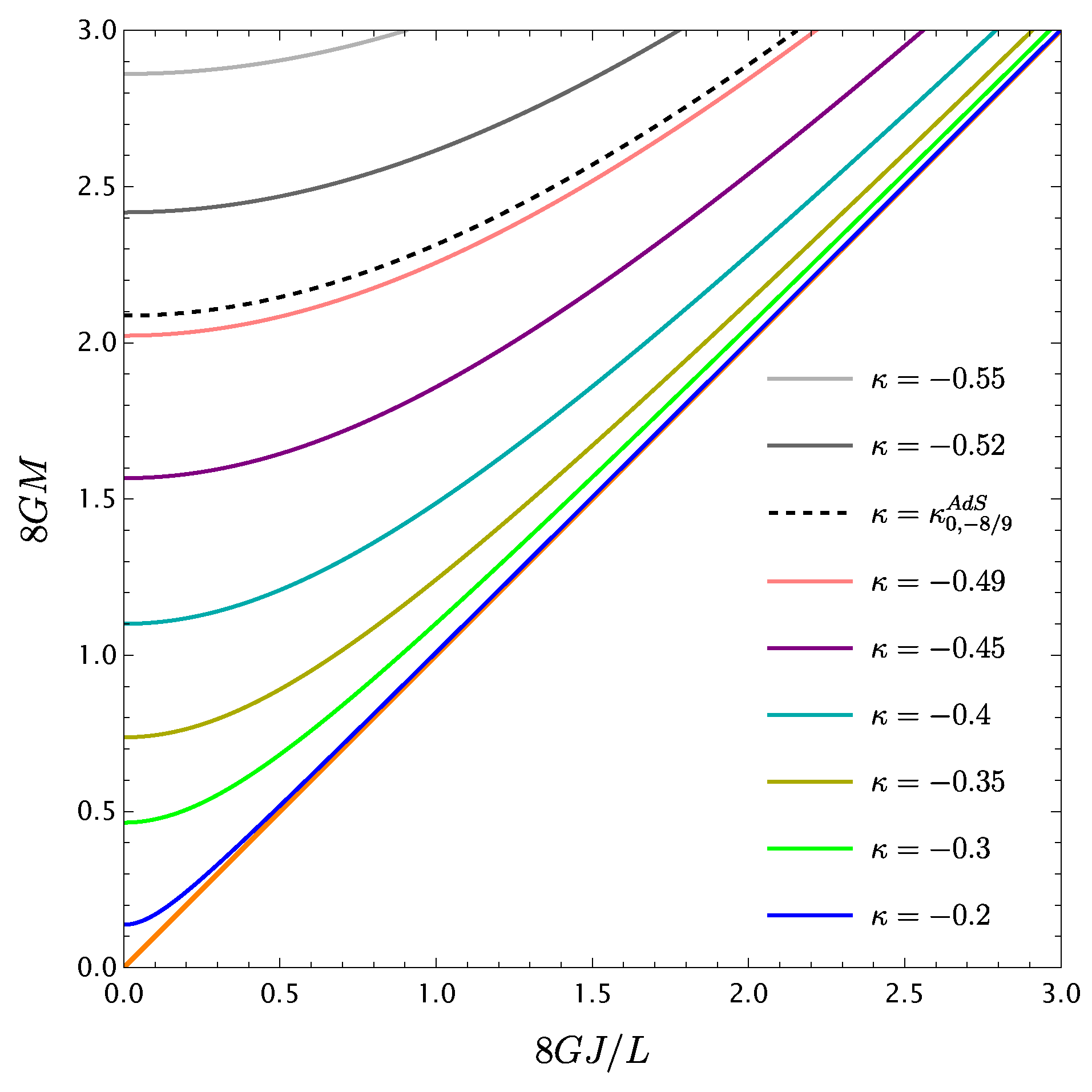}\hspace{1cm}
     \includegraphics[width=0.44\textwidth]{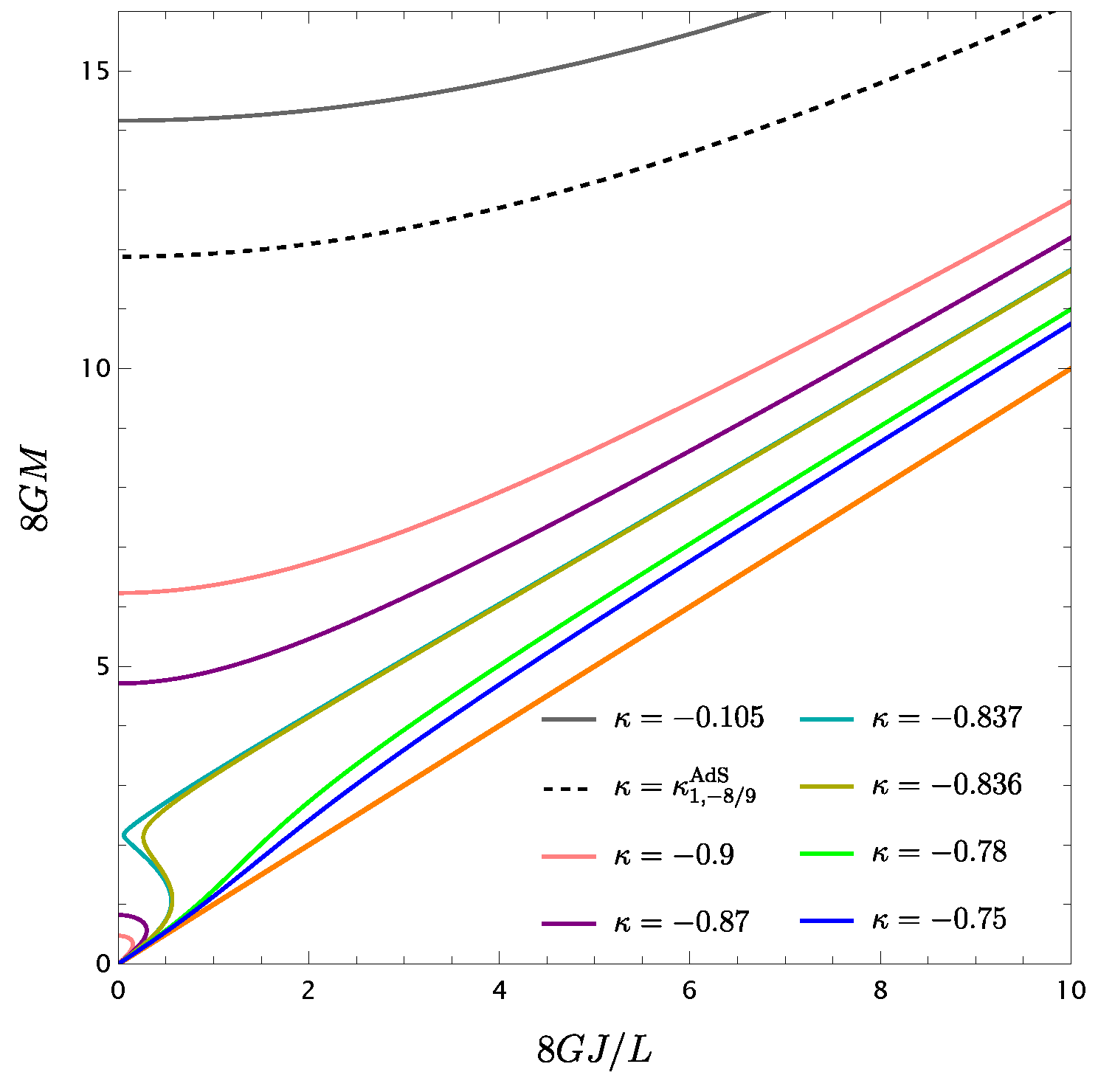}\\
       \includegraphics[width=0.45\textwidth]{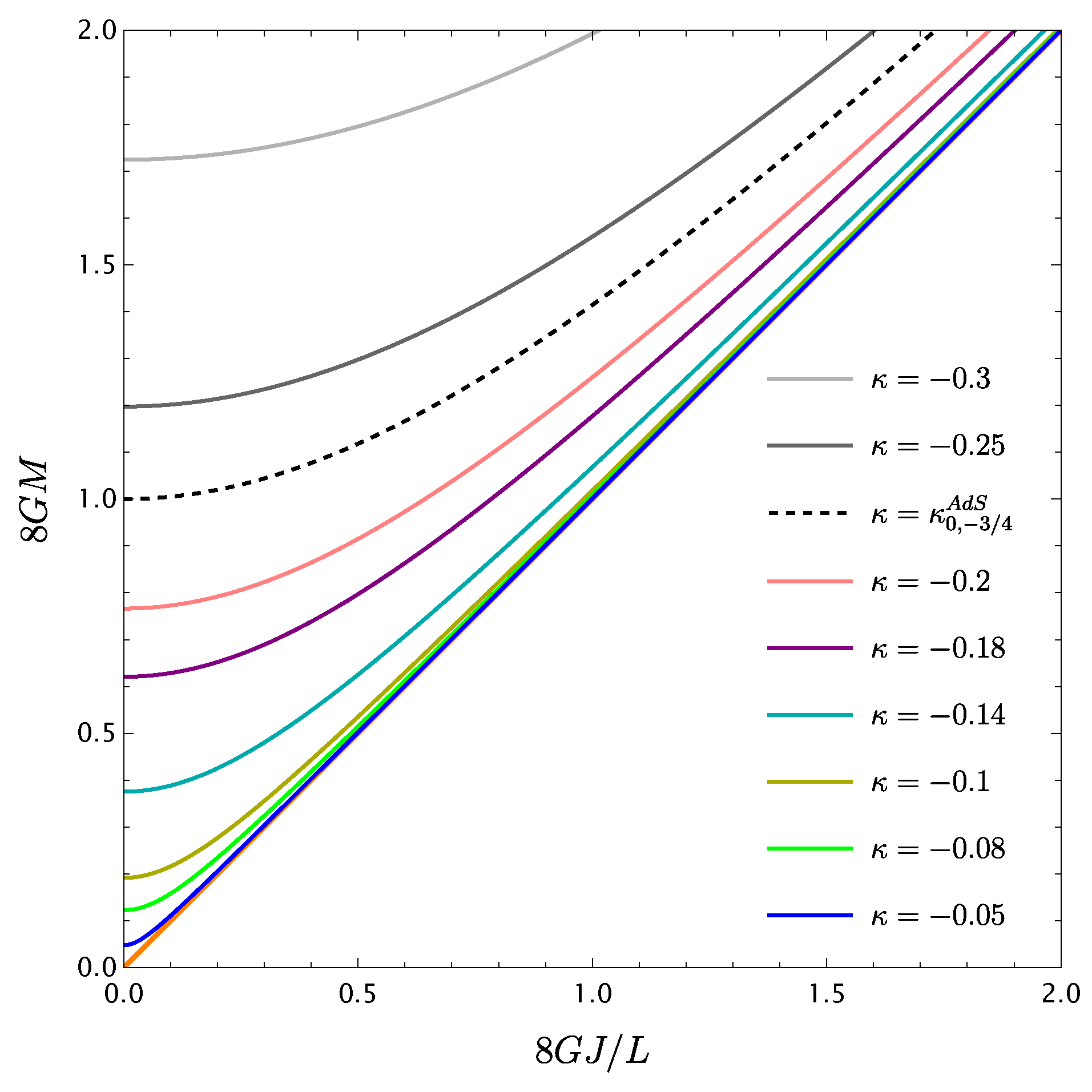}\hspace{1cm}
     \includegraphics[width=0.44\textwidth]{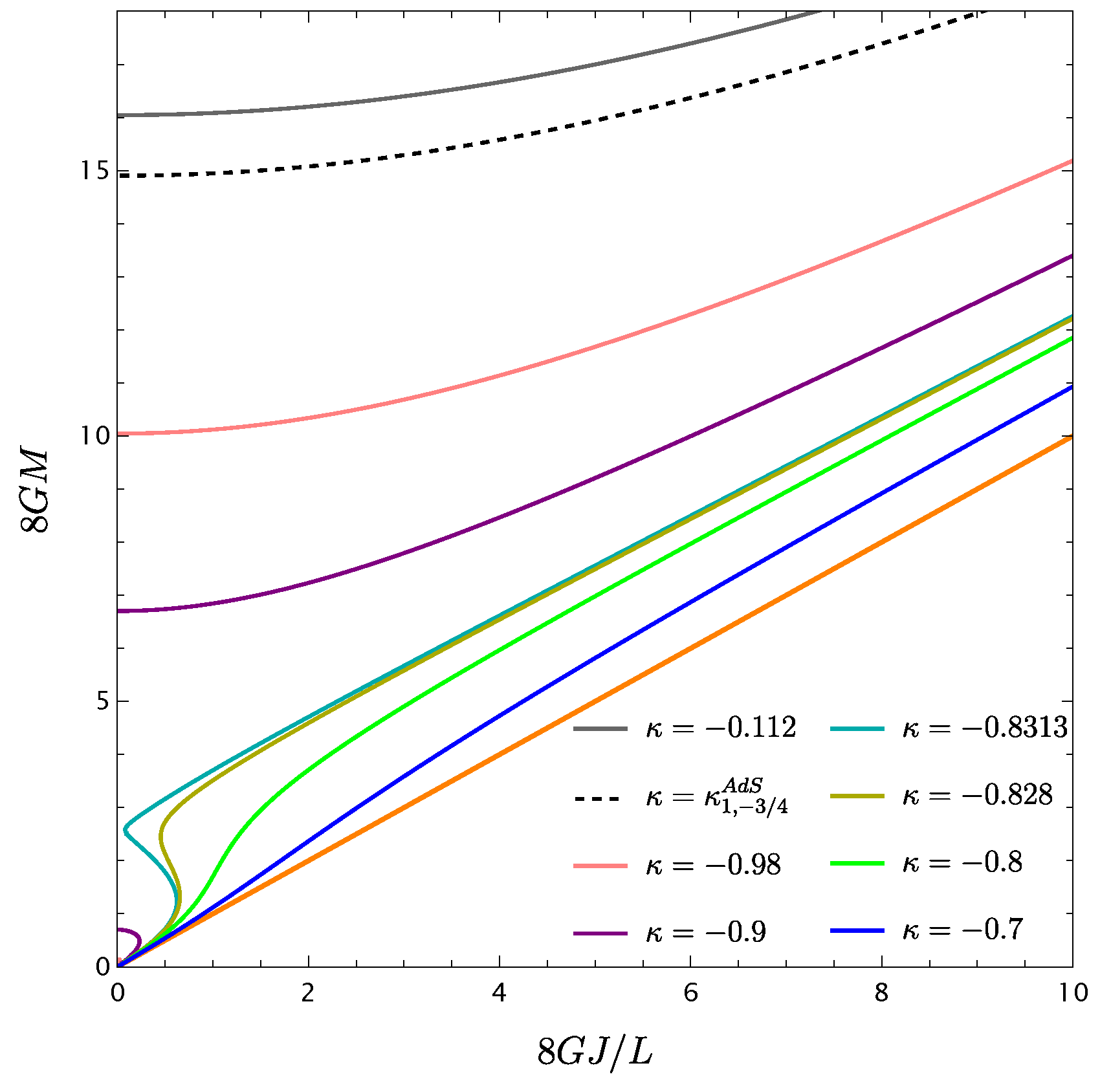}\\
         \includegraphics[width=0.45\textwidth]{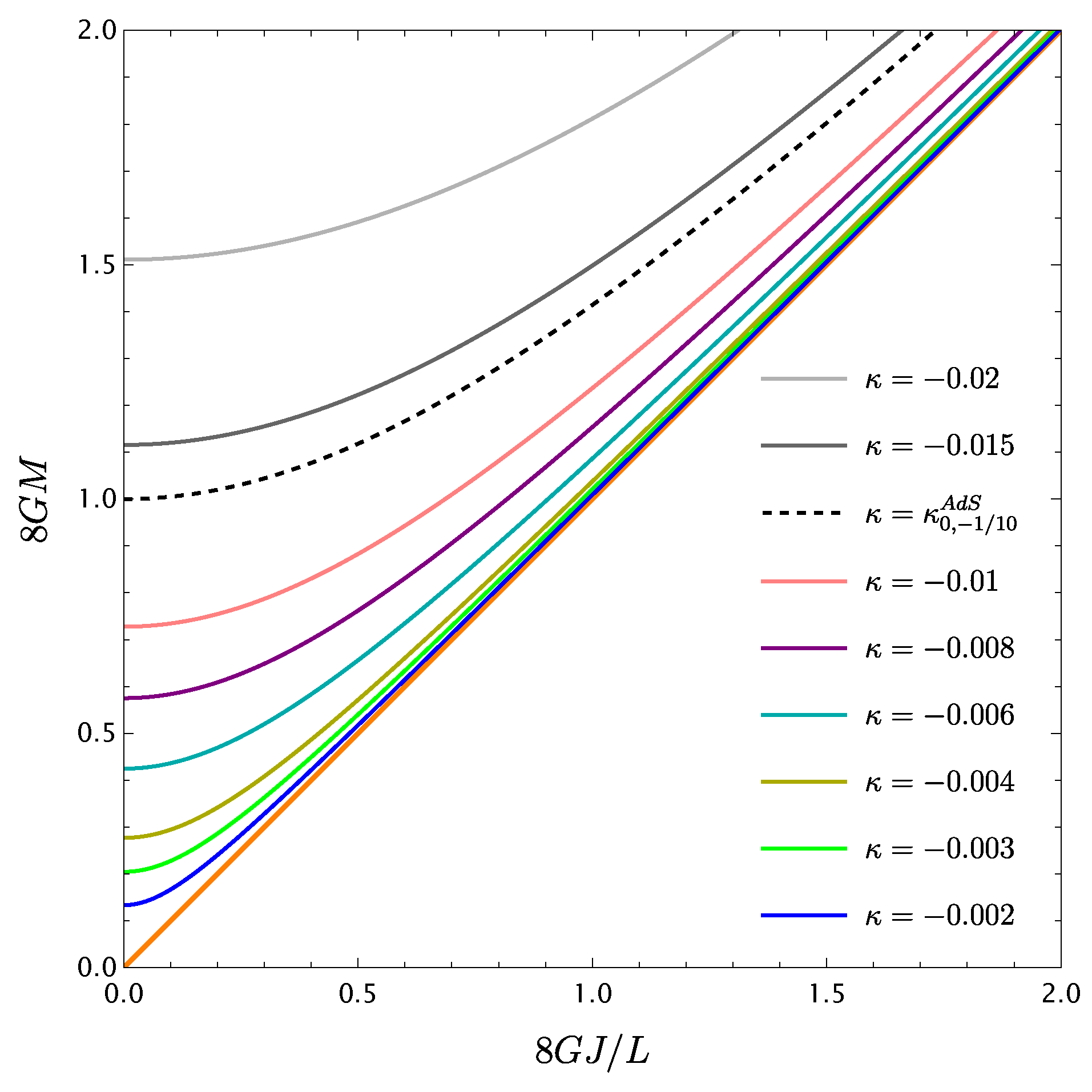}\hspace{1cm}
     \includegraphics[width=0.44\textwidth]{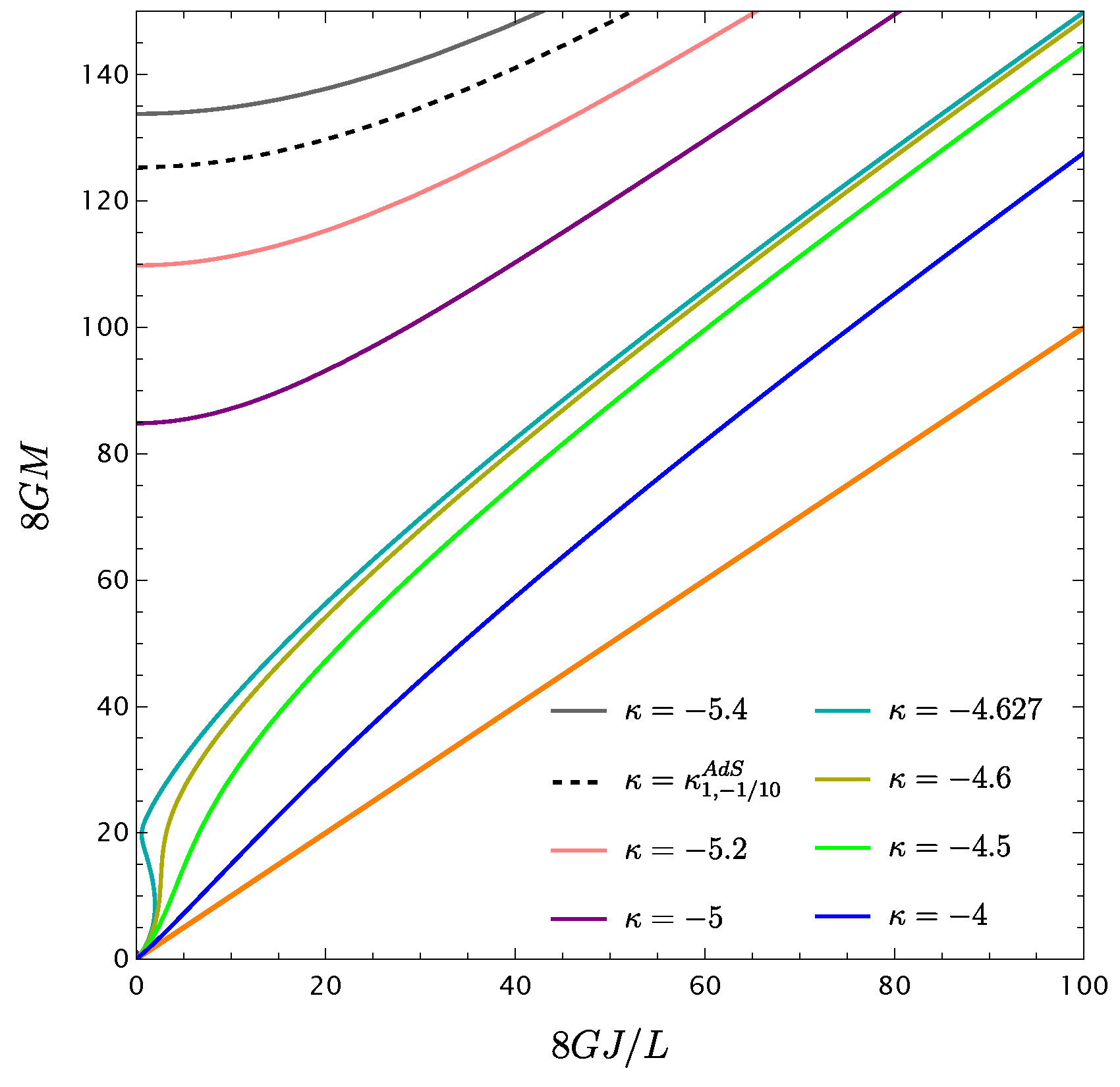}\\
   \caption{Analogue of Fig.~\ref{fig:m0_kappa_onset_curves} (left panel; $m=0$) and Fig.~\ref{fig:m1_kappa_onset_curves} (right panel; $m=1$), but now for scalar masses $\bm{\mu^2L^2= -8/9}$ (top), $\bm{\mu^2L^2= -3/4}$ (middle), and $\bm{\mu^2L^2= -1/10}$ (bottom).}
    \label{fig:m01_kappa_onset_curves-OtherMasses}
\end{figure}

\begin{figure}[ht]
\vskip -1.0cm
    \centering
    \includegraphics[width=0.9\linewidth]{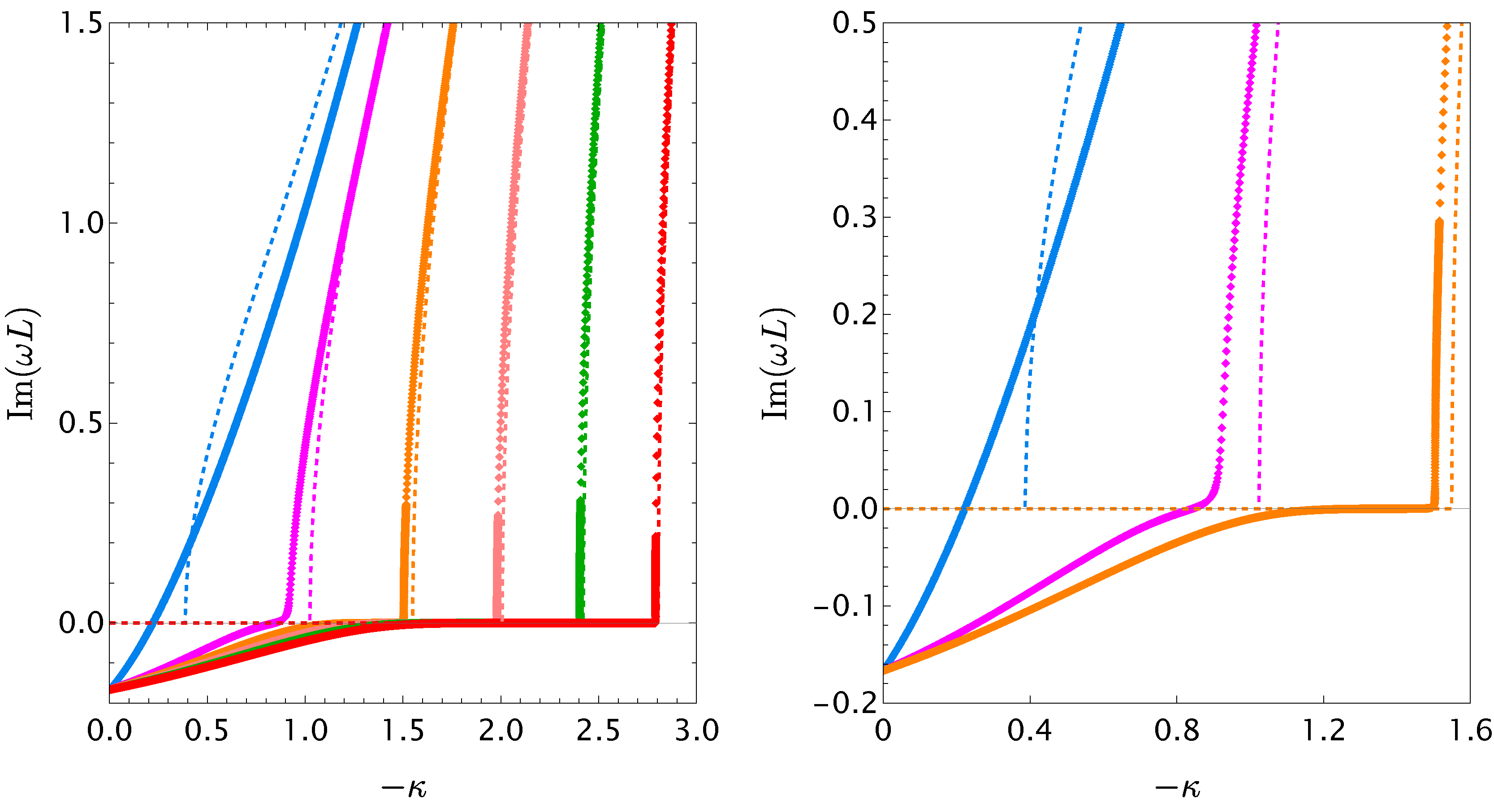}
    \includegraphics[width=0.9\linewidth]{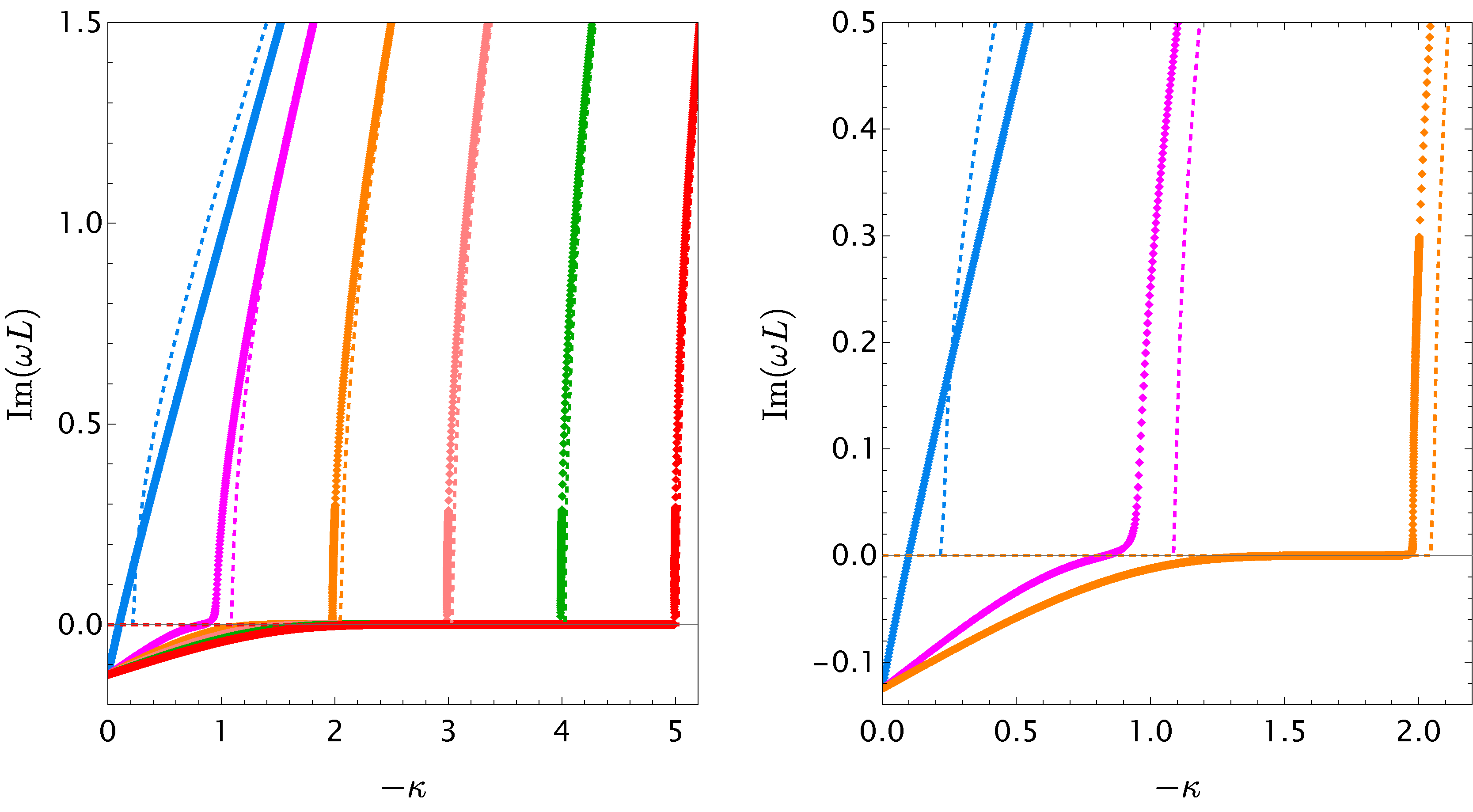}
     \includegraphics[width=0.9\linewidth]{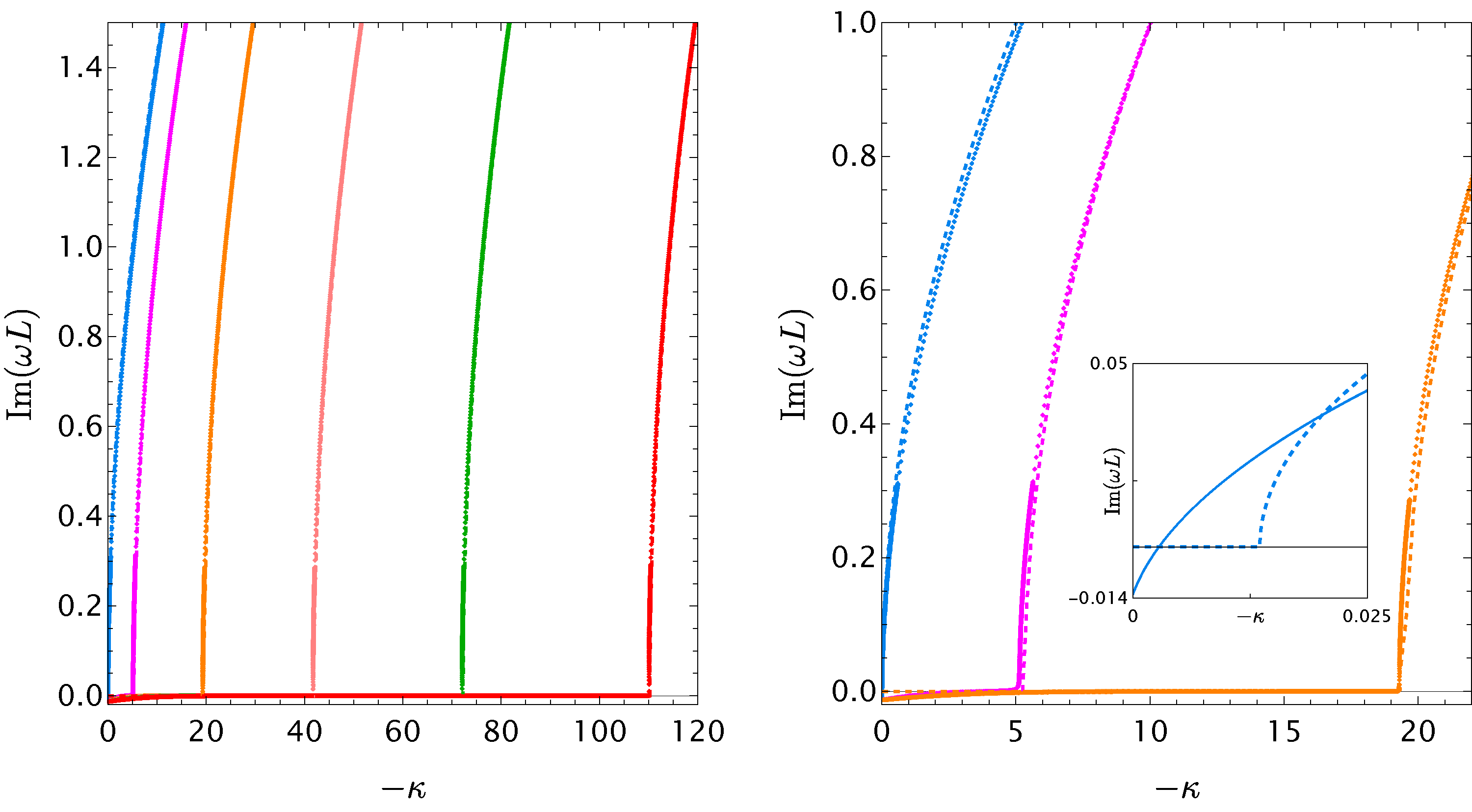}
    \caption{\footnotesize Analogue of Fig.~\ref{fig:m0tom5_btz_VS_ads_instability}, now for scalar masses $\bm{\mu^2L^2=-8/9}$ (top), $\bm{\mu^2L^2=-3/4}$ (middle), and $\bm{\mu^2L^2=-1/10}$ (bottom). We show $\text{Im}(\omega L)$ as a function of~$\kappa$ for $n=0$ co-rotating modes in global AdS$_3$ (dashed lines) and for BTZ black holes (diamonds) with $\{\hat{M},\hat{J}\}=\{5/16,1/4\}$. Curves correspond to $m=0$ (blue), $m=1$ (purple), $m=2$ (orange), $m=3$ (pink), $m=4$ (green), and $m=5$ (red). The right panel is a zoom of the left, showing only the $m=0,1,2$ modes.}
    \label{fig:m0tom5_btz_VS_ads_instability-OtherMasses}
\end{figure}

In the main text, we claimed that the plots presented in section~\ref{sec:BTZresults} for a scalar field with mass $\mu^2L^2=-15/16$ illustrate universal properties - within the mass range $-1<\mu^2L^2<0$ - of double-trace perturbations in BTZ. In this appendix, we provide a selection of results (see Figs.~\ref{fig:BTZw-kappa:m0-OtherMasses}-\ref{fig:m0tom5_btz_VS_ads_instability-OtherMasses}) for the scalar field masses $\mu^2L^2=\{-8/9,-3/4,-1/10\}$ that support these claims. These figures are direct counterparts of the figures for $\mu^2L^2=-15/16$ shown in the main text, and we use the same colour scheme throughout. Consequently, our descriptions here will be brief, as the universal features to be emphasized were already discussed in the main-text analogues. Occasionally, we comment on a feature that is less universal but still worth highlighting (further minor differences can be spotted by direct inspection). Finally, whenever a plot describes perturbations of a specific BTZ black hole, we choose the same parameters $\{\hat{M},\hat{J}\}=\{5/16,1/4\}$ - that is, a BTZ black hole with $\{y_-,y_+\}=\{1/4,1/2\}$ - which matches the choice used in the partner figures of the main text.

Figure~\ref{fig:BTZw-kappa:m0-OtherMasses} presents the analogue of Fig.~\ref{fig:BTZw-kappa:m0}: the quasinormal frequencies of $m=0$ modes as functions of the double-trace parameter $\kappa$ for the BTZ background with $\{\hat{M},\hat{J}\}=\{5/16,1/4\}$, now for scalar masses $\mu^2L^2=-8/9$ (top panel), $\mu^2L^2=-3/4$ (middle panel), and $\mu^2L^2=-1/10$ (bottom panel).

The top panel ($\mu^2L^2=-8/9$) of Fig.~\ref{fig:BTZw-kappa:m0-OtherMasses} is qualitatively very similar to the case $\mu^2L^2=-15/16$ shown in Fig.~\ref{fig:BTZw-kappa:m0} and discussed in detail in the main text; we therefore refrain from further comments and refer the reader to that discussion.

In the middle panel ($\mu^2L^2=-3/4$), we highlight two features not present in Fig.~\ref{fig:BTZw-kappa:m0}. First, some Dirichlet (green squares) and Neumann (red discs) boundary conditions yield identical frequencies, which results in certain Dirichlet$_{\text{\tiny Left}}$ - Neumann-Dirichlet$_{\text{\tiny Right}}$ trajectories appearing as horizontal lines. Second, the $n=0$ mode becomes unstable for $\kappa \lesssim \kappa_{\text{crit}} \sim -0.83$. There exists a window of positive $\kappa$ in which the $n=0$ and $n=2$ modes share the same negative value of ${\rm Im}\,\hat{\omega}$, while their corresponding ${\rm Re}\,\hat{\omega}$ values are symmetric (see the middle-right panel).

In the bottom panel ($\mu^2L^2=-1/10$), we note that for $\kappa\gtrsim -0.09$, the $n=0$ (blue) and $n=1$ overtones share the same ${\rm Im}\,\hat{\omega}$ but have opposite ${\rm Re}\,\hat{\omega}$ (see the right panel). Near $\kappa\simeq 5\times 10^{-4}$ the two modes bifurcate in the imaginary part and merge in the real part: the $n=0$ mode (blue curve) then becomes unstable for $\kappa\lesssim -0.09$, while $n=1$ maintains ${\rm Im}\,\hat{\omega}<0$, and both modes satisfy ${\rm Re}\,\hat{\omega}=0$ in this regime.

In Fig.~\ref{fig:BTZw-kappa:m1-OtherMasses} we present the analogue of 
Fig.~\ref{fig:BTZw-kappa:m1}, \ie the $m=1$ frequency complex plane for a BTZ black hole with $\{\hat{M},\hat{J}\}=\{5/16,1/4\}$, now shown for scalar masses 
$\mu^2L^2=-8/9$ (top-left panel), $\mu^2L^2=-3/4$ (top-right panel), and 
$\mu^2L^2=-1/10$ (middle panel).
All cases in Fig.~\ref{fig:BTZw-kappa:m1-OtherMasses} - with a single exception discussed below — are qualitatively very similar to the $\mu^2L^2=-15/16$ case of Fig.~\ref{fig:BTZw-kappa:m1} discussed in detail in the main text. For this reason, we refrain from repeating that discussion and instead refer the reader to the main text. The purpose of the present figure is to show that the behaviour described there is universal across scalar field masses in the double-trace window $-1<\mu^2L^2<0$.

The middle panel of Fig.~\ref{fig:BTZw-kappa:m1-OtherMasses}, corresponding to 
$\mu^2L^2=-1/10$, departs from this standard pattern in several ways that merit comment.  
The $n=0$ instability (magenta curve) remains present—this is a fundamental and robust feature. However, for this mass one observes that the $n=0$ co-rotating curve appears to extend to arbitrarily large ${\rm Re}(\omega L)$ as $\kappa\to +\infty$, rather than terminating at a Dirichlet (green square) point. The behaviour of this mode is more clearly visible in the bottom-right panel of 
Fig.~\ref{fig:BTZw-kappa:m1-OtherMasses}.  

The $n=0$ counter-rotating mode (dark red curve) also shows distinctive behaviour:  
as $\kappa\to +\infty$ it seems to extend to arbitrarily large negative ${\rm Re}(\omega L)$ with ${\rm Im}(\omega L)\to 0$, whereas as $\kappa\to -\infty$ the curve oscillates around the axis ${\rm Re}(\omega L)=0$ (possibly an infinite number of times as ${\rm Im}(\omega L)\to -\infty$). This behaviour is best seen in the bottom-left panel of Fig.~\ref{fig:BTZw-kappa:m1-OtherMasses}.  

Finally, for this scalar mass, each of the $n\geq 1$ co-rotating modes traces out a closed loop (and similarly for the $n\geq 1$ counter-rotating modes), rather than forming the ``continuous zig-zag chain'' that merges different overtones in the plots for other masses.  Within the broader context of our analysis, these differences for masses closer to the unitarity bound constitute only minor variations. The essential properties of the $n=0$ unstable modes remain universal.

In Fig.~\ref{fig:3D_kappa_onset-OtherMasses} we present the analogue of 
Fig.~\ref{fig:3D_kappa_onset}, \ie the instability onset values $\kappa^{\hbox{\tiny BTZ}}_{m,\hat{\mu}^2}(\hat{M},\hat{J})$ (red surface) and $\kappa^{\hbox{\tiny AdS}}_{m,\hat{\mu}^2}$ (flat grey surface) for $n=0$ double-trace perturbations with $m=0$ (left panels) and $m=1$ (right panels), now shown for scalar masses $\mu^2L^2=-8/9$ (top row), $\mu^2L^2=-3/4$ (middle row), and $\mu^2L^2=-1/10$ (bottom row). All cases displayed in Fig.~\ref{fig:3D_kappa_onset-OtherMasses} are qualitatively very similar to the $\mu^2L^2=-15/16$ case of Fig.~\ref{fig:3D_kappa_onset}, which was discussed in detail in the main text. We therefore do not repeat that discussion here and instead refer the reader to the 
corresponding analysis in the main text.

It is, however, worth emphasising two universal properties that persist for all scalar masses $-1<\hat{\mu}^2<0$.  First, the $m=0$ BTZ onset surface (left panels) always lies above the corresponding $m=1$ surface (right panels), and in fact above the onset surfaces for any $m\geq 1$. Second, the same ordering holds for the AdS$_3$ onset planes: the $m=0$ AdS onset plane always lies above the $m\geq 1$ planes, independently of the scalar mass within the double-trace window.

The purpose of Fig.~\ref{fig:3D_kappa_onset-OtherMasses} is to demonstrate explicitly that the behaviour described in the main text is universal across the entire mass range 
$-1<\mu^2L^2<0$ relevant for double-trace deformations.

In Fig.~\ref{fig:m01_kappa_onset_curves-OtherMasses}, we present the analogue of 
Fig.~\ref{fig:m0_kappa_onset_curves} (left panels, $m=0$) and Fig.~\ref{fig:m1_kappa_onset_curves} (right panels, $m=1$), now for scalar masses 
$\mu^2L^2=-8/9$ (top), $-3/4$ (middle), and $-1/10$ (bottom).  
Each plot shows the instability–onset curves in the $\{\hat{J},\hat{M}\}$ plane for several values of the double–trace parameter $\kappa$ listed in the legend.  
For a fixed $\kappa$, BTZ black holes located between the corresponding onset curve and the extremal line $M=J/L$ (orange) are unstable to $m=0$ (left) or $m=1$ (right) modes.  
The dashed black curve marks the AdS$_3$ onset value $\kappa^{\hbox{\tiny AdS}}_{m,\hat{\mu}^2}$: for $\kappa$ above this threshold, AdS$_3$ is linearly stable while BTZ may be stable or unstable depending on $\{\hat{J},\hat{M}\}$; for $\kappa$ below it (grey/light-grey curves), AdS$_3$ is unstable, and BTZ may again be either stable or unstable depending on its parameters.

All cases in Fig.~\ref{fig:m01_kappa_onset_curves-OtherMasses} are qualitatively equivalent to the $\mu^2L^2=-15/16$ results of Figs.~\ref{fig:m0_kappa_onset_curves} and \ref{fig:m1_kappa_onset_curves}. We therefore refrain from additional comments and refer the reader to the discussion in the main text. These plots further confirm that the behaviour described there is universal for all scalar masses in the double–trace window $-1<\mu^2L^2<0$.

In Fig.~\ref{fig:m0tom5_btz_VS_ads_instability-OtherMasses} we present the 
analogue of Fig.~\ref{fig:m0tom5_btz_VS_ads_instability}, showing 
$\mathrm{Im}(\omega L)$ as a function of~$\kappa$ for $n=0$ co-rotating modes in 
global AdS$_3$ (dashed curves) and in a BTZ black hole (diamonds) with 
$\{\hat{M},\hat{J}\}=\{5/16,1/4\}$. Results are shown for scalar masses 
$\mu^2L^2=-8/9$ (top), $-3/4$ (middle) and $-1/10$ (bottom). Each plot contains 
the modes $m=0$ (blue), $m=1$ (purple), $m=2$ (orange), $m=3$ (pink), 
$m=4$ (green) and $m=5$ (red). The right panel is a zoom of the left, focusing on 
$m=0,1,2$.

All cases behave qualitatively as in the reference mass 
$\mu^2L^2=-15/16$ shown in Fig.~\ref{fig:m0tom5_btz_VS_ads_instability}. Hence 
the conclusions drawn there apply equally here. In particular, the BTZ onset for 
$n=0$ shifts to increasingly negative $\kappa$ as $m$ increases, so any BTZ black 
hole unstable to a given $m$-mode is automatically unstable to all lower modes 
(including $m=0$). Moreover, when several instabilities coexist, the 
$m=0$ mode has the largest ${\rm Im}\,\hat{\omega}$. These features are universal 
throughout the double-trace mass range $-1<\mu^2L^2<0$.

\section{Reproducing results in the literature}\label{app:ReproduceLiterature} 
Some aspects of the BTZ instability to scalar field perturbations with double-trace (Robin) boundary conditions were studied in \cite{Iizuka:2015vsa,Ferreira:2017cta,Dappiaggi:2017pbe}, but only for the $m\geq 1$ sector. These references make no mention of the $m=0$ instability, which is in fact the most relevant: only BTZ black holes that are already unstable to $m=0$ perturbations can also be unstable to $m\geq 1$ perturbations. That is, for values of $\kappa$ where BTZ is unstable to $m\geq 1$, it is necessarily already unstable to $m=0$. Moreover, for a given $\kappa$ where BTZ is unstable to $m=1$ (and thus to $m=0$), the instability is stronger in the $m=0$ sector in the sense that ${\rm Im} \,\hat{\omega}|_{m=0}>{\rm Im} \,\hat{\omega}|_{m=1}$. These properties are illustrated in Fig.~\ref{fig:m0tom5_btz_VS_ads_instability}, which refers to a BTZ with particular $\{M,J/L\}$ and $\mu^2 L^2\in(-1,0)$, but captures typical double-trace properties for $m=1$. 

For context, note that the AdS$_3$ instability discussed in \cite{Dappiaggi:2016fwc,Ferreira:2017cta,Dappiaggi:2017pbe} corresponds to the Wald-Ishibashi instability of global AdS$_d$ ($d=3$) to double-trace boundary conditions, although this connection is not explicitly made in those references. As discussed in the main text, all BTZ black holes that are unstable to $m\geq 1$ perturbations have a $\kappa$ for which AdS$_3$ is unstable to $m=0$. There is, however, a window of $\kappa$ for which AdS$_3$ is stable and BTZ black holes can be stable or unstable to $m=0$ perturbations (some of these can then be stable to $m\geq 1$ modes, but others are also unstable to $m=1$ or higher $m$ modes): see Figs.~\ref{fig:3D_kappa_onset}-\ref{fig:m1_kappa_onset_curves}.

Since we have also studied $m\geq 1$ modes, we can check our $m=1$ results against \cite{Dappiaggi:2017pbe} (we also recover Figure 1 of \cite{Iizuka:2015vsa}). The best non-trivial test is reproducing Figure~1 of \cite{Dappiaggi:2017pbe}, showing only the imaginary part of the frequency, as the real parts are similar across several modes and cluster closely. This corresponds to a BTZ with $y_+ = r_+/L =5$, $y_- = r_-/L = 3$, scalar mass $\mu^2 L^2=-0.65$, and co-rotating modes with $m=1$ and six radial overtones $n=0,1,2,3,4,5$. 

\begin{figure}[ht]
    \centering
    \includegraphics[width=0.6\linewidth]{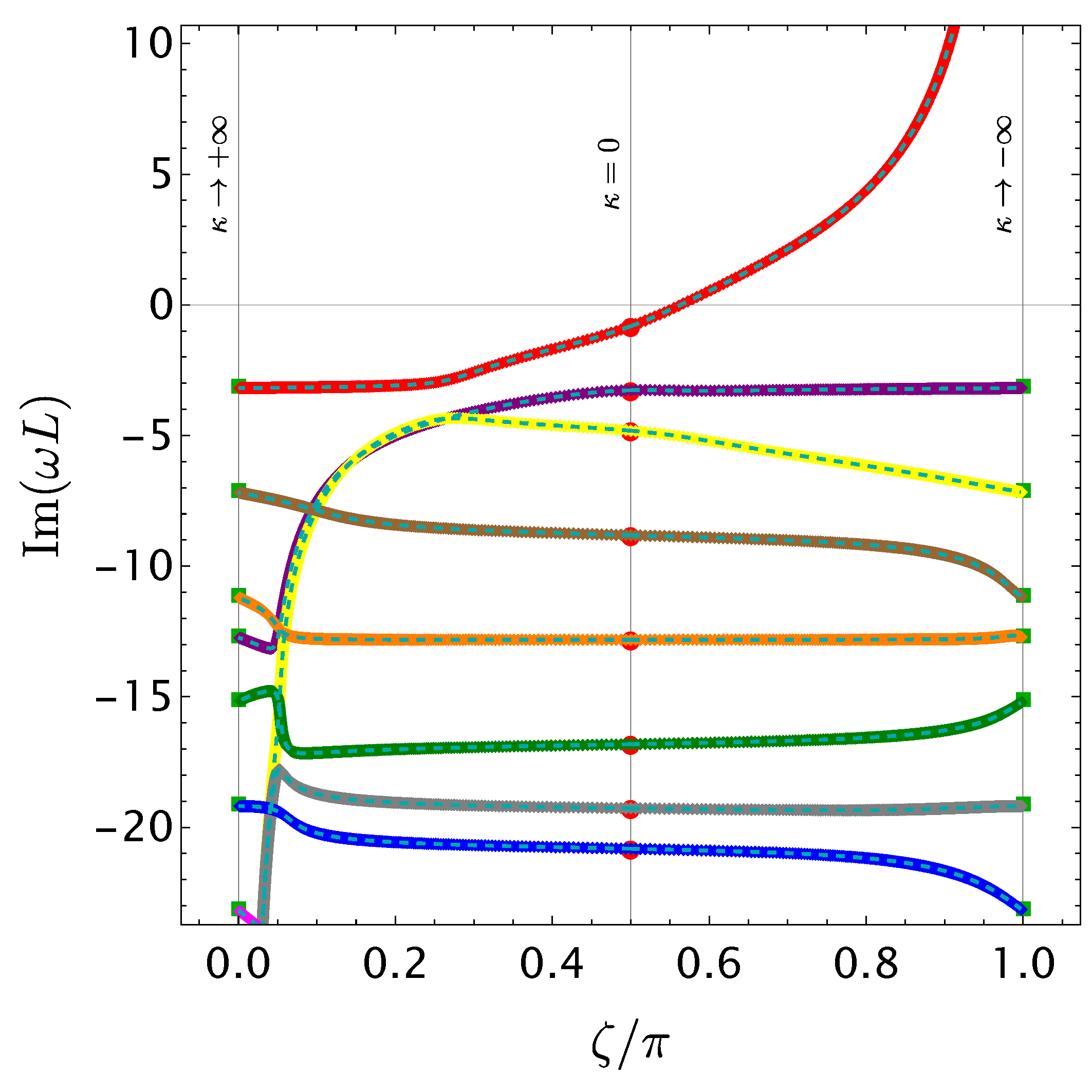}
    \caption{Reproducing the results of \cite{Dappiaggi:2017pbe} to validate our numerical computations. Shown is ${\rm Im}(\omega L)$ as a function of the double-trace parameter $\zeta/\pi$ (or $\kappa$) for a BTZ black hole with $r_+/L =5$, $r_-/L = 3$, and scalar mass $\mu^2 L^2=-0.65$, for co-rotating modes with $m=1$ and six radial overtones $n=0,1,2,3,4,5,6$ ($n=0$ is the top curve, $n=6$ the bottom). Green squares at $\zeta/\pi=0,1$ correspond to Dirichlet frequencies, red discs at $\zeta/\pi=1/2$ to Neumann frequencies. The dashed cyan curves show the results from figure 1 of \cite{Dappiaggi:2017pbe}.}
    \label{fig:ReproduceLiterature}
\end{figure}

In Fig.~\ref{fig:ReproduceLiterature}, we plot the imaginary part of the frequency for these six modes as a function of the double-trace parameter $\zeta/\pi$ used in \cite{Dappiaggi:2017pbe} ($n=0$ is the top curve, $n=5$ the bottom). The relation between this parameter and our double-trace parameter $\kappa$ in \eqref{2xTrace:BC} and \eqref{FGasympExp} is given by
\begin{equation}
\zeta= \text{arccot}\left(\kappa/(y_+^2 - y_-^2)^{\sqrt{1+\hat{\mu}^2}}\right)  
\qquad \Rightarrow
\begin{cases}
 \zeta/\pi = 0  \quad & \longleftrightarrow \quad  \kappa\to +\infty \,,\\
 \zeta/\pi  = \frac{1}{2} \quad   & \longleftrightarrow \quad  \kappa =0  \,,\\
 \zeta/\pi = 1  \quad  & \longleftrightarrow \quad  \kappa\to -\infty \,.
\end{cases}
\end{equation}

The dependence of $\zeta$ on both $\kappa$ and $y_\pm$ arises because \cite{Dappiaggi:2017pbe} use Robin BCs differing from the double-trace BCs \eqref{2xTrace:BC} by a factor $(y_+^2 - y_-^2)^{\sqrt{1+\hat{\mu}^2}}$. The green squares at $\zeta/\pi=0,1$ correspond to Dirichlet frequencies $\hat{\omega}^{\hbox{\tiny BTZ}}_{n} |_{\hbox{\tiny Dir}}$ \eqref{BTZ:wDir}, while the red discs at $\zeta/\pi=1/2$ correspond to Neumann frequencies $\hat{\omega}^{\hbox{\tiny BTZ}}_{n} |_{\hbox{\tiny Neu}}$ \eqref{BTZ:wNeu}. The curves with coloured points from our numerical code (section~\ref{sec:Numerical-Method}) coincide with the cyan curves from \cite{Dappiaggi:2017pbe}, which use Mathematica's \texttt{FindRoot} on the double-trace quantization condition \eqref{eqn:BTZ_Gamma_criterion}. This confirms the accuracy of our results.

\section{Near-horizon limit of BTZ and of its Klein-Gordon equation}\label{sec:AppNHanalysis}
It is well established that some instabilities of AdS black holes (with extremal configurations) can be inferred by studying the behaviour of perturbation equations (of any spin, not only scalar fields) in the near-horizon limit of the AdS black hole \cite{Gubser:2008px,Hartnoll:2008kx,Dias:2010ma,Durkee:2010ea,Dias:2011tj,Hollands:2014lra,Dias:2016pma,Dias:2019fmz,Dias:2020ncd}. Concretely, in this limit the perturbation equation reduces to a Klein-Gordon equation for a scalar field in AdS$_2$ with an effective near-horizon mass. If this effective mass violates the AdS$_2$ Breitenlohner-Freedman (BF) bound, one expects the black hole to be unstable. This provides a sufficient, but not necessary, condition for an instability and identifies its origin. Motivated by past successes, we perform this near-horizon analysis for extremal BTZ but find it inconclusive for understanding the BTZ instability.

The extremal BTZ black hole is given by \eqref{eqn:BTZ_metric} with $r_- \to r_+$, so that $\Omega_+^{\rm ext}= 1/L$. To obtain the near-horizon geometry, we zoom in around the horizon via the coordinate transformations:
\begin{equation}\label{NHtransf}
 r \to r_+ +   \varepsilon L_{\hbox{\tiny AdS$_2$}}^2 \rho, \qquad
t = \frac{\tau}{\varepsilon}\,, \qquad
\phi \to \tilde{\phi} + \Omega^{\rm ext}_+\, \frac{\tau}{\varepsilon},
\end{equation}
where $L_{\hbox{\tiny AdS$_2$}}$ is the AdS$_2$ radius. Taking the limit $\varepsilon \to 0$ yields the near-horizon geometry~\cite{Coussaert:1994tu,Azeyanagi:2008dk,Compere:2012jk}:
\begin{equation} \label{NHgeometry}
 {\rm d}s_{\rm NH}^2= L_{\hbox{\tiny AdS$_2$}}^2 \left(- \rho^2{\rm d}\tau^2 + \frac{{\rm d}\rho^2}{\rho^2}\right) + r_+^2\left({\rm d}\tilde{\phi} +\, \frac{L_{\hbox{\tiny AdS$_2$}}}{r_+}\, \rho\, {\rm d}\tau\right)^2, \quad \hbox{with} \quad  L_{\hbox{\tiny AdS$_2$}}^2\equiv \frac{L}{2}.
\end{equation}
This is a warped AdS$_3$ geometry, i.e., a circle fibred over AdS$_2$, with isometry $SL(2,\mathbb{R}) \times U(1)$. It remains a solution of 3-dimensional Einstein-AdS$_3$ theory, while the AdS$_2$ factor alone solves 2-dimensional Einstein-AdS equations, $R_{\mu\nu}=-L_{\hbox{\tiny AdS$_2$}}^{-2} g_{\mu\nu}$.

Applying this near-horizon limit to the BTZ Klein-Gordon equation \eqref{eqn:BTZ_KG_equation} for $\Phi(t,r,\phi) = e^{-i\omega t}e^{i\,m\phi}\psi(r)$, together with the frequency transformation $\omega \to \omega_{\rm ext} + \varepsilon \, \widehat{\omega}$ and $\omega_{\rm ext} \equiv m\, \Omega_+^{\rm ext}$, we obtain:
\begin{equation}\label{NHlimKG}
 \frac{d}{d\rho}\Big(\rho^2\psi'(\rho)\Big) - \left(\mu^2 L_{\hbox{\tiny AdS$_2$}}^2 -2m \frac{L_{\hbox{\tiny AdS$_2$}} }{r_+} \frac{\widehat{\omega}}{\rho} - \frac{\widehat{\omega}^2}{\rho^2}\right)\psi(\rho) = 0. 
\end{equation}
A Frobenius analysis of \eqref{NHlimKG} gives
\begin{equation}
\label{BFdecaysAdS2}
\psi{\bigl |}_{\rho\to\infty}\simeq \widehat{a} \, \rho^{\,-\widehat{\Delta}_-} +
\cdots + \widehat{b} \,\rho^{\,-\widehat{\Delta}_+}+\cdots\,, \quad
 \hbox{with} \quad \widehat{\Delta}_\pm^{(\rm s)}=\frac{1}{2}\pm \sqrt{ \frac{1}{4}+\mu_{\rm eff}^2 L_{\hbox{\tiny AdS$_2$}}^2}\;,
\end{equation}
which identifies the effective near-horizon mass $\mu_{\rm eff}$. Instability would occur if $\mu_{\rm eff}^2 L_{\hbox{\tiny AdS$_2$}}^2 < -1/4$.

In the BTZ case, however, $\mu_{\rm eff} = \mu$, so that AdS$_2$ BF violation would require $\mu^2 L_{\hbox{\tiny AdS$_2$}}^2 = \mu^2 L^2/4 \leq -1/4$, i.e., $\mu^2 L^2 \leq -1$. This violates the AdS$_3$ BF bound \eqref{2xTrace:rangeMass}, which is impossible. Hence, unlike higher-dimensional rotating AdS black holes \cite{Gubser:2008px,Hartnoll:2008kx,Dias:2010ma,Durkee:2010ea,Dias:2011tj,Hollands:2014lra,Dias:2016pma,Dias:2019fmz,Dias:2020ncd}, there is no mass window in BTZ where the AdS$_3$ BF bound is satisfied but the AdS$_2$ BF bound is violated. Therefore, this near-horizon analysis does not provide insight into the BTZ instability.

\bibliographystyle{JHEP}
\bibliography{refs_btz}

\end{document}